\documentstyle[12pt]{article}
\newcommand{\unity}{{\rm \setlength{\unitlength}{1em}
                     \begin{picture}(0.75,1)
                     \put(0,0){1}
                     \put(0.34,0){\line(0,1){0.65}}
                     \end{picture}
                   }}

\expandafter\ifx\csname amssym.def\endcsname\relax \else\endinput\fi
\expandafter\edef\csname amssym.def\endcsname{%
       \catcode`\noexpand\@=\the\catcode`\@\space}
\catcode`\@=11
\def\undefine#1{\let#1\undefined}
\def\newsymbol#1#2#3#4#5{\let\next@\relax
 \ifnum#2=\@ne\let\next@\msafam@\else
 \ifnum#2=\tw@\let\next@\msbfam@\fi\fi
 \mathchardef#1="#3\next@#4#5}
\def\mathhexbox@#1#2#3{\relax
 \ifmmode\mathpalette{}{\m@th\mathchar"#1#2#3}%
 \else\leavevmode\hbox{$\m@th\mathchar"#1#2#3$}\fi}
\def\hexnumber@#1{\ifcase#1 0\or 1\or 2\or 3\or 4\or 5\or 6\or 7\or 8\or
 9\or A\or B\or C\or D\or E\or F\fi}

\font\tenmsa=msam10
\font\sevenmsa=msam7
\font\fivemsa=msam5
\newfam\msafam
\textfont\msafam=\tenmsa
\scriptfont\msafam=\sevenmsa
\scriptscriptfont\msafam=\fivemsa
\edef\msafam@{\hexnumber@\msafam}
\mathchardef\dabar@"0\msafam@39
\def\dashrightarrow{\mathrel{\dabar@\dabar@\mathchar"0\msafam@4B}}
\def\dashleftarrow{\mathrel{\mathchar"0\msafam@4C\dabar@\dabar@}}

\def\ulcorner{\delimiter"4\msafam@70\msafam@70 }
\def\urcorner{\delimiter"5\msafam@71\msafam@71 }
\def\llcorner{\delimiter"4\msafam@78\msafam@78 }
\def\lrcorner{\delimiter"5\msafam@79\msafam@79 }
\def\yen{{\mathhexbox@\msafam@55 }}
\def\checkmark{{\mathhexbox@\msafam@58 }}
\def\circledR{{\mathhexbox@\msafam@72 }}
\def\maltese{{\mathhexbox@\msafam@7A }}

\font\tenmsb=msbm10
\font\sevenmsb=msbm7
\font\fivemsb=msbm5
\newfam\msbfam
\textfont\msbfam=\tenmsb
\scriptfont\msbfam=\sevenmsb
\scriptscriptfont\msbfam=\fivemsb
\edef\msbfam@{\hexnumber@\msbfam}
\def\Bbb#1{{\fam\msbfam\relax#1}}
\def\widehat#1{\setbox\z@\hbox{$\m@th#1$}%
 \ifdim\wd\z@>\tw@ em\mathaccent"0\msbfam@5B{#1}%
 \else\mathaccent"0362{#1}\fi}
\def\widetilde#1{\setbox\z@\hbox{$\m@th#1$}%
 \ifdim\wd\z@>\tw@ em\mathaccent"0\msbfam@5D{#1}%
 \else\mathaccent"0365{#1}\fi}
\font\teneufm=eufm10
\font\seveneufm=eufm7
\font\fiveeufm=eufm5
\newfam\eufmfam
\textfont\eufmfam=\teneufm
\scriptfont\eufmfam=\seveneufm
\scriptscriptfont\eufmfam=\fiveeufm

\csname amssym.def\endcsname
\expandafter\ifx\csname pre amssym.tex at\endcsname\relax \else \endinput\fi
\expandafter\chardef\csname pre amssym.tex at\endcsname=\the\catcode`\@
\catcode`\@=11
\newsymbol\boxdot 1200
\newsymbol\boxplus 1201
\newsymbol\boxtimes 1202
\newsymbol\square 1003
\newsymbol\blacksquare 1004
\newsymbol\centerdot 1205
\newsymbol\lozenge 1006
\newsymbol\blacklozenge 1007
\newsymbol\circlearrowright 1308
\newsymbol\circlearrowleft 1309
\undefine\rightleftharpoons
\newsymbol\rightleftharpoons 130A
\newsymbol\leftrightharpoons 130B
\newsymbol\boxminus 120C
\newsymbol\Vdash 130D
\newsymbol\Vvdash 130E
\newsymbol\vDash 130F
\newsymbol\twoheadrightarrow 1310
\newsymbol\twoheadleftarrow 1311
\newsymbol\leftleftarrows 1312
\newsymbol\rightrightarrows 1313
\newsymbol\upuparrows 1314
\newsymbol\downdownarrows 1315
\newsymbol\upharpoonright 1316
 
\newsymbol\downharpoonright 1317
\newsymbol\upharpoonleft 1318
\newsymbol\downharpoonleft 1319
\newsymbol\rightarrowtail 131A
\newsymbol\leftarrowtail 131B
\newsymbol\leftrightarrows 131C
\newsymbol\rightleftarrows 131D
\newsymbol\Lsh 131E
\newsymbol\Rsh 131F
\newsymbol\rightsquigarrow 1320
\newsymbol\leftrightsquigarrow 1321
\newsymbol\looparrowleft 1322
\newsymbol\looparrowright 1323
\newsymbol\circeq 1324
\newsymbol\succsim 1325
\newsymbol\gtrsim 1326
\newsymbol\gtrapprox 1327
\newsymbol\multimap 1328
\newsymbol\therefore 1329
\newsymbol\because 132A
\newsymbol\doteqdot 132B
 
\newsymbol\triangleq 132C
\newsymbol\precsim 132D
\newsymbol\lesssim 132E
\newsymbol\lessapprox 132F
\newsymbol\eqslantless 1330
\newsymbol\eqslantgtr 1331
\newsymbol\curlyeqprec 1332
\newsymbol\curlyeqsucc 1333
\newsymbol\preccurlyeq 1334
\newsymbol\leqq 1335
\newsymbol\leqslant 1336
\newsymbol\lessgtr 1337
\newsymbol\backprime 1038
\newsymbol\risingdotseq 133A
\newsymbol\fallingdotseq 133B
\newsymbol\succcurlyeq 133C
\newsymbol\geqq 133D
\newsymbol\geqslant 133E
\newsymbol\gtrless 133F
\newsymbol\sqsubset 1340
\newsymbol\sqsupset 1341
\newsymbol\vartriangleright 1342
\newsymbol\vartriangleleft 1343
\newsymbol\trianglerighteq 1344
\newsymbol\trianglelefteq 1345
\newsymbol\bigstar 1046
\newsymbol\between 1347
\newsymbol\blacktriangledown 1048
\newsymbol\blacktriangleright 1349
\newsymbol\blacktriangleleft 134A
\newsymbol\vartriangle 134D
\newsymbol\blacktriangle 104E
\newsymbol\triangledown 104F
\newsymbol\eqcirc 1350
\newsymbol\lesseqgtr 1351
\newsymbol\gtreqless 1352
\newsymbol\lesseqqgtr 1353
\newsymbol\gtreqqless 1354
\newsymbol\Rrightarrow 1356
\newsymbol\Lleftarrow 1357
\newsymbol\veebar 1259
\newsymbol\barwedge 125A
\newsymbol\doublebarwedge 125B
\undefine\angle
\newsymbol\angle 105C
\newsymbol\measuredangle 105D
\newsymbol\sphericalangle 105E
\newsymbol\varpropto 135F
\newsymbol\smallsmile 1360
\newsymbol\smallfrown 1361
\newsymbol\Subset 1362
\newsymbol\Supset 1363
\newsymbol\Cup 1264
 
\newsymbol\Cap 1265
 
\newsymbol\curlywedge 1266
\newsymbol\curlyvee 1267
\newsymbol\leftthreetimes 1268
\newsymbol\rightthreetimes 1269
\newsymbol\subseteqq 136A
\newsymbol\supseteqq 136B
\newsymbol\bumpeq 136C
\newsymbol\Bumpeq 136D
\newsymbol\lll 136E
 
\newsymbol\ggg 136F
 
\newsymbol\circledS 1073
\newsymbol\pitchfork 1374
\newsymbol\dotplus 1275
\newsymbol\backsim 1376
\newsymbol\backsimeq 1377
\newsymbol\complement 107B
\newsymbol\intercal 127C
\newsymbol\circledcirc 127D
\newsymbol\circledast 127E
\newsymbol\circleddash 127F
\newsymbol\lvertneqq 2300
\newsymbol\gvertneqq 2301
\newsymbol\nleq 2302
\newsymbol\ngeq 2303
\newsymbol\nless 2304
\newsymbol\ngtr 2305
\newsymbol\nprec 2306
\newsymbol\nsucc 2307
\newsymbol\lneqq 2308
\newsymbol\gneqq 2309
\newsymbol\nleqslant 230A
\newsymbol\ngeqslant 230B
\newsymbol\lneq 230C
\newsymbol\gneq 230D
\newsymbol\npreceq 230E
\newsymbol\nsucceq 230F
\newsymbol\precnsim 2310
\newsymbol\succnsim 2311
\newsymbol\lnsim 2312
\newsymbol\gnsim 2313
\newsymbol\nleqq 2314
\newsymbol\ngeqq 2315
\newsymbol\precneqq 2316
\newsymbol\succneqq 2317
\newsymbol\precnapprox 2318
\newsymbol\succnapprox 2319
\newsymbol\lnapprox 231A
\newsymbol\gnapprox 231B
\newsymbol\nsim 231C
\newsymbol\ncong 231D
\newsymbol\diagup 231E
\newsymbol\diagdown 231F
\newsymbol\varsubsetneq 2320
\newsymbol\varsupsetneq 2321
\newsymbol\nsubseteqq 2322
\newsymbol\nsupseteqq 2323
\newsymbol\subsetneqq 2324
\newsymbol\supsetneqq 2325
\newsymbol\varsubsetneqq 2326
\newsymbol\varsupsetneqq 2327
\newsymbol\subsetneq 2328
\newsymbol\supsetneq 2329
\newsymbol\nsubseteq 232A
\newsymbol\nsupseteq 232B
\newsymbol\nparallel 232C
\newsymbol\nmid 232D
\newsymbol\nshortmid 232E
\newsymbol\nshortparallel 232F
\newsymbol\nvdash 2330
\newsymbol\nVdash 2331
\newsymbol\nvDash 2332
\newsymbol\nVDash 2333
\newsymbol\ntrianglerighteq 2334
\newsymbol\ntrianglelefteq 2335
\newsymbol\ntriangleleft 2336
\newsymbol\ntriangleright 2337
\newsymbol\nleftarrow 2338
\newsymbol\nrightarrow 2339
\newsymbol\nLeftarrow 233A
\newsymbol\nRightarrow 233B
\newsymbol\nLeftrightarrow 233C
\newsymbol\nleftrightarrow 233D
\newsymbol\divideontimes 223E
\newsymbol\varnothing 203F
\newsymbol\nexists 2040
\newsymbol\Finv 2060
\newsymbol\Game 2061
\newsymbol\mho 2066
\newsymbol\eth 2067
\newsymbol\eqsim 2368
\newsymbol\beth 2069
\newsymbol\gimel 206A
\newsymbol\daleth 206B
\newsymbol\lessdot 236C
\newsymbol\gtrdot 236D
\newsymbol\ltimes 226E
\newsymbol\rtimes 226F
\newsymbol\shortmid 2370
\newsymbol\shortparallel 2371
\newsymbol\smallsetminus 2272
\newsymbol\thicksim 2373
\newsymbol\thickapprox 2374
\newsymbol\approxeq 2375
\newsymbol\succapprox 2376
\newsymbol\precapprox 2377
\newsymbol\curvearrowleft 2378
\newsymbol\curvearrowright 2379
\newsymbol\digamma 207A
\newsymbol\varkappa 207B
\newsymbol\Bbbk 207C
\newsymbol\hslash 207D
\undefine\hbar
\newsymbol\hbar 207E
\newsymbol\backepsilon 237F
\catcode`\@=\csname pre amssym.tex at\endcsname
\newtheorem{Lemma}{Lemma}[section]
\newtheorem{Theorem}[Lemma]{Theorem}
\newtheorem{Corollary}[Lemma]{Corollary}
\newtheorem{Proposition}[Lemma]{Proposition}
\newtheorem{Definition}[Lemma]{Definition}
\newtheorem{Assumption}[Lemma]{Assumption}
\newtheorem{Example}[Lemma]{Example}
\newtheorem{Remark}[Lemma]{Remark}

\begin{document}
\begin{titlepage}
\vspace*{60pt}
\centerline{\LARGE \bf Integrable systems
and}
\vskip 20pt
\centerline{\LARGE \bf Riemann surfaces of
infinite genus}
\vskip 60pt
\centerline{\large Martin U.
Schmidt\footnote{Supported by
Forschungsinstitut f{\"u}r Mathematik
(ETH Z{\"u}rich) and by DFG,
SFB~288 ``Differentialgeometrie und
Quantenphysik''} }
\vskip 5pt
\centerline{Institut f{\"u}r
Theoretische Physik}
\vskip 2pt
\centerline{Freie Universit{\"a}t
Berlin}
\vskip 2pt
\centerline{Arnimallee 14}
\vskip 2pt
\centerline{D-14195 Berlin}
\vskip 60pt
\tableofcontents
\centerline{\bf {Abstract.}}

To the spectral curves of smooth periodic solutions of the
$n$-wave equation the points with infinite energy are added.
The resulting spaces are considered as generalized
Riemann surfcae. In general the genus is equal to infinity,
nethertheless these Riemann surfaces are similar to compact
Riemann surfaces. After proving a Riemann Roch Theorem we can
carry over most of the constructions of the finite gap
potentials to all smooth periodic potentials. The symplectic
form turns out to be closely related to Serre duality.
Finally we prove that all non-linear PDE's, which belong to
the focussing case of the non-linear Schr\"odinger equation,
have global solutions for arbitrary smooth periodic
inital potantials.

\end{titlepage}

\section{Introduction} \label{Section0}
The investigation of the  Korteweg-de Vries equation
initiated the development of many new ideas
on integrable systems\footnote{We commend
the review \cite{DKN}.}. A large class of
integrable systems was discovered, which
turned out to be closely related to
the theory of Riemann surfaces. In 1976 McKean and
Trubowitz \cite{MKT} succeeded to establish
a one to one correspondence between
periodic smooth solutions of the  Korteweg-de Vries
equation and divisors of Riemann
surfaces of infinite genus. In 1980
Adler and van Moerbeke \cite{AvM} and
afterwards Reyman and Semenov-Tian-Shansky
\cite{RST} generalized ideas of Krichever
\cite{Kr}. In the finite dimensional
case they showed that the Lagrangian
submanifolds are connected components of
the Picard group\footnote{The Picard
group of a Riemann surface
is defined to be the set of all equivalence
classes of holomorphic line bundles together
with the multiplication induced by the tensor
product of line bundles. In this article
we will often use divisors to describe
line bundles. This is possible if the line
bundle admits a meromorphic cross
section. For Riemann surfaces in the
correct sense this is always true (see
e.g.\cite[29.17]{Fo}). We do not know
whether this is true for the generalized Riemann
surfaces considered in this article. But
the line bundles we are interested in
admit always meromorphic cross
sections.} of compact Riemann
surfaces. Moreover, the hamiltonian
flows turned out to be given by the
action of one dimensional subgroups of
the Picard group on these
Lagrangian submanifolds
(see e.g. \cite{RST}). It was clear
that the same is true at least for the
so called algebraic geometric solutions
of the corresponding infinite
dimensional systems. In consideration of
the results of \cite{MKT} and \cite{MKT2}
it is natural
to expect that this correspondence could be
generalized to non-algebraic geometric solutions.
Unfortunately there does not exist a
Riemann surface with an infinite dimensional
Picard group. Either the Riemann surface
is compact and the corresponding Picard
group is finite dimensional, or the
Riemann surface is not compact and the
corresponding Picard group is trivial.

McKean and Trubowitz overcame this
problem by using another space
associated to a Riemann surface instead
of a connected component of the Picard
group. The first homology group has a natural
embedding into the dual space of the
holomorphic forms. This embedding may be
described by the integration of the
holomorphic forms over representative
1-cycles. For compact Riemann surfaces
the Jacobian variety,
the connected component of the identity
of the Picard group, is isomorphic to
the quotient of the dual space of the
holomorphic forms divided by the first
homology group. This quotient turns out
to be a compact abelian group.
For Riemann surfaces of infinite genus
the natural generalization of this
quotient is the dual of the Hilbert
space of holomorphic forms divided by
the closure of the first homology group.
In fact, in \cite{MKT} it is proven,
that the real part of this quotient is a
compact abelian group. Moreover, this
group can be identified with all
divisors corresponding to the potentials
out of the isospectral sets. For compact
Riemann surfaces it is possible to
define the solutions directly as
functions on these compact abelian
groups: Multivalued holomorphic functions on
these groups are called theta functions.
The Its Matveev formula \cite{IM} gives
solutions of the  Korteweg-de Vries equation in terms
of these theta functions. McKean and
Trubowitz were able to construct such
theta functions on the infinite
dimensional quotient space of square
integrable holomorphic forms divided by
the first homology group of some Riemann
surface of infinite genus. In \cite{MKT2}
the Its Matveev formula is proven even
in this situation. Now in general
for Riemann surfaces of infinite genus
there are two possibilities:
Either we use these theta
functions\footnote{To the authors
knowledge this
is done in \cite{FKT}.},
or we characterize those
divisors, which correspond to the
potentials. In the first approach the
correspondence between line bundles and
potentials may be omitted. We choose the
second approach, which refers back to
the Picard group: Due to Abel's Theorem
for compact Riemann surfaces the elements of the
quotient space of holomorphic forms
divided by the first homology group are in one to one
correspondence with the equivalence
classes of divisors of degree zero.
The methods of \cite{MKT}
and \cite{MKT2}
suggest a modification of the equivalence
relation in the case of non-compact
Riemann surfaces of infinite genus. Only
those meromorphic functions have a
divisor equivalent to the zero divisor
in this modified sense, which
are bounded in some sense near infinity.

This paper gives a slightly different
and more geometric approach to the
Picard group of Riemann surfaces of
infinite genus. Similar to the algebraic
geometric case we add to the spectral
curve some points corresponding to the
value \begin{math} \lambda =\infty \end{math}
of the spectral parameter. In every
neighbourhood of the form \begin{math}
|\lambda |>1/\epsilon \end{math} of
these points there
are in general infinitely many
branchpoints. Hence the resulting
space is not a Riemann surface in the
correct sense. Now we define a base of
neighbourhoods of such points, such that
in some neighbourhood of these points
there are no branchpoints. Together with
the usual topology of the spectral curve
this gives a topology on the resulting
space. Typically these neighbourhoods
are of the following form: \begin{displaymath}
\left\{ \lambda ^{-1}\in {\Bbb C}\left| \
|\lambda ^{-1}|<\epsilon ,\
|\lambda ^{-1}-a_{n}|>\frac{c_{n}}{\epsilon }
\mbox{ for all } n\in {\Bbb N}
\right. \right\} ,\end{displaymath}
with some \begin{math} \epsilon >0\end{math},
some sequence \begin{math} (a_{n})_{n\in {\Bbb N}
}\end{math}, which converges to zero,
and some positive sequence \begin{math}
(c_{n})_{n\in {\Bbb N}}\end{math}, which
converges much faster to zero, than
the sequence \begin{math} (|a_{n}|)_{n\in {\Bbb N}
}\end{math}. \\
\begin{picture}(460,220)(-220,-120)

\put(-120,0){\circle*{18} }
\put(-81,0){\circle*{12} }
\put(-54,0){\circle*{8} }
\put(-36,0){\circle*{5.3} }
\put(-24,0){\circle*{3.5} }
\put(-16,0){\circle*{2.3} }
\put(-10.5,0){\circle*{1.5} }
\put(-7,0){\circle*{1} }

\put(120,0){\circle*{18} }
\put(81,0){\circle*{12} }
\put(54,0){\circle*{8} }
\put(36,0){\circle*{5.3} }
\put(24,0){\circle*{3.5} }
\put(16,0){\circle*{2.3} }
\put(10.5,0){\circle*{1.5} }
\put(7,0){\circle*{1} }
\put(-50,-115){Figure \begin{math}
1\end{math}}
\end{picture}

\noindent
In figure 1
we draw a schematic picture of such a
neighbourhood in the \begin{math}
1/\lambda \end{math}-plane.
Now we call a function \begin{math} f\end{math}
on such a set holomorphic, if
\begin{description}
\item[(i)] \begin{math} f\end{math} is
holomorphic on the interior of this set
with respect to the usual topology of \begin{math}
{\Bbb C}\end{math},
\item[(ii)] \begin{math} f\end{math} and
all derivatives of \begin{math} f\end{math}
extend to continuous functions on the
whole set with respect to the topology
defined by these neighbourhoods.
\end{description} This definition makes
it possible to carry over almost all
concepts of the theory of Riemann
surfaces to these generalized Riemann
surfaces. For example the sheaves of
holomorphic and meromorphic functions,
and the sheaves of holomorphic and
meromorphic forms are defined in a
natural way, respectively. Furthermore,
the concept of
holomorphic line bundles and of divisors
may be generalized to this situation.

These generalized Riemann surfaces turn out
to be similar to compact Riemann
surfaces, although they are not compact.
For example all global holomorphic
functions are constant, and their Picard
groups are very large. A large part of the
article will be devoted to an analysis
of these Picard groups. It
turns out that the smooth periodic
solutions of the generalized nonlinear Schr{\"o}\-din\-ger equation
are in one to one
correspondence with some part of the
Picard groups.

In some sense the nonlinear Schr{\"o}\-din\-ger equation is the
simplest case of those integrable
systems, which may be described by some
Lax equation or some zero curvature
equation\footnote{The nonlinear
Schr{\"o}\-din\-ger equation is the standard example of
\cite{FT}.}. In this article we restrict
attention to the generalization of the
nonlinear Schr{\"o}\-din\-ger equation
to semisimple Lie groups introduced in
\cite{FK}. It will not suffice to look
at the nonlinear Schr{\"o}\-din\-ger equation
only, because a simple count of dimensions
in the infinite dimensional
case suggests that the systems corresponding to
higher groups than \begin{math} SL(2,{\Bbb C}
)\end{math} are not
integrable\footnote{Compare with
\cite{BS}.}. Indeed,
the number of series of integrals are
equal to the rank of the group and the
number of fields is equal to the
dimension of the group minus the rank of
the group. This article shows that the
generalized nonlinear Schr{\"o}\-din\-ger equation is quite
similar to the nonlinear Schr{\"o}\-din\-ger equation in the
case of periodic boundary
conditions\footnote{On the line this might be different.
Then all Riemann surfaces are singular.
For higher groups all isospectral sets
may decompose into uncountable many
components with respect to the action of
the Picard group. Hence there may be
additional integrals of motion (see
Example~\ref{Example9.2}). Due to
Proposition~\ref{Proposition9.1} this is
impossible for \begin{math} GL(2,{\Bbb C})\end{math}
and \begin{math} SL(2,{\Bbb
C})\end{math}.}.
We believe that almost all methods and
results of this paper may be generalized
to other integrable systems, as for example
the generalized sine Gordon and the sinh
Gordon equation. The case of the two
dimensional Kadomtsev-Petviashvili equation seems to be more
complicated. Some of the ideas
of Appendix~\ref{Appendixb} may be
helpful.

It is possible to modify these methods
in order to cover all potentials, which
are elements of some Sobolev spaces,
instead of the Fr\'{e}chet space of smooth
periodic potentials. For this purpose
the topology of the Riemann surface will
have to be changed. Furthermore, condition
(ii) of the definition of holomorphic
functions on these neighbourhoods
described above has to be replaced by
\begin{description}
\item[(ii)'] \begin{math} f\end{math}
and all derivatives up to some fixed
order extend to continuous functions on
the whole set.
\end{description}
In this article only the smooth case is
covered. Then the resulting Riemann
surfaces seem to be more similar to
usual Riemann surfaces.

Let us now give a short summary of the
article. In \cite{HSS} it is shown that
the Floquet matrix\footnote{This matrix
is the same as the monodromy matrix.} may be diagonalized
formally by a matrix valued formal power
series in \begin{math}
\lambda ^{-1}\end{math}. In the second section
we prove that this power series is an
asymptotic expansion of a holomorphic
diagonalization. The domain of this
asymptotic expansion is chosen to be as large
as possible.

This domain is used in the
third section to define the topology of
the completed spectral curve. The
resulting space is a Riemann surface in
the sense described above. Moreover, we
prove that all global holomorphic
functions are constant
(Lemma~\ref{Lemma2.1}) and that the
total residue of meromorphic
differential forms is zero
(Theorem~\ref{Theorem2.1}). Finally the
divisor of a meromorphic function is
shown to be an infinite sum of finite
divisors of degree zero
(Corollary~\ref{Corollary2.1}). With the
exception of a finite part of the divisor
each of these finite divisors are located
in one of the domains, which are
excluded from the neighbourhoods of
points corresponding to \begin{math} \lambda =\infty
\end{math} as described above. Hence the
degree of a divisor turns out to be a
sequence of degrees of finite divisors
indexed by the excluded domains.

In the
fourth section the eigen vectors of the
monodromy are shown to define holomorphic
line bundles over this Riemann surface. It
turns out that the Riemann surfaces
corresponding to some potentials are
singular. This case is treated separately
in the tenth section. The degree of the
eigen bundle is determined in
Theorem~\ref{Theorem3.1}. Finally it is
shown that the Riemann surface together
with the eigen bundle completely
determines the potential.

In order to
classify all line bundles, which
are equivalent to the eigen bundle of
some potential, a Riemann-Roch Theorem
for the generalized Riemann surfaces
is proven in the fifth section. A
condition on divisors is given, such
that the space of global sections is
finite dimensional and the space of
holomorphic forms with values in the
dual line bundle is finite
dimensional, too (Theorem~\ref{Theorem4.3}). In a
general situation all integral divisors
of a specified degree fulfil this
condition
(Corollary~\ref{Corollary4.1}). Hence
the topology defined above essentially
determines the location properties of
these divisors.

Now the classification
of all eigen bundles of the potentials is
given in Theorem~\ref{Theorem5.1}.
Furthermore the one to one correspondence
between potentials and divisors defines
a homeomorphism with respect to a
suitable topology on the set of these divisors.

The seventh section is a short
excursion on Darboux
coordinates\footnote{In Example~\ref{Example9.1}
it is shown that in general they are
not global coordinates.}. These Darboux
coordinates are given by the values of
the spectral parameter \begin{math} \lambda \end{math}
and the logarithm of the values of the Floquet
multiplier \begin{math} \mu \end{math}
(eigenvalues of the monodromy) at all
points of the divisor, which describes
the eigen bundles corresponding to the
potentials. In case of the  Korteweg-de Vries equation
this was proven in \cite[Theorem 2.8 and
Theorem 3.5]{PT}. This book gives a
comprehensive picture of these
coordinates. In the finite dimensional
case the analogous result was recently proven in
\cite{AHH}.

To the authors knowledge the main
result of the eighth section is new
even in the finite dimensional case.
Theorem~\ref{Theorem7.1}
shows that the symplectic form is given
by the Serre duality. Loosely speaking
the tangent space may be
identified with the direct sum of the first
cohomology group of the sheaf of
holomorphic functions on the spectral
curve and the space of holomorphic forms
on the spectral curve. The first summand
is naturally isomorphic to the Lie algebra
of the Picard group and therefore also
isomorphic to the tangent space along
the corresponding Lagrangian
submanifold. The second summand is
isomorphic to any maximal isotropic
subspace of the tangent space
transversal to the tangent space along
the Lagrangian submanifold. Then the
symplectic form defines a non-degenerate
pairing between these two spaces.
Theorem~\ref{Theorem7.1} shows that this
pairing is the same as the pairing given
by Serre duality\footnote{For compact Riemann
surfaces Serre duality is isomorphic to
a natural symplectic form of the Riemann
surface. In fact, the wedge product and
integration over the Riemann surface
together defines a natural symplectic
form on the first cohomology group of
the deRham complex and the decomposition
into the direct sum of the
space of hoplomorphic and
antiholomorphic forms is a decomposition
into Lagrangian subspaces with respect
to this symplectic form. Due to Dolbeault's
Theorem this decomposition of the first
cohomology group of the deRham complex
is isomorphic the direct sum
of the space of holomorphic forms and
the first cohomology group of the sheaf
of holomorphic functions
and this isomorphism transforms
the symplectic form into
Serre duality.}. This relation might be
the central point of the connection
between Riemann surfaces and integrable
systems. Theorem~\ref{Theorem7.1}
proves in addition that these systems are
completely integrable in a weak sense:
the tangent space
along the Lagrangian submanifolds is
shown to be maximal isotropic with
respect to the symplectic form.

In the ninth section we formulate a reality
condition. In case of the nonlinear Schr{\"o}\-din\-ger equation
this yields the self-focussing nonlinear Schr{\"o}\-din\-ger
equation\footnote{The methods of \cite{MKT} do
not cover this case.}. Summing up
Corollary~\ref{Corollary8.1} and
Corollary~\ref{Corollary9.1} leads to the
following picture: All isospectral sets
are homoemorphic to finite unions of
groups of the form \begin{displaymath} \left(
{\Bbb R}/{\Bbb Z}\right) ^{{\cal I}_{
\mbox{\it \scriptsize effective} }}\mbox{ times a
finite dimensional abelian Lie group.}
\end{displaymath} Here \begin{math} {\cal I}_{
\mbox{\it \scriptsize effective} }\end{math}
is some countable set. Moreover, all tangent
vectors along the isospectral set
correspond to a hamiltonian flow on this
isospectral set. In particular all partial
differential equations, which describe
such hamiltonian flows corresponding to local
integrals of motion (see \cite{HSS} and
\cite{Sch}), are shown to have
global solutions\footnote{By pure analytic methods
this is proven in \cite{Bo} for some
of these partial differential
equations.}.

In Appendix A we show
that the formal power series, which
diagonalizes the monodromy, is Borel
summable if and only if the potential is
analytic. Hence the asymptotic expansion
of the eigen vectors completely
determines the potential (and therefore
also the meromorphic eigenvector
function) if and only if the potential
is analytic.

In Appendix B we include
another reality condition in our
approach. In case of the nonlinear Schr{\"o}\-din\-ger equation
this gives the non-focussing nonlinear Schr{\"o}\-din\-ger
equation\footnote{The methods of \cite{MKT} can
be carried over to this case.}. It turns
out that the two reality conditions are
related to the two covering maps of the
Riemann surface induced by the spectral
parameter and the Floquet multiplier,
respectively.

In some footnotes we mark those parts of
the article, which may be passed over, if
only a rough comprehension is intended.

\section{An asymptotic expansion} \label{Section1}
Let \begin{math} p\end{math}  be a diagonal matrix
\begin{math} p=diagonal(p_{1},\ldots ,p_{n})\end{math},
such that all \begin{math} p_{i}\end{math}
are distinct and let \begin{math} q(x)\end{math}
be a smooth function into
the \begin{math} n \times n \end{math}-matrices, which is periodic
with period 1.  Let us
first consider the fundamental solution of the auxiliary
problem
\begin{displaymath} L\cdot g(x,\lambda ,q)=0,\; \;
g(0,\lambda ,q)=\unity ,\end{displaymath}
with the Lax operator
{\begin{math} L=\frac{d}{dx}
+q(x)+\lambda p\end{math}. }
The following well known lemma will give
us a power series expansion of \begin{math} g\end{math}.
\begin{Lemma} \label{Lemma1.1} Let \begin{math} g(x)\end{math}  be the
unique fundamental solution of the homogenous differential
equation
\begin{displaymath} \left( \frac{d}{dx} +a(x)\right)
g(x)=0,\; \; g(0)=\unity .\end{displaymath}
Then {\begin{math} f(x)=g(x)\left(\int_{0}^{x}
g^{-1}(t)b(t)dt+f_{0}\right) \end{math} }
is the unique solution of the inhomogeneous differential
equation
\begin{displaymath} \left( \frac{d}{dx} +a(x)\right)
f(x)=b(x),\; \; f(0)=f_{0}.\end{displaymath}
\end{Lemma}
We omit the easy proof.
\hspace*{\fill } \begin{math} \Box \end{math}

\noindent
Now we make an ansatz for the fundamental
solution \begin{math} g(x)\end{math}
with \begin{math} a(x)=q(x)+p\lambda \end{math}.
\begin{equation} g(x,\lambda ,q)=\sum_{0}^{\infty }
\gamma _{n}(x,\lambda ,q) \label{g1.2} \end{equation}
\begin{displaymath} \mbox{with} \; \;
\gamma _{0}(x,\lambda )=\exp (-x\lambda p)
\; \; \mbox{and} \; \; \left(
\frac{d}{dx} +\lambda p\right)
\gamma _{n+1}(x,\lambda ,q)=-q(x)\gamma
_{n}(x,\lambda ,q).\end{displaymath}
Due to Lemma~\ref{Lemma1.1} we obtain the recursion
relation \begin{displaymath} \gamma _{n+1}(x,\lambda
,q)=-\int_{0}^{x} \gamma _{0}(x-t,\lambda )q(t)\gamma
_{n}(t,\lambda ,q)dt,\end{displaymath}
and finally the explicit solution
\begin{math} \gamma _{n}(x,\lambda
,q)=\end{math} \begin{equation} =
(-1)^{n}\int_{0\leq t_{1}\leq ...\leq t_{n}\leq x} \gamma
_{0}(x-t_{n},\lambda )q(t_{n})\gamma _{0}(t_{n}-t_{n-1},\lambda
)...q(t_{1})\gamma _{0}(t_{1},\lambda )dt_{1}...dt_{n}.\label{g1.3}
\end{equation}
On the space of potentials \begin{math} q\end{math}  we make
use of the natural scalar product \\
{\begin{math} \left\langle \tilde{q} ,q\right\rangle =
\int_{0}^{1}\mbox{tr} \left( \tilde{q}
(t)q^{*}(t)\right) dt\end{math} }
and the corresponding norm \begin{math} \| q \| \end{math}.
Although, in the end  we are only interested in smooth periodic
potentials, we consider the fundamental solution \begin{math}
g\end{math}  for arbitrary \begin{math}
q\end{math} in the Hilbert space \begin{math}
{\cal H}\end{math}
 corresponding to the above scalar product. The space of \begin{math}
n\times n \end{math}-matrices is endowed with the Banach
norm of operators of the Hilbert space
\begin{math} {\Bbb C}^{n}\end{math}.
\begin{Theorem} \label{Theorem1.1} The formal power series (\ref{g1.2})
for \begin{math} g(x,\lambda ,q)\end{math}  converge
uniformly on bounded subsets of \begin{math} [0,1]\times {\Bbb C}\times
{\cal H}\end{math}  to the unique solution of the auxiliary
problem. For each fixed \begin{math} x\end{math}
 and \begin{math} q,\; g(x,\cdot ,q)\end{math}  is an entire
function. \end{Theorem}
This result compares with Theorem 1 of \cite{PT} in the case
of Hill's equation.\\
Proof: For \begin{math} \gamma _{0}(x,\lambda )\end{math}
we start with the obvious bound
\begin{displaymath} \| \gamma _{0}(x,\lambda )\|
=\exp \left( x\cdot \sup \left\{ \Re(\lambda p_{i}) |
i=1,\ldots ,n\right\} \right) .
\end{displaymath}
Inserting this estimate into (\ref{g1.3}) gives
\begin{displaymath} \| \gamma _{n}(x,\lambda ,q)\| \leq \| \gamma
_{0}(x,\lambda )\| \frac{1}{n!} \left( \int_{0}^{x} \|
q(t)\| dt\right) ^{n} \end{displaymath}
For the last factor we give a bound in terms of \begin{math}
\| q\| \end{math} :
\begin{displaymath} \int_{0}^{x} \|q(t)\| dt \leq \sup \left\{
|\left\langle
q,\tilde{q} \right\rangle | \; \left| \| \; \tilde{q} \|
^{2}=x\right. \right\} =\| q\| \sqrt{x} .\end{displaymath}
In fact this is true because on the space of \begin{math} n\times
n\end{math}-matrices we have
\begin{displaymath} \| A\| \leq \sup \left\{ |\mbox{tr} (AB)|\; \left|
\; \mbox{tr} (BB^{*})=1\right. \right\} .\end{displaymath}
Inserting this bound into (\ref{g1.2})  finally gives
\begin{displaymath} g(x,\lambda ,q)\leq \exp \left( x\cdot
\sup \left\{ \Re (\lambda p_{i})\; |\; i=1,\ldots ,n\right\}
+\| q\| \sqrt{x}\right) .\end{displaymath}
This shows the convergence of \begin{math} g(x,\lambda ,q)\end{math}.
The uniqueness is well known. All summands of the expansion (\ref{g1.2})
are entire functions with respect to \begin{math} \lambda \end{math}
for fixed \begin{math} x\end{math}  and
\begin{math} q\end{math}.  Hence \begin{math} g(x,\cdot
,q)\end{math}  is entire, too.
\hspace*{\fill } \begin{math} \Box \end{math}

\noindent
Now let \begin{math} \tilde{q} \end{math}  be a small
perturbation of \begin{math} q:\end{math}
{\begin{math} \|q-\tilde{q} \|<\epsilon \end{math}. }
Then formula (\ref{g1.3})  implies
\begin{displaymath} \| \gamma _{n}(x,\lambda ,q)-\gamma
_{n}(x,\lambda ,\tilde{q} )\|\leq n\epsilon
\sqrt{x} \| \gamma
_{0}(x,\lambda )\| \frac{1}{n!} \left(
(\| q\| +\epsilon )\sqrt{x} \right)
^{n-1}\leq \end{displaymath}
\begin{displaymath} \leq \epsilon
\sqrt{x} \| \gamma _{0}(x,\lambda
)\| \frac{1}{(n-1)!} \left( (\| q\| +\epsilon
)\sqrt{x} \right)
^{n-1}.\end{displaymath}
This proves the
\begin{Corollary} \label{Corollary1.1}
If \begin{math} q\end{math}  and \begin{math} \tilde{q} \end{math}
 satisfy \begin{math} \| q-\tilde{q} \| <\epsilon
\end{math}, the following estimate holds:
\begin{displaymath} \| g(x,\lambda ,q)-g(x,\lambda ,\tilde{q}
)\| \leq \epsilon \sqrt{x} \|
\gamma _{0}(x,\lambda )\| \exp \left( (\| q\|
+\epsilon )\sqrt{x} \right).\end{displaymath} \end{Corollary}
\hspace*{\fill } \begin{math} \Box \end{math}

\noindent
Let us now recall, how the fundamental solution transforms
under gauge transformations. If \begin{math} g(x)\end{math}
is the fundamental solution of
\begin{displaymath} \left( \frac{d}{dx} +a(x)\right)
g(x)=0,\; \; g(0)=\unity \end{displaymath}
and if \begin{math} h(x)\end{math}  is a differentiable
function into \begin{math} GL(n,{\Bbb C})\end{math},
then \begin{math}
\tilde{g} (x)=h^{-1}(x)g(x)h(0)\end{math}
is the fundamental
solution of \begin{displaymath}
\left( \frac{d}{dx} +\tilde{a} (x)\right) \tilde{g} (x)=0,\;
\; \tilde{g} (0)=\unity  \; \; \mbox{with}
\; \; \tilde{a}
(x)=Ad(h^{-1}(x))a(x)+h^{-1}(x)\frac{dh(x)}{dx} .\end{displaymath}
In the case, in which both \begin{math} a(x)\end{math}  and \begin{math}
h(x)\end{math}  are periodic with period 1,
the Floquet matrix \begin{math} g(1)\end{math}  transforms
to \begin{math} \tilde{g} (1)=Ad(h^{-1}(0))g(1)\end{math}.
This shows that whenever a periodic \begin{math} h(x)\end{math}
 `diagonalizes' the operator \begin{math} \frac{d}{dx} +a(x)\end{math},
 which means that \begin{math} \tilde{a} (x)\end{math}  is
a diagonal matrix for all \begin{math} x\end{math},  then \begin{math}
h(0)\end{math}  diagonalizes the Floquet matrix \begin{math}
g(1)\end{math}. The next theorem from \cite{HSS}  presents
a formal diagonalization of the foregoing Lax operator. The
rest of this sections concerns the
analytic content of this formal diagonalization.
\begin{Theorem} \label{Theorem1.2}
Let \begin{math} q\end{math}  be a smooth periodic
potential. Then there exist two series
\begin{displaymath} \begin{array}{ll}
a_{1}(x), a_{2}(x),\ldots
& \mbox{ of offdiagonal matrices and} \\
b_{0}(x), b_{1}(x), \ldots
& \mbox{ of diagonal
matrices, respectively,} \end{array}
\end{displaymath}
such that \begin{math} a_{m+1}(x)\end{math}  and \begin{math}
b_{m}(x)\end{math}  are differential polynomials in \begin{math}
q(x)\end{math}  with derivatives of order \begin{math}
m\end{math}  at most and the following
equality
for formal power series of \begin{math}
\lambda ^{-1}\end{math}  holds:
\begin{equation} L\left( \unity +\sum_{m=1}^{\infty } a_{m}(x)\lambda
^{-m}\right) =\left( \unity +\sum_{m=1}^{\infty } a_{m}(x)\lambda
^{-m}\right) \left( \frac{d}{dx} +p\lambda +\sum_{m=0}^{\infty
} b_{m}(x)\lambda ^{-m}\right) \label{g1.6} .\end{equation}
In the particular case where \begin{math} q\end{math}  is only
\begin{math} M\end{math}  times differentiable, there exists
an \begin{math} \epsilon >0\end{math}  and \begin{math} c>
0\end{math},  such that for all \begin{math} |\lambda |^{-1}<
\epsilon ,\; x\in [0,1] \end{math}
\begin{equation} \left\| \left( \unity +\sum_{m=1}^{M+1} a_{m}(x)\lambda
^{-m}\right) ^{-1}L\left( \unity +\sum_{m=1}^{M+1} a_{m}(x)\lambda
^{-m}\right) -\frac{d}{dx} +p\lambda +\sum_{m=0}^{M}
b_{m}(x)\lambda ^{-m}\right\| <
\frac{c}{|\lambda |^{M+1}} .\label{g1.7}
\end{equation} \end{Theorem}
Proof: We solve inductively the ansatz (\ref{g1.6})
 in all powers of \begin{math} \lambda ^{-1}\end{math}.  For
the power \begin{math} \lambda ^{0}\end{math}  we obtain the
equation
\begin{equation} q(x)=[a_{1}(x),p]-b_{0}(x).\label{g1.7a} \end{equation}
Since all diagonal entries of \begin{math}
p\end{math}  are distinct,
\begin{math} ad(p)\end{math}  is invertible on the space of
all offdiagonal matrices and the equation has a unique
solution. For the power \begin{math} \lambda ^{-M}\end{math}
 we obtain the equation
\begin{equation} \frac{da_{M}(x)}{dx}
+q(x)a_{M}(x)-\sum_{m=1}^{M}
a_{m}(x)b_{M-m}(x)=[a_{M+1}(x),p]+b_{M}(x).\label{g1.7b} \end{equation}
These equations give inductively a unique solution of (\ref{g1.6})
 with the desired properties. For fixed \begin{math} M\end{math}
 it is obvious that for all \begin{math} \lambda \end{math}
 out of some neighbourhood of infinity and for all \begin{math}
x\in [0,1]\; \; \left( \unity +\sum_{m=1}^{M} a_{m}(x)\lambda
^{-m}\right) \end{math}  has a uniformly bounded inverse. On
the other hand the solution of
(\ref{g1.6}) implies that
there exists a \begin{math} \tilde{c} \end{math},  such
that for the same domain
\begin{displaymath} \left\| \left(
\unity +\sum_{m=1}^{M+1} a_{m}(x)\lambda
^{-m}\right) ^{-1}L\left( \unity +\sum_{m=1}^{M+1} a_{m}(x)\lambda
^{-m}\right) -\frac{d}{dx} +p\lambda +\sum_{m=0}^{M}
b_{m}(x)\lambda ^{-m}\right\| <\frac{\tilde{c} }{|\lambda |^{M+1}} .
\end{displaymath}  This completes the proof of the theorem.
\hspace*{\fill } \begin{math} \Box \end{math}

\noindent
It was shown in \cite{HSS} that these formal power series
are convergent, if and only if
\begin{math} q(x)\end{math}  is an
algebraic geometric potential. In
Appendix~\ref{Appendixa}  we
present a proof that these power series are Borel
summable,
if and only if \begin{math} q(x)\end{math}  is an analytic potential.
But for general smooth potentials these power series seem
not to be convergent in any sense.
Formally these power
series diagonalize the Floquet matrix at \begin{math} \lambda
=\infty \end{math}.  We are now going to prove the main
statement of this section that indeed these power series
are an asymptotic expansion of the diagonalization of the
Floquet matrix near \begin{math} \lambda =\infty \end{math}.
Later on we will see that in some sense it is even a Taylor
expansion of this diagonalization. Let us first define the
domain of the asymptotic expansion.
\begin{Definition} \label{Definition1.1}
For any \begin{math} l\in {\Bbb N}_{0},\; \epsilon
\in {\Bbb R}^{+} \end{math}  let \begin{math}
O_{l,\epsilon }\end{math}  be the set
\begin{displaymath} O_{l,\epsilon }=\left\{ \lambda \in {\Bbb C}
\; \left| \;
|\lambda |^{-1}<\epsilon ,\; |\lambda -k\pi
\sqrt{-1} |>\frac{1}{\epsilon (|k|\pi )^{l}}
\mbox{ for all }
k\in {\Bbb Z}\setminus \{ 0\} \right. \right\}
\footnote{For \begin{math} l=0\end{math}
we will always assume \begin{math}
\epsilon >2/\pi \end{math}. This ensures
that all these sets contain circles
around infinity.\label{footnote1}
}.\end{displaymath}
For a fixed \begin{math} M\end{math}  times continuously
differentiable potential \begin{math} q\end{math}  let \begin{math}
\exp (p_{i}(\lambda ))\end{math}  be the \begin{math} i\end{math}-th
eigenvalue of the Floquet matrix of the Lax
operator on the right hand side of (\ref{g1.6}):
\begin{equation} p_{i,M}(\lambda )=-p_{i}\lambda
-\sum_{m=0}^{M} \lambda ^{-m}\int_{0}^{1} \left(
b_{m}(x)\right) _{ii}dx.\label{g1.8}
\end{equation}
Finally let \begin{math} U_{M,\epsilon }\end{math}
be the set given by
\begin{displaymath} U_{M,\epsilon }=\left\{ \lambda \in {\Bbb C}
\; \left| \;
\frac{p_{i,M}(\lambda )-p_{j,M}(\lambda )}{2} \in O_{M,\epsilon
} \; \; \mbox{for all} \; i\neq j\in \{ 1,\ldots ,n\}
\right. \right\}.\end{displaymath}
\end{Definition}
The polynomials \begin{math} p_{i}(\lambda )\end{math}  and
the sets \begin{math} U_{M,\epsilon }\end{math}
depend on the potential \begin{math} q\end{math}.
But for large \begin{math} \lambda \end{math},  which
corresponds to small \begin{math} \epsilon ,\; p_{i}(\lambda
)\end{math}  is nearly equal to \begin{math} p_{i}\lambda \end{math}.
 Hence \begin{math} U_{M,\epsilon }\end{math}  consists of
all sufficient large \begin{math} \lambda \end{math}  with the
exception of small domains near \begin{math}
\lambda =\frac{2n\pi \sqrt{-1} }{p_{i}-p_{j}} ,\;
n\in {\Bbb Z}\end{math}.
With the help of (\ref{g1.8})  it would be easy to give an
asymptotic expansion of the exact
localization of
these
excluded domains. Later on we will see that in the case of
Hill's equation, this reproduces the well known asymptotic
expansion of the periodic and antiperiodic eigenvalues
(compare e.g. \cite{MW}, \cite{Ho} .)
\begin{Theorem} \label{Theorem1.3}
Let \begin{math} q\end{math} be an
\begin{math} M\end{math}  times continuously
differentiable periodic potential.
Then three is a holomorphic matrix valued function
\begin{math} h\end{math} on some \begin{math}
U_{l,\epsilon }\end{math}, which diagonalizes
the Floquet matrix \begin{math}
g(1,\lambda ,q)\end{math}
for all \begin{math} l\leq
M\end{math}.  Moreover this matrix
valued function can be chosen to be of
the form \begin{math}
h=\unity +\end{math} a holomorphic offdiagonal matrix
valued function. Then this function \begin{math}
h\end{math} and
the eigenvalues \begin{math} \mu
_{1},\ldots ,\mu _{n}\end{math}  of the Floquet matrix
may be expanded
asymptotically and uniformly on some \begin{math}
U_{l,\epsilon }\end{math}:
\begin{displaymath} \left\|
h(\lambda )-\left( \unity +\sum_{m=1}^{M-l}
a_{m}(0)\lambda ^{-m}\right) \right\| <\frac{\delta }{|\lambda
|^{M-l+1}} \footnote{Actually we will
prove that for all \begin{math} 1\leq l\leq
M\end{math} and all \begin{math} \delta
>0\end{math} there exists an \begin{math}
\epsilon >0\end{math}, such that this estimate
holds uniformly on \begin{math} U_{l,\epsilon
}\end{math}. Furthermore, for \begin{math}
l=0\end{math} this estimate holds with
some \begin{math} \delta >0\end{math}
uniformly on \begin{math} U_{0,4/\pi }\end{math}
(compare with
footnote~\ref{footnote2}).} ,\end{displaymath}
\begin{displaymath} |\mu _{i}-\exp (p_{i,M}(\lambda ))|<
\frac{c}{|\lambda |^{M+1}} |\exp (p_{i,M}(\lambda
))| \; \; i=1,\ldots ,n,\end{displaymath}
with some \begin{math} \delta ,c>0\footnote{The
rest of this section may be passed over.
It contains the proof of this theorem.}
\end{math}. \end{Theorem}
In order to prove this theorem we need some lemmata.
\begin{Lemma} \label{Lemma1.2}
Let \begin{math} B\end{math}  be a diagonal matrix \begin{math}
B=\mbox{\it diagonal} (\beta _{1},\ldots ,\beta _{n})\end{math}
such that \begin{math} \beta _{1}\end{math}
is the eigenvalue of maximal length,
which implies \begin{math} |\beta
_{1}|=\| B\|. \end{math} In addition
consider a matrix \begin{math} A\end{math}
 of the form \begin{math} A=\left( \begin{array}{cc}
a & b\\
c & d\end{array} \right) \end{math}  with respect to the
decomposition \begin{math} {\Bbb C}^{n}={\Bbb C}\oplus
{\Bbb C}^{n-1}\end{math}.
Moreover we assume the following estimates
to be valid for some small \begin{math}
\epsilon >0,\; 0<\delta \leq \frac{1}{4} :\end{math}
\begin{displaymath} \| A-B\| \leq \epsilon \| B\| \; \; \mbox{and}
\; \; |\beta _{1}-\beta _{i}|\geq \frac{\epsilon }{\delta }
|\beta _{1}|\mbox{ for all \begin{math}
i\not= 1\end{math}. } \end{displaymath}
Then there exists exactly one eigenvalue \begin{math} \alpha
\end{math}  of \begin{math} A\end{math}  obeying the
estimate \begin{math} |\alpha -\beta _{1}|\leq \epsilon
|\beta _{1}|\end{math}.  Furthermore there exists
exactly one
matrix of the form \begin{math} \left( \begin{array}{cc}
1 & w\\
v & \unity \end{array} \right)
\end{math},  which satisfies \begin{displaymath}
\left( \begin{array}{cc}
a & b\\
c & d\end{array} \right) \left( \begin{array}{cc}
1 & w\\
v & \unity \end{array} \right) =\left( \begin{array}{cc}
1 & w\\
v & \unity \end{array} \right) \left( \begin{array}{cc}
\alpha  & 0\\
0 & cw+d\end{array} \right) .\end{displaymath}
Finally the
assumptions imply the following inequalities:
\begin{displaymath} \|
v\| \leq \frac{\delta }{1-2\delta } ,\; \; \| w\| \leq
\frac{\delta }{1-2\delta } \; \; \mbox{and} \; \;
\left\| \left( \begin{array}{cc}
\alpha  & 0\\
0 & cw+d\end{array} \right) -B\right\| \leq \epsilon
\frac{1-\delta }{1-2\delta }
\| B\| .\end{displaymath} \end{Lemma}
Proof: Set \begin{math} A(z)=B+z(A-B)\end{math}.
Now we make the ansatz \begin{math}
A(z)v(z)=\alpha (z)v(z), \end{math}  with \begin{displaymath}
\alpha (z)=\beta _{1}+\sum_{k=1}^{\infty }
\alpha _{k}z^{k},\; v(z)=\sum_{k=0}^{\infty }
v_{k}z^{k},\; v_{0}=\left( \begin{array}{c}
1 \\
0 \\
\vdots \end{array} \right) .\end{displaymath}
For the \begin{math}
K\end{math}-th power of \begin{math} z\end{math}
we obtain  the equation: \begin{math}
Bv_{K}+(A-B)v_{K-1}=\sum_{k=0}^{K}\alpha _{k}v_{K-k}\end{math}.
 All these equations have a unique
solution with \begin{math}
v_{0}^{t}v_{K}=0\end{math}
for all \begin{math} K>0,\; \alpha _{K}=
v_{0}^{t}(B-A)v_{K-1}\end{math}
 and \begin{math}
(B-\beta _{1}\unity )v_{K}=(B-A)v_{K-1}+\sum_{k=1}^{K}
\alpha _{k}v_{K-k}\end{math}.
 Using the assumptions we can give
bounds for \begin{math}
|\alpha _{K}|\end{math}  and \begin{math} \|
v_{K}\| \end{math}: \begin{displaymath}
|\alpha _{K}|\leq \epsilon |\beta
_{1}|\cdot \| v_{K-1}\| \mbox{ and } \end{displaymath}
\begin{displaymath} \| v_{K}\| \leq
\frac{\delta }{\epsilon |\beta _{1}|} \left(
\epsilon |\beta _{1}|\cdot \| v_{K-1}\| +\sum_{k=1}^{K-1}
\epsilon |\beta _{1}|\cdot \| v_{k-1}\|
\cdot \| v_{K-k}\| \right)
\leq \delta \sum_{k=0}^{K-1} \| v_{k}\|
\cdot \| v_{K-k-1}\| .\end{displaymath}
The equation \begin{math}
\gamma (z)=\delta z\gamma ^{2}(z)+1 \end{math}  has one
solution \begin{math} \gamma
(z)=1/(2\delta z)\left( 1-\sqrt{1-4\delta z} \right)
\end{math}  with \begin{math} \gamma (0)=1\end{math}.
The Taylor expansion of
this function converges on the domain \begin{math}
|z|\leq 1/(4\delta ) \end{math}  and is
bounded on this domain by \begin{math}
|\gamma (z)|\leq 2\end{math}.  This proves
that \begin{math} \alpha =\alpha (1)\end{math}  and \begin{math}
v(1)\end{math}  exist and are bounded by
\begin{displaymath} |\alpha -\beta _{1}|\leq
2\epsilon |\beta _{1}|,\; \; \left\|
v(1)-v_{0}\right\| \leq \frac{1}{2\delta }
\left( 1-2\delta -\sqrt{1-4\delta } \right) .\end{displaymath}
Let us now improve these estimates.
Since \begin{math} \alpha \end{math}  is an
eigenvalue of \begin{math} A\end{math}
we have \begin{displaymath} \inf \left\{ \left.
|\alpha -\beta _{i}|\; \right| \; i=1,\ldots ,n\right\} \leq
\epsilon |\beta _{1}|\end{displaymath} Together with \begin{math}
|\alpha -\beta _{1}|\leq 2\epsilon |\beta _{1}|\end{math}
 and \begin{math}
|\beta _{1}-\beta _{i}|\geq 4\epsilon |\beta _{1}|
\end{math}  for all \begin{math} i=2,\ldots ,n\end{math}  this
implies \begin{math} |\alpha -\beta _{1}|\leq
\epsilon |\beta _{1}|\end{math}.
 The following ansatz \begin{displaymath} \left( \begin{array}{cc}
a & b\\
c & d\end{array} \right) \left( \begin{array}{cc}
1 & w\\
v & \unity \end{array} \right) =\left( \begin{array}{cc}
1 & w\\
v & \unity \end{array} \right) \left( \begin{array}{cc}
\alpha   & 0\\
0 & cw+d\end{array} \right) \end{displaymath} is equivalent to the
three equations \begin{displaymath} \begin{array}{cc}
a+bv=\alpha , & aw+b=wcw+wd,\\
c+dv=v\alpha \end{array} \end{displaymath}
Since \begin{math} \alpha \end{math}  is an eigenvalue
of \begin{math} A,\; \; v=(\alpha \unity -d)^{-1}c\end{math}
is a solution of the two equations
involving \begin{math} w\end{math}.  The third equation is
equivalent to \begin{displaymath} \left( \begin{array}{cc}
a & b\\
c & d\end{array} \right) \left( \begin{array}{cc}
1 & w\\
0 & \unity \end{array} \right) =\left( \begin{array}{cc}
1 & w\\
0 & \unity \end{array} \right) \left( \begin{array}{cc}
\alpha   & 0\\
c & cw+d\end{array} \right) \; \; \mbox{and to} \end{displaymath}
\begin{displaymath} \left( \begin{array}{cc}
1 & -w\\
0 & \unity \end{array} \right) \left( \begin{array}{cc}
a & b\\
c & d\end{array} \right) =\left( \begin{array}{cc}
\alpha   & 0\\
c & cw+d\end{array} \right) \left( \begin{array}{cc}
1 & -w\\
0 & \unity \end{array} \right) .\end{displaymath}
Again since \begin{math} \alpha \end{math}  is an
eigenvalue, \begin{math} w=b(d-\alpha
\unity )^{-1}\end{math}
is a solution. The assumptions also guarantee the following three
estimates: \begin{displaymath}
\| d-\mbox{\it diagonal} (\beta _{2},\ldots ,\beta _{n})\|
\leq  \epsilon
|\beta _{1}|,\end{displaymath}
\begin{displaymath}
\left\| \mbox{\it diagonal} (\beta _{2},\ldots ,\beta
_{n})v-\beta _{1}v\right\|
\geq \frac{\epsilon }{\delta} \| \beta
_{1}v\| \mbox{ for all \begin{math} v\in
{\Bbb C}^{n-1}\end{math}, } \end{displaymath}
\begin{displaymath} \| \beta _{1}\unity -\alpha \unity \| \leq
\epsilon |\beta _{1}|.\end{displaymath}
\begin{displaymath} \mbox{This implies }
\| dv-\alpha
v\| \geq \frac{\epsilon |\beta _{1}|(1-2\delta )}{\delta }
\| v\| \mbox{
and furthermore }\| (d-\alpha
\unity )^{-1}\| \leq  \frac{\delta }{\epsilon (1-2\delta )|\beta
_{1}|} .\end{displaymath}
\begin{displaymath} \mbox{Therefore }\| v\| \leq
\frac{\delta }{1-2\delta } \; \; \mbox{ and} \; \; \| w\| \leq
\frac{\delta }{1-2\delta } \; \; \mbox{and
finally} \end{displaymath}
\begin{displaymath} \left\| \left( \begin{array}{cc}
\alpha  & 0\\
0 & cw+d\end{array} \right) -B\right\| \leq
\epsilon |\beta _{1}|+\left\| \left( \begin{array}{cc}
0 & 0\\
0 & cw\end{array} \right) \right\| \leq \epsilon |\beta
_{1}|\left( 1+\frac{\delta }{1-2\delta } \right) .\end{displaymath}
All other eigenvalues \begin{math} \alpha ' \end{math}
of \begin{math} A\end{math}
 are also eigenvalues
of \begin{math} cw+d\end{math}.
 Hence they obey the estimate \begin{math}
\inf \left\{ \left. |\alpha '-\beta _{i}|\right|
\; i=2,\ldots ,n\right\} \leq \frac{3}{2}
\epsilon |\beta _{1}|\end{math},  which is a
contradiction to \begin{math}
|\alpha '-\beta _{1}|\leq \epsilon |\beta _{1}|\end{math}.
This proves that \begin{math} \alpha \end{math}
 is unique.
\hspace*{\fill } \begin{math} \Box \end{math}
\begin{Lemma} \label{Lemma1.3}
Let \begin{math} B\end{math}  be the diagonal matrix \begin{math}
B=\mbox{\it diagonal} (\beta _{1},\ldots
,\beta _{n})\end{math}  and \begin{math}
A\end{math}  another \begin{math} n\times n\end{math}-matrix
such that the following estimates hold for small
\begin{math} \epsilon ,\delta \in {\Bbb R}^{+}\end{math}
and all \begin{math} i=1,\ldots
,n\end{math}:
\begin{displaymath} \left\| \bigwedge ^{i}(A)-\bigwedge
^{i}(B)\right\|
\leq \epsilon \left\| \bigwedge ^{i}(B)\right\| \mbox{ and }
|\beta _{i}-\beta _{j}|\geq
\frac{\epsilon }{\delta } \sup \{ |\beta _{i}|,|\beta _{j}|\}
.\end{displaymath}
Then there exists an invertible matrix \begin{math} h\end{math}
 of the form \begin{math} \unity
+\mbox{\it offdiagonal} \end{math}
satisfying \begin{math} Ah=h\cdot
\mbox{\it diagonal} (\alpha
_{1},\ldots ,\alpha _{n})\end{math}  such that the following
inequalities are valid: \begin{displaymath}
\| h-\unity \| =o(\delta )\; \; \mbox{and}
\; \; |\alpha _{i}-\beta _{i}|=
o(\epsilon )|\beta _{i}|.\end{displaymath}
\end{Lemma}
Proof: First we rearrange lines and columns in a way,
such that \begin{math} \prod_{j=1}^{i} \beta _{j}\end{math}  is
an eigenvalue of maximal length of \begin{math} \bigwedge
^{i}(B)\end{math}.  Then we make use of
the last lemma and, furthermore claim
that \begin{displaymath} \left\|
\bigwedge ^{i}(cw+d)-\bigwedge ^{i}\left(
\mbox{\it diagonal} (\beta _{2},\ldots ,\beta _{n})\right)
\right\| =o(\epsilon )\left\| \bigwedge
^{i}\left(
\mbox{\it diagonal} (\beta _{2},\ldots ,\beta _{n})\right)
\right\| .\end{displaymath}  In order
to prove this claim we note that
the matrix \begin{displaymath}
\bigwedge ^{i+1}
\left( (1-wv)\left( \begin{array}{cc}
1 & w\\
v & \unity \end{array} \right) ^{-1}\right)
=\bigwedge ^{i+1}\left( \begin{array}{cc}
1 & -w\\
-v & (1-wv)\unity +vw\end{array} \right)
\mbox{ has the form }
\left( \begin{array}{cc}
\unity & \cdot \\
\cdot  & \cdot \end{array} \right)
\end{displaymath}
with respect to the natural
decomposition \begin{math} \bigwedge
^{i+1}\left( {\Bbb C}\oplus {\Bbb C}^{n-1}\right)
\simeq \bigwedge ^{i}\left( {\Bbb C}^{n-1}\right)
\oplus \bigwedge ^{i+1}\left( {\Bbb C}^{n-1}\right)
\end{math}.  Hence the matrix \begin{displaymath}
\bigwedge ^{i+1}\left( \begin{array}{cc}
1 & -w\\
-v & (1-wv)\unity +vw\end{array} \right)
\left( \bigwedge ^{i+1}(A)-\bigwedge ^{i+1}(B)\right)
\end{displaymath} has the form
\begin{displaymath} \left( \begin{array}{cc}
\alpha \bigwedge ^{i}(cw+d)-\beta
_{1}\bigwedge ^{i}\left(
\mbox{\it diagonal} (\beta _{2},\ldots ,\beta
_{n})\right) & \cdot \\
\cdot & \cdot\end{array} \right) .\end{displaymath}
Finally the bound \begin{displaymath}
\left\| \bigwedge ^{i}(cw+d)-\bigwedge
^{i}\left( \mbox{\it diagonal} (\beta
_{2},\ldots ,\beta _{n})\right) \right\| \leq
\end{displaymath}
\begin{displaymath} \leq \frac{1}{|\beta _{1}|}
\left\| \bigwedge ^{i+1}\left( \begin{array}{cc}
1 & -w\\
-v & (1-wv)\unity +vw\end{array} \right)
\right\| \cdot \left\|
\bigwedge ^{i+1}(A)-\bigwedge ^{i+1}(B)
\right\| +\frac{|\alpha -\beta _{1}|}{|\beta
_{1}|} \left\| \bigwedge
^{i}(cw+d)\right\| \end{displaymath}
and the last lemma proves the claim. The
inductive use of this claim and the last
lemma now proves the lemma.
\hspace*{\fill } \begin{math} \Box \end{math}

\noindent
The assumptions of Lemma~\ref{Lemma1.3}
contain two
estimates. In the application we have in
mind the first is just the content of
Corollary~\ref{Corollary1.1}, but to
ensure the second estimate we need two further lemmata.
\begin{Lemma} \label{Lemma1.4}
For all \begin{math} \delta >0\end{math}
and \begin{math} l\in {\Bbb N}\end{math}  there
exists an
\begin{math} \epsilon >0\end{math},  such that for all \begin{math}
\lambda \in O_{l,\epsilon }\end{math} \begin{displaymath}
|\sinh (\lambda )|\geq \frac{1}{\delta
|\lambda |^{l}} \exp \left( |\Re (\lambda )|\right)
\footnote{This estimate is proven in
\cite[Lemma~2.1]{PT} for \begin{math}
l=0,\delta =4\end{math} and \begin{math}
\epsilon =4/\pi \end{math}.\label{footnote2} }
.\end{displaymath}
\end{Lemma}
Proof: First we use the following bound from below:
\begin{math} |\sinh (\lambda )|\geq \exp \left( |\Re
(\lambda )|\right) |\sin \left( \Im (\lambda )\right)
|/2\end{math}.
To each \begin{math} \delta >0\end{math}
 there exists an \begin{math} \epsilon >0\end{math},
such that for all \begin{math}
\sqrt{-1} \Im (\lambda )\in O_{l,\epsilon }\;
\; |\sin \left( \Im (\lambda )\right)
|\geq 2\delta ^{-1}|\lambda |^{-l}\end{math},  and therefore
the above estimate also holds. On the other hand
\begin{math} |\sinh (\lambda )|\end{math}
is again bounded from below by \begin{displaymath}
|\sinh (\lambda )|\geq \frac{1}{2}
\left( 1-\exp \left( -2|\Re (\lambda )|\right)
\right) \exp \left( |\Re (\lambda )|\right) .\end{displaymath}
Hence the proposed estimate holds also in the domain
\begin{math} \left( 1-\exp \left( -2|\Re
(\lambda )|\right) \right) \geq 2\delta ^{-1}
|\lambda |^{-l}\end{math}.  The union of
the
domains of both kinds clearly contains some \begin{math}
O_{l,\epsilon }\end{math}.
\hspace*{\fill } \begin{math} \Box \end{math}
\begin{Lemma} \label{Lemma1.5}
Let \begin{math} \beta _{1}(\lambda )\end{math}  and \begin{math}
\beta _{2}(\lambda )\end{math}  be two functions of the form
\begin{displaymath} \beta _{1}(\lambda )=\exp \left(
p_{1}\lambda +polynomial(\lambda ^{-1})\right) =\exp \left(
p_{1}(\lambda )\right) ,\end{displaymath}
\begin{displaymath} \beta _{2}(\lambda )=\exp \left(
p_{2}\lambda +polynomial(\lambda ^{-1})\right) =\exp \left(
p_{2}(\lambda )\right) \end{displaymath}
Then for all \begin{math} \delta >0\end{math}
and \begin{math} l\in {\Bbb N}\end{math} there exists an
\begin{math} \epsilon >0\end{math},  such that for all \begin{math}
\frac{p_{1}(\lambda )-p_{2}(\lambda )}{2} \in O_{l,\epsilon }\end{math}
\begin{displaymath} |\beta _{1}(\lambda )-\beta _{2}(\lambda )|\geq
\frac{1}{\delta |\lambda |^{l}} \sup \{
|\beta _{1}(\lambda )|,|\beta _{2}(\lambda )|\} .
\end{displaymath} \end{Lemma}
Proof: We have \begin{displaymath}
\exp \left( p_{1}(\lambda )\right) -\exp \left(
p_{2}(\lambda )\right) =
2\exp \left( \frac{p_{1}(\lambda )+p_{2}(\lambda )}{2} \right)
\sinh \left( \frac{p_{1}(\lambda )-p_{2}(\lambda )}{2} \right)
,\end{displaymath} \begin{displaymath}
\sup \left\{ |\exp \left( p_{1}(\lambda )\right)
|,|\exp \left( p_{2}(\lambda )\right)
|\right\} \left| \exp \left(
-\frac{p_{1}(\lambda )+p_{2}(\lambda )}{2} \right)
\right| =\exp \left( \left| \Re \left(
\frac{p_{1}(\lambda )-p_{2}(\lambda )}{2}
\right) \right| \right) .\end{displaymath}
Hence the last lemma gives for each \begin{math} \delta >0\end{math}
an \begin{math} \epsilon >0\end{math},
such that for all \begin{math}
\frac{p_{1}(\lambda )-p_{2}(\lambda )}{2} \in O_{l,\epsilon }\end{math}
\begin{displaymath} |\exp \left( p_{1}(\lambda )\right) -\exp \left(
p_{2}(\lambda )\right) |\geq \frac{\exp \left( \sup \left\{
\Re \left( p_{1}(\lambda )\right) ,\Re
\left( p_{2}(\lambda )\right) \right\}
\right) }{\delta } \left| \frac{p_{1}(\lambda
)-p_{2}(\lambda )}{2} \right|
^{-l}.\end{displaymath}
For \begin{math} p_{1}-p_{2}\end{math}  not being zero,
there surely exists some neighbourhood of \begin{math}
\lambda =\infty\end{math}  and some \begin{math} c>0\end{math}
such that
\begin{displaymath} \left|
\frac{p_{1}(\lambda )-p_{2}(\lambda )}{2}
\right| ^{-l}\geq c|\lambda |^{-l}.\end{displaymath}
This proves the
lemma.
\hspace*{\fill } \begin{math} \Box \end{math}

\noindent
Now we are ready to give the Proof of Theorem~\ref{Theorem1.3}:
We want to apply Lemma~\ref{Lemma1.3}.
Two estimates are assumed in this lemma.
These estimates are shown to be
fulfilled one after another.
First we make an observation. If \begin{math} g\end{math}
is the fundamental solution of the differential equation
\begin{displaymath} \left( \frac{d}{dx} +a(x)\right)
g(x)=0,\; \; g(0)=\unity ,\end{displaymath}
then \begin{math} \bigwedge ^{i}(g)\end{math}
is the fundamental
solution of the differential equation \begin{displaymath}
\left( \frac{d}{dx} +d\bigwedge ^{i}\left( a(x)\right) \right)
\bigwedge ^{i}\left( g(x)\right) =0,\; \; \bigwedge ^{i}\left(
g(0)\right) =\unity .\end{displaymath}
Here \begin{math} d\bigwedge ^{i}(A)\end{math}
is defined to be \begin{math} (A\wedge  \unity \wedge ...\wedge
\unity )+...+(\unity \wedge
\unity \wedge ...\wedge A)\end{math}.  We also
remark that if \begin{math} \|A-B\| \leq \epsilon \end{math},
then \begin{math} \| d\bigwedge ^{i}(A)-d\bigwedge ^{i}(B)\|
\leq i\epsilon \end{math}.  If we take the
two Lax operators of inequality (\ref{g1.7})
in Theorem~\ref{Theorem1.2}  and the
\begin{math} i\end{math}-th
exterior powers, respectively, these
inequalities ensure the validity of the
assumption of Corollary~\ref{Corollary1.1}
 with \begin{math} \epsilon =c|\lambda
|^{-M-1}\end{math}.
Corollary~\ref{Corollary1.1}  for its part ensures the first
estimate of the assumption in Lemma~\ref{Lemma1.3}  with \begin{math}
\epsilon =c|\lambda |^{-M-1}\end{math}  for some
\begin{math} c>
0\end{math} (not necessarily the same as in
Theorem~\ref{Theorem1.2}). Now the second
estimate of the assumption of Lemma~\ref{Lemma1.3}
 is guaranteed by Lemma~\ref{Lemma1.5},
if \begin{math} \delta \end{math}
 in Lemma~\ref{Lemma1.3}  is changed to \begin{math}
\delta c|\lambda |^{l-M-1}\end{math} and
the same \begin{math} \epsilon \end{math}
as in Theorem~\ref{Theorem1.3}.  Now the first estimate
of Lemma~\ref{Lemma1.3}  ensures that the matrix, which
diagonalizes the first Lax operator in the inequality (\ref{g1.7})
 exists and is uniformly bounded by \begin{math}
\delta c|\lambda |^{l-M-1}\end{math} with
some constant \begin{math} c\end{math}
 (not necessary the same as above) on the
domain of Lemma~\ref{Lemma1.5}.
The second estimate of Lemma~\ref{Lemma1.3}  proves the
second estimate of Theorem~\ref{Theorem1.3}.
\hspace*{\fill } \begin{math} \Box \end{math}

\section{The Riemann surface} \label{Section2}
In this section we will introduce the Riemann surface
corresponding to the Floquet matrix. For the moment it is
defined as the curve given in terms of the eigenvalue equation
of the Floquet matrix \begin{displaymath}
R(\lambda ,\mu )=\det \left( \mu \unity
-g(1,\lambda ,q)\right) =0. \end{displaymath}
After normalization this is an open \begin{math}
n\end{math}-fold covering of the complex plane
\begin{math} \lambda \in {\Bbb C}\end{math} and therefore
an open smooth Riemann surface (see for
example Theorem~8.9 \cite{Fo} ). We
intend to establish a one to one
correspondence between some line bundles
and potentials. This construction has already
been carried out
in the case of algebraic geometric
potentials \cite{Kr} , \cite{RST} .
On open Riemann surfaces all line bundles
are trivial (Theorem~30.3 \cite{Fo} ),
so we have to impose some decay
condition near \begin{math}
\lambda =\infty \end{math}. We want to do this
with the help of the asymptotic
expansion of the last section. Indeed,
similar to the algebraic geometric case,
we may add covering points of \begin{math}
\lambda =\infty \end{math}  to the Riemann
surface, if we concede them an exceptional
position. In this section we want to
provide a first impression of this strange
behaviour of the Riemann surface near \begin{math}
\lambda =\infty \end{math}. It can be
summarized in the statement that almost
everything holds as in the context of
compact Riemann surfaces.
\begin{Definition} \label{Definition2.1}
 From now on we fix a smooth periodic
potential \begin{math} q\end{math}.
Theorem~\ref{Theorem1.3}  shows that
for an arbitrary \begin{math} l\in {\Bbb N}_{0}\end{math}
 over some \begin{math} U_{l,\epsilon }\end{math}
 the Riemann surface is an unbranched \begin{math}
n\end{math}-fold covering, and that
there is a natural way to index the
sheets with the numbers \begin{math}
i=1,\ldots ,n\end{math}  corresponding to
the eigenvalue \begin{math} \mu _{i}\end{math}
 of the Floquet matrix. Each of these
sheets can of course uniquely be extended
to those small excluded domains of \begin{math}
U_{l,\epsilon }\end{math},  where only other
sheets have branchpoints. To each
extended sheet over a domain near \begin{math}
\lambda =\infty \end{math}  we add the point \begin{math}
\lambda =\infty \end{math}  and take the
extended sets of \begin{math}
U_{l,\delta },\; \; \delta \leq \epsilon
\end{math},  as
a base of neighbourhoods of this new
point: For each \begin{math} i=1,\ldots ,n\end{math}
 these neighbourhoods are the liftings
into the \begin{math}
i\end{math}-th sheet of the sets \begin{displaymath}
\left\{ \lambda \in {\Bbb C}\left|
\frac{p_{i,l}(\lambda )-p_{j,l}(\lambda )}{2} \in
O_{l,\delta }\; \; \mbox{for all }\; \; j\neq
i\right. \right\} \bigcup \left\{ \infty
\right\} .\end{displaymath} (Compare
with Definition~\ref{Definition1.1}).
This base of neighbourhoods of the \begin{math}
n\end{math}  covering points over \begin{math}
\lambda =\infty \end{math}  together with the
usual topology over \begin{math} {\Bbb C}\end{math}
defines the topology \begin{math}
\tau _{l}\footnote{For \begin{math} l=0\end{math}
we assume that \begin{math}
\epsilon>2/\pi \end{math} (compare with
footnote~\ref{footnote1}).}\end{math}.  Because \begin{math}
O_{l,\epsilon }\end{math}  is contained in \begin{math}
O_{m,\epsilon }\end{math} for \begin{math}
\epsilon <1,\; \; l\leq m\end{math},  these
topologies are ordered: \begin{math}
\tau _{0}\supset \tau _{1}\supset ...\; \; \end{math}.
Finally we define \begin{math}
\tau _{\infty }\end{math}  as the
finest topology, which is coarser than all these topologies.
This \begin{math} n\end{math}-fold
covering over \begin{math} {\Bbb C}\cup \{ \infty
\} \end{math} together with the topology
\begin{math} \tau _{\infty }\end{math}
will be called the Riemann surface \begin{math}
Y\end{math}.  Taking the sets \begin{math}
U_{l,\epsilon }\cup \{ \infty \} \end{math} as
a base of neighbourhoods of \begin{math}
\lambda =\infty \end{math}  there is also a
sequence of topologies \begin{math}
\tau _{0}\supset \tau _{1}\supset ...\supset
\tau _{\infty }\end{math}  on \begin{math}
{\Bbb C}\cup \{ \infty \} \end{math}.  The space \begin{math}
{\Bbb C}\cup \{\infty \} \end{math}  together
with the topology \begin{math} \tau _{\infty
}\end{math}  will be called \begin{math}
X. \end{math}  \end{Definition}
\begin{Remark} \label{Remark2.1a}
We will see in the next section that
the Riemann surface \begin{math} Y\end{math}
 may be an unbranched covering over some
of these excluded domains. For the
analytic investigation of the Riemann
surface \begin{math} Y\end{math}
it would be appropriate not to
exclude such domains from the
neighbourhoods of the covering points of
infinity. But it could happen that a
singular Riemann surface, namely with
singularities in such unbranched
coverings over excluded domains,
corresponds to the potential \begin{math}
q\end{math}  in a way, which is established in the
next section. The topology defined above
fits in the most singular case, namely
the algebraic curve defined by the
equation \begin{math} R(\lambda ,\mu)=0\end{math}.
\end{Remark}
Obviously \begin{math} X\end{math}  and \begin{math}
Y\end{math}  are Hausdorff
spaces, but no topological manifolds.
The covering map \begin{math} \pi :Y\mapsto
{\Bbb P}_{1}\end{math}  is continuous, but in general
the map \begin{math} \pi :Y\mapsto X\end{math}  is
not continuous. A function on \begin{math}
X\end{math}  and \begin{math} Y\end{math},
 respectively is called holomorphic, if
it is holomorphic on \begin{math} {\Bbb C}\end{math}
and \begin{math} \pi ^{-1}({\Bbb C})\end{math},
 respectively and if all derivatives of \begin{math}
f\end{math}  can be
extended continuously to the whole of \begin{math} X\end{math}
 and \begin{math} Y\end{math},
respectively. Analogously meromorphic
functions are meromorphic functions on \begin{math}
{\Bbb C}\end{math}  and \begin{math} \pi
^{-1}({\Bbb C})\end{math},  respectively, such
that on some neighbourhood of infinity
and \begin{math} \pi ^{-1}(\infty )\end{math}
 respectively \begin{math} \lambda ^{-l}f\end{math}
 is holomorphic for some \begin{math}
l\in {\Bbb N}\end{math}.  Now it should be
evident how to define the sheaves \begin{math} {\cal O}_{X}\end{math}
 and \begin{math} {\cal O}_{Y}\end{math}  of
holomorphic functions and \begin{math}
{\cal M}_{X}\end{math}  and \begin{math}
{\cal M}_{Y}\end{math}  of meromorphic
functions.
\begin{Lemma} \label{Lemma2.1}
For each open set \begin{math} U\end{math}
 of \begin{math} {\Bbb P}_{1}\end{math}, the
sections of the sheaf \begin{math}
{\cal O}_{X}\end{math}  restricted to \begin{math}
U\end{math}  coincide with the
holomorphic functions on \begin{math} U\end{math}
 in the usual sense. In particular \begin{math}
H^{0}(X,{\cal O}_{X})={\Bbb C}\end{math}  and \begin{math}
H^{0}(Y,{\cal O}_{Y})={\Bbb C}\end{math}. \end{Lemma}
Proof: By the maximum modulus Theorem \cite{Co}
 the sections of \begin{math} {\cal O}_{X}\end{math}
 restricted to some \begin{math} U\end{math}
 and all their derivatives are continuous
with respect to the topology of \begin{math}
{\Bbb P}_{1}\end{math}.  Hence they are
holomorphic in the usual sense. All
holomorphic functions on \begin{math} {\Bbb
P}_{1}\end{math}  are constant, so that \begin{math}
H^{0}(X,{\cal O}_{X})={\Bbb C}\end{math}
follows. Due to the same argument all the
elementary symmetric functions of a
holomorphic function on \begin{math} Y\end{math}
 with respect to the covering \begin{math}
\pi :Y\mapsto {\Bbb P}_{1}\end{math}  \cite{Fo}
 have to be constant and therefore the
holomorphic functions on \begin{math} Y\end{math}
 also have to be constant.
\hspace*{\fill } \begin{math} \Box \end{math}

\noindent
This is the first analogy of our Riemann
surface with compact Riemann surfaces.
Before we proceed to further
analogies, let us join on to the last
section.
\begin{Lemma} \label{Lemma2.2}
A function holomorphic on some \begin{math}
U_{l,\epsilon }\end{math}  can be
extended to a function holomorphic
on \begin{math}
U_{l,\epsilon }\cup \{ \infty \} \end{math},
iff it can be expanded asymptotically
uniformly on some \begin{math} U_{L,\delta }\end{math}
 for each \begin{math} L>l\end{math}  in
the following manner
\begin{displaymath} \left|
f(\lambda )-\sum_{m=0}^{M}a_{m}\lambda ^{-m}\right|
<\frac{c_{M}}{|\lambda |^{M+1}} \; \; \mbox{for all}
\; \; M\in {\Bbb N}.\end{displaymath}
\end{Lemma}
Proof: If \begin{math} f\end{math}  is
holomorphic on some \begin{math}
U_{l,\epsilon }\end{math},  it has by definition a
Taylor expansion at \begin{math} \lambda =\infty
\end{math}.  This Taylor expansion obviously
gives an asymptotic expansion of the
desired form. Conversely let \begin{math} f\end{math}
 have such an asymptotic
expansion. For each \begin{math} \lambda _{0}\in
U_{L,\delta '}\end{math}  with \begin{math}
\delta '<\delta \end{math}  there is a ball \begin{math}
B(\lambda _{0}^{-1},c|\lambda
_{0}|^{-L-2})=\left\{ \lambda \; \left|
\; |\lambda ^{-1}-\lambda _{0}^{-1}|<c|\lambda
_{0}|^{-L-2}\right. \right\}
\end{math}  inside \begin{math}
U_{L,\delta }\end{math}  with some \begin{math}
c>0\end{math}  depending on \begin{math}
L\end{math}  and \begin{math} \delta \end{math}
 but not on \begin{math}
\lambda _{0}\end{math}.  Now with the help of
Cauchy's estimate \cite{Co}  the
asymptotic expansion leads to an
asymptotic expansion of \begin{math} f'=
\frac{df}{d\lambda ^{-1}} \end{math}  on \begin{math}
U_{L,\delta '}\end{math} of the form \begin{displaymath}
\left|
f'(\lambda )-\sum_{m=0}^{M}ma_{m}\lambda ^{-m+1}\right|
<\frac{c_{M}}{c|\lambda |^{M-L+3}} \end{displaymath}
and therefore \begin{displaymath}
\left|
f'(\lambda )-\sum_{m=0}^{M}ma_{m}\lambda ^{-m+1}\right|
<\frac{\tilde{c} _{M}}{|\lambda |^{M-L+3}} \end{displaymath}
with some new constants \begin{math}
\tilde{c} _{M}\end{math}.  This proves that \begin{math}
f'(\lambda )\end{math}  is continuous with
respect to the topology \begin{math}
\tau _{L}\end{math}  and has an asymptotic
expansion similar to \begin{math}
f\end{math}.  The repeated application
of this claim to higher and higher
derivatives shows that all derivatives
can continuously be extended with respect
to all topologies \begin{math} \tau _{L}\end{math}.
This proves the claim.
\hspace*{\fill } \begin{math} \Box \end{math}

\noindent
There are similar statements about the
holomorphic functions on \begin{math} Y\end{math}
with completely analogous proofs.

We emphasize that the inverse
of a meromorphic function is not
necessarily also a
meromorphic function. In fact, the
inverse of a meromorphic function would
be meromorphic, if we could exclude that
this meromorphic function has a zero of
infinite order at infinity or at some
of the covering points of infinity,
respectively. However, this is not
possible and we will provide an example
of a function with these properties
at the end of this section.
Now it is quite obvious that the divisor
of an invertible meromorphic function consists of a
finite part far away from infinity, of
sequences of finite divisors inside
the small excluded domains of the
neighbourhood of infinity and of a
finite contribution at infinity. More
precisely, to each meromorphic function
there exists a neighbourhood of infinity and
\begin{math} \pi ^{-1}(\infty )\end{math}
 respectively, such that the restriction
of this function to this
neighbourhood has poles or zeros
at most at infinity. Let us choose a
subset of this neighbourhood out of the
base given above in
Definition~\ref{Definition2.1},
defining a sequence of excluded domains. In the
sequel we will often make some
statements about these excluded domains of
this base of neighbourhoods. For this
purpose we abbreviate the multiple index
\begin{math} (i,j,k)\end{math}  used in
Definition~\ref{Definition2.1} by \begin{math}
\iota :\end{math} \begin{displaymath} \iota \in
{\cal I},\; \; \mbox{with} \; \; {\cal I}=\left\{
(i,j,k)\; \left| \; k\in {\Bbb Z},\; \;
i<j\in \{ 1,\ldots ,n\} \right. \right\} .\end{displaymath}
Moreover, the absolute value of \begin{math}
\iota \end{math}  is defined as \begin{math}
|\iota |=|(i,j,k)|=|k|\end{math}.  This
index will be used as follows. Given any
neighbourhood \begin{math} U_{l,\epsilon }\end{math}
as in Definition~\ref{Definition1.1},
the index \begin{math} \iota =(i,j,k)\in {\cal I}\end{math}
 specifies the excluded domain of \begin{math}
U_{l,\epsilon }\end{math}  defined as the union
of the \begin{math} i\end{math}-th and \begin{math}
j\end{math}-th sheet in \begin{math} Y\end{math}
over the set of \begin{math}
\lambda \end{math}'s satisfying \begin{displaymath}
\left| \frac{p_{i,l}(\lambda )-p_{j,l}(\lambda )}{2}
-k\pi \sqrt{-1}\right| \leq \frac{1}{\epsilon (|k|\pi
)^{l}} .\end{displaymath}  By
definition, if \begin{math}
k=0\end{math}, the excluded domains form
the whole of \begin{math} Y\end{math}.
Excluded domains having non empty
intersection will be identified. Thus
the set \begin{math} {\cal I}\end{math}  does
not really label the excluded domains.
For the sake of simplicity we do not use a
more precise notation.
The actual precise meaning will be clear
from the context.

Holomorphic and meromorphic
differential forms on \begin{math} X\end{math}
 and \begin{math} Y\end{math}, respectively, are
defined near infinity as holomorphic
and meromorphic functions, respectively,
times \begin{math} d\lambda ^{-1}\end{math}
and otherwise as usual.
\begin{Theorem} \label{Theorem2.1}
For each meromorphic differential form,
the sequence \begin{math} (r_{\iota })_{\iota \in
{\cal I}}\end{math}  of the
corresponding total residues of
the excluded domains defines an element of
the Fr\'{e}chet space \begin{displaymath}
l_{{\cal I}}^{\infty }=\left\{ (\alpha _{\iota })_{\iota \in
{\cal I}}\; \left| \; \sum_{\iota \in
{\cal I}}|\alpha _{\iota }|\cdot |\iota
|^{l}<\infty \; \; \mbox{for all}
\; \; l\in {\Bbb N}_{0}\right. \right\} \end{displaymath}
Hence the total residue is well defined
and is equal to zero.\end{Theorem}
Proof: The sequence of residues of every
form \begin{math} \omega =\left(
\sum_{m=M_{1}}^{M_{2}} a_{m}\lambda ^{-m}\right)
d\lambda \end{math}  is asymptotically equal
to zero for arbitrary integers \begin{math}
M_{1}\leq M_{2}\end{math}.  If we use
the Cauchy integral representation of the
total residue, the asymptotic expansion
of the meromorphic differential form
analogous to the asymptotic expansion of Lemma~\ref{Lemma2.2}
shows the convergence of the sequence
of total residues to be faster than any
inverse power of \begin{math} |\lambda |\end{math}.
Since the absolute value of the index
\begin{math} \iota \end{math}
 is bounded by some constant times
the absolute value of \begin{math} \lambda \end{math}
 inside the excluded domain with index \begin{math}
\iota :\; \; |\iota |<c|\lambda |\end{math},  the claim
about the asymptotic behaviour follows.
The last statement is now obvious.
\hspace*{\fill } \begin{math} \Box \end{math}
\begin{Corollary} \label{Corollary2.1}
Let \begin{math} f\end{math}  be a
meromorphic function, whose inverse is
also a meromorphic function. Then the
sequence of total degrees \begin{math}
(d_{\iota })_{\iota \in {\cal I}}\end{math}  of the
excluded domains is asymptotically
equal to zero: \begin{math} d_{\iota }=0\end{math}
 for all \begin{math} |\iota |\geq N\end{math},
with some positive integer \begin{math}
N\end{math}.
Therefore the total degree of \begin{math}
f\end{math}  is well defined and is
equal to zero.\end{Corollary}
Proof: Due to the assumption on \begin{math}
f,\; \; f^{-1}df\end{math}  is a
meromorphic differential form. Therefore the
residue of this form is a integer
multiple of \begin{math} 2\pi \sqrt{-1}\end{math}
and the claim follows from the
preceeding theorem.
\hspace*{\fill } \begin{math} \Box \end{math}
\begin{Remark} \label{Remark2.1}
As a consequence of this corollary
the equivalence classes of divisors
decompose into components in analogy to
the connected components of the Picard
group of compact Riemann surfaces.
But they are not only
labeled by the total degree as in the
case of compact Riemann surfaces: For
every divisor, which in general
is defined as a
cross section of the multiplicative sheaf \begin{math}
{\cal M}^{*}_{X}/{\cal O}^{*}_{X}\end{math}  and \begin{math}
{\cal M}^{*}_{Y}/{\cal O}^{*}_{Y}\end{math}
respectively, there exists a
neighbourhood of infinity and all
covering points of infinity,
respectively, such that the restriction
to this neighbourhood may have only
contributions at infinity and all
covering points of infinity
respectively. Hence we can associate to
each divisor a sequence of total degrees
in the excluded domains indexed by \begin{math}
\iota \in {\cal I}:\end{math}  \begin{displaymath}
\deg :\mbox{ divisors } \rightarrow
\mbox{integer valued sequences indexed by}
\; \; \iota \in {\cal I},\; \; D\mapsto
\left( \deg _{\iota }(D)\right) _{\iota \in
{\cal I}}.\end{displaymath}
If the absolute value of \begin{math}
\iota \end{math}  is small, \begin{math}
\deg _{\iota }\end{math}  depends on the
choice of \begin{math} U_{l,\epsilon }\end{math}
defining the excluded domain with index
\begin{math} \iota \end{math}.
Moreover, as
mentioned before, there corresponds only
one degree to excluded domains
having a non empty intersection, namely
the total degree of the union of these
excluded domains.
If \begin{math} \deg _{\iota }(D)=\deg _{\iota }(D')\end{math}
for all \begin{math} |\iota |\geq N\end{math},
 with some positive integer \begin{math}
N\end{math},  the degrees of \begin{math}
D\end{math}  and \begin{math} D'\end{math}
 are called asymptotically equal.
If in addition \begin{math} \sum_{|\iota |\leq K}
\deg _{\iota }(D)-\deg _{\iota }(D')=0\end{math}
for all \begin{math} K\geq N\end{math}  we
call the degrees of two such
divisors asymptotically and totally equal.
Hence the degrees of two linear
equivalent divisors \begin{math} D\end{math}
 and \begin{math} D'\end{math}
are asymptotically and totally equal.
Consequently the
components of the equivalence classes of
divisors are labeled by equivalence classes
of integer valued sequences indexed by \begin{math}
\iota \in {\cal I}\end{math}, which are
asymptotically and totally equal.
\end{Remark}
Having exhibited some analogies between our
Riemann surfaces \begin{math} X\end{math}
 and \begin{math} Y\end{math}  and
ordinary compact Riemann surfaces, we
will turn to the question whether there
actually exist meromorphic functions with
infinitely many poles. Before we answer
this question in the next section for
the surface \begin{math} Y\end{math},
we will now give the answer for the
surface \begin{math} X\footnote{These
two examples may be passed over.}\end{math}.
\begin{Example} \label{Example2.1}
Let \begin{math} (\alpha _{i})_{i\in {\Bbb N}}\end{math}
 and \begin{math} (\beta _{i})_{i\in {\Bbb N}}\end{math}
 be two sequences of \begin{math} {\Bbb C}
\subset X={\Bbb P}_{1}\end{math}
 satisfying the following conditions:
\begin{description}
\item[(i)] \begin{math} |\alpha _{i+1}
|\geq |\alpha _{i}|,\; \; |\beta _{i+1}|\geq
|\beta _{i}|\end{math}.
\item[(ii)] Both sequences have no accumulation
points in \begin{math} X\end{math}.
\item[(iii)] For all large \begin{math} i,\;
\; \alpha _{i}\end{math}  and \begin{math}
\beta _{i}\end{math}  are contained in a
common excluded domain.
\item[(iv)] \begin{math}
|\alpha _{i}|^{l}>i,\; \; |\beta _{i}|^{l}>i\end{math}
 for \begin{math} i\geq I\end{math},
with some positive integers \begin{math}
I\end{math}  and \begin{math} l\end{math}.
\end{description}
Due to (ii) each excluded domain contains
only a finite number of \begin{math}
\alpha 's\end{math}  and \begin{math} \beta 's\end{math}.
Condition (iv) ensures that this
number is bounded by some power of
\begin{math} |\lambda |\end{math}
 with \begin{math} \lambda \end{math}  being a
point in the excluded domain. In
consideration of the topology of \begin{math}
X\end{math}  these conditions
ensure that the sequence \begin{math}
(\alpha _{i}-\beta _{i})_{i\in {\Bbb N}}\end{math}
defines an element of the Fr\'{e}chet space \begin{math}
l_{\infty }\end{math}.  Here \begin{math}
l_{\infty }\end{math}  is \begin{displaymath}
l_{\infty }=\left\{ (\alpha _{\iota })_{\iota \in
{\Bbb N}}\; \left| \; \sum_{\iota \in
{\Bbb N}}|\alpha _{\iota }|\cdot |\iota
|^{l}<\infty \; \; \mbox{for all}
\; \; l\in {\Bbb N}_{0}\right. \right\} \end{displaymath}
 Now with help of the well known test of
convergence for infinite products \cite{Co}
 it is straightforward to prove that
\begin{math} f(\lambda )=\prod_{i\in {\Bbb N}
}(\lambda -\alpha _{i})/(\lambda -\beta _{i})\end{math}  is a
meromorphic function on \begin{math} X\end{math}.
\end{Example}
\begin{Example} \label{Example2.2}
For all \begin{math} \iota \in {\cal I}\end{math}
let \begin{math} \alpha _{\iota }(\lambda
)/\beta _{\iota }(\lambda )\end{math}
be a rational function on \begin{math}
{\Bbb P}_{1}\end{math}, which has a zero at
infinity and poles only inside the image
under \begin{math} \pi \end{math}
of the excluded domain corresponding to
the index \begin{math} \iota \end{math}
of some \begin{math} U_{l,\epsilon }\end{math}
 for all \begin{math} l\in {\Bbb N}\end{math}.
Let \begin{math} \| \cdot \|_{l,\epsilon ,\iota }\end{math}
be the supremum norm on the boundary of
the domain excluded from \begin{math}
U_{l,\epsilon }\end{math}  and
corresponding to the
index \begin{math} \iota \end{math}.  Then
it is quite easy to prove that
\begin{displaymath} f(\lambda )=\sum_{\iota \in
{\cal I}} \frac{\alpha _{\iota }(\lambda )}{\beta _{\iota }(\lambda )}
\end{displaymath} is a meromorphic
function on \begin{math} X\end{math},
whenever \begin{displaymath} \sum_{\iota \in
{\cal I}} \left\| \frac{\alpha _{\iota
}(\lambda )}{\beta _{\iota }(\lambda )}
\right\| _{l,\epsilon ,\iota }<\infty \; \; \mbox{for all}
\; \; l\in {\Bbb N}\; \; \mbox{with some} \;
\; \epsilon >0\; \; \mbox{depending on} \; \;
l.\end{displaymath}  \begin{displaymath}
\mbox{This even implies
that for all \begin{math} l,k\in {\Bbb
N} \end{math}  and the same \begin{math}
\epsilon \end{math}  as before }\sum_{\iota \in
{\cal I}} \left\| \frac{\alpha _{\iota
}(\lambda )}{\beta _{\iota }(\lambda )}
\right\| _{l,\epsilon ,\iota } |\iota
|^{k}<\infty .\end{displaymath} \end{Example}
Finally we want to present the promised
counterexample of a meromorphic function,
whose inverse is not a  meromorphic
function. Let \begin{math} g(x)\end{math}
 be a smooth periodic function, with a
Taylor expansion at \begin{math} x=0\end{math}
 identically equal to zero. Then the
coefficients of the Fourier expansion \begin{math}
g(x)=\sum_{j\in {\Bbb Z}} \gamma _{j}\exp \left( 2\pi
j \sqrt{-1} x\right) \end{math}  satisfy
the relations \begin{math} \sum_{j\in {\Bbb Z}
}j^{l} \gamma _{j}=0\end{math}  for all \begin{math}
l\in {\Bbb N}\end{math}.  Now we define \begin{displaymath}
f(\lambda )=\sum_{j\in {\Bbb Z}} \frac{\gamma _{j}\lambda }{\lambda -j}
=\sum_{j\in {\Bbb Z}}\frac{\gamma _{j}}{1-j/\lambda } ,\end{displaymath}
which according to Example~\ref{Example2.2}
 is a meromorphic function on some
appropriately chosen \begin{math} X\end{math}.
But all derivatives of \begin{math} f\end{math}
 at infinity vanish: \begin{displaymath}
f^{(l)}(\lambda )=\sum_{j\in {\Bbb Z}}
\frac{(j)^{l}\gamma _{j}l!}{(1-j/\lambda )^{l+1}} .\end{displaymath}
Hence the inverse of \begin{math} f\end{math}
 is not meromorphic. On the other hand \begin{math} f\end{math}
 is not identically zero, if \begin{math}
g\end{math}  is not identically zero.
With the choice \begin{displaymath}
g(x)=\exp \left( \frac{1}{(z-1)^{2}} \right)
,\; \; z=\exp \left( 2\pi \sqrt{-1} x\right)
\end{displaymath}  we obtain a
meromorphic function \begin{math} f\neq
0\end{math},  whose inverse is not
meromorphic.

\section{The dual eigen bundle} \label{Section3}
In this section we will associate a
line bundle \begin{math} E(q)\end{math}
over the Riemann surface \begin{math}
Y\end{math} to the
potential \begin{math} q\end{math}.
This correspondence \begin{math} q\mapsto
E(q)\end{math} will
be injective, which means that
although there are many potentials,
which correspond to the same Riemann
surface \begin{math} Y\end{math},  there is only one, which
in addition corresponds to a given line bundle.
Moreover, we will be able even to reconstruct the
potential from the line bundle. This
inversion is called the inverse
scattering map or the inverse spectral
transform. Indeed, in the
case of Hill's equation for example, the line bundle
can be described in terms of the
corresponding Dirichlet or Neumann
eigenvalues; both of them essentially
define equivalent divisors corresponding to
the line bundle (for the precise
statements see e.g. \cite{MKT}).

The matrix \begin{math} h(\lambda )\end{math}
in Theorem~\ref{Theorem1.3}  combines
all eigenvectors of the Floquet matrix.
Therefore it can also be used to
describe the eigen bundle. But in case
the potential has a trivial Taylor
expansion at \begin{math} x=0\end{math},
 the offdiagonal part of \begin{math}
h\end{math}  has a trivial asymptotic
expansion at infinity. This would face
us with the problem of poles of infinite order. To
avoid this we conjugate the Lax operator
and therefore also the Floquet matrix
with the matrix \begin{displaymath}
h_{0}=\frac{1}{\sqrt{n} } \left(
\begin{array}{ccccc}
1 & 1 & . & . & 1\\
1 & s & . & . & s^{n-1}\\
. & . & . & . & .\\
. & . & . & . & .\\
1 & s^{n-1} & . & . & s^{(n-1)(n-1)}\end{array}
\right) ,\; \; \; \mbox{ \begin{math} s\end{math}
 a primitive \begin{math} n\end{math}-th
root of unity,} \end{displaymath}
\begin{displaymath} \mbox{i.e. } L \mapsto
Ad(h_{0})L,\; \; g(1,\lambda ,q)\mapsto
F(\lambda ,q)=Ad(h_{0})g(1,\lambda ,q)\; \; \mbox{and}
\; \; h(\lambda )\mapsto h_{0}h(\lambda ).\end{displaymath}
\begin{Lemma} \label{Lemma3.1} Both
equations \begin{equation}
\pi ^{*}\left( (F(\lambda ,q)\right) v=v\mu  \;
\; \mbox{and} \label{g3.1a} \end{equation}
\begin{equation} w\pi ^{*}\left( F(\lambda ,q)\right)
=\mu w\label{g3.1b} \end{equation}  have
solutions being vector valued
meromorphic functions \begin{math} v=\left(
\begin{array}{c}
v_{1}\\
\vdots \\
v_{n}\end{array} \right) ,\;
w=(w_{1},\ldots ,w_{n})\end{math}  on \begin{math}
Y. \end{math}  These solutions are unique
under the additional normalization
condition \begin{math}
v_{i}=1\end{math}  and \begin{math} w_{i}=1\end{math}
 for any fixed \begin{math} i=1,\ldots ,n\end{math}.
\end{Lemma}
Proof: From the definition of the
Riemann surface we know that \begin{math}
\det \left( \mu \unity -F(\lambda ,q)\right) =0\end{math}.
Therefore for any \begin{math} \lambda \in {\Bbb C}\end{math}
there exist solutions \begin{math} v\end{math}
 and \begin{math} w\end{math}  as above
with \begin{displaymath}
\left( \mu \unity -F(\lambda ,q)\right) v=0\; \; \mbox{and}
\; \; w\left(\mu \unity -F(\lambda ,q) \right)
=0.\end{displaymath}
 Explicitly such solutions \begin{math}v\end{math}
 and \begin{math} w\end{math}  can
be given in terms of polynomials of the
entries of \begin{math} F(\lambda ,q)\end{math}
 and \begin{math} \mu \end{math}.
Therefore at least formally we can explicitly
write down solutions of the desired
form in terms of rational functions in
the entries of \begin{math} F(\lambda ,q)\end{math}
 and \begin{math} \mu \end{math}.  If the
denominator does not vanish identically,
this gives a meromorphic solution on \begin{math}
\pi ^{-1}({\Bbb C})\end{math}.  In the proof
of Theorem~\ref{Theorem1.3} it was
shown that there exist values of \begin{math}
\lambda \end{math},  for which all eigenvalues
are distinct. This implies the uniqueness
of the solutions of the desired form.
Furthermore Theorem~\ref{Theorem1.3} gives
 a quotient of two asymptotic expansions
for the solutions. The leading term of
the denominator is given by some entry
of \begin{math} h_{0}\end{math}  and is
therefore not zero. Hence we obtain an
asymptotic expansion near infinity of
the solutions, guaranteeing the formal
solutions on \begin{math} \pi ^{-1}({\Bbb C})\end{math}
 to be meromorphic. Now Lemma~\ref{Lemma2.2}
proves the claim.
\hspace*{\fill } \begin{math} \Box \end{math}

\noindent
This lemma shows that equations (\ref{g3.1a})
and (\ref{g3.1b}) define holomorphic
maps from \begin{math} Y\end{math}  to \begin{math}
{\Bbb P}_{n-1}\end{math}  and therefore
holomorphic line bundles on \begin{math}
Y\end{math}.  The line bundles described
by \begin{math} v\end{math}  and \begin{math}
w\end{math}  are called the
eigen bundle and the transposed
eigen bundle of \begin{math} q\end{math}
and denoted by \begin{math}
E(q)\end{math}  and \begin{math}
E^{t}(q)\end{math}  respectively. It is more
convenient to work with the
corresponding dual bundles
\begin{math} E^{*}(q)\end{math}  and \begin{math}
E^{t^{*}}(q)\end{math}.  We want to
use divisors to describe these
line bundles: Set \begin{math} D(q)\end{math}
 to be the negative divisor of \begin{math}
v\end{math}  with \begin{math} v_{1}=1\end{math}
and \begin{math} D^{t}(q)\end{math}
to be the negative divisor of \begin{math}
w\end{math}  with \begin{math} w_{1}=1\end{math}.
 They are both integral divisors.
In general a divisor of a matrix valued
meromorphic function is defined by the
highest order of the poles and by the lowest order of the
zeros of all its entries respectively.
Finally let us introduce the sheaf \begin{math}
{\cal O}_{q}\end{math}  of holomorphic functions
on \begin{math} {\Bbb P}_{1}\end{math}
with values in the space of
the \begin{math} n\times n\end{math}-matrices,
which commute with \begin{math}
F(\cdot ,q)\end{math}.  Outside of the
branchpoints, each diagonalization
of \begin{math} F\end{math}  also
diagonalizes every section of \begin{math}
{\cal O}_{q}\end{math}.  Hence the eigenvalues
of sections of \begin{math} {\cal O}_{q}\end{math}
 can be considered as holomorphic
functions on \begin{math} Y\end{math},
at least outside the branchpoints,
and with the help of \cite[Theorem 8.2]{Fo}
everywhere on \begin{math} Y\end{math}.
This establishes a sheaf homomorphism \begin{math}
\varepsilon _{q}:{\cal O}_{q}\rightarrow \pi _{*}({\cal
O}_{Y}) \end{math}
 of sheaves of rings on \begin{math} {\Bbb
P}_{1}\end{math}. The map \begin{math}
\varepsilon _{q}\end{math}
is injective, because the
diagonalization of \begin{math} F\end{math}
 can be used to invert \begin{math}
\varepsilon _{q}\end{math}
 outside of the branchpoints.
\begin{Definition} \label{Definition3.1}
The potential \begin{math} q\end{math}
is called non-singular, if \begin{math}
\varepsilon _{q}\end{math}
 is an isomorphism of sheaves and
singular, if \begin{math} \varepsilon _{q}\end{math}
is not surjective. If \begin{math}
q\end{math}  is singular the image of \begin{math}
{\cal O}_{q}\end{math}  defines a singular
Riemann surface \begin{math} Y'\end{math}
 with normalization \begin{math} p:Y\rightarrow
Y'\end{math},  such that the following
diagram commutes\\
\begin{picture}(460,80)(0,0)
\put(210,55){\begin{math} Y\end{math} }
\put(280,55){\begin{math} Y'\end{math} }
\put(245,15){\begin{math} {\Bbb P}_{1}\end{math}
} \put(245,65){\begin{math} p\end{math} }
\put(215,35){\begin{math} \pi \end{math}
} \put(275,35){\begin{math} \pi '\end{math}
}\put(215,50){\vector(1,-1){25}}
\put(280,50){\vector(-1,-1){25}}
\put(230,60){\vector(1,0){40} }
\end{picture}

\noindent
and the induced map \begin{math}
\varepsilon _{q}':{\cal O}_{q}\rightarrow
\pi '_{*}({\cal O}_{Y'})\end{math}  is an
isomorphism of sheaves. Here \begin{math}
{\cal O}_{Y'}\end{math}  is the structure
sheaf of \begin{math} Y'\end{math}.
\end{Definition}
The commutative diagram  ensures that \begin{math}
Y'\end{math}  can only have covering
points over the same base point as
multiple points. This is in fact the only
possibility for singular points: a
branchpoint of the eigenvalues of \begin{math}
F\end{math},  which is removed on \begin{math}
Y\end{math}.  But we want to emphasize
that there may  exist
branchpoints of the eigenvalues of \begin{math}
F\end{math},  which also are removed on \begin{math}
Y'\end{math}.  Hence we have in general
three branching divisors \begin{math}
b_{\mbox{\it \scriptsize analytic} }\leq
b_{\mbox{\it \scriptsize effective} }\leq
b_{\mbox{\it \scriptsize algebraic} }\end{math}
on \begin{math} Y\end{math}.  Here the
analytic branching divisor is defined by
the branching order of the map \begin{math}
\pi \end{math}.  The effective branching divisor
is in some sense the branching divisor
of the singular Riemann surface \begin{math}
Y'\end{math}. It will be defined in
Section~\ref{Section9}.
Finally the algebraic branching
divisor is defined by the zeros of
\begin{math} \frac{\partial R(\lambda
,\mu )}{\partial \mu } \end{math}.
The algebraic branching divisor is equal
to the effective branching divisor
corresponding to the most singular
curve, which is defined by \begin{math}
R(\lambda ,\mu )=0\end{math}.
 From now, unless stated explicitly, we restrict
ourselves to the simplest case \begin{math}
b_{\mbox{\it \scriptsize analytic} }=
b_{\mbox{\it \scriptsize effective} }=
b_{\mbox{\it \scriptsize algebraic} }\end{math}.
The general case will be
treated in Section~\ref{Section9}.

The global sections \begin{math}
v_{1}=1,v_{2},\ldots ,v_{n}\end{math}  and
\begin{math} w_{1}=1,w_{2},\ldots ,w_{n}\end{math}
 define sheaf homomorphisms\\ \begin{math}
\phi _{v}:{\cal O}_{{\Bbb P}_{1}}^{n}\rightarrow
\pi
_{*}({\cal O}_{D(q)})\footnote{For all
divisors \begin{math} D\end{math} the
sheaf \begin{math} {\cal O}_{D}\end{math}
is defined to be the sheaf of
meromorphic functions \begin{math} f\end{math}
satisfying locally \begin{math} (f)\geq
-D\end{math}.},(f_{1},\ldots ,f_{n})\mapsto
\sum_{i=1}^{n} f_{i}v_{i}\\
\phi _{w}:{\cal O}_{{\Bbb P}_{1}}^{n}\rightarrow
\pi
_{*}({\cal O}_{D^{t}(q)}),(f_{1},\ldots ,f_{n})\mapsto
\sum_{i=1}^{n} f_{i}w_{i}\end{math}.
\begin{Theorem} \label{Theorem3.2}
\begin{math} \phi _{v}\end{math}  and \begin{math}
\phi _{w}\end{math}  are isomorphisms. In
particular \begin{math}
v_{1},\ldots ,v_{n}\end{math}  and \begin{math}
w_{1},\ldots ,w_{n}\end{math}  span
the space of global sections of \begin{math}
{\cal O}_{D(q)}\end{math}  and
\begin{math} {\cal O}_{D^{t}(q)}\end{math},
 respectively. Moreover, \begin{math}
v_{1},\ldots ,v_{n}\end{math}  and \begin{math}
w_{1},\ldots ,w_{n}\end{math}  are
uniquely defined by \begin{math} D(q)\end{math}
 and \begin{math} D^{t}(q)\end{math},
respectively and by their values at all
covering points of infinity.\end{Theorem}
Proof: The proof is given for \begin{math}
v\end{math}.  It is the same for \begin{math}
w\end{math}. Clearly the image of \begin{math} {\phi }_{v}\end{math}
is an \begin{math} {\cal O}_{q}\end{math}
 module. Hence it is also an \begin{math}
\pi _{*}({\cal O}_{Y})\end{math}  module. But \begin{math}
{\cal O}_{D(q)}\end{math}  is generated by \begin{math}
v_{1},\ldots ,v_{n}\end{math}  as an \begin{math}
{\cal O}_{Y}\end{math}  module. Therefore \begin{math}
\phi _{v}\end{math}  is surjective. At each
point \begin{math} \lambda _{0}\end{math}
with distinct eigenvalues \begin{math}
\mu _{1},\ldots ,\mu _{n}\end{math}  of \begin{math}
F(\lambda _{0},q)\end{math},  the values of \begin{math}
v\end{math}  constitute a basis of \begin{math}
{\Bbb C}^{n}\end{math}.  Hence \begin{math}
\phi _{v}\end{math}  is injective in a neighbourhood of
such a point. Then \begin{math}
\phi _{v}\end{math}  is an isomorphism. The
values of \begin{math} v\end{math}  at
all covering points of infinity are
the columns of \begin{math} h_{0}\end{math}.
There they also form a basis. This verifies the last claim.
\hspace*{\fill } \begin{math} \Box \end{math}

\noindent
The dependence on the values at infinity
suggests a modification of the notion of linear
equivalence: Modified principal
divisors are defined as divisors of meromorphic
functions, which are equal to \begin{math}
1\end{math}  at all covering points of
infinity. This also excludes
meromorphic functions with zeros of
infinite order. Geometrically this
corresponds to the identification of all
the covering points of infinity and
therefore makes \begin{math} Y\end{math}
 singular. But for reasons of simplicity
we use the notion of modified linear
equivalence. A line bundle in this
modified sense is a line bundle on
\begin{math} Y\end{math}  together with
an identification of all its fibers over
the covering points of infinity.
Theorem~\ref{Theorem3.2}
shows that \begin{math}
D(q)\end{math}  is the unique integral
divisor, which is modified equivalent to
\begin{math} D(q)\end{math},  or
equivalently, that \begin{math} E^{*}(q)\end{math}
 has a unique cross section, which
respects the identification over all
covering points of infinity. The \begin{math}
v_{i}'s\end{math}  are uniquely
determined as the quotients of
cross sections with specified values at
the covering points divided by this
unique cross section.
\newtheorem{cLemma}[Lemma]{Counting
Lemma} \begin{cLemma} \label{Lemma3.2}
In the sense of
Remark~\ref{Remark2.1}
the branching divisor \begin{math} b\end{math}
 has asymptotically and
totally the degree \begin{math}
(2,2,\ldots )\end{math}.  \end{cLemma}
We want to remind the reader that
although the number of branchpoints is
of course infinite, this has a precise
meaning: All but finitely many
branchpoints are pairwise located in
excluded domains indexed by
\begin{math} \iota  \in {\cal I}\end{math}
with the exception of finitely many
indices. Furthermore the number of
excluded branchpoints is twice the
number of excluded indices.

Proof: The branchpoints are the zeroes
of the discriminant of \begin{math}
R(\lambda ,\mu )\end{math},
 considered as a polynomial in \begin{math}
\mu \end{math}.  In \cite{PT} it is shown
that Lemma~\ref{Lemma1.4}  is true even
for \begin{math} \delta =4,l=0\end{math}  and
\begin{math} \epsilon =4/\pi \end{math}.  Hence
Lemma~\ref{Lemma1.5}  is also true for \begin{math}
\delta =2,l=0\end{math}  and \begin{math}
\epsilon =4/\pi :\end{math}  For \begin{math}
\lambda \in O_{0,\epsilon }\end{math} \begin{displaymath}
\left| \exp (p_{1}\lambda )-\exp (p_{2}\lambda )\right| \geq
\frac{1}{2} sup\left\{ \left|
\exp (p_{1}\lambda )\right| ,\left| \exp (p_{2}\lambda )\right|
\right\} .\end{displaymath}  On the
other hand for \begin{math} |\lambda |\end{math}
large enough Theorem~\ref{Theorem1.3}
 gives: \begin{displaymath}
\left| \mu _{1}-\exp (p_{1}\lambda )\right| <\frac{c}{|\lambda |}
\left| \exp (p_{1}\lambda )\right| \; \; \mbox{and}
\; \; \left| \mu _{2}-\exp (p_{2}\lambda )\right|
<\frac{c}{|\lambda |} \left| \exp (p_{2}\lambda )\right|
.\end{displaymath}  These two estimates give for \begin{math}
|\lambda |\end{math}  large enough and \begin{math}
\lambda \in O_{0,\epsilon }:\end{math}  \begin{displaymath}
\left|
(\mu _{1}-\mu _{2})^{2}-\left(
\exp (p_{1}\lambda )-\exp (p_{2}\lambda )\right) ^{2}\right|
<\left| \exp (p_{1}\lambda )-\exp (p_{2}\lambda )\right|
^{2}.\end{displaymath}  The same is of
course true for \begin{math}
\mu _{i},\mu _{j},\; \; i\neq j\in \{
1,\ldots ,n\} \end{math}.  Now Rouch\'{e}'s
Theorem \cite{Co}  proves the claim.
\hspace*{\fill } \begin{math} \Box \end{math}

\noindent
Any two solutions of (\ref{g3.1a}) and
(\ref{g3.1b}) can be used to define a
matrix valued meromorphic function on the
Riemann surface \begin{math} Y:\; \; P=v(w\cdot
v)^{-1}w
\end{math}.  If \begin{math} f\end{math}
 and \begin{math} g\end{math}  are
meromorphic functions on \begin{math} Y\end{math},
 and \begin{math} \tilde{v} =vf\end{math}
 and \begin{math} \tilde{w} =gw\end{math},
 then \begin{math} \tilde{v} (\tilde{w}
\cdot \tilde{v} )^{-1}\tilde{w}
=v(w\cdot v)^{-1}w \end{math}.
 Therefore \begin{math} P\end{math}
 does not depend on the special choice of \begin{math}
v\end{math}  and \begin{math} w\end{math}.
In the definition of \begin{math} P\end{math}
 we may in particular assume that locally
\begin{math} v\end{math}  and \begin{math}
w\end{math}  have no poles and zeros.
This function \begin{math} P\end{math}
has some nice properties:
\begin{Lemma} \label{Lemma3.3}
\begin{description}
\item[(i)] \begin{math} P^{2}=P\end{math}
\item[(ii)] \begin{math} P\pi
^{*}(F)=\pi ^{*}(F)P=\mu P\end{math}
\item[(iii)] The sum over the sheets of \begin{math}
P\end{math}  is equal to \begin{math}
\unity \end{math}.
\item[(iv)] The divisor of \begin{math}
P\end{math}  is \begin{math} -b\end{math}.
\end{description} \end{Lemma}
Proof: (i) and (ii) directly follow from
the definition. Now let \begin{math}
v^{1},\ldots ,v^{n}\end{math}  and \begin{math}
w^{1},\ldots ,w^{n}\end{math}  be the values
of \begin{math} v\end{math}  and \begin{math}
w\end{math},  respectively at all the
covering points of some \begin{math}
\lambda _{0}\end{math}  not being a
branchpoint: set \begin{math}
\mu _{1},\ldots ,\mu _{n}\end{math}  for the
distinct eigenvalues of \begin{math}
F(\lambda _{0},q)\end{math}.  The following
relations ensure the \begin{math}
v's\end{math}  and the \begin{math}
w's\end{math}  to form two bases of \begin{math}
{\Bbb C}^{n}:Fv^{i}=v^{i}\mu _{i},\; \;
w^{i}F=\mu _{i}w^{i}\end{math}.  This
implies \begin{math} w^{i}v^{j}=0\end{math}
 if \begin{math} i\neq j\end{math},  and
the \begin{math} v's\end{math}  and
the \begin{math} w's\end{math}  are
therefore up to a factor dual bases of \begin{math} {\Bbb C}
^{n}\end{math}.  Hence we have \begin{math}
\sum_{i=1}^{n}
v^{i}(w^{i}v^{i})^{-1}w^{i}=\unity \end{math},
 which proves (iii). The negative
divisor of \begin{math} P\end{math}
must be integral, because \begin{math} P\end{math}
 only can have poles. In fact we
may assume that \begin{math} v\end{math}
 and \begin{math} w\end{math}  have
neither poles nor zeroes, and only the
denominator \begin{math} wv\end{math}
can vanish. The proof of
(iii) shows that the poles of \begin{math}
P\end{math}  are exactly the branchpoints
and even more precisely, that the divisor of
\begin{math} P\end{math}  is \begin{math}
-b\end{math}. Indeed considering
the proof of (iii) we know that for \begin{math}
y,y'\end{math}  in a small
neighbourhood of some branchpoint \begin{math}
y_{0}\end{math},  the number of zeros of
the function \begin{math} w(y)v(y')\end{math},
 if we fix either \begin{math} y\end{math}
 or \begin{math} y'\end{math},  is equal
to the branching order. This proves (iv).
\hspace*{\fill } \begin{math} \Box \end{math}
\begin{Theorem}\label{Theorem3.1}
The divisors \begin{math}
D(q)\end{math}  and \begin{math}
D^{t}(q)\end{math}  have asymptotically
and totally degree \begin{math}
(1,1,\ldots )\end{math}  in the sense of
Remark~\ref{Remark2.1}. Moreover, if
\begin{math} v\end{math}  and \begin{math}
w\end{math}  are solutions of (\ref{g3.1a})
and (\ref{g3.1b}) with \begin{math}
v_{1}=1=w_{1}\end{math},  the following
equation for divisors holds:
\begin{equation} D(q)+D^{t}(q)+(w\cdot
v)=b.\label{g3.2} \end{equation}
Finally, the function \begin{math} wv\end{math}
induces an equivalence relation in the
modified sense. \end{Theorem}
Proof: The divisor of \begin{math}
v\cdot w\end{math}  is equal to \begin{math}
-D(q)-D^{t}(q)\end{math}.  Hence
equation (\ref{g3.2}) is a direct
consequence of (iv) from
Lemma~\ref{Lemma3.3}. At each covering
point of infinity, \begin{math} w\cdot v\end{math}
 is equal to \begin{math} n\end{math}.
This proves the last statement.
Now we claim that \begin{math} D(q)\end{math}
 and \begin{math} D^{t}(q)\end{math}
have asymptotically and totally the same
degree. With this claim  the first statement of the
theorem follows from Corollary~\ref{Corollary2.1}
 and the counting
Lemma\footnote{Implicit in
Section~\ref{Section4} we will give
another proof. In fact, the proof of
the implication (i)\begin{math}
\Rightarrow \end{math} (ii) of Theorem~\ref{Theorem4.3}
implies the first statement.}. In order to
prove the claim we need another lemma.
By \begin{math} \pi \end{math}  we also
denote the induced homomorphism from
the group of divisors of \begin{math} Y\end{math}
into the group of divisors of \begin{math}
X\end{math}.
\begin{Lemma} \label{Lemma3.5}
Let \begin{math} g\in GL(n,{\cal M}_{X})\end{math}
 be the unique solution of \begin{math}
gv=w^{t}\end{math}.  Then one has the following
relation between divisors: \begin{displaymath} \left(
\det (g)\right) =\pi \left( D(q)\right)
-\pi \left( D^{t}(q)\right) .\end{displaymath}
\end{Lemma}
Proof: Due to Theorem~\ref{Theorem3.2}
the support of the divisor of \begin{math}
\det (g)\end{math}  is contained in the
image of the union of the support of \begin{math}
D(q)\end{math}  and \begin{math}
D^{t}(q)\end{math}  under \begin{math}
\pi \end{math}.  For every meromorphic
function \begin{math} f\end{math}  on \begin{math}
Y\end{math},  there exists a unique \begin{math}
n\times n\end{math}-matrix valued
function \begin{math} g_{f}\end{math}
on \begin{math} X\end{math},
 such that \begin{math} g_{f}v=vf\end{math}
 and \begin{math} g_{f}w^{t}=w^{t}f\end{math},  and
the determinant of \begin{math} g_{f}\end{math}
 is equal to the \begin{math}
n\end{math}-th elementary symmetric
function of \begin{math} f\end{math}
with respect to the covering map \begin{math}
\pi \end{math}. Theorem~\ref{Theorem3.2}
 even shows that the supports of the
divisors of \begin{math} \det (g)\det (g_{f})\end{math}
 and \begin{math} \det ^{-1}(g_{f})\det (g)\end{math}
 are contained in the image
of the union of the support of \begin{math}
(f)+D(q)\end{math}  and \begin{math}
D^{t}(q)\end{math}  and the union of the
support of \begin{math} D(q)\end{math}
and \begin{math} (f)+D^{t}(q)\end{math}
respectively. This proves the claim.
\hspace*{\fill } \begin{math} \Box \end{math}

\noindent
Completion of the proof of
Theorem~\ref{Theorem3.1}:
The function \begin{math} g\end{math}
of the last lemma is a meromorphic
function on \begin{math} X\end{math},
which is equal to \begin{math}
h_{0}^{-1}h_{0}^{-1^{t}}\end{math}
 at infinity. Hence there exists a
neighbourhood of infinity, such that \begin{math}
|\det (g)-1|<1\end{math}  in this
neighbourhood. Now Rouch\'{e}'s Theorem \cite{Co}
 implies that \begin{math} D(q)\end{math}
 and \begin{math} D^{t}(q)\end{math}
have asymptotically and totally the same
degree. Indeed, if there exist disjoint excluded
domains, whose images under \begin{math}
\pi \end{math}  have non empty
intersection, the \begin{math} n\end{math}-fold
covering \begin{math} \pi \end{math}
decomposes near these excluded domains
into coverings containing actual only one
excluded domain. The application of the
last lemma to each of these coverings
proves the claim even in this case.
\hspace*{\fill } \begin{math} \Box \end{math}

\noindent
Due to equation (\ref{g3.2}) the
equivalence class of one of the divisors
\begin{math} D(q)\end{math}  and \begin{math}
D^{t}(q)\end{math}  determines the other
equivalence class. In the remainder of this
section we present a way to reconstruct
the potential \begin{math} q\end{math}
from the divisor \begin{math} D(q)\end{math}.
 This will be done in three steps:
\begin{description}
\item[step 1:] Theorem~\ref{Theorem3.2}
shows that the sections \begin{math}
v_{1}=1,v_{2},\ldots ,v_{n}\end{math}  of the
sheaf \begin{math} {\cal O}_{D(q)}\end{math}
are uniquely determined by \begin{math}
D(q)\end{math}.
\item[step 2:] Theorem~\ref{Theorem1.2}
and Theorem~\ref{Theorem1.3}  enable us
to obtain the complete Taylor expansion
of \begin{math} q(x)\end{math}  at the
point \begin{math} x=0\end{math}  from
the asymptotic expansion of \begin{math}
v\end{math}.
\item[step 3:] For each \begin{math}
x\in {\Bbb R}/{\Bbb Z}\end{math},  let \begin{math}
T_{x}\end{math}  be the shift by \begin{math}
x\end{math}  on the space of potentials:
\begin{math} (T_{x}q)(y)=q(x+y)\end{math}.
The Riemann surface \begin{math} Y\end{math}
 will not change under the shift and
the line bundle \begin{math}
E(T_{x}q)\end{math} will be described
in terms of the
line bundle \begin{math} E(q)\end{math}.
\end{description}
The first step has already been carried
out.

In the second step we have to restrict
the space of potentials:
\begin{Assumption} \label{Assumption1}
The diagonal part of the potential
is equal to a fixed constant. \end{Assumption}
This fits with the group
theoretical treatment. In fact, with this
assumption the space of potentials forms
a coadjoint orbit and therefore possesses
a natural symplectic structure. This
observation and
a corresponding hamiltonian formulation
is well known and can be found in e.g.\cite{FT},
\cite{Sch}.

Equation (\ref{g1.7a}) gives a formula
for \begin{math} q(0)\end{math} in terms
of the asymptotic expansion of \begin{math}
v\end{math}.  One may obtain similar
formulas for the higher order
coefficients of the Taylor expansion of \begin{math}
q\end{math}  with the help of (\ref{g1.7b}).
Some details may be found in
Appendix~\ref{Appendixa}.

Finally we turn to the third step.
First we have to ensure that the shift \begin{math}
T_{x}\end{math}  does not change the
Riemann surface. For this purpose let \begin{math}
g(x)\end{math}  be the fundamental
solution of the differential equation \begin{math}
\left( \frac{d}{dx} +a(x)\right)
g(x)=0,\; \; g(0)=\unity \end{math}  with
some periodic \begin{math} a(x)\end{math}.
Then the Floquet matrix of \begin{math}
T_{x}a\end{math}  is given by \begin{math}
g(x+1)g^{-1}(x)=Ad(g(x))g(1)\end{math}.
Hence the eigenvalues of the Floquet
matrices of all the \begin{math} T_{x}a\end{math}
 coincide. In particular to all the
potentials \begin{math} T_{x}q\end{math}
 there corresponds the same Riemann
surface. The main tool of this step is
given by Floquet theory \cite{Fr}:

If the Floquet matrix \begin{math} g(1)\end{math}
 of the above differential equation has
a logarithm: \begin{math} g(1)=\exp (\alpha
)\end{math},
 then the gauge transformation with the
periodic differentiable matrix \\ \begin{math}
h(x)=g(x)\exp (-x\alpha )\end{math}  transforms
\begin{math} a(x)\end{math}  into the
constant \begin{math} \alpha \end{math}.  For
the application of this observation in
the present context we need a lemma.
\begin{Lemma} \label{Lemma3.4} On some
neighbourhood of all covering points of
infinity, \begin{math} \ln (\mu )\end{math}
is a meromorphic function.\end{Lemma}
Proof: For \begin{math} \epsilon \end{math}
small enough, the second estimate of
Theorem~\ref{Theorem1.3}  implies an
asymptotic expansion uniformly on \begin{math}
U_{M,\epsilon }:\end{math}  \begin{math} \left|
\ln (\mu _{i})-p_{i,M}(\lambda )\right|
<c|\lambda |^{-M-1}\end{math}  with some
constant depending on \begin{math} M\end{math}.
 Of course only one branch of \begin{math}
\ln (\mu )\end{math}  satisfies this
estimate. Together with Lemma~\ref{Lemma2.2}
 this guarantees the existence of the
meromorphic function \begin{math}
\ln (\mu )\end{math}.
\hspace*{\fill } \begin{math} \Box \end{math}

\noindent
Now an easy application of  Floquet theory
provides solutions of the equations (\ref{g3.1a})
and (\ref{g3.1b}) for the
potential \begin{math} T_{x}q\end{math} of
the form: \begin{displaymath} v(x,q)=\pi ^{*}\left(
h_{0}g(x,\cdot ,q)h_{0}^{-1}\right) v(q)\mu
^{-x}\end{displaymath}
\begin{displaymath} w(x,q)=\mu ^{x}w(q)
\pi ^{*}\left( h_{0}g^{-1}(x,\cdot ,q)h_{0}^{-1}\right)
.\end{displaymath}
Here \begin{math}
\mu ^{x}\end{math}  is an
abbreviation of \begin{math} \exp \left(
x\cdot \ln (\mu )\right) \end{math}.
Now we claim that for all \begin{math}
x\in {\Bbb R}\end{math} these
solutions are holomorphic vector valued
functions on the neighbourhood described
in the foregoing lemma, which take the same
values as \begin{math} v(q)\end{math}
and \begin{math} w(q)\end{math},
respectively, at all
covering points of infinity.
Let \begin{math} v(T_{x}q)\end{math} and
\begin{math} w(T_{x}q)\end{math} be the
corresponding normalized solutions:
\begin{displaymath}
v(T_{x}q)=\frac{v(x,q)}{v_{1}(x,q)} \end{displaymath}
\begin{displaymath} w(T_{x},q)=\frac{w(x,q)}{w_{1}(x,q)}
.\end{displaymath} Then the functions \begin{math}
v_{1}(x,q)\end{math} and \begin{math}
w_{1}(x,q)\end{math} are the unique solutions of
the differential equations \begin{displaymath}
\frac{\partial v_{1}(x,q)}{\partial x} =-\left( \left( \pi
^{*}(h_{0}(p\lambda +q(x))h_{0}^{-1})v(T_{x}q)\right)
_{1}+\ln (\mu )\right) v_{1}(x,q),\
\ v_{1}(0,q)=1 \mbox{ and } \end{displaymath}
\begin{displaymath} \frac{\partial w_{1}(x,q)}{\partial x}
=\left( \left( w(T_{x}q)\pi
^{*}(h_{0}(p\lambda +q(x))h_{0}^{-1})\right) _{1}
+\ln (\mu )\right) w_{1}(x,q),\
\ w_{1}(0,q)=1,\mbox{ respectively} .\end{displaymath}
These solutions are given by \begin{displaymath}
v_{1}(x,q)=\exp -\left( \int_{0}^{x} \left( \left( \pi
^{*}(h_{0}(p\lambda +q(t))h_{0}^{-1})v(T_{t}q)\right)
_{1}+\ln (\mu )\right) dt \right) \mbox{ and }
\end{displaymath} \begin{displaymath}
w_{1}(x,q)=\exp \left( \int_{0}^{x}
\left( \left( w(T_{t}q)\pi
^{*}(h_{0}(p\lambda +q(t))h_{0}^{-1})\right)
_{1}+\ln (\mu )\right) dt\right) ,\mbox{respectively}
.\end{displaymath} Now the asymptotic
expansion of
Theorem~\ref{Theorem1.3}, whose
coefficients are calculated in
Theorem~\ref{Theorem1.2}, shows that
both functions under the integrals of are
holomorphic functions on the
neighbourhood described in the previous lemma.
Furthermore, due to Assumption~\ref{Assumption1}
they vanish at all covering points of
infinity. This proves the claim.

The definition of these solutions of the
equations (\ref{g3.1a}) and (\ref{g3.1b})
for the potential \begin{math} T_{x}q\end{math}
may be transformed to \begin{displaymath}
\pi ^{*}(h_{0}g(x,\cdot
,q)h_{0}^{-1})v(q)=v(x,q)\mu ^{x}\mbox{and }
\end{displaymath} \begin{displaymath}
w(x,q)\pi ^{*}(h_{0}g(x,\cdot
,q)h_{0}^{-1})\mu ^{x}w(q),\mbox{ respectively}
.\end{displaymath}
It is well known that \begin{math}
\det \left( g(x,\lambda ,q)\right) =\exp \left(
\int_{0}^{x}\mbox{tr} \left( p\lambda +q(t)\right) dt\right)
\end{math}.  Hence the map \begin{math}
\lambda \rightarrow g(x,\cdot ,q)\end{math}
defines an entire
function from \begin{math} {\Bbb C}\end{math}
to \begin{math} GL(n,{\Bbb C})\end{math}.
Thus the left hand sides define
line bundles on \begin{math} \pi ^{-1}({\Bbb C}
)\end{math}  isomorphic to the
restriction of \begin{math}
E(q)\end{math}  and \begin{math}
E^{t}(T_{x}q)\end{math}, respectively.
Due to Lemma~\ref{Lemma3.4}, the
function \begin{math} \mu ^{x}\end{math}
is the cocycle of a holomorphic line bundle over \begin{math}
Y\end{math}  with respect
to the covering \begin{math} \pi
^{-1}({\Bbb C})\end{math}  and the
neighbourhood given by
Lemma~\ref{Lemma3.4}.
Let us denote this
line bundle by \begin{math} L(x)\end{math}.
Then the foregoing claim implies
that \begin{displaymath}
E(q)\simeq E(T_{x}q)\otimes L(x),
\end{displaymath} \begin{displaymath}
E^{t}(T_{x}q)\simeq E^{t}(q)\otimes
L(x).\end{displaymath}
After Theorem~\ref{Theorem3.2}  we
introduced a modified notion for
line bundles. In order to finish the
reconstruction of the potential out of
the line bundle in the modified sense, we
remark that the claim proves these isomorphisms
even to be isomorphisms in the
modified sense. Since obviously \begin{math}
L^{*}(x)\simeq L(-x)\end{math},  this proves the
\begin{Theorem} \label{Theorem3.3}
There are two isomorphisms of
line bundles in the modified sense for
all \begin{math} x\in {\Bbb R}:\end{math} \begin{displaymath}
E^{*}(T_{x}q)\simeq E^{*}(q)\otimes L(x),\end{displaymath}
\begin{displaymath}
E^{t^{*}}(T_{x}q)\simeq
E^{t^{*}}(q)\otimes L(-x).\end{displaymath}
\end{Theorem}
\hspace*{\fill } \begin{math} \Box \end{math}

\noindent
For all \begin{math} x\in {\Bbb Z},\; \; L(x)\end{math}
 is trivial, because \begin{math} \mu \end{math}
 and \begin{math} 1/\mu \end{math}  are holomorphic
functions on \begin{math} \pi ^{-1}({\Bbb C})\end{math}.
 The line bundles on the right hand sides are
well defined for all \begin{math} x\in
{\Bbb C}\end{math},  but the left hand
sides in general make no sense for \begin{math}
x\not\in {\Bbb R}\end{math}.  Only for
analytic potentials the
line bundles on the left hand
side are defined for some \begin{math}
x\not\in {\Bbb R}\end{math}.  This
observation provides an
idea of the largeness even of the
components of the Picard group
\begin{math} H^{1}\left( Y,{\cal O}^{*}_{Y}\right)
\end{math}.
 From Remark~\ref{Remark2.1} we
recall that the Picard group has
as many components, as there
are classes of asymptotically and totally
equivalent integer valued
sequences indexed by \begin{math} \iota \in
{\cal I}\end{math}.  Now we are only interested in
a small part of one component, namely
the one corresponding to the sequence \begin{math}
(1,1,\ldots )\end{math}.  The next section
will provide more details.

\section{The Riemann-Roch Theorem} \label{Section4}
Our next goal is to classify the
divisors, which correspond to some
potential \begin{math} q\end{math}.  For
this purpose we need as the main tool
an appropriate version of the
Riemann-Roch Theorem adapted to the present
situation. This
is of interest in its own right and the
only subject of this section.

We start with a discussion, which will
serve as a motivation.
On a compact Riemann surface \begin{math}
Y\end{math}  the Riemann-Roch Theorem
(see e.g. \cite{Fo})
is given by the relation \begin{displaymath}
\dim H^{0}(Y,{\cal O}_{D})-\dim
H^{1}(Y,{\cal O}_{D})=1-g+\deg (D),\end{displaymath}
where \begin{math} g\end{math}  is the
genus of \begin{math} Y\end{math},  and \begin{math}
D\end{math}  is any divisor on \begin{math}
Y\end{math}.  If \begin{math} Y\end{math}
is an \begin{math} n\end{math}-fold
covering of the Riemann sphere, with
branching divisor \begin{math} b\end{math},
 the genus can be calculated with the
help of the Riemann-Hurwitz formula
(see e.g. \cite{Fo}): \begin{displaymath} g=\frac{\deg (b)}{2}
-n+1.\end{displaymath} In our case \begin{math}
\deg (b)\end{math}  is infinite, and the
meaning of the Riemann-Roch Theorem is not clear.
But if there is a finite interpretation
of the expression \begin{math}
\deg (D)-\deg (b)/2\end{math}, then the
Riemann-Roch Theorem still makes sense.
Now Theorem~\ref{Theorem3.1}
shows that the divisors, we are
interested in, are exactly of this kind.
Also in Theorem~\ref{Theorem3.2} the
dimension of \begin{math} H^{0}\left(
Y,{\cal O}_{D(q)}\right) \end{math}  is
calculated to be equal to \begin{math}
n\end{math}.  If we insert this into the
Riemann-Roch Theorem, we would obtain \begin{math}
\dim H^{1}(Y,{\cal O}_{D(q)})
=0\end{math}. At this point the
first difficulty arises: Equation
(\ref{g3.2}) shows that there exists an
exact sequence of
homomorphisms of sheaves \begin{displaymath}
0\rightarrow {\cal O}_{D(q)}\rightarrow {\cal O}_{b}\rightarrow
{\cal O}_{D^{t}(q)}/{\cal O}_{Y}\rightarrow 0.\end{displaymath}
If \begin{math} \dim
H^{1}(Y,{\cal O}_{D(q)})\end{math}
were to be equal to zero, the exact cohomology
sequence of the short foregoing exact
sequence would imply the short exact
sequence:\begin{displaymath}
0\rightarrow H^{0}(Y,{\cal O}_{D(q)})\rightarrow
H^{0}(Y,{\cal O}_{b})\rightarrow H^{0}\left(
Y,{\cal O}_{D^{t}(q)}/{\cal O}_{Y}\right) \rightarrow
0.\end{displaymath}  This means that to
any sequence of complex numbers indexed
by the points of the divisor \begin{math} D^{t}(q)\end{math},
 there exists a form which is almost holomorphic
and takes these values at the
index points. (i.e. only with poles of order at most two at
all covering points of some element of \begin{math}
{\Bbb P}_{1}\end{math})
This can't be true (see
Theorem~\ref{Theorem8.3}). The
contradiction is related to the conclusion of the
last section. Indeed, as we explained,
we are only
interested in an `admissible' part of one
component of the Picard group.
Therefore we have to
restrict the first cohomology groups to
some `admissible' parts. One way to avoid
this difficulty is to rephrase the
classical formulation of the Riemann-Roch
Theorem: \begin{displaymath} \dim
H^{0}(Y,{\cal O}_{D})-\dim
H^{0}(Y,\Omega _{-D})
=1-g+\deg (D),\end{displaymath}  which due to Serre duality (see
e.g. \cite{Fo})
is equivalent to the up-to-date version. There
is an alternative approach, which in some sense
compactifies the Riemann surface \begin{math}
Y\end{math}:  We use the inverse image
topology of the covering map to
calculate the sheaf cohomology.
For any sheaf \begin{math} {\cal F}\end{math}
on \begin{math} Y\end{math} define
\begin{math}
\tilde{H} ^{i}(Y,{\cal F})= H^{i}({\Bbb
P}_{1},\pi _{*}({\cal F}))\end{math}. Clearly
\begin{math} \tilde{H} ^{0}(Y,{\cal
F})=H^{0}(Y,{\cal F})\end{math}.
The behaviour of \v{C}ech cohomology under
refinement (see e.g. \cite{Fo})  shows that \begin{math}
\tilde{H} ^{1}(Y,{\cal F})\end{math}  is
in fact a subgroup of \begin{math}
H^{1}(Y,{\cal F})\end{math}.
\begin{Definition} \label{Definition4.1}
A divisor \begin{math}
D\end{math}  on \begin{math} Y\end{math}
is called of Riemann-Roch
type, if
\begin{description}
\item[(i)] the support of \begin{math} D\end{math}
 is contained in \begin{math} \pi
^{-1}({\Bbb C})\end{math}.
\item[(ii)] for some neighbourhood \begin{math}
U\end{math}  of infinity in \begin{math}
{\Bbb P}_{1}\end{math},  there exist
cross sections \begin{math}
u_{1},\ldots ,u_{n}\end{math}  of \begin{math}
{\cal O}_{D}\end{math}  on \begin{math} \pi
^{-1}(U)\end{math},  such that the map\\
\begin{math} \phi _{u}:{\cal O}_{{\Bbb P}_{1}}^{n}\rightarrow \pi
_{*}({\cal O}_{D}),\; \; (f_{1},\ldots ,f_{n})\mapsto
\sum_{i=1}^{n} f_{i}u_{i}\end{math}  is
an isomorphism of sheaves on \begin{math}
U\end{math}. \end{description} \end{Definition}
The restriction of (i) may be dropped,
but then we have to pay attention to the
possibility of zeros of infinite order.
On the other hand the condition (i) fits with
the modification of linear equivalence
in the last section. From this definition
and Lemma~\ref{Lemma2.1}  it is quite
obvious that for each
divisor of Riemann-Roch
type, \begin{math} \pi _{*}({\cal O}_{D})\end{math}
 is isomorphic to the sheaf of sections
of a holomorphic vector bundle on \begin{math}
{\Bbb P}_{1}\end{math}.  In view of
the Riemann-Roch Theorem for holomorphic
vector bundles this suggests
the following:
\newtheorem{RTheorem}[Lemma]{Riemann-Roch Theorem}
\begin{RTheorem} Each divisor of Riemann-Roch
type \begin{math} D\end{math}  has asymptotic
degree \begin{math} (1,1,\ldots )\end{math}
in the sense of
Remark~\ref{Remark2.1}. Hence \begin{math}
\deg (D)-\deg (b)/2\end{math}  is a well
defined integer. Moreover both modified
cohomology groups are finite dimensional
and the following formula is valid:
\begin{displaymath} \dim \tilde{H}
^{0}(Y,{\cal O}_{D})-\dim
\tilde{H} ^{1}(Y,{\cal O}_{D})
=n+\deg (D)-\frac{\deg (b)}{2} .\end{displaymath}
\end{RTheorem}
Proof: In condition (ii) we can
impose that \begin{math} u_{i}\end{math}
 has the same value as \begin{math} v_{i}\end{math}
 of Lemma~\ref{Lemma3.1}, with \begin{math}
v_{1}=1\end{math}, at all covering points of
infinity.  The same argument as
used in the proof of Theorem~\ref{Theorem3.1}
 shows that the divisors \begin{math} D\end{math}
and \begin{math} D(q)\end{math}  have
asymptotically the same degree. If in
addition
the \begin{math} u_{i}\end{math}  are
even meromorphic on the whole of \begin{math}
Y\end{math},  so that the total
degrees of the divisors of the
vector valued function \begin{math}
(u_{1},\ldots ,u_{n})\end{math}  and of \begin{math}
D(q)\end{math}  coincide. But \begin{math}
\pi _{*}({\cal O}_{D})\end{math}  defines a
holomorphic vector bundle on \begin{math}
{\Bbb P}_{1}\end{math},  which always has \begin{math}
n\end{math}  independent meromorphic
sections \cite{Gu}. Hence we can even
assume that the \begin{math} u_{i} \end{math}
are meromorphic on \begin{math} {\Bbb
P}_{1}\end{math}.  Now the claim follows
from the Riemann-Roch Theorem for
holomorphic vector bundles \cite{Gu}.
\hspace*{\fill } \begin{math} \Box \end{math}

\noindent
 From the proof of Lemma~\ref{Lemma3.3}
we conclude that \begin{math} \pi
_{*}({\cal O}_{D^{t}(q)})\end{math}  is
natural isomorphic to the sheaf of
sections of the vector bundle dual to the
vector bundle associated
to \begin{math} \pi
_{*}({\cal O}_{D(q)})\end{math}.  More
generally, for each
divisor \begin{math} D\end{math}  of Riemann-Roch
type \begin{math} b-D\end{math}
 is also of Riemann-Roch
type and the associated vector bundles
are dual to each other. On the other
hand the
sheaf of holomorphic forms on \begin{math}
Y\end{math}  is isomorphic to the
tensor product of \begin{math} {\cal O}_{b}\end{math}
 with the pullback of the sheaf of
holomorphic forms on \begin{math} {\Bbb
P}_{1}\end{math}.  Now the Serre duality for
holomorphic vector bundles on \begin{math}
{\Bbb P}_{1}\end{math}  \cite{Gu}  implies
Serre duality for divisors of Riemann-Roch
type on \begin{math}
Y:\end{math}
\newtheorem{STheorem}[Lemma]{Serre
duality Theorem}
\begin{STheorem} For each
divisor of Riemann-Roch
type \begin{math} D\; \; \Omega _{-D}\end{math}
 corresponds also to a
divisor of Riemann-Roch
type \begin{math} D'\sim b-D-2\pi
^{-1}(\lambda )\end{math}  with some \begin{math}
\lambda \in {\Bbb P}_{1}\end{math}.  Moreover there is
a natural non-degenerate pairing between
\begin{math} \tilde{H} ^{0}(Y,{\cal O}_{D})\end{math}
 and \begin{math} \tilde{H}
^{1}(Y,\Omega _{-D})\end{math}. \end{STheorem}
\hspace*{\fill } \begin{math} \Box \end{math}

\noindent
The definition of divisors of Riemann-Roch
type essentially postpones the main
problem, namely to find sufficiently
many divisors of Riemann-Roch type.
In the rest of this section we try to
characterize some divisors of Riemann-Roch
type through their location. In particular we
are looking for sufficient geometric
conditions which in addition are easy to handle.

Let us choose an arbitrary \begin{math}
U_{l,\epsilon }\end{math}.
For each \begin{math} \iota \in {\cal I}\end{math}
we define \begin{math} \pi :Y_{l,\epsilon ,\iota }\mapsto
{\Bbb P}_{1} \end{math}
to be the \begin{math} n\end{math}-fold
covering, which is obtained from \begin{math} Y\end{math}
after removing all branching points
outside of the domain excluded from \begin{math}
U_{l,\epsilon }\end{math}  with
index \begin{math} \iota \end{math}. Let \begin{math}
D_{l,\epsilon ,\iota }\end{math}  be the divisor
of \begin{math} Y_{l,\epsilon ,\iota }\end{math},  whose
support is contained in the excluded
domain with index \begin{math} \iota \end{math},
and whose restriction to this excluded
domain coincides with the restriction of
\begin{math} D\end{math}  to this
excluded domain. For large \begin{math}
|\iota |\end{math} this definition does
not depend on the choice of \begin{math}
U_{l,\epsilon }\end{math}.  If we are only
interested in such \begin{math} \iota \end{math},
 we will sometimes omit the indices \begin{math}
l\end{math}  and \begin{math} \epsilon \end{math}.
\begin{Definition} \label{Definition4.2}
A divisor \begin{math} D\end{math}  of \begin{math}
Y\end{math}  is called admissible, if
\begin{description}
\item[(i)] the support of \begin{math} D\end{math}
 is contained in \begin{math} \pi
^{-1}({\Bbb C})\end{math}.
\item[(ii)] the direct image sheaf \begin{math} \pi
_{*}({\cal O}_{D_{\iota }}) \end{math}
 is isomorphic to \begin{math}
{\cal O}_{{\Bbb P}_{1}}^{n}\end{math}
for all \begin{math} \iota \in
{\cal I}\end{math},  which correspond to
excluded domains in the inverse image of
some neighbourhood of infinity under the
covering map \begin{math} \pi\end{math}.
\end{description}
\end{Definition}
The Riemann-Roch Theorem and the Riemann-Hurwitz
formula together ensure \begin{math}
\deg (D_{\iota })\end{math}  to be equal to
half the branching order of \begin{math}
Y_{\iota }\end{math}.  Hence all admissible
divisors have asymptotic degree \begin{math}
(1,1,\ldots )\end{math}.  Let
\begin{math} D\end{math} be any admissible
divisor. For all sufficiently large
\begin{math} \iota \end{math}  there exist \begin{math}
n\end{math}  unique cross sections \begin{math}
v_{1,\iota },\ldots ,v_{n,\iota }\end{math}  of \begin{math}
{\cal O}_{D_{l,\epsilon ,\iota }}\end{math},
which take the same
values as \begin{math} v_{1},\ldots
,v_{n}\end{math}, with the normalization \begin{math}
v_{1}=1\end{math}, at the
covering points of infinity.  For all \begin{math}
\lambda \end{math} outside the
domain excluded from \begin{math}
U_{l,\epsilon }\end{math}  with index \begin{math}
\iota \end{math}  we define the \begin{math}
n\times n\end{math}-matrix \begin{math}
h_{l,\epsilon ,\iota }(D,\lambda ) \end{math}  as \begin{math}
(h_{l,\epsilon ,\iota }(D,\lambda ))_{i,j}=\end{math}  the
value of \begin{math} v_{i,\iota }\end{math}
at the covering point in the \begin{math}
j\end{math}-th sheet over \begin{math}
\lambda \end{math}.  Therefore at infinity \begin{math}
h_{l,\epsilon ,\iota }(D,\lambda )\end{math}  is equal to \begin{math}
h_{0}\end{math}.  Now we can state the
main theorem of this section.
\begin{Theorem} \label{Theorem4.3}
The following statements are equivalent:
\begin{description}
\item[(i)] The divisor \begin{math} D\end{math}
 is of Riemann-Roch type.
\item[(ii)] The divisor \begin{math} D\end{math}
is admissible and
\begin{math} \sum_{\iota } \left\|
h_{l,\epsilon ,\iota }(D,\cdot )-h_{0}\right\|
_{l,\epsilon ,\iota }<\infty \end{math} for all \begin{math}
l\in {\Bbb N}\end{math}  with some \begin{math}
\epsilon \end{math}  depending on \begin{math}
l\end{math}.
\end{description} \end{Theorem}
We defined the norm \begin{math} \| \cdot \|
_{l,\epsilon ,\iota }\end{math}  in
Example~\ref{Example2.2} as the
supremum norm on the boundary of the
domain excluded from \begin{math}
U_{l,\epsilon }\end{math}  with index \begin{math}
\iota \end{math}.  The proof\footnote{The proof of this
theorem may be passed over to the first
paragraph in front of equation
(\ref{g4.5}).} of this
theorem is divided into two steps
concerning the two implications. In each
step we will first state some lemmata,
which are needed in the following main
part of the actual step. Let us begin with
the implication (i) \begin{math}
\Rightarrow \end{math} (ii).
\begin{Lemma} \label{Lemma4.1a}
Let \begin{math} D(\lambda _{0},R)\end{math}
be the circle \begin{math} \left\{ \lambda \in
{\Bbb C}\left| \; |\lambda -\lambda _{0}|=R\right. \right\}
\end{math}  and let \\ \begin{math}
g:D(\lambda _{0},R)\rightarrow GL(n,{\Bbb C}),\; \lambda \mapsto
g(\lambda )\end{math}  be an analytic map such
that \begin{math} \| g(\lambda )-\unity \| <1\end{math}
 for all \begin{math} \lambda \in D(\lambda _{0},R)\end{math}.
This map \begin{math} g\end{math}
defines a cocycle  of the covering \begin{math}
\left\{ \lambda \in {\Bbb P}_{1}\left| \;
|\lambda -\lambda _{0}|\leq
R\right. \right\} \cup \left\{ \lambda \in {\Bbb P}
_{1}\left| \; |\lambda -\lambda _{0}|\geq R\right. \right\}
\end{math}. Then the corresponding holomorphic vector bundle on
\begin{math} {\Bbb P}_{1}\end{math} is trivial. Moreover, the
trivialization can be chosen to be in
accordance with
a given identification of the fiber over
infinity with \begin{math} {\Bbb C}^{n}\end{math}.
\end{Lemma}
Proof: We remind the reader of the
biholomorphic map of \begin{math} {\Bbb P}
_{1}\end{math},  which transforms \begin{math}
D(\lambda _{0},R)\end{math}  into \begin{math}
D(0,1)\end{math}  and fixes infinity.
Let \begin{math} H\end{math}  be the
Hilbert space \begin{math}
L^{2}(S^{1},{\Bbb C}^{n})\end{math}. We
have the decomposition \begin{math}
H=H_{+}\oplus H_{-}\end{math}
into boundary values of
holomorphic maps from the interior of \begin{math}
D(0,1)\end{math}  to \begin{math} {\Bbb C}
^{n}\end{math}  and holomorphic maps
defined outside of \begin{math} D(0,1)\end{math},
 which are equal to zero at infinity,
respectively. In
\cite[Proposition~(6.3.1)]{PS} it is shown
that \begin{math} LGL(n,{\Bbb C})\end{math}
is a subgroup of \begin{math}
GL_{res}(H_{+}\oplus H_{-})\end{math}.
Moreover it is proven in
\cite[Theorem~(8.1.2) and
Proposition~(8.4.1)]{PS} that \begin{math}
g\end{math}  admits a Birkhoff
factorization \begin{math}
g=g_{+}g_{-}\end{math},  if \begin{math}
g=\left( \begin{array}{cc}
a & b\\
c & d\end{array} \right) \end{math}  can
be written as \begin{math} \left( \begin{array}{cc}
a-bd^{-1}c & b\\
0 & d\end{array} \right)
\left( \begin{array}{cc}
\unity  & 0\\
d^{-1}c & \unity \end{array} \right)
\end{math}
 with respect to the decomposition \begin{math}
H=H_{+}\oplus H_{-}\end{math}.  Hence
the vector bundle is trivial, if \begin{math}
d\end{math}  is invertible. The
assumption implies \begin{math} \sup \left\{
\| g(\lambda )-\unity \| \left| \; \lambda \in D(0,1)\right.
\right\} <1\end{math}  and therefore \begin{math}
\left\| \left( \begin{array}{cc}
a & b\\
c & d\end{array} \right) -\unity \right\| <1\end{math}.
 This proves that \begin{math} d\end{math}
 is invertible. In the Birkhoff
factorization above \begin{math} g_{-}\end{math}
 is equal to \begin{math} \unity \end{math}
at infinity, and the induced trivialization of
the vector bundle is in accordance with a given
identification of the fiber over
infinity with \begin{math} {\Bbb C}^{n}\end{math}.
\hspace*{\fill } \begin{math} \Box \end{math}

\noindent
Proof of the implication (i) \begin{math}
\Rightarrow \end{math}  (ii) of
Theorem~\ref{Theorem4.3}:
Let \begin{math} D\end{math}  be a
divisor of Riemann-Roch type. Then there
exist \begin{math} n\end{math}
cross sections \begin{math}
v_{1}(D),\ldots ,v_{n}(D)\end{math}
 of \begin{math} {\cal O}_{D}\end{math} over
the inverse image of a neighbourhood of
infinity under \begin{math} \pi \end{math},
which induce an
isomorphism between the direct image
sheaf \begin{math} \pi _{*}({\cal O}_{D}) \end{math}
 and \begin{math} {\cal O}_{{\Bbb P}_{1}}^{n}\end{math}.
They can be chosen to take the same values as \begin{math}
v_{1},\ldots ,v_{n}\end{math} at the covering points of
infinity. On some neighbourhood of infinity in \begin{math}
X\end{math}  these cross sections define
a holomorphic \begin{math} n\times
n\end{math}-matrix valued function: \begin{math}
h(D,\lambda )_{i,j}=\end{math}  the value of \begin{math}
v_{i}(D)\end{math}  at the covering
point in the \begin{math} j\end{math}-th
sheet over \begin{math} \lambda \end{math}.
Let \begin{math} D(\lambda _{\iota },R_{\iota })\end{math}
be any circle around a domain excluded from \begin{math}
U_{l,\epsilon }\end{math}  with index \begin{math}
\iota \end{math},  such that all other
excluded domains sit outside of this
circle. The holomorphic vector bundle of
Lemma~\ref{Lemma4.1a} defined by the
cocycle \begin{math} g(\lambda
)=h(D,\lambda )\end{math}
corresponds to the direct image sheaf \begin{math}
\pi_{*}({\cal O}_{D_{l,\epsilon ,\iota }})\end{math}.
More precisely, the Birkhoff
factorization of \begin{math} h(D,\cdot
)\end{math}  with respect to the circle \begin{math}
D(\lambda _{\iota },R_{\iota })\end{math}
 is given by \begin{math}  h(D,\cdot )=\left(
h(D,\cdot )h_{l,\epsilon ,\iota
}^{-1}(D,\cdot )h_{0}\right) \cdot \left(
h_{0}^{-1}h_{l,\epsilon ,\iota
}(D,\cdot )\right) \end{math}.
Then Lemma~\ref{Lemma4.1a}
proves that the divisor \begin{math} D\end{math}
 is admissible, since \begin{math} h_{0}^{-1}h(D,\lambda )\end{math}
is continuous and equal to the identity
matrix at infinity. Moreover,
\begin{math} h(D,\lambda )\end{math}  has an
asymptotic expansion: \begin{displaymath}
\left\| h(D,\lambda )-\left(
h_{0}+\sum_{m=1}^{M} a_{m}\lambda ^{-m}\right)
\right\| <\frac{c_{M}}{|\lambda |^{M+1}} \mbox{ for all
\begin{math} M\in {\Bbb N}\end{math}
uniformly on \begin{math} U_{l,\epsilon }
\end{math} }\end{displaymath}
If \begin{math} \epsilon \end{math} is
small enough, then \begin{math}
h_{0}+\sum_{m=1}^{M} a_{m}\lambda ^{-m} \end{math}
is a holomorphic map from \begin{math} U_{l,\epsilon
} \end{math} into \begin{math} GL(n,{\Bbb C})\end{math}
and furthermore \begin{math}
\left\| \left( h_{0}+\sum_{m=1}^{M} a_{m}\lambda ^{-m}
\right) ^{-1}\right\| \leq 2\end{math}  on
this set. Thus we have \begin{displaymath}
\left\| \left( h_{0}+\sum_{m=1}^{M} a_{m}\lambda ^{-m}
\right) ^{-1}h(D,\lambda )-\unity \right\|
<\frac{2c_{M}}{|\lambda |^{M+1}} \mbox{ for all
\begin{math} M\in {\Bbb N}\end{math}
uniformly on \begin{math} U_{l,\epsilon }.
\end{math} }\end{displaymath}
The proof of Lemma~\ref{Lemma4.1a}
even shows that \begin{displaymath} \left\|
g_{-}(\lambda )-\unity \right\| \leq \frac{\left\|
g(\lambda )-\unity \right\| }{1-\left\| g(\lambda )-\unity \right\|
} \mbox{ for all \begin{math} \lambda \in
D(\lambda _{0},R)\end{math}. } \end{displaymath}
Hence we obtain the estimate \begin{displaymath}
\left\| h_{0}^{-1}h_{l,\epsilon ,\iota }(D,\lambda )-\unity \right\|
_{l,\epsilon ,\iota }\leq \frac{\tilde{c} _{M}}{|\lambda |^{M+1}}
\end{displaymath} for all \begin{math} M\in {\Bbb N}\end{math}
and for all indices \begin{math} \iota \end{math},
which correspond to domains excluded
from \begin{math} U_{l,\epsilon }\end{math}.
 This is true for all \begin{math} l\in
{\Bbb N}\end{math}  with some \begin{math}
\epsilon \end{math}  depending on \begin{math}
l\end{math}. This implies (ii).

Now we want to prove the implication (ii)
\begin{math} \Rightarrow \end{math}
(i). Let us first explain the strategy of the
proof. Let \begin{math} D\end{math} be a divisor with the assumed
properties. We will see that there
exist unique cross sections \begin{math}
v_{1}(D),\ldots ,v_{n}(D)\end{math}  of \begin{math}
{\cal O}_{D}\end{math}  over \begin{math} \pi
^{-1}(U)\end{math}, with some
neighbourhood \begin{math} U\end{math}
of infinity on \begin{math} {\Bbb P}_{1}
\end{math} , which takes the same values as \begin{math}
v_{1},\ldots ,v_{n}\end{math} at
infinity. Their elementary symmetric functions with
respect to the covering map \begin{math}
\pi \end{math}  \cite{Fo}  are of the
form \begin{math} c+\sum
\alpha _{i}(\lambda )/\beta _{i}(\lambda )\end{math}  as in
Example~\ref{Example2.2}  at the end of
Section 2. This implies that
\begin{math} D\end{math}  is of Riemann-Roch
type. To establish this claim we will use
perturbative methods. If we remove all
the branchpoints and simultaneously all
the points of the divisor \begin{math}
D\end{math}  the resulting surface is isomorphic
to \begin{math} ({\Bbb P}_{1})^{n}\end{math}
 and \begin{math} {\cal O}_{D}\end{math}
becomes isomorphic to \begin{math} \left(
{\cal O}_{{\Bbb P}_{1}}\right)
^{n}\end{math}.  Thus the claim is trivial
in this case. In a first step we deform
the Riemann surface \begin{math} Y\end{math}
 to the compact Riemann surface \begin{math}
Y_{\iota }\end{math} over some neighbourhood of the
branchpoints out of one excluded domain.
By Definition~\ref{Definition4.2}
the space of cross sections
of \begin{math} {\cal O}_{D_{\iota }}\end{math}  is
isomorphic to the space of cross sections
of \begin{math} \left( {\cal O}_{{\Bbb P}_{1}}\right)
^{n}\end{math}. These deformations of
the Riemann surface \begin{math} Y\end{math}
perturb each other. In a second step we
present a way to fit together all these
deformations. The
main tool will be provided by the elementary symmetric
functions, because they determine
functions on \begin{math} Y\end{math}
in terms of functions on \begin{math} {\Bbb
P}_{1}\end{math}.  It will turn out that
these perturbations deform the
isomorphism \begin{math}
\pi_{*}({\cal O}_{D_{\iota }})\simeq {\cal O}_{{\Bbb P}_{1}}\end{math}
 into another isomorphism \begin{math}
\pi_{*}({\cal O}_{D})\simeq {\cal O}_{{\Bbb P}_{1}}\end{math}.

Let us omit the index \begin{math}
\iota \end{math} in the first step.
In the next lemma \begin{math}
\pi :Y\rightarrow {\Bbb P}_{1}\end{math}
is assumed to be a compact \begin{math}
n\end{math}-fold covering of \begin{math}
{\Bbb P}_{1}\end{math}  and \begin{math} D\end{math}
 is assumed to be a divisor of \begin{math}
Y\end{math},  such that \begin{math}
\pi _{*}\left( {\cal O}_{D}\right) \end{math}
is isomorphic to \begin{math} {\cal O}_{{\Bbb P}
_{1}}^{n}\end{math}.
If we impose that a cross section of \begin{math}
{\cal O}_{D}\end{math} takes the values \begin{math}
x_{1},\ldots ,x_{n}\end{math} at the covering
points of infinity,  this
cross section is uniquely defined. Then
the elementary symmetric functions of this
cross section are rational
functions, with poles only at the
base points of the integral part of the
divisor. The values at infinity are of
course the usual elementary symmetric
functions of \begin{math}
x_{1},\ldots ,x_{n}\end{math}.  Hence
for every choice of \begin{math}
x_{1},\ldots ,x_{n}\end{math} we
have rational
functions \begin{math}
c_{1}(\lambda ,x_{1},\ldots ,x_{n}),\ldots ,c_{n}(\lambda
,x_{1},\ldots ,x_{n})
\end{math}  on \begin{math} {\Bbb P}_{1}\end{math},
such that \begin{equation}
f^{n}+f^{n-1}c_{1}(\lambda
,x_{1},\ldots ,x_{n})+...+c_{n}(\lambda ,x_{1},\ldots ,x_{n})=0
\label{g4.2} \end{equation}  defines the
unique cross section \begin{math} f\end{math}
 of \begin{math} {\cal O}_{D}\end{math},
which takes the values \begin{math}
x_{1},\ldots ,x_{n}\end{math} at the covering points of
infinity.  Each of these
elementary symmetric functions has the
form \begin{equation}
c_{i}(\lambda ,x_{1},\ldots ,x_{n})=e_{i}(x_{1},\ldots ,x_{n})+
\frac{\alpha _{i}(\lambda ,x_{1},\ldots ,x_{n})}{\beta (\lambda )}
,\label{g4.3} \end{equation}  where \begin{math}
e_{i}(x_{1},\ldots ,x_{n})\end{math}  is the
usual \begin{math} i\end{math}-th
elementary symmetric function of \begin{math}
x_{1},\ldots ,x_{n},\; \; \beta (\lambda )\end{math}  is the
polynomial whose zeroes are given by the
base points of the integral part of the
divisor and \begin{math}
\alpha _{i}(\lambda ,x_{1},\ldots ,x_{n})\end{math}  are
polynomials in \begin{math} \lambda \end{math}
 of degree less than the degree of \begin{math}
\beta (\lambda )\end{math},  whose coefficients are
homogenous polynomials in \begin{math}
x_{1},\ldots ,x_{n}\end{math}.
\newtheorem{DLemma}[Lemma]{Deformation
Lemma} \begin{DLemma} \label{Lemma4.2}
Let \begin{math} Y\end{math}  and \begin{math}
D\end{math}  fulfil the conditions given above.
Let \begin{math} U\end{math}
 be an open subset of \begin{math} {\Bbb
P}_{1}\end{math},  which contains a ball
of radius \begin{math} R>0\end{math}
around all the base points of the
integral part of the divisor. Moreover,
let \begin{math} B\end{math}  be a ball
contained in the intersection of all
balls with radius \begin{math} R/2\end{math}
 around these base points, and let \begin{math}
\| \cdot \| _{\partial B}\end{math} be the
supremum norm on the boundary of \begin{math}
B\end{math}.  Moreover, choose some
values \begin{math} x_{1},\ldots ,x_{n}\end{math},
 such that \begin{math} |x_{i}-x_{j}|>c>0\end{math}
 for all \begin{math} i\neq j\end{math}.
Assume \begin{math} \| \alpha _{i}(\cdot
,x_{1},\ldots ,x_{n})/\beta \| _{\partial B}\end{math}
to be small enough for all \begin{math}
i=1,\ldots ,n\end{math}, and let
\begin{math} f_{1},\ldots ,f_{n}\end{math}
be functions on \begin{math} U\end{math},  such that
\begin{math} |f_{i}-x_{i}|<\epsilon \end{math},
with some \begin{math} \epsilon
<c/4\end{math}. Then there exists a unique cross section \begin{math}
\tilde{f} \end{math}  of \begin{math}
{\cal O}_{D}\end{math}  over \begin{math} \pi
^{-1}(U)\end{math},  such that the
regular parts of  the elementary
symmetric functions \begin{math} \tilde{c}
_{i}\end{math} of \begin{math} \tilde{f}
\end{math}  with respect to the covering
\begin{math} \pi \end{math}  are equal to the
usual elementary symmetric functions of \begin{math}
f_{1},\ldots ,f_{n}:\end{math}  \begin{displaymath}
\tilde{c}
_{i}=e_{i}(f_{1},\ldots ,f_{n})+\frac{\tilde{\alpha }
_{i}}{\beta } ,\; \; \mbox{with some polynomial \begin{math}
\tilde{\alpha } _{i}\end{math} of degree
smaller than the degree of \begin{math}
\beta \end{math}. }
\end{displaymath} On the \begin{math}
i\end{math}-th sheet the section \begin{math}
\tilde{f} \end{math}  is moreover
assumed to be nearly equal to \begin{math}
f_{i}\end{math} away from the ball \begin{math}
B\end{math}. In this sense \begin{math}
\tilde{f} \end{math} is a deformation of \begin{math}
f_{1},\ldots ,f_{n}\end{math}  on the
trivial covering \begin{math} {\Bbb
P}_{1}^{n}\rightarrow {\Bbb P}_{1}\end{math}.
\end{DLemma}
In order to prove this lemma we need two
more lemmata.
\begin{Lemma} \label{Lemma4.3}
Let \begin{math} U\end{math}  be an open
subset of \begin{math} {\Bbb P}_{1}\end{math},
which contains a ball of radius \begin{math}
R>0\end{math} around some points \begin{math}
\lambda _{1},\ldots ,\lambda _{n}\in {\Bbb C}\end{math}.  Let
\begin{math} Hol(A)\end{math}  be the
Banach space of holomorphic functions on
\begin{math} U\end{math},  which extend
continuously to the closure \begin{math}
A\end{math}  of \begin{math} U\end{math}.
Then the linear map \begin{displaymath}
Hol(A)\rightarrow Hol(A),\; \; g\mapsto
reg\left( \frac{g}{\prod_{l=1}^{d}
(\lambda -\lambda _{l})} \right) \; \;
\mbox{is bounded in norm by \begin{math}
(2/R)^{d}\end{math}.} \end{displaymath}
\end{Lemma}
Proof: We use induction in \begin{math}
d\end{math}.  It is quite obvious that
the regular part of \begin{math}
g(\lambda )(\lambda -\lambda _{0})^{-1}\prod_{l=1}^{d}
(\lambda -\lambda _{l})^{-1}\end{math}  is equal to
the regular part of \begin{math} \tilde{g}
(\lambda )\prod_{l=1}^{d}(\lambda -\lambda _{l})^{-1}\end{math},
 where \begin{math} \tilde{g}
(\lambda )=(g(\lambda )-g(\lambda _{0}))/(\lambda -\lambda
_{0})\end{math}.
This function \begin{math} \tilde{g} \end{math}
 extends to a continuous
function on \begin{math} A\end{math}.
With the help of the maximum modulus
Theorem \cite{Co} we obtain the bound \begin{math}
\| \tilde{g} \| \leq \| g\| 2/R\end{math}.
This proves the start of the induction
and also each induction step.
\hspace*{\fill } \begin{math} \Box \end{math}
\begin{Lemma} \label{Lemma4.4}
In the situation of the preceeding lemma
let \begin{math} B\end{math}  be a ball
containing the points \begin{math}
\lambda _{1},\ldots ,\lambda _{d}\end{math},  which
for his part is
contained in the intersection of all
balls with radius \begin{math} R/3\end{math}
 around the points \begin{math}
\lambda _{1},\ldots ,\lambda
_{d}\end{math}. Again let \begin{math}
\| \cdot \| _{\partial B}\end{math} be the
supremum norm on the boundary of \begin{math}B\end{math}.
Moreover, let \begin{math} \alpha \end{math}
 be a polynomial of degree less than \begin{math} d,
\end{math} and \begin{math}
\beta (\lambda )=\prod_{l=1}^{d}(\lambda -\lambda _{l})\end{math}.
 Then the linear map \begin{displaymath}
Hol(A)\rightarrow Hol(A),\; \; g\mapsto
reg\left( \frac{g(\lambda )\alpha (\lambda
)}{\beta (\lambda )} \right) \; \;
\mbox{is bounded in norm by \begin{math} 2\| \alpha
/\beta \| _{\partial
B}\end{math}. }\end{displaymath}
\end{Lemma}
Proof: The rational function \begin{math}
\alpha (\lambda )/\beta (\lambda )\end{math}  can be written as
\begin{displaymath} \frac{\alpha (\lambda
)}{\beta (\lambda )}
=\sum_{j=1}^{d}
\frac{\gamma _{j}}{\prod_{l=j}^{d}(\lambda -\lambda _{l})}
\mbox{ with } \gamma _{j}=\frac{1}{2\pi \sqrt{-1} }
\int_{\partial  B} \frac{\alpha (\lambda
)}{\beta (\lambda )}
\prod_{l=j+1}^{d} (\lambda -\lambda _{l})d\lambda .\end{displaymath}
Here \begin{math} |\gamma _{j}|\end{math}  is
bounded by \begin{math} \left\| \alpha
/\beta \right\|
_{\partial B}\left( R/3\right) ^{d-j+1}\end{math}.
 The preceeding lemma now gives the estimate \begin{math}
\left\| g\alpha /\beta \right\| \leq \| g\| \left\| \alpha
/\beta \right\|
_{\partial B}\sum_{j=1}^{d}
(2/3)^{d-j+1}\leq 2\| g\| \left\| \alpha
/\beta \right\| _{\partial
B}\end{math}.
\hspace*{\fill } \begin{math} \Box \end{math}

\noindent
Proof of the Deformation
Lemma~\ref{Lemma4.2}: Let \begin{math}
u_{1},\ldots ,u_{n}\end{math}  be the unique
cross sections of \begin{math} {\cal O}_{D}\end{math},
 whose matrix of values at all covering
points of infinity is equal to the
identity matrix. For all \begin{math}
\tilde{f} _{1},\ldots ,\tilde{f} _{n}\in
Hol(A)^{n}\end{math}  let \begin{math}
\tilde{c} _{1},\ldots ,\tilde{c} _{n}\end{math}
be the elementary symmetric functions with
respect to the covering \begin{math} \pi
\end{math}  of \begin{math} \tilde{f}
=\sum_{i=1}^{n} \tilde{f} _{i}u_{i}\end{math}.
Due to the last lemma the map \begin{displaymath}
Hol(A)^{n}\rightarrow Hol(A)^{n},\; \; (\tilde{f}
_{1},\ldots ,\tilde{f} _{n})\mapsto \left(
reg(\tilde{c} _{1}),\ldots ,reg(\tilde{c} _{n})\right) \end{displaymath}
is holomorphic. If \begin{math} \tilde{f}
_{1},\ldots \tilde{f} _{n}\end{math}  are
all constant, they are mapped onto the
ordinary elementary symmetric functions
of these constants. But on the open subset of
\begin{math} Hol(A)^{n}\end{math}  with \begin{math}
|\tilde{f} _{i}-x_{i}|<2\epsilon \end{math},
the map \begin{displaymath} (\tilde{f}
_{1},\ldots ,\tilde{f} _{n})\mapsto
\mbox{the usual elementary symmetric functions of }
(\tilde{f} _{1},\ldots ,\tilde{f} _{n})\end{displaymath}
is invertible, if \begin{math}
c\end{math}  is larger than \begin{math}
4\epsilon \end{math}.  According to the last
lemma the difference of these two maps
is small. Hence we can apply the
inverse function Theorem \cite{PT}, and
the above map is invertible for \begin{math}
|\tilde{f} _{i}-x_{i}|<2\epsilon \end{math}.
This proves the claim.
\hspace*{\fill } \begin{math} \Box \end{math}

\noindent
Now let \begin{math} D\end{math}  be any
admissible divisor of \begin{math} Y\end{math}.
As we mentioned before, the surfaces \begin{math}
Y_{l,\epsilon ,\iota }\end{math} and the
divisors \begin{math}
D_{l,\epsilon ,\iota }\end{math}  fit into the
situation described in the deformation Lemma
whenever \begin{math} |\iota |\end{math}  is
large enough. In Example~\ref{Example2.2}
we introduced the supremum norm on the
boundary of these excluded domains \begin{math}
\| \cdot \| _{l,\epsilon ,\iota }\end{math}.  Let \begin{math}
C_{\iota }(D,l,\epsilon )>0\end{math}  be the smallest
constant, such that \begin{displaymath}
\left\| \frac{\alpha _{i,\iota } (\cdot
,x_{1},\ldots ,x_{n})}{\beta _{\iota
}(\cdot )}\right\|
_{l,\epsilon ,\iota }\leq C_{\iota
}(D,l,\epsilon )\; \; \mbox{for all }
|x_{1}|\leq 1,\ldots ,|x_{n}|\leq 1\; \; \mbox{and}
\; \; i=1,\ldots ,n.\end{displaymath}
\begin{Proposition} \label{Proposition4.1}
Let \begin{math} D\end{math}  be an
admissible divisor of \begin{math}
Y\end{math}. Then \begin{math}
D\end{math} is of Riemann-Roch
type, if \begin{math}
\sum_{\iota } C_{\iota }(D,l,\epsilon )<\infty \end{math}
for all \begin{math} l\in {\Bbb N}\end{math}
with some \begin{math} \epsilon >0\end{math}
depending on \begin{math} l\end{math}. \end{Proposition}
Proof: Let \begin{math}
x_{1},\ldots ,x_{n}\end{math} be
complex numbers, such that \begin{math}
|x_{i}-x_{j}|>c\end{math}. First we show that there exists
a unique cross section of \begin{math}
{\cal O}_{D}\end{math}  over the inverse
image under \begin{math} \pi \end{math}
of some neighbourhood
of infinity on \begin{math} {\Bbb P}_{1}\end{math},
which takes the values \begin{math}
x_{1},\ldots ,x_{n}\end{math}  at all
covering points of infinity. For the
elementary symmetric functions of this
cross section we make
the ansatz: \begin{displaymath}
\tilde{c} _{i}(\lambda ,x_{1},\ldots ,x_{n})=e_{i}(x_{1},\ldots ,x_{n})+
\sum_{\iota \in {\cal I}'} \frac{\alpha _{i,\iota }(\lambda
)}{\beta _{\iota }(\lambda )}
\; \; \mbox{(compare with (\ref{g4.3}) and
Example~\ref{Example2.2}).}
\end{displaymath}
The sum contains only those indices,
which correspond to excluded domains
over the neighbourhood of infinity on \begin{math}
{\Bbb P}_{1}\end{math}.  We choose two
neighbourhoods \begin{math} U_{l,\delta }\end{math},
and \begin{math} U_{\tilde{l} ,\tilde{\delta }
}\end{math},  such that for each index \begin{math}
\iota \in {\cal I}'\end{math}  in
the deformation Lemma the closure of
\begin{math} U\end{math}  can be chosen to be
equal to the image under \begin{math}
\pi \end{math}  of the excluded domain
with index \begin{math} \iota \end{math}
 of \begin{math} U_{l,\delta
}\end{math}. Also we choose \begin{math}
B\end{math}  to be equal to the image
under \begin{math} \pi \end{math}  of
the excluded domain with index \begin{math}
\iota \end{math}  of \begin{math} U_{\tilde{l}
,\tilde{\delta } }\end{math}.
Taking \begin{math} \tilde{l} >l\end{math}  for
granted, this can be always attained by
reducing the neighbourhood of infinity
on \begin{math} {\Bbb P}_{1}\end{math} and
thereby the corresponding set of indices \begin{math}
{\cal I}'\end{math}  of excluded domains over
this neighbourhood. Now we introduce the Banach space of
sequences of polynomials \begin{math}
\alpha _{i,\iota }(\lambda )\end{math}  indexed by \begin{math}
i=1,\ldots ,n\end{math}  and \begin{math}
\iota \in {\cal I}'\end{math},  such that the
degree of \begin{math} \alpha _{i,\iota }\end{math}
 is less than the degree \begin{math}
d_{\iota }\end{math} of \begin{math}
\beta _{\iota }(\lambda )\end{math} from equation
(\ref{g4.3}). The norm is given by
\begin{displaymath}  \left\|
(\alpha _{i,\iota }(\cdot ))\right\| =
\sum_{\iota \in {\cal I}'} \sup \left\{ \left. \left\|
\frac{\alpha _{i,\iota }}{\beta _{\iota }} \right\| _{l,\delta
,\iota }\; \right| i=1,\ldots ,n\right\} .\end{displaymath}
With the help of the deformation Lemma
we can now define a map on an open
subset of this Banach space into this
Banach space:\begin{displaymath} \left\{
(\alpha _{i,\iota })_{(i,\iota )\in \{ 1,\ldots ,n\} \times
{\cal I}'}\; \left| \; \| (\alpha _{i,\iota })\|\leq
\epsilon \right. \right\} \rightarrow \left\{
(\alpha _{i,\iota })_{(i,\iota )\in \{ 1,\ldots ,n\} \times
{\cal I}'}\right\} ,\; \; (\alpha _{i,\iota })\mapsto
(\tilde{\alpha } _{i,\iota }),\end{displaymath}
such that for all \begin{math} \kappa \in {\cal I}'\end{math}
the section \begin{math} \tilde{f} _{\kappa }\end{math}
of \begin{math} {\cal O}_{D}\end{math}
over the excluded domain with
index \begin{math} \kappa \end{math},
defined by
the elementary symmetric functions \begin{math} \tilde{c}
_{j,\kappa }=c_{j,\kappa }+\tilde{\alpha } _{j,\kappa
}/\beta _{\kappa } \end{math}
is the deformation of the section \begin{math}
f_{\kappa }\end{math} of \begin{math}
{\cal O}_{{\Bbb P}_{1}}^{n}\end{math} over the same
domain defined by the elementary
symmetric functions \begin{displaymath}
c_{j,\kappa
}=e_{j}(x_{1},\ldots ,x_{n})+\sum_{\iota
\in {\cal I}'\setminus \{
\kappa \}} \frac{\alpha _{j,\iota
}}{\beta _{\iota }} .\end{displaymath}
Now Lemma~\ref{Lemma4.4} gives the
estimate \begin{displaymath}
\left\| \frac{\tilde{\alpha }
_{j,\kappa }}{\beta_{\kappa }} \right\| _{l,\delta
}\leq C_{\kappa }(D,\tilde{l}
,\tilde{\delta } )\cdot
\mbox{constant }\end{displaymath} The
constant depends only on \begin{math}
c\end{math}  and \begin{math}
x_{1},\ldots ,x_{n}\end{math}.
The assumption on the divisor implies
that \begin{math} \sum_{\kappa \in {\cal I}'}
C_{\kappa }(D,\tilde{l} ,\tilde{\delta }
)<\infty \end{math}. Hence the image of
this map is contained in the domain, if
the neighbourhood of infinity and the
corresponding set of indices \begin{math}
{\cal I}'\end{math} is small enough. The same
argument for arbitrary \begin{math}
l\in {\Bbb N}\end{math} shows that the
elementary symmetric functions \begin{math}
\tilde{c} _{j,\kappa }\end{math}  are
meromorphic functions of the form given
in Example~\ref{Example2.2}.
Again Lemma~\ref{Lemma4.4} shows that \begin{displaymath}
\left\| \frac{\tilde{\alpha } _{j,\kappa
}}{\beta _{\kappa }}
-\frac{\tilde{\alpha '} _{j,\kappa
}}{\beta _{\kappa }} \right\|
_{l,\delta }\leq \left\| ( \alpha _{i,\iota }) -(\alpha '_{i,\iota })
\right\| C_{\kappa }(D,\tilde{l} ,\tilde{\delta }
)\cdot \mbox{constant } \end{displaymath}
Again the constant depends only
on \begin{math} c\end{math}  and \begin{math}
x_{1},\ldots ,x_{n}\end{math}.
This map is a contraction, if the neighbourhood
of infinity and the corresponding set of
indices \begin{math} {\cal I}'\end{math}
is small enough. Due to Picard's fix point principle
it has a unique fixed point. For such a fixed point
the elementary symmetric functions \begin{math}
\tilde{c} _{j,\kappa }\end{math}  do
not depend on \begin{math} \kappa .
\end{math} Therefore
they define a section of \begin{math}
{\cal O}_{D}\end{math}  over some
neighbourhood of infinity of the desired
form. In the proof of the deformation
lemma we showed that near every
excluded domain this cross section can be
written as \begin{math} \tilde{f}
=\sum_{i=1}^{n}\tilde{f} _{i,\kappa
}u_{i,\kappa }\end{math},
where \begin{math} u_{1,\kappa
},\ldots ,u_{n,\kappa } \end{math}
defines an isomorphism of sheaves \begin{math}
{\cal O}_{{\Bbb P}_{1}}^{n}\simeq \pi _{*}({\cal O}_{D})\end{math}
 near this excluded domain and \begin{math}
|\tilde{f} _{i}-x_{i}|<2\epsilon \end{math}.
Hence we can choose the values \begin{math}
x_{1},\ldots ,x_{n}\end{math}  for \begin{math}
n\end{math}  different cross sections \begin{math}
u_{1},\ldots ,u_{n}\end{math} of \begin{math}
{\cal O}_{D}\end{math}  over some
neighbourhood of infinity in a way, such
that condition (ii) of Definition~\ref{Definition4.1}
is satisfied.
\hspace*{\fill } \begin{math} \Box \end{math}

\noindent
Proof of the implication (ii) \begin{math}
\Rightarrow \end{math} (i) of
Theorem~\ref{Theorem4.3}: It is quite
obvious that the condition (ii) is
equivalent to the assumptions of
Proposition~\ref{Proposition4.1}. This
concludes the proof of
Theorem~\ref{Theorem4.3}.
\hspace*{\fill } \begin{math} \Box \end{math}

\noindent
In the case that \begin{math} n=2\end{math}
 and that \begin{math} D\end{math}  is an
integral divisor, the elementary
symmetric functions in
equation~(\ref{g4.2}) of the form given in
equation~\ref{g4.3}
 can be calculated explicitly: Let \begin{math}
\lambda _{1}\end{math}  and \begin{math}
\lambda _{2}\end{math}  be the two base points
of the two branchpoints. Furthermore let
the divisor be specified by its
base point \begin{math} \lambda _{0}\end{math}
and by a choice of the branch \begin{math}
\sqrt{(\lambda _{0}-\lambda _{1})(\lambda _{0}-\lambda _{2})}\end{math}.
Let \begin{math} f\end{math} be the unique cross
section, which takes the values \begin{math} x_{1}\end{math}
and \begin{math} x_{2}\end{math} at the covering points of
infinity. It is given by \begin{eqnarray}
f^{2}-\left(
x_{1}+x_{2}+\frac{x_{1}-x_{2}}{\lambda -\lambda _{0}}
\sqrt{(\lambda _{0}-\lambda _{1})(\lambda
_{0}-\lambda _{2})}\right)
f & + &\nonumber \\
x_{1}x_{2}-\frac{x_{1}-x_{2}}{\lambda -\lambda _{0}} \left(
\frac{x_{1}-x_{2}}{4}
(2\lambda _{0}-\lambda _{1}-\lambda _{2})+\frac{x_{1}+x_{2}}{2}
\sqrt{(\lambda _{0}-\lambda _{1})(\lambda _{0}-\lambda _{2})}\right)
& = & 0.\label{g4.5} \end{eqnarray}
Now it is easy to see that the condition of
Proposition~\ref{Proposition4.1}  is
fulfilled in this case. For excluded
domains, which contain only two
branchpoints, the compact \begin{math}
n\end{math}-fold covering \begin{math}
\pi :Y_{\iota }\rightarrow {\Bbb P}_{1}\end{math}
 decompose into one two-fold covering
and \begin{math} n-2\end{math}  copies
of \begin{math} {\Bbb P}_{1}\end{math}.
Hence we can estimate \begin{math} C_{\iota
}(D,l,\epsilon )\end{math} with
the help of equation~(\ref{g4.5}) even in this case.
\begin{Corollary} \label{Corollary4.1}
If all the excluded domains
corresponding to one \begin{math}
U_{l,\epsilon }\end{math}  have asymptotically
no overlap, then all integral divisors of
asymptotic degree \begin{math}
(1,1,\ldots )\end{math} with support
inside of \begin{math} \pi ^{-1}({\Bbb C})\end{math}
are of Riemann-Roch
type.\end{Corollary}
\hspace*{\fill } \begin{math} \Box \end{math}

\noindent
The case that all domains excluded from
\begin{math} U_{l,\epsilon }\end{math}
have asymptotically no overlap is the
generic case. Otherwise the Riemann
surfaces \begin{math} Y_{\iota }\end{math}
 are more complicated and may have
divisors of degree equal to half the
branching order, such that the
corresponding direct image sheaf \begin{math}
\pi _{*}({\cal O}_{D})\end{math}  is not
isomorphic to \begin{math} {\cal O_{{\Bbb
P}_{1}}}^{n}\end{math}.

Formula (\ref{g4.5}) shows that the
coefficients of the singular parts of
the elementary symmetric functions are
bounded by \begin{math} 4\sup \{
|\lambda _{0}-\lambda _{1}|,|\lambda _{0}-\lambda _{2}|\} \end{math}.
 This observation was crucial in the
last proof. These coefficients are
bounded on the domain, where all
base points of the divisor and the
branchpoints are elements of a bounded
subset of \begin{math} {\Bbb C}\end{math},
iff \begin{math} \pi _{*}({\cal
O}_{D})\simeq {\cal O}_{{\Bbb P}_{1}}^{n} \end{math}
were to hold also
for singular surfaces \begin{math} Y\end{math}.
 But this is no more valid even in the
case that \begin{math} n=2\end{math}
and that the support of \begin{math} D\end{math}
 contains more than one point: If \begin{math}
Y\end{math}  is the singular Riemann
surface of two copies of \begin{math} {\Bbb
P}_{1}\end{math}  connected by one double
point, and if the divisor consists of two
points out of one copy of \begin{math}
{\Bbb P}_{1}\end{math}  with multiplicity \begin{math}
1\end{math}  and one point out of the
other copy of \begin{math} {\Bbb P}_{1}\end{math}
 with multiplicity \begin{math} -1\end{math},
\begin{math} \pi _{*}({\cal
O}_{D})\end{math} is not isomorphic to \begin{math}
{\cal O}_{{\Bbb P}_{1}}^{n} \end{math}.
There exist similar counterexamples for \begin{math}
n>2\end{math} and \begin{math} D\end{math}
 integral. These counterexamples suggest
a more restrictive assumption on the
location of the divisor: The statement
\begin{math} \pi _{*}({\cal O}_{D})\simeq
{\cal O}_{{\Bbb P}_{1}}^{n} \end{math}
remains true, if this
assumption guarantees that in the limit
of singular Riemann surfaces some of
the points of the divisors are deformed
into the singular points. Then it can be proven that
the divisors are of Riemann-Roch
type.

\section{The Jacobian variety} \label{Section5}
In this section we want to investigate the set of all
equivalence classes of line bundles, which submit to the
Riemann-Roch Theorem. The following definition is suggested
from the previous section.
\begin{Definition} \label{Definition5.1}
Let the Jacobian variety \begin{math} \mbox{\it Jacobian} (Y)\end{math}
of \begin{math} Y\end{math} be the set of all equivalence
classes in the modified sense of integral divisors of
asymptotic and total degree \begin{math}
(1,1,\ldots )\end{math}
 in the sense of Remark~\ref{Remark2.1}, which are of
Riemann-Roch type. \end{Definition}
For every divisor \begin{math} D\end{math}  of asymptotic and
total degree \begin{math} (1,1,\ldots )\end{math}, which
is of Riemann-Roch type there exists at least one
cross section \begin{math}
f\end{math}  of \begin{math} {\cal O}_{D}\end{math},  which takes
the value 1 at all covering points of infinity.
Hence all such divisors are
equivalent in the modified sense to some integral divisor.
Moreover, if there exists only one section \begin{math} f\end{math}
 of \begin{math} {\cal O}_{D}\end{math},
which is equal to \begin{math}
1\end{math}  at all covering points of infinity, \begin{math}
D\end{math}  is equivalent in the modified sense to exactly
one integral divisor. Such divisors are called non-special
in the modified sense.
\begin{Lemma} \label{Lemma5.1}
For all representatives \begin{math} D\end{math}
of elements of \begin{math} \mbox{\it Jacobian} (Y)\end{math}  the
following statements are equivalent:
\begin{description}
\item[(i)] \begin{math} [D]\in \mbox{\it Jacobian} (Y)\end{math}  is
non-special in the modified sense.
\item[(ii)] \begin{math} \tilde{H} ^{1}\left(
Y,{\cal O}_{D-\pi ^{-1}(\infty )}\right) =0\end{math}.
\item[(iii)] \begin{math} \pi _{*}\left( {\cal O}_{D}\right)
\simeq {\cal O}_{{\Bbb P}_{1}}^{n}\end{math}. \end{description}
\end{Lemma}
Proof: (i)\begin{math} \Leftrightarrow \end{math}
(ii): The Riemann-Roch Theorem implies \begin{displaymath}
\dim \tilde{H} ^{0}\left( Y,{\cal O}_{D-\pi ^{-1} (\infty
)}\right) =\dim \tilde{H} ^{1}\left(
Y,{\cal O}_{D-\pi ^{-1} (\infty )}\right) .\end{displaymath}
For all elements \begin{math} f\end{math}  of \begin{math}
\tilde{H} ^{0}\left( Y,{\cal O}_{D-\pi ^{-1} (\infty )}\right)
,\ \ 1+f\end{math}  is a cross section of \begin{math}
{\cal O}_{D}\end{math},  which is equal to \begin{math}
1\end{math}  at all covering points of infinity
and all these cross sections are of the form \begin{math}
1+f\end{math},  with \begin{math} f\in H^{0}\left(
Y,{\cal O}_{D-\pi ^{-1} (\infty )}\right) \end{math}.
Hence (i) and (ii) are equivalent.

(ii)\begin{math} \Leftrightarrow \end{math}
(iii): \begin{math}
\pi _{*}({\cal O}_{D})\end{math}  is isomorphic to \begin{math}
{\cal O}_{D_{1}}\oplus \ldots \oplus {\cal O}_{D_{n}}\end{math},
 with some divisors \begin{math} D_{1},\ldots
,D_{n}\end{math}  of \begin{math} {\Bbb
P}_{1}\footnote{This shows the Birkhoff
factorization (see e.g.
\cite[Proposition~(8.11.5)]{PS}).} \end{math}
 such that \begin{math} \deg (D_{1})+\ldots +\deg
(D_{n})=0\end{math}.  Then we have \begin{math}
\dim \tilde{H} ^{1}\left( Y,{\cal O}_{D-\pi ^{-1} (\infty
)}\right) =\sum_{i=1}^{n} \max \{0,\deg
(D_{i})\}\end{math}.  Hence is (ii)
equivalent to \begin{math} \deg (D_{i})=0\end{math}
 for all \begin{math} i=1,\ldots ,n\end{math},
which is equivalent to (iii).
\hspace*{\fill } \begin{math} \Box \end{math}

\noindent
Theorem~\ref{Theorem4.3}  suggests to endow \begin{math}
\mbox{\it Jacobian} (Y)\end{math}  with the topology induced by the
infinite sums \begin{math} \sum_{\iota } \|
h_{l,\epsilon ,\iota }(D,\cdot
)-h_{l,\epsilon ,\iota }(D',\cdot )\| _{l,\epsilon ,\iota }\end{math}.
More precisely, only for large \begin{math} |\iota |,\ \
h_{l,\epsilon ,\iota }(D,\cdot )\end{math}  is well defined.
Hence these sums define semidefinite metrics on
the set of all representatives of elements of \begin{math}
\mbox{\it Jacobian} (Y)\end{math}.  Each representative \begin{math}
D\end{math}  decomposes into a finite collection
of points of \begin{math} Y\end{math}  and
sequences of divisors in the excluded domains. For
the finite collection of points we use the
natural topology induced by the topology of \begin{math}
Y\end{math}.  For the sequences of divisors we use
the topology defined by all the sums \begin{math}
\sum_{\iota } \| h_{l,\epsilon ,\iota }(D,\cdot
)-h_{l,\epsilon ,\iota }(D',\cdot )\| _{l,\epsilon ,\iota }\end{math}
with \begin{math} l\in {\Bbb N}\end{math}  and some \begin{math}
\epsilon \end{math}  depending on \begin{math} l\end{math}.
These together define a topology on the set of
all representatives of \begin{math} \mbox{\it Jacobian} (Y)\end{math}.
 Finally the topology of \begin{math} \mbox{\it Jacobian} (Y)\end{math}
 is given by the quotient topology, which is the
finest topology such that the map, which maps
the representatives of elements of \begin{math}
\mbox{\it Jacobian} (Y)\mapsto \end{math} elements of \begin{math}
\mbox{\it Jacobian} (Y)\end{math}, is continuous.
\begin{Lemma} \label{Lemma5.2}
For all integral divisors \begin{math} D\end{math}
 of asymptotic degree \begin{math} (1,1,\cdot )\end{math}
the following statements are equivalent:
\begin{description}
\item[(i)] \begin{math} \sum_{\iota } \|
h_{l,\epsilon ,\iota }(D,\cdot )-h_{0}\|
_{l,\epsilon ,\iota } <\infty \end{math}
 for all \begin{math} l\in {\Bbb N}\end{math}  with some
\begin{math} \epsilon \end{math}  depending on \begin{math}
l\end{math}.
\item[(ii)] \begin{math} \sum_{\iota }\| h_{l,\epsilon ,\iota }(D,\cdot
)-h_{0}\| _{l,\epsilon ,\iota } |\iota |^{k}<\infty \end{math}
for any fixed \begin{math} l\end{math}  and \begin{math}
\epsilon \end{math}  and all \begin{math} k\in {\Bbb N}\end{math}.
\end{description} \end{Lemma}
Proof: All entries of \begin{math}
h_{l,\epsilon ,\iota }(D)\end{math}  are
meromorphic functions on the compact \begin{math}
n\end{math}-fold covering \begin{math}
\pi :Y_{l,\epsilon ,\iota }\rightarrow {\Bbb P}_{1}\end{math}
 and at the covering points of infinity
equal to the corresponding entries of \begin{math}
h_{0}\end{math}.  We will use the
elementary symmetric functions \begin{math}
c_{1},\ldots ,c_{n}\end{math}  defined
by (\ref{g4.2}) and of the form (\ref{g4.3}).
In our case the degree \begin{math} d\end{math}
 of \begin{math} \beta (\lambda )\end{math}  is at
most \begin{math} n(n-1)/2\end{math}.
Now we need another lemma.
\begin{Lemma} \label{Lemma5.3}
Let \begin{math} B\subset {\Bbb C}\subset
{\Bbb P}_{1}\end{math}  be a ball containing the
points \begin{math} \lambda _{1},\ldots
,\lambda _{d}\end{math},  which furthermore is
contained in the intersection of all
balls of radius \begin{math} r\end{math}
 around the points \begin{math}
\lambda _{1},\ldots ,\lambda _{d}\end{math}.
Moreover let \begin{math} A\subset{\Bbb C}
\subset {\Bbb P}_{1}\end{math}  be a ball
containing the balls with radius \begin{math}
R\end{math}  around the points \begin{math}
\lambda _{1},\ldots ,\lambda _{d}\end{math}. Let \begin{math}
\| \cdot \| _{\partial B}\end{math}  and \begin{math}
\| \cdot \| _{\partial A}\end{math}  be
the supremum norm on the boundary of \begin{math}
B\end{math}  and \begin{math} A\end{math}
 respectively. Finally let \begin{math}
\alpha (\lambda )\end{math}  be a polynomial of
degree less than the degree of \begin{math}
\beta (\lambda )=\prod_{i=1}^{d} (\lambda -\lambda _{i})\end{math}.
Then we have the estimate \begin{displaymath}
\left\| \frac{\alpha }{\beta } \right\| _{\partial
A}  \leq \left\| \frac{\alpha }{\beta } \right\|
_{\partial B}\frac{1-(r/R)^{d}}{1-(r/R)}
\frac{r}{R} .\end{displaymath} \end{Lemma}
The proof uses the same methods as the
proof of Lemma~\ref{Lemma4.4}
\hspace*{\fill } \begin{math} \Box \end{math}

\noindent
Conclusion of the proof of
Lemma~\ref{Lemma5.2}: The foregoing
lemma gives an estimate \begin{displaymath}
C'\sum_{\iota } \left\|
h_{\iota }(D,\cdot )-h_{0}\right\|
_{l',\epsilon ',\iota }\leq \sum_{\iota } \left\|
h_{\iota } (D,\cdot )-h_{0}\right\|
_{l,\epsilon ,\iota }\
|\iota |^{k}\leq C''\sum_{\iota } \left\|
h_{\iota }(D,\cdot )
-h_{0}\right\| _{l+k,\epsilon '',\iota }\end{displaymath}
with some constants \begin{math} C',l',\epsilon '\end{math}
 and \begin{math} C'',\epsilon ''\end{math}
depending only on \begin{math}
l,k,\epsilon \end{math}  and \begin{math}
p\end{math}.
\hspace*{\fill } \begin{math} \Box \end{math}

\noindent
This lemma shows that we can use as
well the sequences from
(ii)in the same fashion as
we used the sequences from \begin{math}
(i)\end{math}  to define the topology of
\begin{math} \mbox{\it Jacobian} (Y)\end{math}.

Now let \begin{math} \mbox{\it Isospectral} (Y)\end{math}  be
the subspace of the Fr\'{e}chet space \begin{math}
{\cal H}^{\infty }\end{math}  of smooth
periodic potentials \begin{math} \tilde{q}
\end{math},  such that \begin{math} \det\left(
\mu \unity -F(\lambda ,\tilde{q} )\right) =R(\lambda ,\mu
)=\det \left( \mu \unity -F(\lambda ,q)\right) \end{math}.
In Section~\ref{Section3} we defined a
map: \begin{displaymath} D(\cdot ):
\mbox{\it Isospectral} (Y)\rightarrow
\mbox{\it Jacobian} (Y),\tilde{q} \mapsto [D(\tilde{q}
)].\end{displaymath} We were able to
lift the shift to a flow on the image of
\begin{math} D:(x,[D(\tilde{q} )])\mapsto
[D(T_{x}\tilde{q} )]\end{math}.  This
lifting of the shift was defined by the
property that the action on the
corresponding line bundles is given by
the tensor product with \begin{math}
L(x)\end{math}. In order to extend this
action to the whole of \begin{math}
\mbox{\it Jacobian} (Y)\end{math}, let us describe the
sheaves \begin{math} \pi _{*}({\cal O}_{D})\end{math}
for all \begin{math} [D]\in \mbox{\it Jacobian} (Y)\end{math}
by cocycles:

Let \begin{math} {\cal U}\end{math} be
an open covering of \begin{math} X\end{math}
 of the form \begin{displaymath} {\cal
U}=\{ U\} \cup \{ U_{\iota } | \iota \in {\cal I}\} ,
\mbox{ such that } \end{displaymath}
\begin{description}
\item[Covering (i)] \begin{math} U\end{math}  is
an open neighbourhood of infinity and \begin{math}
\pi ^{-1}(U)\end{math} is an unbranched \begin{math}
n\end{math}-fold covering of \begin{math}
U\end{math}.
\item[Covering (ii)] \begin{math} U_{\iota }\end{math}
contains the domain excluded from \begin{math}
U\end{math} with index \begin{math}
\iota \end{math}.
\item[Covering (iii)] \begin{math} U_{\iota }\cap
U_{\iota '}=\emptyset \end{math} if \begin{math}
\iota \neq \iota '\end{math}.
\end{description}
Then we define the set of cocycles \begin{math}
C^{1}({\cal U},\pi _{*}({\cal O}^{*}_{Y}))\end{math}
to be the set of holomorphic functions
\begin{displaymath} g_{\iota }:U\cap U_{\iota }\rightarrow
GL(n,{\Bbb C}),\lambda \mapsto g_{\iota }(\lambda ),\mbox{ such that }
\end{displaymath} \begin{description}
\item[Cocycle (i)] for all \begin{math} \iota \in {\cal I}\end{math}
there exists some holomorphic map \begin{displaymath}
g_{+,\iota }:U_{\iota }\rightarrow GL(n,{\Bbb C}),\mbox{ such that }
g_{+,\iota }g_{\iota }=g_{\iota }\mbox{\it diagonal}
(\mu _{1},\ldots ,\mu _{n}),\end{displaymath}
where \begin{math}
\mu _{i}\end{math} is the restriction of \begin{math}
\mu \end{math} to the \begin{math}
i\end{math}-sheet over \begin{math}
U\cap U_{\iota }\end{math}.
\item[Cocycle (ii)] For all \begin{math} \iota \in
{\cal I} \ \det (g_{\iota })\end{math}
has trivial winding number around the
excluded domain with index \begin{math}
\iota \end{math} (the total residue of the
form \begin{math} \det
^{-1}(g_{\iota })d\det (g_{\iota })\end{math} on
the set \begin{math} U_{\iota }\end{math} is
zero).
\item[Cocycle (iii)] There exists some \begin{math}
K>0\end{math}, such that for all \begin{math}
|\iota |\geq K \ g_{\iota }\end{math} extends to
a holomorphic function on \begin{math}
U\cup X\setminus U_{\iota }\end{math}, which
is equal to \begin{math} h_{0}\end{math}
at \begin{math} \lambda =\infty \end{math}.
\item[Cocycle (iv)] \begin{math} \sum_{\iota } \|
g_{\iota }-h_{0}\| _{l,\epsilon ,\iota }<\infty \end{math}
for all \begin{math} l\in {\Bbb N}\end{math}
with some \begin{math} \epsilon \end{math}
depending on \begin{math} l\end{math}.
\end{description}
Moreover, we endow \begin{math} C^{1}({\cal
U},\pi _{*}({\cal O}^{*}_{Y}))\end{math} with the
topology defined by the metrics \begin{displaymath}
d\left( (g_{\iota })_{\iota \in {\cal I}},(\tilde{g}
_{\iota })_{\iota \in {\cal I}}\right) =\sum_{\iota } \|
g_{\iota }-\tilde{g} _{\iota }\| _{l,\epsilon ,\iota } \end{displaymath}
for all \begin{math} l\in {\Bbb N}\end{math}
with some \begin{math} \epsilon \end{math}
depending on \begin{math} l\end{math}.
Condition (i) ensures that the cocycles
describe holomorphic vector bundles over \begin{math}
X\end{math}, which are \begin{math}
\pi _{*}({\cal O})\end{math} modules. Such
vector bundles are the direct images of
line bundles over \begin{math}
Y\end{math}. Due to condition (i) the
asymptotic and total degrees of these
line bundles are equal to \begin{math}
(1,1,\ldots )\end{math} in the sense of
Remark~\ref{Remark2.1}. Finally
condition (iii) is analogous to the
assumption (ii) on admissible divisors.
\begin{Proposition} \label{Proposition5.1}
The line bundles defined by these
cocycles of \begin{math} C^{1}({\cal
U},\pi _{*}({\cal O}^{*}_{Y}))\end{math} are
equivalent in the modified sense to some
element of \begin{math} \mbox{\it Jacobian} (Y)\end{math}.
Moreover, the corresponding map \begin{math}
C^{1}({\cal U},\pi _{*}({\cal O}^{*}_{Y}))\rightarrow
\mbox{\it Jacobian} (Y)\end{math} is continuous.
\end{Proposition}
\begin{Theorem} \label{Theorem5.1}
The action of the tensor product with \begin{math}
L(x)\end{math} on holomorphic
line bundles induces a continuous action
of \begin{math} {\Bbb R}\end{math} on \begin{math}
\mbox{\it Jacobian} (Y)\end{math},
which is denoted by \begin{displaymath}
{\Bbb R}\times \mbox{\it Jacobian} (Y)\rightarrow
\mbox{\it Jacobian} (Y),(x,[D])\mapsto
T_{x}[D].\end{displaymath} \end{Theorem}
The proof\footnote{This
proof may be passed over to
Theorem~\ref{Theorem5.2}.} of this proposition is similar
to the proof of
Proposition~\ref{Proposition4.1}.
First we need a lemma analogous to the
deformation Lemma.
\begin{Lemma} \label{Lemma5.1a}
Let \begin{math} g\end{math}  and \begin{math}
\tilde{g} \end{math} be analytic maps
from the circle \begin{math} \{\lambda \in {\Bbb C}|\
|\lambda -\lambda _{0}|=R\}\rightarrow GL(n,{\Bbb C}
)\end{math}, such that \begin{math}
\| g-\unity \| <1\end{math} and \begin{math} \|
\tilde{g} -\unity \| <1\end{math},
respectively. Here the norm denotes the
supremum norm on the circle. Due to
Lemma~\ref{Lemma4.1a} both elements of
the Loop group admit a Birkhoff
factorization \begin{math} g=g_{+}g_{+}\end{math}
and \begin{math} \tilde{g} =\tilde{g}
_{-}\tilde{g} _{+}\end{math},
respectively. Then the following
estimate holds: \begin{displaymath}
\| g_{-}-\tilde{g} _{-}\| \leq
\frac{2\| g-\tilde{g} \| }{(1-\|
g-\unity \| )(1-\| \tilde{g} -\unity \| )} .\end{displaymath}
\end{Lemma}
The proof uses the same arguments as the
proof of Lemma~\ref{Lemma4.1a}.
\hspace*{\fill }\begin{math} \Box \end{math}

\noindent
Proof of
Proposition~\ref{Proposition5.1}: Let \begin{math}
f=(f_{1},\ldots ,f_{n})\end{math} be a \begin{math}
n\end{math}-tuple of invertible
holomorphic functions on \begin{math}
U\end{math}. These functions together
define a section of the sheaf \begin{math}
{\cal O}^{*}_{Y}\end{math} over \begin{math}
\pi ^{-1}(U)\end{math}, where the
restriction of this section to the \begin{math}
i\end{math}-th sheet is given by \begin{math}
f_{i}\end{math} for all \begin{math}
i=1,\ldots ,n\end{math}. By abuse of
notation this section is denoted by \begin{math}
f=(f_{1},\ldots ,f_{n})\end{math}.
Furthermore, let \begin{math}
(\tilde{g} _{+,\iota })_{\iota \in {\cal I}}\end{math}
be any sequence of holomorphic
functions \begin{displaymath}
\tilde{g} _{+,\iota }:U_{\iota }\rightarrow
GL(n,{\Bbb C}),\lambda \mapsto \tilde{g}
_{+,\iota }(\lambda ).\end{displaymath}
Then the line bundles corresponding to two elements \begin{math}
(g_{\iota })_{\iota \in {\cal I}}\end{math}  and \begin{math}
(\tilde{g} _{\iota })_{\iota \in {\cal I}}\end{math}
of \begin{math} C^{1}({\cal
U},\pi _{*}({\cal O}^{*}_{Y}))\end{math} are
equivalent in the modified sense, if \begin{displaymath}
\tilde{g} _{+,\iota }\tilde{g} _{\iota }=g_{\iota }\mbox{\it diagonal}
(f_{1},\ldots ,f_{n})\mbox{ for all }
\iota \in {\cal I}.\end{displaymath}
Due to assumption (iii) on the cocycles
for large \begin{math} |\iota |\end{math}
the Birkhoff factorization around the
open sets \begin{math} U_{\iota }\end{math}
determines \begin{math} \tilde{g} _{\iota }\end{math}
as a function depending only on
\begin{math} g_{\iota }\end{math} and \begin{math}
f=(f_{1},\ldots ,f_{n})\end{math} :
\begin{displaymath} (h_{0}^{-1}\tilde{g}
_{+,\iota }h_{0})(h_{0}^{-1}\tilde{g}
_{\iota })=h_{0}^{-1}g_{\iota }\mbox{\it diagonal}
(f_{1},\ldots ,f_{n}).\end{displaymath}
In order to prove the first statement of
the proposition
it suffices to show that the line
bundles corresponding to the cocycles of
\begin{math} C^{1}({\cal
U},\pi _{*}({\cal O}^{*}_{Y}))\end{math} are
equivalent in the modified sense to the
line bundles corresponding to some
integral divisor \begin{math} D\end{math}
of asymptotic degree \begin{math}
(1,1,\ldots )\end{math} over some open
neighbourhood \begin{math} U_{-}\end{math}
of \begin{math} \lambda =\infty \end{math} of \begin{math}
{\Bbb P}_{1}\end{math}. Now we claim that to
each cocycle \begin{math} (g_{\iota })_{\iota \in
{\cal I}}\end{math} there exists an
invertible holomorphic function \begin{math}
f=(f_{1},\ldots ,f_{n})\end{math} on \begin{math}
\pi ^{-1}(U)\end{math}, which is equal to \begin{math}
1\end{math}  at all covering points of
infinity, and an element \begin{math}
[D]\end{math}  of \begin{math}
\mbox{\it Jacobian} (Y)\end{math}, such that the Birkhoff
factorization of \begin{math}
h_{0}^{-1}g_{\iota }\mbox{\it diagonal}
(f_{1},\ldots ,f_{n})\end{math} is given
by \begin{math} \tilde{g}
_{+,\iota }h_{0}^{-1}h_{\iota }(D)\end{math}
whenever \begin{math} |\iota |\end{math} is
large enough. The function \begin{math}
h_{\iota }(D)\end{math} was defined ahead of
Theorem~\ref{Theorem4.3}. For this
purpose we introduce the Banach space of
sequences \begin{math} (f_{\iota })_{\iota \in
{\cal I}'}=(f_{1,\iota },\ldots ,f_{n,\iota }))_{\iota \in
{\cal I}'}\end{math} of holomorphic
functions \begin{displaymath}
f_{\iota }:\pi ^{-1}\left( U\cup (X\setminus U_{\iota })\right)
\rightarrow {\Bbb C},\end{displaymath} which are
equal to \begin{math} 0\end{math} at all
covering points of infinity, and the
norm \begin{displaymath}
\left\| (f_{\iota })_{\iota \in {\cal I}'}\right\| =
\sum_{\iota \in {\cal I}'} \left\| f_{\iota }\right\|
_{l,\epsilon ,\iota }<\infty \mbox{ with some fixed
\begin{math} l\end{math} and \begin{math}
\epsilon \end{math}.} \end{displaymath}
The set \begin{math} {\cal I}'\end{math}
contains the indices corresponding to
all excluded domains over some
neighbourhood of infinity of \begin{math}
{\Bbb P}_{1}\end{math}.
For all elements \begin{math} (f_{\iota })_{\iota \in
{\cal I}'}\end{math} of this Banach space, the
infinite product \begin{math}
\prod_{\iota } (1+f_{\iota })\end{math} defines
a holomorphic function on \begin{math}
U\cap U_{l,\epsilon }\end{math}. With the help of the
Birkhoff factorization we define a map of an open subset
of this Banach space into the Banach
space \begin{displaymath}
\left\{ (f_{\iota })_{\iota \in {\cal I}'} | \| (f_{\iota })_{\iota \in
{\cal I}'}\| <\epsilon  \right\} \rightarrow
\left\{ (f_{\iota })_{\iota \in {\cal I}'} \right\}
,(f_{\iota })_{\iota \in {\cal I}'}\mapsto
(\tilde{f} _{\iota })_{\iota \in {\cal I}'}:\end{displaymath}
\begin{displaymath} \mbox{Let } \tilde{g}
_{+,\iota }(h_{0}^{-1}\tilde{g} _{\iota })
=h_{0}^{-1}g_{\iota }\mbox{\it diagonal}
\left( \prod_{\iota '\neq \iota } (1+f_{1,\iota '}),\ldots ,
\prod_{\iota '\neq \iota } (1+f_{n,\iota '})\right) .\end{displaymath}
be the Birkhoff factorization around the set
\begin{math} U_{\iota }\end{math}.
Then \begin{math} \tilde{f} _{\iota }\end{math}
is defined as \begin{displaymath} \tilde{f}
_{\iota }=(\tilde{f} _{1,\iota },\ldots ,\tilde{f}
_{n,\iota })=\left( \frac{1}{(\tilde{g} _{\iota
})_{1,1}} -1,\ldots ,\frac{1}{(\tilde{g} _{\iota
})_{1,n}} -1\right) \end{displaymath}
for all \begin{math} \iota \in {\cal
I}'\end{math}. Now the foregoing
lemma shows that the image of this map is
contained in its domain, if \begin{math} {\cal I}'\end{math}
is chosen small enough, and furthermore,
that this map is a contraction. Hence it has an
unique fixed point \begin{math} (f_{\mbox{\it \scriptsize fix}
,\iota })_{\iota \in {\cal I}'}\end{math}. Then there
exists an integral divisor \begin{math}
D\end{math} of asymptotic
degree \begin{math} (1,1,\ldots )\end{math}, such
that for all \begin{math} \iota \in {\cal I}'\
h_{\iota }(D)\end{math} solves the following Birkhoff
factorization: \begin{displaymath}
\tilde{g} _{+,\iota }h_{0}^{-1}h_{\iota }(D)=
g_{\iota }\prod_{\iota \in {\cal I}'} (1+
f_{\mbox{\it \scriptsize fix} ,\iota }).\end{displaymath}
In fact, the solution \begin{math}
h_{\iota }(D)\end{math} of this Birkhoff
factorization diagonalizes some \begin{math} \tilde{g}
_{+,\iota }\end{math} and furthermore, the
first row of this matrix valued
holomorphic function is equal to the
first row of \begin{math}
h_{0}\end{math}. Then the \begin{math}
i\end{math}-th column of this function is
equal to the restriction of a
meromorphic vector valued function \begin{math}
v(D)\end{math}  with \begin{math}
v(D)_{1}=1\end{math} to the \begin{math}
i\end{math}- sheet of the Riemann
surface \begin{math} Y_{\iota }\end{math}
over \begin{math} {\Bbb P}_{1}\setminus
U_{\iota }\end{math}. This proves the claim.
In order to prove the second statement
of the proposition it suffices to
show that the fixed point of the map
above depends continuously on the
cocycle \begin{math} (g_{\iota })_{\iota \in
{\cal I}}\end{math}. This is a consequence of
the estimate of the foregoing lemma.
\hspace*{\fill }\begin{math} \Box \end{math}

\noindent
For the proof of Theorem~\ref{Theorem5.1}
we again need a lemma:
\begin{Lemma} \label{Lemma5.4} There
exists some \begin{math} K\geq 0\end{math},
such that for all \begin{math} |\iota |\geq
K,\ \ln (\mu )\end{math} defines a
holomorphic function on the excluded
domain with index \begin{math} \iota \end{math}.
\end{Lemma}
Proof: Obviously \begin{math}
\ln (\mu )\end{math} defines a multi valued
holomorphic function on \begin{math}
\pi ^{-1}({\Bbb C})\end{math}.  In Theorem~\ref{Theorem1.3}
it was shown that for all \begin{math}
\epsilon >0\end{math} there exists some \begin{math}
c>0\end{math} such that \begin{displaymath}
\left| \mu _{i}-\exp (p_{i}\lambda )\right| <\frac{c}{|\lambda |}
| \exp (p_{i}\lambda )|\end{displaymath} for
all \begin{math} \lambda \in U_{0,\epsilon }\end{math}
and \begin{math} i=1,\ldots ,n\end{math}.
This implies the estimate \begin{displaymath}
\left| \ln
(\mu _{i})-(p_{i}\lambda +2n_{i}\pi \sqrt{-1}
)\right| <\frac{c'}{|\lambda |}
\end{displaymath} with some constant \begin{math}
c'>0\end{math} and some integers \begin{math}
n_{1},\ldots ,n_{n}\in {\Bbb Z}\end{math}.
Now let us extend one branch
corresponding to the integers \begin{math}
n_{i}\end{math} and \begin{math} n_{j}\end{math}
of this
function from the boundary of the
excluded domain with index \begin{math}
\iota =(i,j,k)\in {\cal I}\end{math} to the
interior of this excluded domain.
Then the foregoing estimates on
the \begin{math} i\end{math}-th and \begin{math}
j\end{math}-th sheet imply the following
condition:
\begin{displaymath} \left| n_{i}-n_{j}-k\right|
\leq \frac{2}{\epsilon } +\frac{2c'}{|\lambda |} .\end{displaymath}
If \begin{math} |\iota |\end{math} is large
enough this implies \begin{math}
n_{i}-n_{j}=k\end{math}. Due to this
condition \begin{math} \ln (\mu )\end{math}
extends to a single valued holomorphic
function on the excluded domain with
index \begin{math} \iota \end{math}.
\hspace*{\fill } \begin{math} \Box \end{math}

\noindent
Proof of Theorem~\ref{Theorem5.1}: The
foregoing lemma shows that the Birkhoff
factorization of \begin{displaymath}
h_{l,\epsilon ,\iota }(D,\cdot )\mbox{\it diagonal}
(\mu _{1}^{x},\ldots \mu _{n}^{x})\end{displaymath}
is given for large \begin{math} |\iota |\end{math}
by \begin{displaymath}
g_{+}(\cdot )Ad(\exp (2\pi x\sqrt{-1} \mbox{\it diagonal}
(n_{1},\ldots
,n_{n})))(h_{0}^{-1}h_{l,\epsilon ,\iota }(D,\cdot
)),\end{displaymath} with some
holomorphic \begin{displaymath} g_{+}:\mbox{excluded
domain with index \begin{math} \iota \end{math}
} \rightarrow GL(n,{\Bbb C}),\lambda \mapsto
g_{+}(\lambda )\end{displaymath} and \begin{math}
n_{i}-n_{j}=k\end{math}. Then
Proposition~\ref{Proposition5.1} proves the claim.
\hspace*{\fill } \begin{math} \Box \end{math}

\noindent
It is quite obvious that the subset of \begin{math}
\mbox{\it Jacobian} (Y) \end{math} of all
equivalence classes of divisors, which
are non-special in the modified sense,
is an open subset of \begin{math}
\mbox{\it Jacobian} (Y)\end{math}. All equivalence classes
of this subspace have only one integral
representative. Moreover, the topology
of this subspace is given directly by the
topology of the set of representatives
of the elements of \begin{math}
\mbox{\it Jacobian} (Y)\end{math}.
Now let \begin{math} \mbox{\it Jacobian} _{0}(Y)\subset
\mbox{\it Jacobian} (Y)\end{math}
be the subspace \begin{math}
\mbox{\it Jacobian} _{0}(Y)=\end{math}
\begin{displaymath} =\left\{
[D]\in \mbox{\it Jacobian} (Y)\left|
T_{x}[D] \mbox{ is non-special in the modified
sense for all }x\in {\Bbb R}\right. \right\}
.\end{displaymath} This subset is open, too.
In fact, let \begin{math}
[D]\end{math} be any element of \begin{math}
\mbox{\it Jacobian} _{0}(Y)\end{math}. Due to the last
theorem for all \begin{math}
x\in [0,1]\end{math} there exists an open
neighbourhood \begin{math} U_{x}\end{math}
of \begin{math} [D]\end{math} and
an open interval \begin{math}
(x-\epsilon _{x},x+\epsilon _{x})\end{math}, such that
for all \begin{math} (x',[D'])\in
(x-\epsilon _{x},x+\epsilon _{x})\times U_{x}\end{math}
the divisor \begin{math} T_{x'}[D']\end{math}
is non-special in the modified sense.
The open covering \begin{math} \{
(x-\epsilon _{x},x+\epsilon _{x})\ | x\in [0,1]\} \end{math}
of the compact interval \begin{math}
[0,1]\end{math}  has a finite
subcovering. Then the intersection of the
corresponding open neighbourhoods \begin{math}
U_{x}\end{math} is an open neighbourhood
of \begin{math} D\end{math} contained in \begin{math}
\mbox{\it Jacobian} _{0}(Y)\end{math}. Hence this subset
is open.
\begin{Theorem} \label{Theorem5.2}
The map \begin{math} [D(\cdot )]:
\mbox{\it Isospectral} (Y)\rightarrow
\mbox{\it Jacobian} (Y)\end{math}
induces a homeomorphism
between \begin{math} \mbox{\it Isospectral} (Y)\end{math}
and \begin{math} \mbox{\it Jacobian} _{0}(Y)\end{math}.
\end{Theorem}
Proof\footnote{The rest of this section
may be passed over. It contains the
proof of this theorem.}: Theorem~\ref{Theorem3.1} and
Lemma~\ref{Lemma5.1} shows that \begin{math}
D(q)\end{math} is non-special in the
modified sense for all \begin{math}
q\in \mbox{\it Isospectral} (Y)\end{math}
and Theorem~\ref{Theorem3.3}
implies the relation \begin{displaymath}
[D(T_{x}q)]=T_{x}[D(q)].\end{displaymath}
Hence the map \begin{displaymath}
D(\cdot ):\mbox{\it Isospectral} (Y)\rightarrow \mbox{\it Divisors}
(Y),q\mapsto D(q)\end{displaymath} induces another map
denoted by \begin{displaymath}
[D(\cdot )]:\mbox{\it Isospectral} (Y)\rightarrow
\mbox{\it Jacobian} _{0}(Y),q\mapsto
[D(q)].\end{displaymath} Due
to Lemma~\ref{Lemma5.1} there exists a
map \begin{displaymath} {\Bbb R}\times
\mbox{\it Jacobian} _{0}(Y)\rightarrow
\mbox{\it vector valued functions on }
Y,(x,[D]\mapsto v(x,[D])=\left( \begin{array}{c}
v_{1}(x,[D]) \\
\vdots \\
v_{n}(x,[D]) \end{array} \right) ,\end{displaymath}
such that \begin{displaymath}
\phi _{v(x,[D])}:{\cal O}^{n}_{{\Bbb P}_{1}}\rightarrow
\pi _{*}({\cal O}_{T_{x}D}),(f_{1},\ldots
,f_{n})\mapsto \sum_{i}
f_{i}v_{i}(x,[D])\end{displaymath} is an
isomorphism and \begin{math} v(x,[D])\end{math}
takes the same values as the columns of \begin{math}
h_{0}\end{math} at the covering points of infinity.
The proof of
Proposition~\ref{Proposition4.1} and the
preceeding theorem show that all Taylor
coefficients of \begin{math} v(x,[D])\end{math}
at the covering points of infinity are
continuous functions on \begin{math} {\Bbb R}
\times \mbox{\it Jacobian} _{0}(Y)\end{math}, which are
periodic in \begin{math} x\end{math}
with period 1.
Hence there exists a unique continuous
and in \begin{math} x\end{math} periodic
potential \begin{displaymath} q:{\Bbb R}
\times \mbox{\it Jacobian}
_{0}(Y)\rightarrow n\times n\mbox{-matrices}
,(x,[D])\mapsto q(x,[D]),\mbox{ such that} \end{displaymath}
\begin{displaymath}
\pi ^{*}(h_{0}(p\lambda +q(x,[D]))h_{0}^{-1})v(x,[D])-v(x,[D])\ln
(\mu )\end{displaymath} is holomorphic on
the neighbourhood of Lemma~\ref{Lemma3.5}
and at all covering points of infinity
equal to zero. The diagonal part of \begin{math}
q(x,[D])\end{math} does not depend on \begin{math}
x\end{math} and \begin{math} [D]\end{math}.
Now let \begin{math} f(x,[D])\end{math}
be the first entry of this function: \begin{displaymath}
f(x,[D])=\left( \pi ^{*}(h_{0}(p\lambda
+q(x,[D]))h_{0}^{-1})v(x,[D])-v(x,[D])\ln
(\mu )\right) _{1}.\end{displaymath}
Then the function \begin{displaymath}
\pi ^{*}(h_{0}(p\lambda
+q(x,[D]))h_{0}^{-1})v(x,[D])-v(x,[D])(\ln
(\mu )-f(x,[D]))\end{displaymath} extends to a
meromorphic function on \begin{math} Y\end{math}.
Indeed, the function given above
is a multi valued meromorphic function on \begin{math}
Y\end{math}. Then by definition of \begin{math}
f\end{math} this function is single
valued and the first component is equal
to zero. Now we claim that \begin{math}
v(x,[D])\end{math} is differentiable
with respect to \begin{math} x\end{math}
and that the derivative is equal to
\begin{displaymath}
\frac{\partial v(x,[D])}{\partial x}
=\pi ^{*}(h_{0}(p\lambda
+q(x,[D]))h_{0}^{-1})v(x,[D])-v(x,[D])(\ln
(\mu )-f(x,[D])).\end{displaymath}
One way to prove this claim is to
improve Proposition~\ref{Proposition5.1}
and Proposition~\ref{Proposition4.1} and
to show directly that this function is
differentiable with respect to \begin{math}
x\end{math} (see
footnote~\ref{footnote26}). But the proofs of these two
propositions are rather extensive. There
is another less tedious way to prove
this claim: We define
for all \begin{math} x\in {\Bbb R}\end{math}
and all \begin{math} \lambda \in U\setminus \{
\infty \} \end{math} with
some neighbourhood \begin{math} U\end{math}
of infinity in \begin{math} X\end{math}
the function \begin{math} g(x,\lambda )\in
GL(n,{\Bbb C})\end{math}
such that \begin{displaymath}
\pi ^{*}(g(x,\cdot ))v(0,[D])=v(0,[D])\mu ^{-x}.\end{displaymath}
Due to Proposition~\ref{Proposition5.1} this function has a
decomposition \begin{math}
g(x,\cdot)=g_{+}(x,\cdot )g_{-}(x,\cdot
)\end{math} into a continuous function \begin{displaymath}
g_{+}:{\Bbb R}\times {\Bbb C}\rightarrow GL(n,{\Bbb C}
),(x,\lambda )\mapsto g_{+}(x,\lambda ),\end{displaymath}
which is holomorphic with respect to \begin{math}
\lambda \end{math},
times a continuous function \begin{displaymath}
g_{-}:{\Bbb R}\times U\rightarrow GL(n,{\Bbb C}
),(x,\lambda )\mapsto g_{-}(x,\lambda ),\end{displaymath}
which is again holomorphic with respect
to \begin{math} \lambda \end{math} and
equal to \begin{math} \unity \end{math} for
\begin{math} \lambda =\infty \end{math}, such
that\footnote{The function
\begin{math} (\pi ^{*}(g_{-}(x,\cdot
))v(0,[D]))_{1}\end{math} corresponds to
the function \begin{math}
\prod_{\iota }(1+f_{\iota })\end{math}
of Proposition~\ref{Proposition5.1}. }
\begin{displaymath} v(x,[D])=\frac{\pi ^{*}(g_{-}(x,\cdot
))v(0,[D])}{(\pi ^{*}(g_{-}(x,\cdot
))v(0,[D]))_{1}} .
\end{displaymath} Moreover, define
the \begin{math} n\times
n\end{math}-matrix valued function
\begin{displaymath} a:{\Bbb R}\times U\rightarrow
n\times n\mbox{-matrices} ,(x,\lambda )\mapsto
a(x,\lambda )\end{displaymath} such that \begin{displaymath}
a(x,\cdot )v(x,[D])=v(x[D])\ln (\mu ).\end{displaymath}
Then we saw above that this function has
a decomposition \begin{displaymath}
a(x,\cdot )=a_{+}(x,\cdot
)+a_{-}(x,\cdot ),\mbox{ with } \end{displaymath}
\begin{displaymath}
a_{+}(x,\lambda )=h_{0}(p\lambda +q(x,[D]))h_{0}^{-1}
\mbox{ and } \end{displaymath} \begin{displaymath}
\pi ^{*}(a_{-}(x,\lambda ))v(x,[D])=
\pi ^{*}(h_{0}(p\lambda +q(x,[D]))h_{0}^{-1})v(x,[D])-v(x,[D])\ln
(\mu ).\end{displaymath} Then all these
functions satisfy the integral equation \begin{displaymath}
\int_{0}^{x}
g_{+}(t,\lambda )a(t,\lambda )g_{-}(t,\lambda )dt=
g_{+}(x,\lambda )g_{-}(x,\lambda )-
g_{+}(0,\lambda )g_{-}(0,\lambda ),\end{displaymath}
or more generally\footnote{\label{footnote26}
It is easy to see that the left hand
side is differentiable with respect
to \begin{math} x\end{math} at the point
\begin{math} x=y\end{math} and that the
derivative is equal to \begin{math}
a(x,\lambda )\end{math}. On the other
side it is possible to improve Proposition~\ref{Proposition5.1}
and Proposition~\ref{Proposition4.1} and
to show directly that \begin{math} g_{+}\end{math}
and \begin{math} g_{-}\end{math} are
differentiable with respect to \begin{math}
x\end{math}. Then Lemma~\ref{Lemma5a} is
an obvious consequence.} \begin{equation}
\int_{x}^{y}
g_{+}^{-1}(x,\lambda )g_{+}(t,\lambda )a(t,\lambda )
g_{-}(t,\lambda )g_{-}^{-1}(y,\lambda )dt=
g_{+}^{-1}(x,\lambda )g_{+}(y,\lambda )
-g_{-}(x,\lambda )g_{-}^{-1}(y,\lambda
).\label{g5.1} \end{equation}
Now we need a \begin{Lemma} \label{Lemma5a}
Due to the integral
equation~(\ref{g5.1}), \begin{math}
g_{+}(\cdot ,\lambda )\end{math} and \begin{math}
g_{-}(\cdot ,\lambda )\end{math} obey the
integral equations \begin{displaymath}
\int_{0}^{x}
a_{+}(t,\lambda )g_{+}^{-1}(t,\lambda )dt
=\unity -g_{+}^{-1}(x,\lambda )
\mbox{ and } \end{displaymath} \begin{displaymath}
\int_{0}^{x}
a_{-}(t,\lambda )g_{-}(t,\lambda )dt=g_{-}(x,\lambda )-\unity , \mbox{
respectively.} \end{displaymath}
\end{Lemma}
Proof: We define \begin{math}
G_{+,l}(x,y,\lambda )=
g_{+}^{-1}(x,\lambda )\cdot \end{math}
\begin{displaymath} \cdot
\int\limits_{x\leq t_{1}\leq \ldots \leq t_{l}\leq y}
\mbox{\it Ad} (g_{+}(t_{1},\lambda
))(a_{+}(t_{1},\lambda ))
\ldots
\mbox{\it Ad} (g_{+}(t_{l},\lambda
))(a_{+}(t_{l},\lambda ))
dt_{1}\ldots dt_{l}\cdot g_{+}(y,\lambda )
\end{displaymath} and \begin{math}
G_{-,l}(x,y,\lambda )=
g_{-}(x,\lambda )\cdot \end{math}
\begin{displaymath} \cdot
\int\limits_{x\leq t_{1}\leq \ldots \leq t_{l}\leq y}
\mbox{\it Ad} (g_{-}^{-1}(t_{1},\lambda
))(a_{-}(t_{1},\lambda ))
\ldots
\mbox{\it Ad} (g_{-}^{-1}(t_{l},\lambda
))(a_{-}(t_{l},\lambda ))
dt_{1}\ldots dt_{l}\cdot
g_{-}^{-1}(y,\lambda ).\end{displaymath}
Now we claim that for all \begin{math}
L\in {\Bbb N}_{0}\end{math} the following
equation holds: \begin{displaymath}
g_{+}^{-1}(x,\lambda )g_{+}(y,\lambda )
-g_{-}(x,\lambda )g_{-}^{-1}(y,\lambda )=
\end{displaymath} \begin{displaymath}
=\sum_{l=1}^{L} G_{-,l}(x,y,\lambda )
-\sum_{l=1}^{L} (-1)^{l}G_{+,l}(x,y,\lambda )+
\sum_{l=0}^{L} (-1)^{l}\int_{x}^{y}
G_{+,l}(x,t,\lambda )a(t)G_{-,L-l}(t,y,\lambda )dt.
\end{displaymath} In fact, for \begin{math}
L=0\end{math}  this is just the integral
equation (\ref{g5.1}). Furthermore, this
integral equation implies the
equations: \begin{displaymath}
\sum_{l=0}^{L} (-1)^{l}\int_{x}^{y}
G_{+,l}(x,t,\lambda )a(t)G_{-,L-l}(t,y,\lambda )dt=
\end{displaymath} \begin{displaymath}
=\sum_{l=0}^{L+1} (-1)^{l}\int_{x}^{y}
G_{+,l}(x,t,\lambda )a(t)G_{-,L+1-l}(t,y,\lambda )dt
+G_{-,L+1}(x,y,\lambda )-(-1)^{L+1}G_{+,L+1}(x,y,\lambda )
\end{displaymath} for all \begin{math}
L\in {\Bbb N}_{0}\end{math}. The inductive use of
these equations proves the claim. The
same arguments as in Section~\ref{Section1}
gives the bounds: \begin{displaymath}
\| G_{+,l}(x,y,\lambda )\| \leq \frac{C}{l!}
\mbox{ and } \| G_{-,l}(x,y,\lambda )\| \leq \frac{C}{l!}
\end{displaymath} with some constant \begin{math}
C>0\end{math} depending on \begin{math}
x,y,\lambda \end{math}. Hence the sum \begin{displaymath}
\sum_{l=1}^{\infty } G_{-,l}(x,y,\lambda )
-\sum_{l=1}^{\infty } (-1)^{l}G_{+,l}(x,y,\lambda )
\end{displaymath} converges for any
fixed \begin{math} x,y,\lambda \end{math}.
Moreover, the limit of the first sum is
holomorphic with respect to \begin{math}
\lambda \end{math} for all \begin{math} \lambda \in
U\setminus \{\infty \} \end{math} and
equal to zero for \begin{math} \lambda =\infty
\end{math}. The second sum converges to
an entire function with respect to \begin{math}
\lambda \end{math}. Then we have \begin{displaymath}
\sum_{l=1}^{\infty }
G_{-,l}(x,y,\lambda )=\unity -g_{-}(x,\lambda )g_{-}^{-1}(y,\lambda )
\mbox{ for all } \lambda \in U, \mbox{ and }
\end{displaymath} \begin{displaymath}
-\sum_{l=1}^{\infty }
(-1)^{l}G_{+,l}(x,y,\lambda )=g_{+}^{-1}(x,\lambda )g_{+}(y,\lambda )
-\unity  \mbox{ for all } \lambda \in {\Bbb C}, \mbox{ respectively.}
\end{displaymath} Now let \begin{math} \tilde{g}
_{+}(\cdot ,\lambda )\end{math} and \begin{math}
\tilde{g} _{-}(\cdot ,\lambda )\end{math} be
the unique solutions of the integral
equations: \begin{displaymath}
\int_{0}^{x} \tilde{g}
_{+}(t,\lambda )\mbox{\it Ad} (g_{+}(t,\lambda ))
(a_{+}(t,\lambda ))dt
=\unity -\tilde{g} _{+}(x,\lambda )\mbox{ and }
\end{displaymath} \begin{displaymath}
\int_{0}^{x} \tilde{g}
_{-}(t,\lambda )\mbox{\it Ad} (g_{-}^{-1}(t,\lambda ))
(a_{-}(t,\lambda ))dt
=\tilde{g} _{-}(x,\lambda )-\unity , \mbox{ respectively.}
\end{displaymath} These solutions can be
given in terms of the infinite sums
above: \begin{displaymath}
\tilde{g} _{+}(x,\lambda )g_{+}(x,\lambda )=
\sum_{l=0}^{\infty } (-1)^{l}G_{+,l}(0,x,\lambda )
\mbox{ and } \end{displaymath} \begin{displaymath}
\tilde{g} _{-}(x,\lambda )g_{-}^{-1}(x,\lambda )=
\sum_{l=0}^{\infty } G_{+,l}(0,x,\lambda ),
\mbox{ respectively.} \end{displaymath}
We have seen above that both infinite
sums on the right hand side converge to
\begin{math} \unity \end{math}. Hence \begin{math}
\tilde{g} _{+}(x,\lambda )\end{math} is equal
to \begin{math} g_{+}^{-1}(x,\lambda )\end{math}
 and \begin{math} \tilde{g} _{-}(x,\lambda )\end{math}
is equal to \begin{math}
g_{-}(x,\lambda )\end{math}. This proves the
lemma.
\hspace*{\fill } \begin{math} \Box \end{math}

\noindent
Conclusion of the proof of
Theorem~\ref{Theorem5.2}:
Lemma~\ref{Lemma5a} now implies that \begin{math}
g_{+}(x,\lambda )\end{math}  and \begin{math}
g_{-}(x,\lambda )\end{math} are differentiable
with respect to \begin{math} x\end{math}
 and that the derivatives are given by \begin{displaymath}
\frac{\partial g_{+}}{\partial x}
(x,\lambda )=g_{+}(x,\lambda )a_{+}(x,\lambda ) \mbox{ and }
\frac{\partial g_{-}}{\partial x}(x,\lambda )
=a_{-}(x,\lambda
)g_{-}(x,\lambda ),\mbox{ respectively.}
\end{displaymath}
Then we have \begin{displaymath}
(\pi ^{*}(g_{-}(x,\cdot ))v(0,[D]))_{1}
=\exp \left( \int_{0}^{x}
f(t,[D])dt\right).\end{displaymath}
This shows that \begin{displaymath}
\frac{\partial v(x,[D])}{\partial x}
=\pi ^{*}(h_{0}(p\lambda +q(x,[D]))h_{0}^{-1})v(x,[D])-v(x,[D])(\ln
(\mu )-f(x,[D])).\end{displaymath} By
definition \begin{math} q(x,[D])\end{math}
and \begin{math} f(x,[D])\end{math} are
differentiable with respect to \begin{math}
x\end{math} up to the same order as
\begin{math} v(x,[D])\end{math}.
Hence \begin{math} v(x,[D])\end{math},
\begin{math} f(x,[D])\end{math}
and \begin{math} q(x,[D])\end{math} are
smooth functions with respect to \begin{math} x\end{math}.
Due to the differential equation for \begin{math}
g_{+}(x,\lambda )\end{math} we have \begin{displaymath}
g_{+}(x,\lambda )=h_{0}g^{-1}(x,\lambda ,q(\cdot
,[D]))h_{0}^{-1}.\end{displaymath} By
definition of \begin{math} g(x,\lambda )\end{math}
this implies \begin{displaymath}
\pi ^{*}(F(\cdot ,q(\cdot
,[D])))v(0,[D])=v(0,[D])\mu .\end{displaymath}
This shows that \begin{math} [D]\mapsto
q(\cdot ,[D])\end{math} defines a map \begin{math}
\mbox{\it Jacobian} _{0}(Y)\rightarrow
\mbox{\it Isospectral} (Y)\end{math},
such that the composition with \begin{math}
[D(\cdot )]\end{math} is the identity map of \begin{math}
\mbox{\it Jacobian} _{0}(Y)\end{math}. In Section~\ref{Section3}
we have already proven that the
composition of \begin{math} D\end{math}
with this map is the identity map of \begin{math}
\mbox{\it Isospectral} (Y)\end{math}. All Taylor coefficients
of \begin{math} q(x,[D])\end{math} with
respect to \begin{math} x\end{math} can
be given in terms of the Taylor
coefficients of \begin{math} v(x,[D])\end{math}
with respect to \begin{math} \lambda ^{-1}\end{math}
at the covering points of infinity (see
Appendix~\ref{Appendixa}).
Hence the map \begin{displaymath} \mbox{\it Jacobian} _{0}(Y)\rightarrow
\mbox{\it Isospectral} (Y),[D]\mapsto
q(\cdot ,[D])\end{displaymath}
is continuous. Theorem~\ref{Theorem1.3},
the proof of the implication (i)\begin{math}
\Rightarrow \end{math} (ii) of
Theorem~\ref{Theorem4.3} and
Lemma~\ref{Lemma5.1a}
prove that the map \begin{math}
[D(\cdot )]\end{math} is continuous, too.
This completes the proof.
\hspace*{\fill } \begin{math} \Box \end{math}

\noindent
The map \begin{math} q(x,[D])\end{math}
is defined even for all \begin{math}
(x,[D])\in {\Bbb R}\times \mbox{\it Jacobian} (Y)\end{math},
such that \begin{math} T_{x}[D]\end{math}
is non-special in the modified sense.
This observation suggests that at least
to all \begin{math} [D]\in \mbox{\it Jacobian} (Y)\end{math},
such that \begin{math} T_{x}[D]\end{math}
is non-special in the modified sense for
all \begin{math} x\end{math} in an open
dense subset of \begin{math} [0,1]\end{math},
there corresponds a potential with
singularities. Moreover, it would be
natural to establish a relation between
the kind of singularity of the potential
\begin{math} q(\cdot ,[D])\end{math}  at
the point \begin{math} x\end{math} and
the index of speciality in the modified
sense, which is equal to \begin{math} \dim \tilde{H} ^{1}(Y,{\cal
O}_{T_{x}D-\pi ^{-1}(\infty )})\end{math}.

\section{Darboux coordinates} \label{Section6}
Our next goal is to prove that the dynamical system
of the potentials are completely
integrable. In this section we take an excursion to certain
Darboux coordinates.
\begin{Definition} \label{Definition6.1}
Let \begin{math} H^{0}_{\mbox{\it
\scriptsize modified} }(Y,\Omega )\end{math}
 be the vector space of all meromorphic
differential forms, which have only
poles of order at most 1
 at all covering points of infinity.
\end{Definition}
Due to Theorem~\ref{Theorem2.1} the sum
over the residues at the covering points
of infinity of all elements of \begin{math}
H^{0}_{\mbox{\it \scriptsize modified}
}(Y,\Omega )\end{math}  is equal to
zero. Hence \begin{math}
H^{0}_{\mbox{\it \scriptsize modified}
}(Y,\Omega )\end{math} is the space of
all regular differential forms on the
singular Riemann surface obtained by the
identification of all covering points of
infinity of \begin{math} Y\end{math}  to
one multiple point (see e.g. \cite{Se}).

For each \begin{math} q\in {\cal H}^{\infty }\end{math}
let \begin{math} T_{q}{\cal H}^{\infty }\end{math}
 be the Fr\'{e}chet space of tangent vectors
\begin{math} \delta q\end{math}  at the point \begin{math}
q\end{math}. Due to Assumption~\ref{Assumption1}
the diagonal part of \begin{math}
\delta q\end{math}  is equal to zero. Now let \begin{math}
\Omega _{q}\end{math}  be the map:\begin{displaymath}
\Omega _{q}:T_{q}{\cal H}^{\infty }\rightarrow
H^{0}_{\mbox{\it \scriptsize modified}
}(Y,\Omega ), \delta q\mapsto \left( \frac{1}{\mu }
\frac{d\mu }{dq} (\delta q)\right) d\lambda .\end{displaymath}
Here we set \begin{math} \frac{d\lambda }{dq}
=0\end{math}.  Indeed, with this choice
the expression \begin{math} \frac{1}{\mu }
\frac{d\mu }{dq} (\delta q)\end{math}  becomes a
holomorphic function on the complement
of the union of all branchpoints and all
covering points of infinity and due to
Lemma~\ref{Lemma3.5}  a holomorphic
function on some neighbourhood of all
covering points of infinity. Moreover,
it is easy to see that this function is
a global section of \begin{math} {\cal O}_{b}\end{math}
and due to Assumption~\ref{Assumption1}
even a section of \begin{math}
{\cal O}_{b-\pi^{-1}(\infty )}\end{math}.
This shows that \begin{math} \Omega _{q}(\delta q)\end{math}
 is an element of \begin{math}
H^{0}_{\mbox{\it \scriptsize modified}
}(Y,\Omega )\end{math}.  More
generally, \begin{math} \Omega _{q}(\delta q)\end{math}
is equal to \begin{math} \left( \frac{1}{\mu } \frac{d\mu }{dq}
(\delta q)\right) d\lambda -\left( \frac{d\lambda }{dq}
(\delta q)\right) \frac{d\mu }{\mu } \end{math}
for arbitrary \begin{math} \frac{d\lambda }{dq}
\end{math}  and \begin{math} \frac{d\mu }{dq}
\end{math}. In fact, let \begin{math}
R(\lambda ,\mu )=0\end{math}  be the equation
defining the Riemann surface
corresponding to \begin{math} q\end{math}.
For fixed \begin{math} q\end{math}  we
have \begin{displaymath} \frac{\partial
R}{\partial \lambda } d\lambda +\frac{\partial R}{\partial
\mu } d\mu =0\end{displaymath}  and the
derivatives with respect to \begin{math}
\delta q\end{math}  obey \begin{displaymath} \frac{\partial
R}{\partial \lambda } \frac{d\lambda }{dq} +\frac{\partial
R}{\partial \mu } \frac{d\mu }{dq} +\frac{\partial
R}{\partial q} =0.\end{displaymath}
Combining these two equations we obtain \begin{displaymath}
\left( \frac{1}{\mu } \frac{d\mu }{dq}
(\delta q)\right) d\lambda -\left( \frac{d\lambda }{dq}
(\delta q)\right) \frac{d\mu }{\mu } =\left( \frac{1}{\mu }
\frac{d\mu }{dq} (dq)-\frac{d\lambda }{dq}
(\delta q)\frac{-\frac{\partial R}{\partial \lambda }
}{\mu \frac{\partial R}{\partial \mu } } \right)
d\lambda =\end{displaymath} \begin{displaymath}
=-\frac{\frac{\partial R}{\partial q}
(\delta q)}{\mu \frac{\partial R}{\partial \mu } }
d\lambda =\frac{\frac{\partial R}{\partial q}
(\delta q)}{\frac{\partial R}{\partial \lambda } }
\frac{d\mu }{\mu } .\end{displaymath}
As we mentioned before, with Assumption~\ref{Assumption1}
the space of potentials \begin{math}
{\cal H}^{\infty }\end{math} forms a coadjoint orbit and
therefore possesses a natural holomorphic
symplectic structure. Let \begin{math}
\omega \end{math}  be the \begin{math}
2\end{math} form on \begin{math} {\cal H}^{\infty
}\end{math} defined by \begin{displaymath}
\omega (\delta q,\delta \tilde{q} )=\sum_{i\neq j}
\int_{0}^{1}\frac{\delta q_{ij}(x)\delta \tilde{q}
_{ji}(x)}{p_{i}-p_{j}} dx.\end{displaymath}
With Assumption~\ref{Assumption1} it is
quite obvious that \begin{math} \omega \end{math}
is a non-degenerate 2
form and extends to a holomorphic closed
non-degenerate 2
form on the Hilbert space \begin{math}
{\cal H}\end{math}.\footnote{Those readers, which
are not interested in these Darboux
coordinates may pass over the rest of
this section.}
\begin{Lemma} \label{Lemma6.1}
Let \begin{math} q\in {\cal H}^{\infty }\end{math}
 be a potential such that the
corresponding Riemann surface satisfies
the assumption \begin{math} b_{\mbox{\it
\scriptsize analytic} }=b_{\mbox{\it
\scriptsize algebraic} }\end{math}.
 Now let \begin{math}
(\lambda _{1},\mu _{1})+\ldots +(\lambda
_{d},\mu _{d})\end{math}
be the local part of the divisor \begin{math}
D\end{math}  in some small open set of \begin{math}
Y\end{math}.  After reducing this open
set we can always attain \begin{math}
(\lambda _{1},\mu _{1})=\ldots =(\lambda _{d},\mu _{d})\end{math}.
 Then the functions \begin{math}
\sum_{i=1}^{d}
\lambda _{i}^{k}(q)\mu _{i}^{l}(q)\end{math}
extend to holomorphic functions on some
open neighbourhood of \begin{math} q\end{math}
 in the Hilbert space \begin{math} {\cal H}\supset
{\cal H}^{\infty }\end{math}  for all \begin{math}
k,l\in {\Bbb N}_{0}\end{math}. \end{Lemma}
Proof: We know already from Theorem~\ref{Theorem1.1}
that \begin{math} g(x,\cdot ,\cdot )\end{math}
is an entire function on \begin{math} {\Bbb C}
\times {\cal H}\end{math}. Hence \begin{math}
v(\lambda ,\mu ,q)\end{math}  and \begin{math}
w(\lambda ,\mu ,q)\end{math}  defined in Lemma~\ref{Lemma3.1}
with normalization \begin{math}
v_{1}(\lambda ,\mu ,q)=1\end{math}  and \begin{math}
w_{1}(\lambda ,\mu ,q)=1\end{math} are
meromorphic functions on \begin{math} {\Bbb C}
\times {\Bbb C}\times {\cal H}\end{math}.  Then
there exists some \begin{math} 1\leq i\leq
n\end{math}, such that the local part of
the divisor \begin{math} D(\tilde{q} )\end{math}
 is given by the divisor of \begin{math}
1/v_{i}(\cdot ,\cdot ,\tilde{q} )\end{math}
 for all \begin{math} \tilde{q} \end{math}
in some open neighbourhood of \begin{math}
q\end{math}. Hence we have \begin{displaymath}
\sum_{i=1}^{d} \lambda _{i}^{k}(\tilde{q}
)\mu _{i}^{l}(\tilde{q} )=\frac{1}{2\pi \sqrt{-1} }
\int_{\Gamma (\tilde{q} )} \lambda ^{k}\mu
^{l}\frac{\frac{\partial
v_{i}}{\partial \lambda } d\lambda +\frac{\partial
v_{i}}{\partial \mu } d\mu }{v_{i}} ,\end{displaymath}
where \begin{math} \Gamma (\tilde{q} )\end{math}
 is some loop on the Riemann surface \begin{math}
Y(\tilde{q} )\end{math}  corresponding
to the potential \begin{math} \tilde{q} \end{math}
 around the local part of the divisor \begin{math}
D(\tilde{q} )\end{math}  for all \begin{math}
\tilde{q} \end{math} in some open neighbourhood of
\begin{math} q\end{math}.  Now it is easy to see
that the right hand side is holomorphic
on some open neighbourhood of \begin{math}
q\end{math}.  In fact, Corollary~\ref{Corollary1.1}
gives an estimate for \begin{math}
\| F(\lambda ,q)-F(\lambda ,\tilde{q} )\| \end{math}
in terms of \begin{math} \| q-\tilde{q}
\| \end{math}, which ensures that there
exists some \begin{math} \epsilon >0\end{math},
 such that the above given formula is
valid for all \begin{math} \| q-\tilde{q}
\| <\epsilon \end{math}.
\hspace*{\fill } \begin{math} \Box \end{math}

\noindent
With the next theorem we finish our
short excursion to the parametrization
of the space of potentials given by the
values of \begin{math} \lambda \end{math}  and
\begin{math} \mu \end{math}  at all points
of the divisors. It is shown that these
are almost\footnote{In
Example~\ref{Example9.1} this
coordinates fail to be one to one.}
global coordinates and, moreover, are
Darboux coordinates of the symplectic
manifold. The book \cite{PT} describes
from this point of view Hill's equation
as a completely integrable system with
action angle variables, given by these
values of \begin{math} \lambda \end{math}  and
\begin{math} \ln(\mu )\end{math}  at all
points of the divisor. In fact, in this
case the Dirichlet isospectral sets are
exactly the Lagrangian submanifolds
defined by the property that the values
of \begin{math} \lambda \end{math}  at all
points of the divisor are kept fixed.
The next theorem generalizes the `basic'
Theorem~2.8 and Theorem~3.5. It should
be possible to carry over other parts of
the beautiful analysis given in that
book. After this theorem we want to
return to our isospectral sets.
\begin{Theorem} \label{Theorem6.1}
Let \begin{math} q\in{\cal H}^{\infty }\end{math}
 be a potential such that the
corresponding Riemann surface fulfills
the assumption \begin{math} b_{\mbox{\it \scriptsize analytic}
}=b_{\mbox{\it \scriptsize algebraic} }\end{math}.
If \begin{math} \tilde{q} \in {\cal H}^{\infty
}\end{math} is another potential such
that the values of \begin{math} \lambda \end{math}
 and \begin{math} \mu \end{math} at all
points of the divisors \begin{math} D(q)\end{math}
 and \begin{math} D(\tilde{q} )\end{math}
and the corresponding multiplicities are
equal up to permutation of the points,
then \begin{math} q=\tilde{q} \end{math}.
 Moreover these coordinates are almost
Darboux coordinates in the sense that \begin{displaymath}
\omega (\delta q,\delta \tilde{q} )=\sum_{i\in D(q)}
\frac{d\lambda _{i}}{dq}
(\delta q)\frac{d\ln (\mu _{i})}{dq} (\delta \tilde{q}
)-\frac{d\lambda _{i}}{dq}
(\delta \tilde{q} )\frac{d\ln (\mu _{i})}{dq}
(\delta q).\end{displaymath} This means that in
terms of these coordinates the
symplectic form is given by \begin{displaymath}
\omega =\sum_{i} d\lambda _{i}\wedge d\ln(\mu _{i}).\end{displaymath}
\end{Theorem}
Proof\footnote{Those readers, which are
not interested in this proof may jump
to the next section.} : Let \begin{math} R(\lambda ,\mu ,q)=0\end{math}
 and \begin{math} R(\lambda ,\mu ,\tilde{q} )=0\end{math}
 be the two defining equations of the
Riemann surfaces corresponding to \begin{math}
q\end{math}  and \begin{math} \tilde{q}
\end{math}. We claim that \begin{math}
R(\lambda ,\mu ,\tilde{q} )/\left( \mu \frac{\partial
R(\lambda ,\mu ,q)}{\partial \mu } \right) \end{math}
 is a section of the sheaf \begin{math}
{\cal O}_{b-D(q)}\end{math}  over \begin{math}
\pi ^{-1}({\Bbb C})\end{math} of the Riemann surface \begin{math}
Y(q)\end{math}  corresponding to \begin{math}
q\end{math},  and moreover, that this
function is bounded uniformly on \begin{math}
U_{0,\epsilon }\end{math}  for all \begin{math}
\epsilon >0\end{math}  by \begin{math} c/|\lambda |\end{math},
 with some \begin{math} c>0\end{math}.
In fact the divisor of the denominator
is equal to \begin{math} b\end{math}
and \begin{math} R(\lambda .\mu ,\tilde{q} )\end{math}
 is assumed to have zeroes at all the
points of \begin{math} D(q)\end{math}.
Now let \begin{math} \mu _{i}(q)\end{math}
and \begin{math} \mu _{i}(\tilde{q} )\end{math}
 be the solution of \begin{math}
R(\lambda ,\mu ,q)=0\end{math}  and \begin{math}
R(\lambda ,\mu ,\tilde{q} )=0\end{math}  on the \begin{math}
i\end{math}-th sheet of some
neighbourhood of \begin{math} \lambda =\infty
\end{math}. Then on the \begin{math}
j\end{math}-sheet of this neighbourhood
\begin{math}
R(\lambda ,\mu ,\tilde{q} )/\left( \mu \frac{\partial
R(\lambda ,\mu ,q)}{\partial \mu } \right) \end{math}
is equal to \begin{displaymath}
\frac{\mu _{j}(q)-\mu _{j}(\tilde{q}
)}{\mu _{j}(q)} \prod_{i\neq j}
\frac{\mu _{j}(q)-\mu _{i}(\tilde{q}
)}{\mu _{j}(q)-\mu _{i}(q)}
=\frac{\mu _{j}(q)-\mu _{j}(\tilde{q}
)}{\mu _{j}(q)} \prod_{i\neq j} \left(
1+\frac{\mu _{i}(q)-\mu _{i}(\tilde{q}
)}{\mu _{j}(q)-\mu _{i}(q)} \right) .\end{displaymath}
By definition of \begin{math} U_{0,\epsilon }, \left|
\frac{\mu _{i}(q)}{\mu _{j}(q)-\mu _{i}(q)} \right|
\end{math} is bounded on this set. Then
Theorem~\ref{Theorem1.3} shows the
claim. Due to Theorem~\ref{Theorem3.2}
and Theorem~\ref{Theorem3.1} \begin{math}
R(\lambda ,\mu ,\tilde{q} )/\left( \mu \frac{\partial
R(\lambda ,\mu ,q)}{\partial \mu } \right) \end{math}
can be written as \begin{math}
\sum_{i=1}^{n} f_{i}\frac{w_{i}}{wv} \end{math}
with holomorphic entire functions \begin{math}
f_{i}\end{math}  on \begin{math} {\Bbb C}\end{math}.
Moreover, \begin{math} f_{i}\end{math}
are bounded near \begin{math} \lambda =\infty \end{math}
 by \begin{math} c/|\lambda |\end{math}.
Hence they are all zero, and \begin{math}
R(\lambda ,\mu ,\tilde{q} )\end{math}  is zero
on the Riemann surface corresponding to \begin{math}
q\end{math}. Then \begin{math} q\end{math}
 and \begin{math} \tilde{q} \end{math}
correspond to the same Riemann surface
and due to Section~\ref{Section3}  \begin{math}
q\end{math}  and \begin{math} \tilde{q} \end{math}
are equal. Now let \begin{math} q_{t,\tilde{t}
}\end{math} be the potential \begin{math}
q_{t,\tilde{t} }=q+t\delta q+\tilde{t} \delta \tilde{q}
\end{math}.  Let \begin{math} v(q_{t,\tilde{t}
})\end{math}  and \begin{math} w(q_{t,\tilde{t}
})\end{math}  be the solutions of Lemma~\ref{Lemma3.1}
with normalization \begin{math} v_{1}(q_{t,\tilde{t}
})=1=w_{1}(q_{t,\tilde{t} })\end{math}.
The function \begin{math} (\mu _{t,\tilde{t}
})^{x}=\exp (x\ln (\mu _{t,\tilde{t} }))\end{math}
 is of course a multivalued function on \begin{math}
\pi ^{-1}({\Bbb C})\end{math}  of the Riemann
surface corresponding to \begin{math} q_{t,\tilde{t}
}\end{math}. Hence \begin{math}
v(x,t,\tilde{t} )=\pi ^{*}(h_{0}g(x,\lambda ,q_{t,\tilde{t}
})h_{0}^{-1})v(q_{t,\tilde{t} })\mu _{t,\tilde{t}
}^{-x}\end{math}  and \begin{math} w(x,t,\tilde{t}
)=\frac{1}{w(q_{t,\tilde{t} })v(q_{t,\tilde{t}
})} \mu _{t,\tilde{t} }^{x}w(q_{t,\tilde{t}
})\pi ^{*}(h_{0}g^{-1}(x,\lambda ,q_{t,\tilde{t}
})h_{0}^{-1})\end{math} are multivalued
meromorphic solutions of (\ref{g3.1a})
and (\ref{g3.1b}) corresponding to the
potential \begin{math} T_{x}q_{t,\tilde{t}
}\end{math},  respectively. With the
choice \begin{math} \frac{d\mu }{dq} =0\end{math}
 the functions \begin{math} \frac{\partial
v(x,t=0,0)}{\partial t} ,\ \frac{\partial
v(x,0,\tilde{t} =0)}{\partial \tilde{t} } ,\ \frac{\partial
w(x,t=0,0)}{\partial t} \end{math}  and \begin{math}
\frac{\partial w(x,0,\tilde{t} =0)}{\partial \tilde{t} }
\end{math}  become multivalued
meromorphic functions on the Riemann
surface \begin{math} Y\end{math}
corresponding to \begin{math} q\end{math}.
Let us assume that the matrix \begin{math}
p\end{math}  is invertible. Otherwise
the transformation \begin{math} p\mapsto
p+a\unity \end{math} corresponds to the
transformation \begin{math} (\lambda ,\mu )\mapsto
(\lambda ,\mu \exp (-a\lambda ))\end{math} without
change of the Riemann surface. Let \begin{math}
\kappa _{t,\tilde{t} }\end{math}  be the
meromorphic function \begin{math}
-\mu _{t,\tilde{t} }\frac{d\lambda }{d\mu _{t,\tilde{t} }} \end{math}
of the Riemann surface corresponding to
\begin{math} q_{t,\tilde{t} }\end{math}.
Theorem~\ref{Theorem1.3} shows that at
the \begin{math} i\end{math}-th covering
point of infinity \begin{math} \kappa _{t,\tilde{t} }\end{math}
is equal to \begin{math} 1/p_{i}\end{math}.
It is quite obvious that the poles of \begin{math}
\kappa =\kappa _{0,0}\end{math} are the branchpoints of the
covering map \begin{math} \pi ^{-1}({\Bbb C})\rightarrow
{\Bbb C}\end{math}, induced by the
holomorphic function \begin{math} \mu \end{math}.
More precisely, the branching divisor
of this covering map is given by the
divisor \begin{math} b-(\kappa )\end{math}.
Now we need two lemmata.
\begin{Lemma} \label{Lemma6.2}
Let \begin{math} P(\delta q,\delta \tilde{q} )\end{math}
 be the form \begin{displaymath} P(\delta q,\delta \tilde{q}
)=\lambda \frac{d\mu }{\mu } \int_{0}^{1}\left( \frac{\partial
(\kappa _{0,\tilde{t} =0}w(x,0,\tilde{t} =0))}{\partial \tilde{t} }
h_{0}ph_{0}^{-1}\frac{\partial v(x,t=0,0)}{\partial
t} -\right. \end{displaymath} \begin{displaymath}
\left. -\frac{\partial
(\kappa _{t=0,0}w(x,t=0,0))}{\partial t}
h_{0}ph_{0}^{-1}\frac{\partial v(x,0,\tilde{t} =0)}{\partial
\tilde{t} } \right) dx.\end{displaymath} With the
choice \begin{math} \frac{d\mu }{dq} =0\
P(\delta q,\delta \tilde{q} )\end{math} becomes a
meromorphic differential form on the
Riemann surface \begin{math} Y\end{math}
 with poles only at the branchpoints of
the covering map induced by \begin{math}
\mu \end{math}  and the covering points of
infinity. Furthermore, this form does
not depend on the normalization of the
solutions \begin{math} v(0,t,\tilde{t} )\end{math}
 and \begin{math} w(0,t,\tilde{t} )\end{math}
 of (\ref{g3.1a}) and (\ref{g3.1b}),
respectively, whenever \begin{math}
w(0,t,\tilde{t} )v(0,t,\tilde{t} )=1\end{math}.
The sum of residues at the covering
points of infinity of \begin{math}
P(\delta q,\delta \tilde{q} )\end{math}  is equal to
\begin{math} 2\pi \sqrt{-1} \omega (\delta q,\delta \tilde{q} )\end{math}.
\end{Lemma}
Proof: The different values of \begin{math}
v(x,t,\tilde{t} )\end{math}  and \begin{math}
w(x,t,\tilde{t} )\end{math}  are
obtained by multiplication with \begin{math}
\exp (2\pi kx\sqrt{-1} )\end{math} and \begin{math}
\exp (-2\pi kx\sqrt{-1} )\end{math} with \begin{math}
k\in {\Bbb Z}\end{math}, respectively. An
easy calculation shows that \begin{math}
P(\delta q,\delta \tilde{q} )\end{math} does not depend
on \begin{math} k\end{math} and
therefore is single valued. Let \begin{math}
f(t,\tilde{t} )\end{math} be an
arbitrary smooth function with values in
the meromorphic functions on the Riemann
surface corresponding to \begin{math}
q_{t,\tilde{t} }\end{math}  and set \begin{math}
\tilde{v} (x,t,\tilde{t} )=f(t,\tilde{t}
)v(x,t,\tilde{t} )\end{math}  and \begin{math}
\tilde{w} (x,t,\tilde{t} )=1/f(t,\tilde{t}
)w(x,t,\tilde{t} )\end{math}. Then we
have:\begin{displaymath} \int_{0}^{1}\frac{\partial
(\kappa _{0,\tilde{t} =0}\tilde{w}
(x,0,\tilde{t} =0))}{\partial \tilde{t} }
h_{0}ph_{0}^{-1}\frac{\partial \tilde{v}
(x,t=0,0)}{\partial
t} dx-\end{displaymath} \begin{displaymath}
-\int_{0}^{1} \frac{\partial
(\kappa _{t=0,0}\tilde{w} (x,t=0,0))}{\partial t}
h_{0}ph_{0}^{-1}\frac{\partial \tilde{v}
(x,0,\tilde{t} =0)}{\partial
\tilde{t} } dx-\end{displaymath} \begin{displaymath}
-\int_{0}^{1}\frac{\partial
(\kappa _{0,\tilde{t} =0}w(x,0,\tilde{t} =0))}{\partial \tilde{t} }
h_{0}ph_{0}^{-1}\frac{\partial v(x,t=0,0)}{\partial
t} dx+\end{displaymath} \begin{displaymath}
+\int_{0}^{1} \frac{\partial
(\kappa _{t=0,0}w(x,t=0,0))}{\partial t}
h_{0}ph_{0}^{-1}\frac{\partial v(x,0,\tilde{t} =0)}{\partial
\tilde{t} } dx=\end{displaymath} \begin{displaymath}
=\frac{1}{f(0,0)}
\frac{\partial f(t=0,0)}{\partial
t} \frac{\partial }{\partial \tilde{t}
} \int_{0}^{1}
\kappa _{0,\tilde{t} =0}w(x,0,\tilde{t} =0)
h_{0}ph_{0}^{-1}v(x,0,\tilde{t} =0)dx-\end{displaymath}
\begin{displaymath} -\frac{1}{f(0,0)}
\frac{\partial f(0,\tilde{t} =0)}{\partial
\tilde{t} } \frac{\partial }{\partial t} \int_{0}^{1}
\kappa _{t=0,0}w(x,t=0,0)
h_{0}ph_{0}^{-1}v(x,t=0,0)dx.\end{displaymath}
The function \begin{math} \frac{\partial
g(x,\lambda ,q)}{\partial \lambda } \end{math} is a
solution of the differential equation \begin{displaymath}
\left( \frac{d}{dx} +q(x)+\lambda p\right) \frac{\partial
g(x,\lambda ,q)}{\partial \lambda } +pg(x,\lambda ,q)=0,\
\frac{\partial g(0,\lambda ,q)}{\partial \lambda }
=0.\end{displaymath} Due to Lemma~\ref{Lemma1.1}
\begin{math} \frac{\partial g(1,\lambda ,q)}{\partial
\lambda } \end{math} is given by \begin{displaymath}
\frac{\partial g(1,\lambda ,q)}{\partial \lambda }
=g(1,\lambda ,q)\int_{0}^{1}
g^{-1}(x,\lambda ,q)pg(x,\lambda ,q)dx.\end{displaymath}
Hence we have \begin{displaymath} \kappa _{t,\tilde{t}
}\int_{0}^{1} w(x,t,\tilde{t}
)h_{0}ph_{0}^{-1}v(x,t,\tilde{t}
)dx=-\kappa_{t,\tilde{t}
}w(0,t,\tilde{t} )\pi ^{*}\left(
F^{-1}(\lambda ,q_{t,\tilde{t} })\frac{\partial
F(\lambda ,q_{t,\tilde{t} })}{\partial \lambda } \right)
v(0,t,\tilde{t}
)=\end{displaymath} \begin{displaymath}
=-\frac{\kappa _{t,\tilde{t}
}}{\mu _{t,\tilde{t}
}} \frac{d\mu _{t,\tilde{t}
}}{d\lambda } =1.\end{displaymath} This shows
that \begin{math} P(\delta q,\delta \tilde{q} )\end{math}
 does not depend on the normalization of
\begin{math} v(0,t,\tilde{t} )\end{math}
 and \begin{math} w(0,t,\tilde{t} )\end{math},
whenever \begin{math} w(0,t,\tilde{t}
)v(0,t,\tilde{t} )=1\end{math}. Therefore \begin{math}
P(\delta q,\delta \tilde{q} )\end{math} have poles only
at the branchpoints of the covering map
induced by \begin{math} \mu \end{math}
and the covering points of infinity. In
order to prove the last statement we
claim that at all covering points of
infinity \begin{math} P(\delta q,\delta \tilde{q} )\end{math}
has the same residues as the form \begin{displaymath}
\lambda d\lambda \int_{0}^{1}
\frac{\partial w(x,t=0,0)}{\partial
t} h_{0}ph_{0}^{-1}\frac{\partial
v(x,0,\tilde{t} =0)}{\partial \tilde{t} }
dx-\end{displaymath} \begin{displaymath}
\lambda d\lambda \int_{0}^{1}
\frac{\partial w(x,0,\tilde{t} =0)}{\partial
\tilde{t} } h_{0}ph_{0}^{-1}\frac{\partial
v(x,t=0,0)}{\partial t}
dx\end{displaymath}
as well with the choice \begin{math} \frac{\partial
\mu }{\partial q} =0\end{math} as with the
choice \begin{math} \frac{\partial \lambda }{\partial
q} =0\end{math}.  The first is true
because \begin{math} \kappa \frac{d\mu }{\mu }
=-d\lambda \end{math}  and \begin{math}
\frac{d\kappa _{t,\tilde{t} }}{dt}
\end{math}  has a zero of order \begin{math}
2\end{math}  at all covering points of
infinity. The second is true because the
difference of \begin{math} \frac{\partial
v}{\partial t} \end{math} with the
choice \begin{math} \frac{d\mu }{dq} =0\end{math}
minus \begin{math} \frac{\partial v}{\partial
t} \end{math} with the choice \begin{math}
\frac{d\lambda }{dq} =0\end{math} is equal to \begin{math}
\frac{dv}{d\lambda } \frac{\partial v}{\partial
t} \end{math}, where \begin{math} \frac{\partial
\lambda }{\partial t} \end{math} is taken with
the choice \begin{math} \frac{d\mu }{dq}
=0\end{math}  and similar statements
about \begin{math} \frac{\partial v}{\partial
\tilde{t} } ,\ \frac{\partial w}{\partial
t} \end{math} and \begin{math} \frac{\partial
w}{\partial
\tilde{t} } \end{math}.  In fact, all
these differences have zeroes of order
al least 3  at
all covering points of infinity. Hence
the sum of the residues at the covering
points of infinity of \begin{math}
P(\delta q,\delta \tilde{q} )\end{math}  is equal to
the sum of residues at the covering
points of infinity of \begin{displaymath}
d\lambda \int_{0}^{1}
\frac{\partial w(x,t=0,0)}{\partial
t} \pi ^{*}(h_{0}p\lambda h_{0}^{-1})\frac{\partial
v(x,0,\tilde{t} =0)}{\partial \tilde{t} }
dx-\end{displaymath} \begin{displaymath}
-d\lambda \int_{0}^{1} \frac{\partial w(x,0,\tilde{t} =0)}{\partial
\tilde{t} }
\pi ^{*}(h_{0}p\lambda h_{0}^{-1})\frac{\partial
v(x,t=0,0)}{\partial t} dx.\end{displaymath}
Let \begin{math} \sum \end{math} denote
the sum over all sheets of the covering
map \begin{math} \pi \end{math}. Then
Lemma~\ref{Lemma3.3} implies \begin{displaymath}
\sum \int_{0}^{1}
\frac{\partial w(x,t=0,0)}{\partial
t} \pi ^{*}(h_{0}p\lambda h_{0}^{-1})\frac{\partial
v(x,0,\tilde{t} =0)}{\partial \tilde{t} }
dx-\end{displaymath} \begin{displaymath}
-\sum \int_{0}^{1} \frac{\partial w(x,0,\tilde{t} =0)}{\partial
\tilde{t} }
\pi ^{*}(h_{0}p\lambda h_{0}^{-1})\frac{\partial
v(x,t=0,0)}{\partial t} dx=\end{displaymath}
\begin{displaymath}
=\int_{0}^{1}tr\left( h_{0}^{-1}
\sum \frac{\partial v(x,0,\tilde{t} =0)}{\partial \tilde{t}
} w(x,0,0)
h_{0}p\lambda h_{0}^{-1}\sum \frac{\partial
v(x,t=0,0)}{\partial t} w(x,0,0)
h_{0}-\right. \end{displaymath} \begin{displaymath}
\left. -h_{0}^{-1}
\sum \frac{\partial v(x,t=0,0)}{\partial
t} w(x,0,0)
h_{0}p\lambda h_{0}^{-1}\sum \frac{\partial
v(x,0,\tilde{t} =0)}{\partial \tilde{t} } w(x,0,0)
h_{0}\right) dx.\end{displaymath}
Now the asymptotic expansions of \begin{math}
v\end{math}  and \begin{math} w\end{math}
in Theorem~\ref{Theorem1.2} and Theorem~\ref{Theorem1.3}
imply that the leading term in the
asymptotic expansion of \begin{math}
\left( h_{0}^{-1}\left(
\sum \frac{\partial v(x,t=0,0)}{\partial
t} w(x,0,0)\right)
h_{0}\right) _{ij}\end{math} and \begin{math}
\left( h_{0}^{-1}\left( \sum \frac{\partial
v(x,0,\tilde{t} =0)}{\partial \tilde{t} } w(x,0,0)\right)
h_{0}\right) _{ij}\end{math} is given by
\begin{math}
1/\lambda \frac{\delta q_{ij}}{p_{j}-p_{i}} \end{math}
and \begin{math} 1/\lambda \frac{\delta \tilde{q}
_{ij}}{p_{j}-p_{i}} \end{math} for \begin{math}
i\neq j\end{math}, respectively.
Hence the sum of the residues at the
covering points of infinity of \begin{math}
P(\delta q,\delta \tilde{q} )d\lambda \end{math} is equal to
\begin{displaymath} -2\pi \sqrt{-1} \sum_{i\neq
j} \int_{0}^{1} \frac{\delta \tilde{q}
_{ij}(x)}{p_{j}-p_{i}}
(p_{j}-p_{i})\frac{\delta q_{ji}(x)}{p_{i}-p_{j}}
dx=2\pi \sqrt{-1} \sum_{i\neq
j}\int_{0}^{1}\frac{\delta q_{ij}(x)\delta \tilde{q}
_{ji}(x)}{p_{i}-p_{j}} dx.\end{displaymath}
This completes the proof of the lemma.
\hspace*{\fill } \begin{math} \Box \end{math}
\begin{Lemma} \label{Lemma6.3}
Let \begin{math} Q(\delta q,\delta \tilde{q} )\end{math}
 be the form \begin{math} Q(\delta q,\delta \tilde{q} )
=\end{math} \begin{displaymath}
=-\frac{\kappa }{\mu } d\mu \int_{0}^{1}
\frac{\partial \lambda }{\partial t}
w(x,0,0)h_{0}ph_{0}^{-1}
\frac{\partial v(x,0,\tilde{t} =0)}{\partial \tilde{t}
} dx+\end{displaymath} \begin{displaymath}
+\frac{\kappa }{\mu } d\mu \int_{0}^{1}
\frac{\partial \lambda }{\partial \tilde{t} }
w(x,0,0)h_{0}ph_{0}^{-1}
\frac{\partial v(x,t=0,0)}{\partial t}
dx.\end{displaymath} With
the choice \begin{math} \frac{d\mu }{dq}
=0\ Q(\delta q,\delta \tilde{q} )\end{math}  is a
single valued meromorphic differential
form on the Riemann surface \begin{math} Y\end{math}
 with poles only at the branchpoints
of the covering map induced       by \begin{math}
\mu \end{math}  and the poles of \begin{math}
v(q)\end{math}. If \begin{math} v(q_{t,\tilde{t}
})\end{math}  has a simple pole of order
1  at \begin{math}
(\lambda _{0}(t,\tilde{t} ),\mu _{0}(t,\tilde{t} ))\end{math}
 and if this point is no branchpoint,
the local residue of the form \begin{math}
Q(\delta q,\delta \tilde{q} )\end{math}  at this
point \begin{math}
(\lambda _{0}(t,\tilde{t} ),\mu _{0}(t,\tilde{t} ))\end{math} is
given by \begin{displaymath} 2\pi \sqrt{-1} \left(
\frac{\partial \lambda _{0}}{\partial t}
\frac{\partial \ln (\mu _{0})}{\partial \tilde{t} }
-\frac{\partial \lambda _{0}}{\partial \tilde{t} }
\frac{\partial \ln (\mu _{0})}{\partial t}
\right) .\end{displaymath} Moreover, if \begin{math}
v(q)\end{math} has no poles and zeroes at the
branchpoints, the total sum of all
residues at the branchpoints of the
covering map induced by \begin{math} \mu \end{math}
of \begin{math} P(\delta q,\delta \tilde{q}
)+Q(\delta q,\delta \tilde{q} )\end{math} is equal to zero.\end{Lemma}
Proof: If we multiply \begin{math}
v(x,t,\tilde{t} )\end{math} and \begin{math}
w(x,t,\tilde{t} )\end{math} by \begin{math}
\exp (2\pi kx\sqrt{-1} )\end{math}  and \begin{math}
\exp (-2\pi kx\sqrt{-1} )\end{math},
respectively, again \begin{math} Q(\delta q,\delta \tilde{q} )
\end{math} does not change. Hence it is
single valued. The statement about the
poles is obvious. If \begin{math}
f(t,\tilde{t} )\end{math} is locally a
holomorphic function such that the
divisor of \begin{math} f(t,\tilde{t} )\end{math}
 is locally equal to \begin{math} D(q_{t,\tilde{t}
})\end{math}, \begin{displaymath} \tilde{Q}
(\delta q,\delta \tilde{q} )=
-\frac{\kappa }{\mu }
d\mu \int_{0}^{1} \frac{\partial \lambda }{\partial t}
\tilde{w} (x,0,0)h_{0}ph_{0}^{-1}
\frac{\partial \tilde{v} (x,0,\tilde{t} =0)
}{\partial \tilde{t}
} dx+\end{displaymath} \begin{displaymath}
+\frac{\kappa }{\mu }
d\mu \int_{0}^{1} \frac{\partial \lambda }{\partial \tilde{t} }
\tilde{w} (x,0,0)h_{0}ph_{0}^{-1}
\frac{\partial \tilde{v} (x,t=0,0)
}{\partial t} dx\end{displaymath}
with \begin{math} \tilde{v}
(x,t,\tilde{t} )=f(t,\tilde{t} )v(x,t,\tilde{t} )\end{math}  and
\begin{math} \tilde{w}
(x,t,\tilde{t} )=w(x,t,\tilde{t} )/f(t,\tilde{t} )\end{math}  is
locally holomorphic. Hence the
residue at \begin{math}
(\lambda _{0}(0,0),\mu _{0}(0,0))\end{math} of \begin{math}
Q(\delta q,\delta \tilde{q} )\end{math} is equal to
the residue of \begin{displaymath}
Q(\delta q,\delta \tilde{q} )-\tilde{Q} (\delta q,\delta \tilde{q}
)=\left( \frac{1}{f(0,0)}
\frac{\partial f(0,\tilde{t} =0)}{\partial \tilde{t} }
\frac{\partial \lambda }{\partial t}
-\frac{1}{f(0,0)} \frac{\partial f(t=0,0)}{\partial t}
\frac{\partial \lambda }{\partial \tilde{t} } \right) \frac{d\mu }{\mu }
.\end{displaymath} This residue is equal
to \begin{displaymath} 2\pi \sqrt{-1}
\left( \frac{\partial \lambda _{0}}{\partial t}
\frac{\partial \ln (\mu _{0})}{\partial \tilde{t} }
-\frac{\partial \lambda _{0}}{\partial \tilde{t} }
\frac{\partial \ln (\mu _{0})}{\partial t} \right)
.\end{displaymath} To prove the last
statement we first note that \begin{math}
P(\delta q,\delta \tilde{q} )+Q(\delta q,\delta \tilde{q} )\end{math}
is equal to \begin{displaymath}
\frac{\partial }{\partial \tilde{t} }
\lambda \kappa _{0,\tilde{t} =0}
\int_{0}^{1}w(x,0,\tilde{t} =0)h_{0}ph_{0}^{-1}
\frac{\partial v(x,t=0,\tilde{t} =0)}{\partial t}
dx\frac{d\mu }{\mu } -\end{displaymath} \begin{displaymath}
-\frac{\partial }{\partial t}
\lambda \kappa _{t=0,0}
\int_{0}^{1}w(x,t=0,0)h_{0}ph_{0}^{-1}
\frac{\partial v(x,t=0,\tilde{t} =0)}{\partial \tilde{t} }
dx\frac{d\mu }{\mu } .\end{displaymath}
Now let \begin{math}
(\lambda _{0}(t,\tilde{t} ),\mu _{0}(t,\tilde{t} ))\end{math}
be a branchpoint of order \begin{math}
b_{0}\end{math}  of the covering map
induced by \begin{math} \mu \end{math}
such that \begin{math} v(x,t,\tilde{t} )\end{math}
 has no pole at \begin{math}
(\lambda _{0},\mu _{0})\end{math}. A local
coordinate on the Riemann surface is
given by \begin{math}
\nu ^{b_{0}+1}=\mu -\mu _{0}\end{math}.  Then
we claim that \begin{math} \frac{\kappa }{\mu }
d\mu \int_{0}^{1}w(x,t,\tilde{t} )
h_{0}ph_{0}^{-1}\frac{\partial v}{\partial t}
(x,t,\tilde{t} )dx\end{math}  has a pole of order \begin{math}
1\end{math} at \begin{math}
(\lambda _{0},\mu _{0})\end{math},  and the
residue of this form is locally equal to
the residue of the form \begin{displaymath}
\frac{1}{2}
\frac{d\mu }{\mu } \frac{d}{dt} \ln \left(
\frac{\partial R(\lambda ,\mu ,q_{t,\tilde{t}
})}{\partial \lambda } \right) ,\end{displaymath}
where again the derivative \begin{math}
\frac{d}{dt} \end{math} is taken with
the choice \begin{math} \frac{\partial \mu }{\partial q}
=0:\end{math} \begin{displaymath} \frac{d}{dt}
\ln \left(
\frac{\partial R(\lambda ,\mu ,q_{t,\tilde{t}
})}{\partial
\lambda } \right)
=\frac{R_{\lambda t}}{R_{\lambda }}
+\frac{R_{\lambda \lambda }}{R_{\lambda }} \frac{d\lambda }{dt}
.\end{displaymath} Here we used the
subscript to denote partial derivatives.
In order to prove this claim we need
some preparation. The divisor of \begin{math}
w(x,t,\tilde{t} )\end{math}  is given by \begin{math}
-b\end{math} minus the divisor of \begin{math}
v(x,t,\tilde{t} )\end{math}.  Due to the
assumption \begin{math}
b_{\mbox{\it \scriptsize effective} }=
b_{\mbox{\it \scriptsize analytic} }\end{math}, \begin{math}
w(x,t,\tilde{t} )\end{math}  has also no poles
and zeroes at \begin{math}
(\lambda _{0},\mu _{0})\end{math}.  Now let \begin{math}
(\lambda ,\mu )\end{math}  and \begin{math}
(\lambda ',\mu )\end{math}  be two different
points of the Riemann surface \begin{math}
Y\end{math} such that  the covering map
induced by \begin{math} \mu \end{math}
project them onto the same point \begin{math}
\mu \in {\Bbb C}\end{math}.  Let \begin{math} v'\end{math}  be the
value of \begin{math} v\end{math}
at \begin{math} (\lambda ',\mu )\end{math}.  Then
we have \begin{displaymath} \int_{0}^{1}
w(x,t,\tilde{t} )h_{0}ph_{0}^{-1}v'(x,t,\tilde{t}
)dx=\end{displaymath} \begin{displaymath}
\frac{1}{\lambda -\lambda '}
\int_{0}^{1}w(x,t,\tilde{t} )h_{0}\left( \left( \frac{d}{dx}
+q(x)+p\lambda \right) -\left( \frac{d}{dx}
+q(x)+p\lambda '\right) \right)
h_{0}^{-1}v'(x,t,\tilde{t} )dx=0.\end{displaymath}
More generally, let the prime denote
the value of the corresponding
function at the point with local coordinates \begin{math}
\nu '\end{math}  and the function without
prime denotes this function at the point
with local coordinates \begin{math} \nu \end{math}.
We can expand the function \begin{math}
\int_{0}^{1}
\kappa w(x,t,\tilde{t} )h_{0}ph_{0}^{-1}v'(x,t,\tilde{t} )dx\end{math}
in a Laurent series in \begin{math} \nu \end{math}
and \begin{math} \nu ':\ \sum_{l\geq
-b_{0},l'\geq 0}
a_{l,l'}\nu ^{l}(\nu ')^{l'}\end{math}.
Since \begin{math} \kappa \end{math} has a
pole of order \begin{math} b_{0}\end{math},
we may assume \begin{math} l\geq -b_{0}\end{math}.
For \begin{math} \nu =\nu '\end{math} this
function is equal to \begin{math} 1:\
\sum_{l+l'=L}a_{l,l'}=\delta
_{L,0}\end{math}; and for \begin{math}
\nu ^{b_{0}+1}=(\nu ')^{b_{0}+1},\nu \neq \nu '\end{math},
this function is equal to zero. With the
help of a finite Fourier transformation we
obtain \begin{math} a_{l,l'}=0,\mbox{ if }
l+l'<0\end{math}  and \begin{math}
a_{-l,l}=1/(b_{0}+1)\mbox{ for }
l=0,1,\ldots ,b_{0}\end{math}.  The form
\begin{math} \int_{0}^{1}
\kappa w(x,t,\tilde{t} )h_{0}ph_{0}^{-1}dv(x,t,\tilde{t} )dx\end{math}
has a Laurent expansion of the form \begin{displaymath}
\sum_{l\geq -b_{0},l'\geq 0}
l'a_{l,l'}\nu ^{l+l'-1}d\nu +\sum_{l\geq
-b_{0},l'\geq 0} da_{l,l'}\nu ^{l+l'}.\end{displaymath}
The second term vanishes and the first
term has a simple pole of the form \begin{math}
\sum_{l=0}^{b_{0}} \frac{l}{b_{0}+1} \frac{d\nu }{\nu }
=\frac{b_{0}}{2} \frac{d\nu }{\nu } \end{math}.
 Since \begin{math}
R_{\lambda }(\lambda ,\mu ,q_{t,\tilde{t} })\end{math}  has a
pole of order \begin{math} b_{0}\end{math}
 at \begin{math} (\lambda _{0},\mu _{0})\end{math},
this form has the same pole as the form \begin{math}
\frac{1}{2} d\ln
(R_{\lambda }(\lambda ,\mu ,q_{t,\tilde{t} })\end{math}.  This
proves the claim. If \begin{math} v(q)\end{math}
 has a no poles and zeroes at the
branchpoints of the covering map induced
by \begin{math} \mu \end{math},  the total
sum of residues at the branchpoints of
the form \begin{math} P(\delta q,\delta \tilde{q}
)+Q(\delta q,\delta \tilde{q} )\end{math}  is equal
to the sum of residues at the covering
points of infinity of the form \begin{displaymath}
\frac{1}{2} \frac{d\mu }{\mu } \left(
\frac{\partial \lambda }{\partial \tilde{t} } \frac{d}{dt} (\ln
R_{\lambda })-\frac{\partial \lambda }{\partial t} \frac{d}{d\tilde{t} } (\ln
R_{\lambda })\right) ,\end{displaymath}
where all derivatives are taken
according to the choice \begin{math} \frac{d\mu }{dq}
=0\end{math}.  At the beginning of this
section we showed that \begin{math} \frac{d\mu }{\mu }
\frac{\partial \lambda }{\partial t} =-\Omega (\delta q)\end{math}
and \begin{math} \frac{d\mu }{\mu } \frac{\partial \lambda }{\partial \tilde{t}
}
=-\Omega (\delta \tilde{q} )\end{math}  are
elements of \begin{math}
H^{0}_{\mbox{\it \scriptsize modified} }(Y,\Omega )\end{math}.
Since \begin{math} \frac{d}{dt}
\ln (\kappa )\end{math}  and \begin{math} \frac{d}{d\tilde{t} }
\ln (\kappa )\end{math}  has zeroes of order
at least 2  at
all covering points of infinity, the
total sum of all residues at the
branchpoints of \begin{math} P(\delta q,\delta \tilde{q}
)+Q(\delta q,\delta \tilde{q} )\end{math}  is also
equal to the sum of residues at the
covering points of infinity of the form \begin{displaymath}
\frac{1}{2} \Omega (\delta \tilde{q} )\frac{d}{dt}
\ln (R_{\mu })-\frac{1}{2}
\Omega (\delta q)\frac{d}{d\tilde{t} }
\ln (R_{\mu }),\end{displaymath}  where
all derivatives  are taken according to
the choice \begin{math} \frac{d\mu }{dq}
=0\end{math}. Now we claim that this is
also true if the derivatives are taken
according to the choice \begin{math} \frac{d\lambda }{dq}
=0\end{math}. In fact, the difference of
\begin{math} \frac{d}{dt} \ln (R_{\mu })\end{math}
 with the choice \begin{math} \frac{d\mu }{dq}
=0\end{math}  minus
\begin{math} \frac{d}{dt} \ln (R_{\mu })\end{math}
with the choice \begin{math} \frac{d\lambda }{dq}
=0\end{math} is equal to \begin{math} \frac{d}{d\lambda }
\ln (R_{\mu })\frac{\partial \lambda }{\partial t} \end{math},
 where the derivative is taken
according the choice \begin{math}
\frac{d\mu }{dq} =0\end{math}. Hence the derivatives
can be taken according to the
choice \begin{math} \frac{d\mu }{dq} =0\end{math}
as well as to the choice
\begin{math} \frac{d\lambda }{dq} =0\end{math}.
Let us now take all derivatives
according to the choice \begin{math} \frac{d\lambda }{dq}
=0\end{math}. \begin{displaymath}
\frac{1}{2} \Omega (\delta \tilde{q} )\frac{d}{dt}
\ln (R_{\mu })-\frac{1}{2}
\Omega (\delta q)\frac{d}{d\tilde{t} }
\ln (R_{\mu })=\end{displaymath} \begin{displaymath}
=\frac{1}{2} d\lambda \left( \frac{1}{\mu } \frac{d\mu }{d\tilde{t} }
\left( \frac{R_{t\mu }}{R_{\mu }}
+\frac{R_{\mu \mu }}{R_{\mu }} \frac{d\mu }{dt} \right)
-\frac{1}{\mu } \frac{d\mu }{dt}
\left( \frac{R_{\tilde{t} \mu }}{R_{\mu }}
+\frac{R_{\mu \mu }}{R_{\mu }} \frac{d\mu }{d\tilde{t} } \right) \right)
=\end{displaymath} \begin{displaymath}
=\frac{1}{2} \Omega (\delta \tilde{q}
)\frac{R_{t\mu }}{R_{\mu }} -\frac{1}{2}
\Omega (\delta q)\frac{R_{\tilde{t} \mu }}{R_{\mu }} .\end{displaymath}
Now let \begin{math} \mu _{i}\end{math} be
the value of \begin{math} \mu \end{math}
in the \begin{math} i\end{math}-th sheet
over \begin{math} \lambda \end{math}. Then \begin{displaymath}
R(\lambda ,\mu )=\prod_{i=1}^{n} (\mu -\mu _{i}),\
R_{\mu }=\sum_{j=1}^{n} \prod_{i\neq j}
(\mu -\mu _{i})\mbox{ and }
R_{\mu t}=\sum_{j=1}^{n} \sum_{k\neq j}
-\frac{d\mu _{k}}{dt} \prod_{i\neq k,j}
(\mu -\mu _{i}).\end{displaymath}  In the \begin{math}
j\end{math}-th sheet \begin{math}
\frac{R_{\mu t}}{R_{\mu }} \end{math} takes
the value \begin{displaymath} \sum_{k\neq
j}\frac{\frac{d\mu _{j}}{dt} +\frac{d\mu _{k}}{dt}
}{\mu _{k}-\mu _{i}} .\end{displaymath} \begin{math}
\left| \frac{d}{dt} \ln (\mu )\right| \end{math}
 is bounded on some set \begin{math}
U_{0,\epsilon }\end{math}  by \begin{math}
c|\lambda |^{-1}\end{math}.  In the beginning
of the proof of Theorem~\ref{Theorem6.1}
 we used the fact that \begin{math} \sup \{
|\mu _{k}|,|\mu _{j}|\} /|\mu _{k}-\mu _{j}|\end{math}
 is bounded on all \begin{math}
U_{0,\epsilon .}\end{math} Hence \begin{math} \left|
\frac{R_{\mu t}}{R_{\mu }} \right| \end{math}
is bounded on all \begin{math} U_{0,\epsilon }\end{math}
by \begin{math} c/|\lambda |\end{math}.  This
implies that the total sum of residues
at all branchpoints of the covering map
induced by \begin{math} \mu \end{math}
of the form \begin{math} P(\delta q,\delta \tilde{q}
)+Q(\delta q,\delta \tilde{q} )\end{math}  converges
to zero in the sense of
Remark~\ref{Remark2.1}, if \begin{math}
v(q)\end{math}  has no poles at this
branchpoints. This completes the proof
of Lemma~\ref{Lemma6.3}
\hspace*{\fill } \begin{math} \Box \end{math}

\noindent
Completion of the proof of
Theorem~\ref{Theorem6.1}: It is quite
obvious that \begin{math} Q(\delta q,\delta \tilde{q}
)\end{math} has no poles at the
covering points of infinity. If \begin{math}
D(q)\end{math}  has no poles at the
branchpoints of the covering map induced
by \begin{math} \mu \end{math}  and
furthermore is composed only of simple
points, the last two lemmata show that \begin{displaymath}
\omega (\delta q,\delta \tilde{q} )=\sum_{i\in D(q)}
\frac{d\lambda _{i}}{dq}
(\delta q)\frac{d\ln (\mu _{i})}{dq} (\delta \tilde{q}
)-\frac{d\lambda _{i}}{dq}
(\delta \tilde{q} )\frac{d\ln (\mu _{i})}{dq}
(\delta q),\end{displaymath} and that this sum
converges. If we look more carefully at
the proof, we see that it is also valid
without this restriction.
\hspace*{\fill } \begin{math} \Box \end{math}

\section{The tangent space of the Jacobian variety} \label{Section7}
It is known that in the case of the
Korteweg-de Vries equation the isospectral sets are not
submanifolds of the space of potentials.
Hence we cannot expect that the
isospectral sets are differentiable
manifolds with a tangent bundle.
Nevertheless we can define a subspace of
the tangent space of the space of
potentials, of tangent vectors along the
isospectral sets. By abuse of notation
this will be called the tangent space.
\begin{Definition} \label{Definition7.1}
Let \begin{math} {\cal L}_{q}\subset T_{q}{\cal H}^{\infty
}\end{math}  be the kernel of the map \begin{math}
\Omega _{q}:T_{q}{\cal H}^{\infty }\rightarrow
H^{0}_{\mbox{\it \scriptsize modified}
}(Y,\Omega )\end{math}. \end{Definition}
In general the map \begin{math} \pi :Y\rightarrow
X\end{math}  is not continuous.
Nevertheless, the holomorphic functions
on \begin{math} \pi ^{-1}(U)\end{math}
are well defined for all open sets \begin{math}
U\end{math}  of \begin{math} X\end{math}.
Hence the direct image sheaf of the
holomorphic functions \begin{math}
\pi _{*}({\cal O}_{-\pi^{-1}(\infty )},Y)
\end{math} is a well
defined sheaf over \begin{math} X\end{math}.
 Moreover, there exists a natural
homomorphism \begin{math}
H^{1}(Y,{\cal O}_{-\pi^{-1}(\infty )})\rightarrow
H^{1}(X,\pi _{*}({\cal O}_{-\pi^{-1}(\infty )})\end{math}.
Indeed, let \begin{math} \tilde{Y} \end{math}
be the set \begin{math} Y\end{math}
with the unique coarsest topology
containing the topology of \begin{math}
Y\end{math}  and the inverse image
of the topology of \begin{math} X\end{math}
 under \begin{math} \pi \end{math}.
Then there exists a natural inclusion \begin{math}
H^{1}(Y,{\cal O}_{-\pi^{-1}(\infty )})\hookrightarrow
H^{1}(\tilde{Y} ,{\cal O}_{-\pi^{-1}(\infty )})\end{math}.
It is quite easy to see that each
covering of \begin{math} \tilde{Y} \end{math}
 has a refinement, such that the first
cohomology group with respect to this
refinement is equal to the first
cohomology group with respect to some
inverse image of a covering of \begin{math}
X\end{math}.  Hence \begin{math}
H^{1}(\tilde{Y} ,{\cal O}_{-\pi^{-1}(\infty )})\end{math}
and \begin{math}
H^{1}(X,\pi _{*}({\cal O}_{-\pi^{-1}(\infty )}))\end{math}  are
isomorphic. This gives the natural
inclusion \begin{math}
H^{1}(Y,{\cal O}_{-\pi^{-1}(\infty )})\hookrightarrow
H^{1}(X,\pi _{*}({\cal O}_{-\pi^{-1}(\infty )}))\end{math}. In
Section~\ref{Section3} we already
indicated that there exists a
isomorphism of sheaves over \begin{math} X:\end{math}
\begin{displaymath}
\varepsilon _{q}^{-1}:\pi _{*}({\cal O})\rightarrow
{\cal O}_{q},f\mapsto \sum Pf,\end{displaymath}
 where \begin{math} \sum \end{math}
denotes the sum over all sheets of the
covering map \begin{math} \pi \end{math}.
Now let \begin{math} M_{n}({\cal O}_{-\infty })\end{math}
 denote the sheaf of rings of \begin{math}
n\times n\end{math}-matrices with values
in \begin{math} {\cal O}_{-\infty }\end{math}
on \begin{math} X\end{math}  and \begin{math}
{\cal O}_{-\infty ,q}\end{math}  the subsheaf
of all matrices, which commute with \begin{math}
F(\cdot ,q)\end{math}.  Then there
exists a natural inclusion \begin{math}
{\cal O}_{\infty ,q}\hookrightarrow M_{n}({\cal O}_{-\infty
})\end{math}.  Putting all this together
we have the following sequence of
homomorphisms:\begin{displaymath}
H^{1}(Y,{\cal O}_{-\pi^{-1}(\infty )})\hookrightarrow
H^{1}(X,\pi _{*}({\cal O}_{-\pi^{-1}(\infty
)}))\simeq H^{1}(X,{\cal O}_{-\infty
,q})\rightarrow H^{1}(X,M_{n}({\cal O}_{-\infty
})).\end{displaymath}
Let \begin{math} v(q)\end{math}  and \begin{math}
w(q)\end{math}  be the solutions of
(\ref{g3.1a}) and (\ref{g3.1b}) with
normalization \begin{math}
v_{1}(q)=1=w_{1}(q)\end{math}.
Furthermore, let \begin{math} U_{D}\end{math}
 be the complement of the support of the
divisor \begin{math}
D(q)\end{math}  in \begin{math} Y\end{math}.
\begin{Lemma} \label{Lemma7.1} For each
element \begin{math} c\in H^{1}(X,{\cal O}_{-\infty
,q})\end{math}  in the kernel of \begin{displaymath}
H^{1}(X,{\cal O}_{-\infty
,q})\rightarrow H^{1}(X,M_{n}({\cal O}_{-\infty
}))\end{displaymath} there exists a section \begin{math}
f\end{math}  of \begin{math} {\cal O}_{D}\end{math}
 on \begin{math} \pi ^{-1}({\Bbb C})\end{math},
such that under the homomorphism \begin{displaymath}
H^{1}(Y,{\cal O}_{-\pi^{-1}(\infty )})\hookrightarrow
H^{1}(X,\pi _{*}({\cal O}_{-\pi^{-1}(\infty
)}))\simeq H^{1}(X,{\cal O}_{-\infty
,q}) \end{displaymath} the cocycle \begin{math}
[f]\in H^{1}(Y,{\cal O}_{-\pi^{-1}(\infty )})\end{math}
induced by \begin{math} f\end{math}
with respect to the covering \begin{math}
Y=U_{D}\cup \pi ^{-1}({\Bbb C})\end{math} is
mapped onto \begin{math} c\end{math}.
Moreover, there exists a unique vector valued
meromorphic function \begin{math} \delta v\end{math}
 on \begin{math} Y\end{math},  with
poles only at the poles of \begin{math}
v\end{math}  and zeroes at all covering
points of infinity and a section \begin{math}
a_{+}\end{math} of \begin{math} M_{n}({\cal O}_{-\infty
})\end{math}  on \begin{math} {\Bbb C}\end{math},
 such that \begin{displaymath}
vf=\pi ^{*}(a_{+})v-\delta v\mbox{ and }
(\delta v)_{1}=0.\end{displaymath} \end{Lemma}
Proof: Since \begin{math} c\end{math}
is an element of the kernel of \begin{math}
H^{1}(X,{\cal O}_{-\infty
,q})\rightarrow H^{1}(X,M_{n}({\cal O}_{-\infty
}))\end{math} there exists a cochain \begin{math}
b\in C^{0}({\cal U},M_{n}({\cal O}_{-\infty }))\end{math}
 with some covering \begin{math} {\cal
U}\end{math}  of \begin{math} x\end{math},
 such that \begin{math} \delta b\end{math}
is a representative of \begin{math} c\in
C^{1}({\cal U},{\cal O}_{-\infty ,q})\end{math}.
 For any open set \begin{math} U\end{math}
 of \begin{math} {\cal U}\end{math} we
define on \begin{math} \pi ^{-1}(U)\end{math}
 the function \begin{math}
\delta v_{U}=\pi ^{*}(b_{U})v-v(\pi ^{*}(b_{U})v)_{1}\end{math}.
If \begin{math} V\end{math}  is another
element of the covering \begin{math} {\cal
U}\end{math}, on \begin{math}
\pi ^{-1}(U\cap V)\end{math} the function \begin{math}
\delta v_{U}-\delta v_{V}=\pi
^{*}(b_{U}-b_{V})v-v(\pi ^{*}(b_{U}-b_{V})v)_{1}\end{math}.
Since \begin{math} \delta b\end{math}  is an
element \begin{math} C^{1}({\cal U},{\cal O}_{-\infty
,q}),\
\pi ^{*}(b_{U}-b_{V})v=v(\pi ^{*}(b_{U}-b_{V})v)_{1}\end{math}.
Hence \begin{math} \delta v\end{math} is  a
meromorphic function on \begin{math} Y\end{math}
 with poles only at the poles of \begin{math}
v\end{math}  and zeroes at all covering
points of infinity. If \begin{math} b\end{math}
 is an element of \begin{math} C^{0}({\cal
U},{\cal O}_{-\infty ,q})\end{math}  the same
argument shows that \begin{math} \delta v=0\end{math}.
 This shows that \begin{math} \delta v\end{math}
 depends only on the element \begin{math}
c\end{math}  of \begin{math} H^{1}(X,{\cal O}_{-\infty
,q})\end{math}.  If \begin{math}
(\lambda _{0},\mu _{0})\in Y\end{math}  is a pole
of \begin{math} v\end{math}  and \begin{math}
U\end{math}  is an element of the
covering such that \begin{math}
\pi ^{-1}(U)\end{math}  contains this
pole, then the singular part of \begin{math}
(\pi ^{*}(b_{U})v)_{1}\end{math}  does
only depend on \begin{math} c\end{math}.
Indeed, if \begin{math} V\end{math}  is
another element of the covering, which
contains this pole, \begin{math}
\pi ^{*}(b_{U}-b_{V})v=v(\pi ^{*}(b_{U}-b_{V})_{1}\end{math}
and \begin{math} (\pi ^{*}(b_{U}-b_{V})_{1}\end{math}
 is holomorphic, since \begin{math} \delta b\end{math}
 is an element of \begin{math} C^{1}({\cal
U},{\cal O}_{-\infty ,q})\end{math}.  Hence \begin{math}
c\end{math}  defines a unique Mittag
Leffler distribution on \begin{math} Y\end{math},
more precisely, a global section of the
sheaf \begin{math} {\cal O}_{D}/{\cal O}\end{math}.
Due to \cite[Theorem 26.3]{Fo} there
exists a solution\begin{math} f\end{math}
 of this Mittag Leffler distribution on \begin{math}
\pi ^{-1}({\Bbb C})\end{math}.  The function \begin{math}
f\end{math}  on \begin{math} \pi ^{-1}({\Bbb C}
)\end{math}  together with the zero
function on \begin{math} U_{D}\end{math}
 defines an element of \begin{math}
C^{0}\left( \{ U_{D},\pi ^{-2}({\Bbb C})\}
,{\cal O}_{D-\pi ^{-1}(\infty )}\right) \end{math}.
 The coboundary of this element defines
an element of \begin{math}
H^{1}(Y,{\cal O}_{-\pi ^{-1}(\infty )})\end{math}
 denoted by \begin{math} [f]\end{math}.
Now we claim that under the map
\begin{math} H^{1}(Y,{\cal O}_{-\pi^{-1}(\infty )})\hookrightarrow
H^{1}(X,\pi _{*}({\cal O}_{-\pi^{-1}(\infty
)}))\simeq H^{1}(X,{\cal O}_{-\infty
,q})\ [f]\end{math}  is mapped onto \begin{math}
c\end{math}.  A representative of the
image of \begin{math} [f]\end{math}  in \begin{math}
H^{1}(X,\pi _{*}({\cal O}_{-\pi^{-1}(\infty
)}))\end{math}  is given by the
coboundary of the cochain defined by the
zero section of \begin{math}
\pi _{*}({\cal O}_{-\pi^{-1}(\infty )})\end{math} on \begin{math}
{\Bbb P}_{1}\setminus \pi (\mbox{\it support}
D)\end{math}  and
the section \begin{math} f\end{math}  of
\begin{math} \pi _{*}({\cal O}_{D-\pi^{-1}(\infty )})\end{math}
 on \begin{math} {\Bbb C}\end{math}.  Now let
\begin{math} {\cal U}\cap \{ {\Bbb C},{\Bbb P}
_{1}\setminus \pi (\mbox{\it support}
D)\} \end{math} be the
covering \begin{math} \left\{ U\cap{\Bbb C}
,U\cap({\Bbb P}_{1}\setminus \pi (\mbox{\it support}
D))|\ U\in {\cal
U}\right\} \end{math}.  Then we define
the following element of \begin{math}
C^{0}({\cal U}\cap \{ {\Bbb C},{\Bbb P}
_{1}\setminus \pi (\mbox{\it support}
D)\}
,\pi _{*}({\cal O}_{-\pi^{-1}(\infty )}))\end{math}, which
on \begin{math} U\cap {\Bbb C}\end{math} is equal
to \begin{math} f-(\pi ^{*}(b_{U})v)_{1}\end{math}
 and on \begin{math} U\cap {\Bbb P}
_{1}\setminus \pi (\mbox{\it support}
D)\end{math}  equal to
zero. The coboundary of this element is
equal to the image of \begin{math} [f]\end{math}
 under \begin{math}
H^{1}(Y,{\cal O}_{-\pi^{-1}(\infty )})\hookrightarrow
H^{1}(X,\pi _{*}({\cal O}_{-\pi^{-1}(\infty
)}))\end{math} minus the image of \begin{math}
c\end{math}  under the isomorphism \begin{math}
H^{1}(X,{\cal O}_{-\infty
,q})\simeq H^{1}(X,\pi _{*}({\cal O}_{-\pi^{-1}(\infty
)}))\end{math}. This proves the claim.
Finally we note that by construction of \begin{math}
f\end{math}  the function \begin{math}
a_{+}=\sum Pf-\sum \delta vw/wv\end{math}  is a
section of \begin{math} M_{n}({\cal O}_{-\infty
})\end{math}  on \begin{math} {\Bbb C}\end{math},
 and \begin{math} vf=\pi ^{*}(a_{+})v-\delta v\end{math}.
The functions \begin{math} a_{+}\end{math}
 on \begin{math} {\Bbb C}\end{math}  and the
function \begin{math} a_{-}=-\sum
\delta vw/wv\end{math} on \begin{math} {\Bbb P}
_{1}\setminus \pi (\mbox{\it support}
D)\end{math}  define an
element \begin{math} a\end{math}  of \begin{math}
C^{0}\left( \{ {\Bbb C},{\Bbb P}_{1}\setminus
\pi (\mbox{\it support}
D)\} ,M_{n}({\cal O}_{-\infty})\right) \end{math},
such that \begin{math} \delta a\end{math}  is
equal to the image of \begin{math} [f]\end{math}
 under \begin{math}
H^{1}(Y,{\cal O}_{-\pi^{-1}(\infty )})\hookrightarrow
H^{1}(X,\pi _{*}({\cal O}_{-\pi^{-1}(\infty
)}))\simeq H^{1}(X,{\cal O}_{-\infty
,q})\end{math}. This completes the
proof of the lemma.
\hspace*{\fill } \begin{math} \Box \end{math}
\begin{Definition} \label{Definition7.2}
Let \begin{math} H_{q,\mbox{\it
\scriptsize modified} }^{1}(Y,{\cal O})\subset
H^{1}(Y,{\cal
O}_{-\pi^{-1}(\infty )})\end{math}  be the
vector space of all cocycles
in the kernel of the homomorphism \begin{displaymath}
H^{1}(Y,{\cal O}_{-\pi^{-1}(\infty )})\hookrightarrow
H^{1}(X,\pi _{*}({\cal O}_{-\pi^{-1}(\infty
)}))\simeq H^{1}(X,{\cal O}_{-\infty
,T_{x}q})\rightarrow H^{1}(X,M_{n}({\cal O}_{-\infty
}))\end{displaymath} for all \begin{math}
x\in [0,1]\footnote{With the help of
Proposition~\ref{Proposition5.1} and
Theorem~\ref{Theorem5.1} one can
prove that this condition is equivalent
for all \begin{math} x\in
[0,1]\end{math}. Hence we may impose
this condition only for \begin{math}
x=0\end{math}. But in general \begin{math}
H_{q,\mbox{\it
\scriptsize modified} }^{1}(Y,{\cal O})\end{math}
does not coincide for all \begin{math}
q\in \mbox{\it Isospectral}
(Y)\end{math}. This vector space is
the same for all \begin{math}
q\end{math} and \begin{math} \tilde{q}
\end{math}, such that \begin{math} E^{*}(q)\otimes E(\tilde{q}
)\end{math}  is an element of the real
part of the Picard group (Compare with
Theorem~\ref{Theorem8.3}).} \end{math}.  Furthermore let
\begin{math} \tilde{H} ^{1}_{q,\mbox{\it \scriptsize
modified}}(Y,{\cal
O})\subset H^{1}(Y,{\cal O}_{-\pi^{-1}(\infty )})\end{math}
 be the preimage under the homomorphism \begin{displaymath}
H^{1}(Y,{\cal O}_{-\pi^{-1}(\infty )})\hookrightarrow
H^{1}(X,\pi _{*}({\cal O}_{-\pi^{-1}(\infty
)}))\simeq H^{1}(X,{\cal O}_{-\infty
,q})\end{displaymath} of the subspace \begin{math}
H^{1}({\Bbb P}_{1},{\cal O}_{-\infty ,q})\subset
H^{1}(X,{\cal O}_{-\infty ,q})\end{math}.
\end{Definition}
\begin{Lemma} \label{Lemma7.2} \begin{math}
\tilde{H} ^{1}_{q,\mbox{\it \scriptsize modified}}(Y,{\cal O})\end{math}
 is the same for all \begin{math} q\in
\mbox{\it Isospectral} (Y)\end{math}.  Hence we will omit the
index \begin{math} q\end{math}.
Moreover, \begin{math} \tilde{H}
^{1}_{\mbox{\it \scriptsize modified}}(Y,{\cal O})\end{math}  is
contained in \begin{math}
H_{q,\mbox{\it \scriptsize
modified} }^{1}(Y,{\cal O})\end{math}
for all \begin{math} q\in \mbox{\it Isospectral}
(Y)\end{math}. \end{Lemma}
Proof: Since \begin{math} H^{1}({\Bbb
P}_{1},M_{n}({\cal
O}_{-\infty }))=0\end{math}  we
can apply the last lemma. Since \begin{math}
{\Bbb P}_{1}\end{math}  is compact, we can
calculate the cohomology groups with
respect to some finite covering.
Moreover, for each open subset \begin{math}
U\subset {\Bbb C}\subset {\Bbb P}_{1},\
H^{1}(U,{\cal
O}_{-\infty ,q})\simeq H^{1}(\pi ^{-1}(U),{\cal
O})=0\end{math} (see
\cite[26.1]{Fo} ). Hence we can
calculate the cohomology group with
respect to some covering \begin{math} {\Bbb P}
_{1}=U_{+}\cup U_{-}\end{math}, with \begin{math}
U_{+}\subset {\Bbb C}\subset {\Bbb P}_{1}\end{math}
 and some neighbourhood \begin{math}
U_{-}\end{math}  of \begin{math} \infty
\in {\Bbb P}_{1}\end{math}.  To each element \begin{math}
c\in H^{1}({\Bbb P} _{1},{\cal O}_{-\infty ,q})
\end{math}  there exists a cochain \begin{math}
b\in C^{0}\left( \{ U_{+},U_{-}\}
,M_{n}({\cal O}_{-\infty })\right) \end{math},
 such that \begin{math} \delta b\end{math}
is a representative of \begin{math} c\end{math}
in \begin{math} C^{1}\left( \{
U_{+},U_{-}\} {\cal O}_{-\infty ,q}\right)
\end{math}. Similar arguments as in the
proof of the lemma before show that the
coboundary of the cochain \begin{math}
f\in C^{0}\left( \{
\pi ^{-1}(U_{+}),\pi ^{-1}(U_{-})\} ,{\cal
M}\right) \end{math},  with \begin{math}
f_{\pi ^{-1}(U_{+})}=(\pi ^{*}(b_{U_{+}})v)_{1}\end{math}
 and \begin{math}
f_{\pi ^{-1}(U_{-})}=(\pi ^{*}(b_{U_{-}})v)_{1}\end{math}
 is a representative in \begin{math}
C^{1}\left( \{
\pi ^{-1}(U_{+}),\pi ^{-1}(U_{-})\} ,{\cal
O}_{-\pi^{-1}(\infty )}\right) \end{math} of the
preimage of \begin{math} c\end{math}
under the inclusion \begin{math}
H^{1}(Y,{\cal
O}_{-\pi^{-1}(\infty )})\hookrightarrow H^{1}({\Bbb
P}_{1},{\cal
O}_{-\infty ,q})\end{math}.  Now
let \begin{math} \tilde{q} \in \mbox{\it Isospectral}
(Y)\end{math}
be another potential. Then it is obvious
that \begin{math} [f]\end{math}
corresponds also to an element of \begin{math} H^{1}({\Bbb
P}_{1},{\cal
O}_{-\infty ,\tilde{q} })\end{math}.
 This shows that \begin{math} \tilde{H}
^{1}_{q,\mbox{\it \scriptsize modified}}(Y,{\cal O})\end{math} does
not depend on \begin{math} q\end{math}
and furthermore, that \begin{math} \tilde{H}
^{1}_{\mbox{\it \scriptsize
modified}}(Y,{\cal O})\end{math} is
contained in \begin{math}
H^{1}_{q,\mbox{\it \scriptsize modified}}(Y,{\cal O})\end{math}.
\hspace*{\fill } \begin{math} \Box \end{math}

\noindent
For compact Riemann surfaces Serre
duality gives a non-degenerate pairing
between the first cohomology group of
the sheaf of holomorphic functions and
the vector space of holomorphic
differential forms.
With the help of Lemma~\ref{Lemma7.1} we
can write down the corresponding pairing
between \begin{math} H_{q,\mbox{\it
\scriptsize modified}
}^{1}(Y,{\cal O})\end{math}  and \begin{math}
H^{0}_{\mbox{\it \scriptsize
modified}}(Y,\Omega )\end{math}  as an
infinite sum: Let \begin{math}
[f]\in H_{q,\mbox{\it \scriptsize
modified} }^{1}(Y,{\cal O})\end{math}
be represented by a cocycle of the form
defined in Lemma~\ref{Lemma7.1}. Then
for all \begin{math} \omega \in
H_{\mbox{\it \scriptsize modified}
}^{1}(Y,\Omega )\end{math} \begin{displaymath}
\mbox{\it Res} ([f],\omega )=-\sum
\mbox{\it Res}  (f\omega ),\end{displaymath}
where the sum runs over all poles of \begin{math}
f\end{math}  on \begin{math} \pi ^{-1}({\Bbb C}
)\end{math}. This is equal to the sum of
residues of \begin{math} f\omega \end{math}
at all covering points of infinity. But
in general this sum does not converge.
If \begin{math} [f]\end{math} is an
element of \begin{math} \tilde{H}
^{1}_{\mbox{\it \scriptsize modified} }(Y,{\cal O})\end{math}  of the
form defined in Lemma~\ref{Lemma7.2} \begin{displaymath}
\mbox{\it Res} ([f],\omega )=-\sum \mbox{\it Res} \left(
f_{\pi ^{-1}(U_{+})}\omega \right) +\sum \mbox{\it Res} \left(
f_{\pi ^{-1}(U_{-}}\omega \right) ,\end{displaymath}
where the first sum runs over all poles
of \begin{math} f_{\pi ^{-1}(U_{+})}\end{math}
 on \begin{math} \pi ^{-1}(U_{+})\end{math}
 and the second sum runs over all poles
of \begin{math} f_{\pi ^{-1}(U_{-})}\end{math}
on \begin{math} \pi ^{-1}(U_{-})\end{math}.
Since \begin{math} -f_{\pi ^{-1}(U_{+})}
+f_{\pi ^{-1}(U_{-})}\end{math} is
holomorphic on \begin{math}
\pi ^{-1}(U_{+}\cap U_{-})\end{math}, we
can omit the poles on \begin{math}
\pi ^{-1}(U_{+}\cap U_{-})\end{math}.
Then the first sum becomes finite and
the second sum converges due to
Theorem~\ref{Theorem2.1}. This shows
that \begin{math} \mbox{\it Res}  \end{math} is a
well defined pairing between \begin{math}
\tilde{H} ^{1}_{\mbox{\it \scriptsize
modified} }(Y,{\cal O})\end{math}
and \begin{math} H_{\mbox{\it
\scriptsize modified} }^{1}(Y,\Omega )
\end{math}.  Now we
can state the main theorem of this
section: \begin{Theorem} \label{Theorem7.1}
\begin{description}
\item[(i)] There exists an isomorphism of
vector spaces \begin{math} d\Gamma
_{q}: H_{q,\mbox{\it \scriptsize modified} }^{1}(Y,{\cal O})\rightarrow
{\cal L}_{q}\end{math}, which is uniquely
determined by the property that for all \begin{math}
[f]\in H_{q,\mbox{\it \scriptsize
modified} }^{1}(Y,{\cal O})\end{math} \begin{displaymath}
\frac{\partial v}{\partial q}
(d\Gamma _{q}([f]))=\delta v,\end{displaymath} where \begin{math}
\delta v\end{math}  was defined in
Lemma~\ref{Lemma7.1}.
\item[(ii)] For all \begin{math}
[f]\in H_{q,\mbox{\it \scriptsize
modified} }^{1}(Y,{\cal O})\end{math}  and all \begin{math}
\delta q\in T_{q}{\cal H}^{\infty}\end{math}
the following relation holds: \begin{displaymath}
\omega \left( d\Gamma _{q}([f]),\delta q\right) =\frac{1}{2\pi \sqrt{-1} }
\mbox{\it Res} \left( [f],\Omega _{q}(\delta q)\right) \end{displaymath}
and the infinite sum on the right hand
converges.
\item[(iii)] The pairing \begin{math}
\mbox{\it Res} \end{math} between \begin{math}
\tilde{H} ^{1}_{\mbox{\it \scriptsize modified}
}(Y,{\cal O})\end{math} and \begin{math}
H_{\mbox{\it \scriptsize modified}
}^{0}(Y,\Omega )\end{math} is non-degenerate.
\item[(iv)] \begin{math}
{\cal L}_{q}\end{math}  is a maximal isotropic
subspace of \begin{math} T_{q}{\cal H}^{\infty
}\end{math} with respect to the
symplectic form \begin{math} \omega \end{math}.
\end{description}
\end{Theorem}
Proof\footnote{This proof may be passed
over.}: Let \begin{math} [f]\in
H_{q,\mbox{\it \scriptsize modified}
}^{1}(Y,{\cal O})\end{math}
be a cocycle of the form defined in
Lemma~\ref{Lemma7.1} and \begin{displaymath}
a(x)=\sum v(T_{x}q)\frac{f}{w(T_{x}q)v(T_{x}q)}
w(T_{x}q).\end{displaymath} Due to the
proof of Theorem~\ref{Theorem3.3}
\begin{math} a(x)\end{math} is equal to
\begin{math} h_{0}g(x,\cdot
,q)h_{0}^{-1}a(0)h_{0}g^{-1}(x,\cdot
,q)h_{0}^{-1}\end{math}. This implies
that \begin{math} \left[ a(x),\frac{d}{dx}
+h_{0}\left( p\lambda +q(x)\right) h_{0}^{-1}\right]
=0\end{math}.  By definition of \begin{math}
H_{q,\mbox{\it \scriptsize modified}
}^{1}(Y,{\cal O})\end{math}  and due to Lemma~\ref{Lemma7.1}
there exists for all \begin{math} x\in
[0,1]\end{math}  a decomposition
\begin{math} a(x)=a_{+}(x)+a_{-}(x)\end{math}, such
that \begin{math} a_{+}(x)\end{math}  is
an entire function from \begin{math} {\Bbb C}\end{math}
 into the \begin{math} n\times
n\end{math}-matrices and \begin{math} a_{-}(x)\end{math}
 is a meromorphic function on \begin{math}
X\end{math},  which vanishes at \begin{math}
\lambda =\infty \end{math}. Then the
commutator \begin{math}
[a(x),h_{0}(p\lambda +q(x))h_{0}^{-1}] \end{math}
has a decomposition of the same form and
both functions \begin{math}
a_{+}(x)\end{math}  and \begin{math}
a_{-}(x)\end{math}  are smooth with
respect to \begin{math} x\end{math}.
Then we define \begin{displaymath}
\delta q(x)=\left[ h_{0}^{-1}a_{+}(x)h_{0},\frac{d}{dx}
+p\lambda +q(x)\right]
=\left[ \frac{d}{dx}
+p\lambda +q(x),h_{0}^{-1}a_{-}(x)h_{0}\right]
.\end{displaymath} This equation shows
that \begin{math} \delta q(x)\end{math} is an
entire function with respect to \begin{math}
\lambda \end{math} and is holomorphic near \begin{math}
\lambda =\infty \end{math}. Therefore it does
not depend on \begin{math} \lambda \end{math}.
Moreover, it is a smooth function with
respect to \begin{math} x\end{math}. Now
we claim that \begin{math}
\delta v=-\pi ^{*}(a_{-}(0))v(q)\end{math}
satisfies the relation \begin{displaymath}
\pi ^{*}(F(\cdot ,q))\delta v+\pi ^{*}\left( \frac{\partial
F}{\partial q} (\cdot ,q)(\delta q)\right) v=\delta v\mu .\end{displaymath}
Let \begin{math} \delta v(x)\end{math}  be the
multivalued meromorphic function \begin{math}
\delta v(x)=-\pi ^{*}(a_{-}(x)h_{0}g(x,\cdot
,q)h_{0}^{-1})v\mu ^{-x}\end{math}. Since \begin{math}
a(x)\end{math} is periodic with respect
to \begin{math} x\end{math},  we have \begin{math}
\delta v=\delta v(0)=\delta v(1)\end{math}.  Furthermore, \begin{math}
\delta v(x)\end{math} is a solution of the
differential equation \begin{displaymath}
\pi ^{*}\left( \frac{d}{dx}
+h_{0}(p\lambda +q(x))h_{0}^{-1}\right)
\delta v(x)+\delta v(x)\ln (\mu )+\pi ^{*}\left(
h_{0}\delta q(x)g(x,\cdot ,q)h_{0}^{-1}\right)
v\mu ^{-x}=0.\end{displaymath} Due to
Lemma~\ref{Lemma1.1} \begin{math} \delta v(x)\end{math}
is then equal to \begin{displaymath}
\delta v(x)=\pi ^{*}\left( h_{0}g(x,\cdot
,q)h_{0}^{-1}\right) \left( \int_{0}^{x}
\pi ^{*}(h_{0}g^{-1}(t,\cdot
,q)\delta q(t)g(t,\cdot,q)h_{0}^{-1})v+\delta v(0)\right)
\mu ^{-x}.\end{displaymath} For \begin{math}
x=1\end{math}  we obtain the equation \begin{displaymath}
\delta v=\pi ^{*}(F(\cdot ,q))\delta v\mu ^{-1}+\pi ^{*}\left(
\frac{\partial F}{\partial q} (\cdot ,q)(\delta q)\right)
v\mu ^{-1}.\end{displaymath} This proves
the claim. For each \begin{math} \delta q\in
{\cal L}_{q},\ \delta v=\frac{\partial v}{\partial q} (\delta q)\end{math}
is a solution of this equation. With the
normalization \begin{math} (\delta v)_{1}=0\end{math}
this equation has only one meromorphic
solution, since there are many \begin{math}
\lambda \in {\Bbb C}\end{math}, such that \begin{math}
F(\lambda ,q)\end{math}  has \begin{math} n\end{math}
different eigenvalues. This shows that \begin{math}
\delta v\end{math} corresponding to \begin{math}
[f]\end{math} is equal to \begin{math}
\frac{\partial v}{\partial q}
(\delta q)\end{math}.
It is quite obvious that \begin{math} \delta q\end{math}
 is an element of \begin{math} {\cal L}_{q}\end{math}
if and only if \begin{displaymath}
\frac{\partial v}{\partial q}
(\delta q)\mu -\pi ^{*}(F(\cdot ,q))\frac{\partial v}{\partial q}
(\delta q)=\pi ^{*}\left(
\frac{\partial F(\cdot ,q)}{\partial q}
(\delta q)\right) v-v\frac{\partial \mu }{\partial q}
(\delta q)\end{displaymath}
is an element of the form \begin{math}
\pi ^{*}(\delta F)v\end{math}, with some entire
function \begin{math} \delta F:{\Bbb C}\rightarrow
n\times n\end{math}-matrices. Hence we have a map \begin{math}
d\Gamma _{q}:H_{q,\mbox{\it \scriptsize modified}
}^{1}(Y,{\cal O})\rightarrow {\cal L}_{q}\end{math}.
Now we claim that for each
vector valued function \begin{math} \delta v\end{math},
which vanishes at all covering points of
infinity and furthermore satisfies the
relations \begin{displaymath}
\delta v\mu -\pi ^{*}(F(\cdot ,q))\delta v=\pi ^{*}(\delta F)v\mbox{ and }
(\delta v)_{1}=0\end{displaymath} with some
entire function \begin{math} \delta F:{\Bbb C}\rightarrow
n\times n\end{math}-matrices, there
exists one and only one element \begin{math}
c\in H^{1}(X,{\cal O}_{-\infty ,q})\end{math}
in the kernel of \begin{math} H^{1}(X,{\cal
O}_{-\infty ,q})\rightarrow
H^{1}(X,M_{n}({\cal O}_{-\infty }))\end{math},
 such that \begin{math} \delta v\end{math} is
equal to the corresponding meromorphic
function constructed in
Lemma~\ref{Lemma7.1}. The relation
\begin{math} \delta v\mu -\pi ^{*}(F(\cdot
,q))\delta v=\pi ^{*}(\delta F)v\end{math}
is equivalent to the relation \begin{displaymath}
\left[ \sum \frac{\delta vw}{wv} ,F(\cdot ,q)\right]
=\delta F.\end{displaymath} To proceed further we
need \begin{Lemma} \label{Lemma7.1a}
Let the following homomorphism of sheaves
be an isomorphism: \begin{displaymath} {\cal
O}_{{\Bbb C}}^{n}\rightarrow {\cal
O}_{q},(f_{1},\ldots ,f_{n})\mapsto
\sum_{i=1}^{n} F^{i-1}(\cdot ,q)f_{i}.\end{displaymath}
Then for all \begin{math} \lambda _{0}\in {\Bbb C}\end{math}
the centralizer of \begin{math}
F(\lambda _{0},q)\end{math} is spanned by \begin{math}
\unity ,F(\lambda _{0},q),\ldots
,F^{n-1}(\lambda _{0},q)\end{math}. \end{Lemma}
Proof: Let \begin{math} R(\lambda ,\mu )=\det
(\mu \unity -F(\lambda ,q))=0\end{math} be the
eigenvalue equation and let \begin{math}
(\lambda _{0},\mu _{0})\in {\Bbb C}^{2}\end{math}
any solution of this equation. Then the
condition implies that \begin{math}
R(\lambda _{0},F(\lambda _{0},q))
(F(\lambda _{0},q)-\mu _{0}\unity
)^{-1}\end{math} \footnote{The singularity of the function
\begin{math} R(\lambda _{0},\mu )
(\mu -\mu _{0})^{-1}\end{math} at \begin{math}
\mu =\mu _{0}\end{math} may be removed.
Hence this function should be considered
to be a polynomial of degree \begin{math}
n-1\end{math} with respect to \begin{math}
\mu \end{math} as well as with respect to \begin{math}
F(\lambda _{0},q)\end{math}.}
is not equal to zero. This implies
the conclusion of the lemma.
\hspace*{\fill }\begin{math} \Box \end{math}

\noindent
Due to the assumption \begin{math}
b_{\mbox{\it \scriptsize analytic} }=
b_{\mbox{\it \scriptsize effective} }=
b_{\mbox{\it \scriptsize algebraic} }\end{math}
the condition of the foregoing lemma is
fulfilled. Then this lemma and the
equation ahead of this lemma
imply that in a neighbourhood of every
pole of \begin{math} \sum \delta vw/wv
\end{math}
there exists a meromorphic section \begin{math}
f\end{math} of \begin{math} \pi _{*}({\cal
O})\end{math}, such that \begin{math}
\sum Pf-\sum \delta vw/wv \end{math}
is holomorphic on this neighbourhood.
Therefore the singular part of \begin{math}
\sum \frac{\delta vw}{wv} \end{math} defines a Mittag
Leffler distribution on \begin{math}
\pi ^{-1}({\Bbb C})\end{math}.  Now let \begin{math}
f\end{math} be as in Lemma~\ref{Lemma7.1}
a solution of this Mittag Leffler
distribution on \begin{math} \pi ^{-1}({\Bbb C}
)\end{math}. Then \begin{math} vf+\delta v\end{math}
is a vector valued section of the sheaf \begin{math}
{\cal O}_{D(q)}\end{math} over \begin{math}
\pi ^{-1}({\Bbb C})\end{math}. Hence there
exists an entire function \begin{math}
a_{+}:{\Bbb C}\rightarrow n\times
n\end{math}-matrices such that \begin{math}
vf=\pi ^{*}(a_{+})v-\delta v\end{math}. Since \begin{math}
(\delta v)_{1}=0\end{math}, we have \begin{math}
f=(\pi ^{*}(a_{+})v)_{1}\end{math} and \begin{math}
\delta v=\pi ^{*}(a_{+})v-v(\pi ^{*}(a_{+})v)_{1}\end{math}.
If two elements \begin{math}
c\end{math}  and \begin{math} c'\end{math}
of \begin{math} H^{1}(X,{\cal O}_{-\infty
,q})\end{math} correspond to the same element \begin{math}
\delta v\end{math}, the difference of the two
corresponding \begin{math}
0\end{math}-cochains \begin{math} b\end{math}
 and \begin{math} b'\end{math} used in
the proof of Lemma~\ref{Lemma7.1} is a
cochain in \begin{math} C^{0}({\cal U},{\cal
O}_{-\infty ,q})\end{math} and \begin{math}
\delta (b-b')=c-c'\end{math}. Hence they are
equal. This proves the claim. In
particular, we have for all \begin{math}
\delta q\in {\cal L}_{q}\end{math} a cocycle \begin{math}
[f]\in H^{1}(Y,{\cal O}_{-\pi^{-1}(\infty )})\end{math}
as in Lemma~\ref{Lemma7.1}, such that \begin{math}
v(q)f=\pi ^{*}(a_{+})v(q)-\frac{\partial v}{\partial q}
(\delta q)\end{math}, with some entire
function \begin{math} a_{+}:{\Bbb C}\rightarrow
n\times n\end{math}-matrices. Let \begin{math}
v(x,q)\end{math} be the multivalued
solution \begin{math}
v(x,q)=\pi ^{*}(h_{0}g(x,\cdot
,q)h_{0}^{-1})v(q)\mu ^{-x}\end{math} of
(\ref{g3.1a}) to the potential \begin{math}
T_{x}q\end{math}.  Then we have \begin{math}
v(x,q)f=\end{math} \begin{displaymath}
=\pi ^{*}\left( h_{0}g(x,\cdot
,q)h_{0}^{-1}\left( a_{+}+h_{0}\frac{\partial
g(x,\cdot ,q)}{\partial q}
(\delta q)h_{0}^{-1}\right) h_{0}g^{-1}(x,\cdot
,q)h_{0}^{-1}\right) v(x,q)-
\frac{\partial v(x,q)}{\partial q}
(\delta q) \end{displaymath} \begin{displaymath}
\mbox{and }
\pi ^{*}(F(\cdot ,T_{x}q))\frac{\partial v(x,q)}{\partial q}
(\delta q)-\frac{\partial v(x,q)}{\partial q} (\delta q)\mu =\pi ^{*}\left(
\frac{\partial F(T_{x}q)}{\partial q} (T_{x}\delta q)\right)
v(x,q).\end{displaymath} Since the
solutions of the last equations are
unique up to summation of some multiple
of \begin{math} v(x,q),\ \frac{\partial v(x,q)}{\partial q}
(\delta q)\end{math} is equal to \begin{displaymath}
\frac{\partial v(T_{x}q)}{\partial q}
(T_{x}\delta q)v_{1}(x,q)+\frac{\partial v_{1}(x,q)}{\partial q}
(\delta q)v(x,q).\end{displaymath} Similar
arguments as in Theorem~\ref{Theorem3.3}
show that \begin{math} \frac{\partial v(x,q)}{\partial q}
(\delta q)\end{math} vanishes at all covering
points of infinity. Hence we have \begin{displaymath}
v(T_{x}q)f-f_{-}(x)=\pi
^{*}(a_{+}(x))v(T_{x}q)-\frac{\partial
v(T_{x}q)}{\partial q}
(T_{x}\delta q),\mbox{ with } \end{displaymath}
\begin{displaymath}
a_{+}(x)=h_{0}g(x,\cdot ,q)h_{0}^{-1}\left(
a_{+}+h_{0}\frac{\partial g(x,\cdot ,q)}{\partial q}
(\delta q)h_{0}^{-1}\right)
h_{0}g^{-1}(x,\cdot ,q)h_{0}^{-1}\mbox{ and }
\end{displaymath} \begin{displaymath}
f_{-}(x)=-\frac{\partial v_{1}(x,q)}{\partial q}
(\delta q)\frac{v_{1}(T_{x}q)}{v_{1}(x,q)} ,\end{displaymath}
which is a single valued meromorphic
function on \begin{math} Y\end{math} and
vanishes at all covering points of
infinity. The cocycle represented by \begin{math}
f\end{math} is of course the same as the
cocycle represented by \begin{math}
f-f_{-}(x)\end{math}. Therefore
the cocycle corresponding to \begin{math}
\frac{\partial v(T_{x}q)}{\partial q} (T_{x}q)\end{math}
does not depend on \begin{math} x\end{math}.  In
particular it is an element of \begin{math}
H_{q,\mbox{\it \scriptsize modified}
}^{1}(Y,{\cal O})\end{math}.  This
defines a map \begin{displaymath}
{\cal L}_{q}\rightarrow H_{q,\mbox{\it
\scriptsize modified} }^{1}(Y,{\cal O}),\delta q\mapsto
[f],\end{displaymath} such that \begin{math}
\delta v=\frac{\partial v}{\partial q} (\delta q)\end{math}
corresponds to \begin{math} [f]\end{math}
in the sense of Lemma~\ref{Lemma7.1} and
this property determines a unique
isomorphism \begin{math} d\Gamma
_{q}: H_{q,\mbox{\it \scriptsize modified}
}^{1}(Y,{\cal O})\rightarrow
{\cal L}_{q}\end{math}. This proves (i). Now let \begin{math}
\delta q\in T_{q}{\cal H}^{\infty }\end{math}
and \begin{math} [f]\in H_{q,\mbox{\it
\scriptsize modified} }^{1}(Y,{\cal O})\end{math},
where \begin{math} [f]\end{math} is of
the same form as in
Lemma~\ref{Lemma7.1}. Then we define \begin{displaymath}
b_{-}(x)=\sum \frac{\partial v(T_{x}q)}{\partial q}
(T_{x}\delta q)\frac{1}{w(T_{x}q)v(T_{x}q)}
w(T_{x}q)\end{displaymath} \begin{displaymath}
\mbox{ and } a(x)=\sum
v(T_{x}q)\frac{f}{w(T_{x}q)v(T_{x}q)}
w(T_{x}q)=a_{+}(x)+a_{-}(x).\end{displaymath}
This implies \begin{math}
h_{0}d\Gamma ([f])(x)h_{0}^{-1}=\left[ \frac{d}{dx}
+h_{0}(p\lambda +q(x))h_{0}^{-1},a_{-}(x)\right]
\end{math} and \begin{displaymath}
h_{0}\delta q(x)h_{0}^{-1}=\left[ \frac{d}{dx}
+h_{0}(p\lambda +q(x))h_{0}^{-1},b_{-}(x)\right]
+v(T_{x}q)\frac{\frac{\partial \mu }{\partial q}
(\delta q)}{w(T_{x}q)v(T_{x}q)} w(T_{x}q),\end{displaymath}
where the last derivative is taken with respect to
the choice \begin{math} \frac{d\lambda }{dq}
=0\end{math}. \\ Then \begin{math}
2\pi \sqrt{-1} \omega (d\Gamma ([f]),\delta q)\end{math} is equal to
the residue at infinity of the form \begin{displaymath}
d\lambda \int_{0}^{1} tr \left( a_{-}(x)\left(
\frac{d}{2dx} +h_{0}(p\lambda +q(x))h_{0}^{-1}\right)
b_{-}(x)-\right. \end{displaymath} \begin{displaymath}
\left. -b_{-}(x)\left( \frac{d}{2dx}
+h_{0}(p\lambda +q(x)h_{0}^{-1}\right)
a_{-}(x)\right) dx=\end{displaymath} \begin{displaymath}
=d\lambda \int_{0}^{1} tr \left( b_{-}(x)\left(
\frac{d}{2dx} +h_{0}(p\lambda +q(x))h_{0}^{-1}\right)
a_{+}(x)-\right. \end{displaymath} \begin{displaymath}
\left. -a_{+}(x)\left( \frac{d}{2dx}
+h_{0}(p\lambda +q(x)h_{0}^{-1}\right)
b_{-}(x)\right) dx=\end{displaymath} \begin{displaymath}
=-d\lambda \int_{0}^{1} tr
(a_{+}(x)\delta q(x))dx+d\lambda \int_{0}^{1} tr \left(
a_{+}(x)\sum v(T_{x}q)\frac{\frac{\partial \mu
}{\partial q}
(\delta q)}{w(T_{x}q)v(T_{x}q)} w(T_{x}q)\right)
dx.\end{displaymath} The residue at
infinity of the first summand vanishes
and the residue at infinity of the
second summand is equal to the residue
at infinity of \begin{displaymath}
d\lambda \int_{0}^{1} tr \left(
a(x)\sum v(T_{x}q)\frac{\frac{\partial \mu
}{\partial q}
(\delta q)}{w(T_{x}q)v(T_{x}q)} w(T_{x}q)\right)
dx=\sum f\Omega _{q}(\delta q).\end{displaymath}
Therefore \begin{math}
\omega (d\Gamma _{q}([f]),\delta q)\end{math} is equal
to \begin{math} \frac{1}{2\pi \sqrt{-1} }
\mbox{\it Res} ([f],\Omega _{q}(\delta q))\end{math}.
This proves (ii). To each
finite Mittag Leffler distribution on \begin{math}
\pi ^{-1}({\Bbb C})\end{math}, there
exists a solution \begin{math} f\end{math}
on \begin{math} \pi ^{-1}({\Bbb C})\end{math},
and the corresponding cocycle \begin{math}
[f]\end{math} is of course an element of \begin{math}
\tilde{H} ^{1}_{\mbox{\it \scriptsize modified}
}(Y,{\cal O})\end{math}.
Hence \begin{math} \omega
\in H_{\mbox{\it \scriptsize modified}
}^{1}(Y,\Omega )\end{math}
is equal to zero, if \begin{math}
\mbox{\it Res} ([f],\omega )=0\end{math} for all \begin{math}
[f]\in \tilde{H} ^{1}_{\mbox{\it
\scriptsize modified} }(Y,{\cal O})\end{math}.
On the other hand \begin{math} \omega \end{math}
is non-degenerate and therefore \begin{math}
[f]\in \tilde{H} ^{1}_{\mbox{\it \scriptsize modified}
}(Y,{\cal O})\subset
H_{q,\mbox{\it \scriptsize modified}
}^{1}(Y,{\cal O})\end{math} is equal to zero, if \begin{math}
\omega (d\Gamma _{q}([f]),\delta q)=\frac{1}{2\pi \sqrt{-1} }
\mbox{\it Res} ([f],\Omega _{q}(\delta q))=0\end{math} for all
\begin{math} \delta q\in T_{q}{\cal H}^{\infty }\end{math}.
Then \begin{math} \mbox{\it Res} \end{math} is a
non-degenerate pairing between \begin{math}
\tilde{H} ^{1}_{\mbox{\it \scriptsize modified}
}(Y,{\cal O})\end{math}
and \begin{math} H_{\mbox{\it
\scriptsize modified} }^{0}(Y,\Omega )\end{math}. This
implies that \begin{math} \delta q\in {\cal
L}_{q}\end{math}, if and only if \begin{math}
\omega (\delta q,\delta \tilde{q} )=0\end{math} for all \begin{math}
\delta \tilde{q} \in {\cal L}_{q}\end{math}.
This completes the proof of Theorem~\ref{Theorem7.1}
\hspace*{\fill } \begin{math} \Box \end{math}

\noindent
In the finite dimensional case the last
statement of the theorem would imply
that the system is completely
integrable. In the infinite dimensional
case this is only one possibility to
define the meaning of completely
integrable. But there are stronger
definitions, which are more
satisfactory. For example one could
assume the isospectral sets to be
diffeomorphic to some abelian Lie
groups, where the Lie algebras are
isomorphic to the tangent space. For
compact Riemann surfaces the first
cohomology group of the sheaf of
holomorphic functions is the Lie
algebra of the Picard group. If the
isospectral sets were to be one
connected component of the Picard group
this condition would be fulfilled. In
Section~\ref{Section3} we already saw
that this is not the case for Riemann
surfaces of infinite genus. Nevertheless
we can define some subgroup of the
Picard group, which acts on all divisors
of Riemann-Roch type, but this action is
not transitive.
\begin{Definition} \label{Definition7.3}
Let \begin{math} \tilde{H} ^{1}_{\mbox{\it \scriptsize modified}
}(Y,{\cal O}^{*})\end{math} be the subgroup of
the Picard group in the modified sense,
which consists of cocycles of the form \begin{math}
f_{+}/f_{-}\end{math} on \begin{math}
\pi ^{-1}(U_{+}\cap U_{-})\end{math},
where \begin{math} f_{+}\end{math} is
a meromorphic function on \begin{math}
\pi ^{-1}(U_{+})\end{math} and \begin{math}
f_{-}\end{math} is a meromorphic
function on \begin{math} \pi ^{-1}(U_{-})\end{math},
which is equal to 1
at all covering points of infinity, and \begin{math}
{\Bbb P}_{1}=U_{+}\cup U_{-}\end{math} is
some open covering of \begin{math} {\Bbb P}
_{1}\end{math}, such that \begin{math} U_{+}\subset
{\Bbb C}\subset {\Bbb P}_{1}\end{math} and \begin{math}
U_{-}\end{math} is some open
neighbourhood of infinity.\end{Definition}
Similar arguments as in the proof of
Lemma~\ref{Lemma7.2} show that \begin{math}
\tilde{H} ^{1}_{\mbox{\it \scriptsize modified}
}(Y,{\cal O}^{*})\end{math}
is the subgroup of the Picard group in
the modified sense, which is mapped into
\begin{math} H^{1}_{\mbox{\it \scriptsize modified}
}({\Bbb P}_{1},\pi _{*}({\cal
O}^{*}))\simeq H^{1}_{\mbox{\it \scriptsize modified}
}({\Bbb P}_{1},{\cal
O}_{q}^{*})\end{math} for all \begin{math}
q\in \mbox{\it Isospectral} (Y)\end{math}.
Due to Proposition~\ref{Proposition5.1}
the connected component of the identity
of \begin{math} \tilde{H} ^{1}_{\mbox{\it \scriptsize modified}
}(Y,{\cal O}^{*})\end{math} acts on \begin{math}
\mbox{\it Jacobian} (Y)\end{math}, but neither is \begin{math}
\mbox{\it Jacobian} _{0}(Y)\end{math} an invariant
subspace under this action nor is this
action transitive.

The exact sequence of sheaves \begin{displaymath}
0\rightarrow {\Bbb Z}\rightarrow {\cal O}\rightarrow
{\cal O}^{*}\rightarrow 1,\end{displaymath}
induces a homomorphism  \begin{displaymath}
\exp :\tilde{H} ^{1}_{\mbox{\it \scriptsize modified}
}(Y,{\cal O})\rightarrow \tilde{H}
^{1}_{\mbox{\it \scriptsize modified}
}(Y,{\cal O}^{*}).\end{displaymath} In
some sense \begin{math} \tilde{H}
^{1}_{\mbox{\it \scriptsize modified}
}(Y,{\cal O})\end{math} is the Lie
algebra of \begin{math} \tilde{H}
^{1}_{\mbox{\it \scriptsize modified}
}(Y,{\cal O}^{*})\end{math}  and
this map is the exponential map. In the
next section we introduce a reality
condition in order to define a real subgroup
of the Picard group, which acts
transitively on the real part of the
isospectral sets.

\section{A reality condition} \label{Section8}
In this section we will formulate a
reality condition on the space of
potentials. The real part of the Jacobian
variety will turn out to be isomorphic
to a compact abelian group. For the real
part of the dynamical system we will
construct action angle variables. In the
simplest case, the nonlinear
Schr{\"o}\-din\-ger equation, there are two
reality conditions: the non-focussing
and the self focussing nonlinear Schr{\"o}\-din\-ger
equation. The non-focussing nonlinear Schr{\"o}\-din\-ger
equation is an extension of the real
Korteweg-de Vries equation and can be
treated by the methods of \cite{MKT}.
The reality condition, investigated in
this section, in the simplest case turns
out to be the self focussing nonlinear
Schr{\"o}\-din\-ger equation. In the second
Appendix we give a short introduction to
the relation of these two
reality conditions using spectral theory.

\begin{Assumption} \label{Assumption8.1}
In this section we always assume that \begin{math}
p_{1},\ldots ,p_{n}\end{math} are
imaginary.\end{Assumption}
\begin{Definition} \label{Definition8.1}
The Fr\'{e}chet space of smooth periodic
potentials \begin{math} q\end{math},
which satisfies the reality condition \begin{math}
q^{*}=-q\end{math} is denoted by \begin{math}
{\cal H}^{\infty }_{\Bbb R}\end{math}. In
general we will use the subscript \begin{math}
{\Bbb R}\end{math}, to denote the subsets
corresponding to this reality condition.
\end{Definition}
If \begin{math} q\in {\cal H}_{\Bbb R}^{\infty
},\ q(x)+p\lambda \end{math} is an element of
the Lie algebra \begin{math} u(n,{\Bbb C})\end{math},
 for all \begin{math} \lambda \in {\Bbb R}\end{math}.
 It satisfies the relation \begin{displaymath}
(q(x)+p\lambda )^{*}=-(q(x)+p\bar{\lambda } ).\end{displaymath}
Hence \begin{math} g(x,\lambda ,q)\end{math}
and \begin{math} F(\lambda ,q)\end{math}
satisfy the relations \begin{displaymath}
g^{*}(x,\lambda ,q)=g^{-1}(x,\bar{\lambda } ,q)\mbox{ and }
F^{*}(\lambda ,q)=F^{-1}(\bar{\lambda } ,q)\mbox{, respectively.}
\end{displaymath} Then \begin{math}
R(\lambda ,\mu )=\det (\mu \unity -F(\lambda ,q))\end{math}
satisfies the relation:\begin{displaymath} \overline{R\left(
\bar{\lambda }
,\frac{1}{\bar{\mu } } \right) } =
\overline{\det \left(
\frac{1}{\bar{\mu }} \unity -F(\bar{\lambda } ,q)\right)
} =\det \left( \frac{1}{\mu }
\unity -F^{-1}(\lambda ,q)\right) =
\end{displaymath} \begin{displaymath}
=\frac{R(\lambda ,\mu )}{(-\mu )^{n}\det (-F(\lambda ,q))}
=\frac{R(\lambda ,\mu )}{(\mu
)^{n}R(\lambda ,0)} .\end{displaymath}
This shows that \begin{math} R(\lambda ,\mu )=0\end{math}
if and only if \begin{math} R(\bar{\lambda } ,\bar{\mu }
^{-1})=0\end{math}.  Hence the map \begin{displaymath}
\theta :Y\rightarrow Y,(\lambda ,\mu )\mapsto \left( \bar{\lambda }
,\frac{1}{\bar{\mu } } \right)\end{displaymath}
defines an anti-holomorphic involution of
the Riemann surface corresponding to \begin{math}
q\in {\cal H}_{\Bbb R}^{\infty }\end{math}.
Let us assume in this section that \begin{math}
Y\end{math} corresponds to some \begin{math}
q\in {\cal H}_{\Bbb R}^{\infty }\end{math}. We
will see later that this implies the
branching divisor of the covering map \begin{math}
\pi \end{math} to be invariant under the
involution \begin{math} \theta
\end{math}, in other words \begin{math}
\theta (b_{\mbox{\it \scriptsize effective} })=
b_{\mbox{\it \scriptsize effective} }\end{math}.
For every matrix
valued meromorphic function or
meromorphic form \begin{math} f\end{math}
let \begin{math} \theta ^{*}(f)\end{math}
denote the function or form defined by \begin{displaymath}
\theta ^{*}(f)(\lambda ,\mu )=f^{*}(\bar{\lambda } ,\bar{\mu }
^{-1}).\end{displaymath} By abuse of
notation we will denote the natural
isomorphism of the group of divisors of
\begin{math} Y\end{math} induced by \begin{math} \theta \end{math}
 also by \begin{math} \theta \end{math}.
\begin{Theorem} \label{Theorem8.1}
A potential \begin{math} q\in {\cal H}^{\infty }\end{math}
satisfies the reality condition \begin{math}
q^{*}=-q\end{math} if and only if \begin{math}
\theta (D(q))=D^{t}(q)\end{math}. Moreover,
every divisor \begin{math} D\end{math}
of Riemann-Roch type, which obeys the
equivalence relation \begin{math} D+\theta (D)\sim b\end{math}
is non-special in the modified sense. In
particular \begin{displaymath}
\mbox{\it Jacobian} _{\Bbb R}(Y)=
\left\{ [D]\in \mbox{\it Jacobian} (Y)|D+\theta (D)\sim
b\mbox{ in the modified sense} \right\} \end{displaymath}
is contained in \begin{math} \mbox{\it Jacobian} _{0}(Y)\end{math},
and \begin{math} T_{x}\end{math} acts on
this subspace.\end{Theorem}
Proof: For all \begin{math} q\in {\cal H}^{\infty }\end{math}
we have \begin{displaymath}
g^{*}(x,\lambda ,q)=g^{-1}(x,\bar{\lambda } ,-q^{*})\mbox{, and }
F^{*}(\lambda ,q)=F^{-1}(\bar{\lambda } ,-q^{*}).\end{displaymath}
We assumed that \begin{math} R(\lambda ,\mu )=0\end{math}
if and only if \begin{math} R(\bar{\lambda }
,\bar{\mu } ^{-1})=0.
\end{math} Hence \begin{math} \mu \end{math} is an
eigenvalue of \begin{math} F(\lambda ,q)\end{math},
if and only if \begin{math} \bar{\mu }
^{-1}\end{math} is an eigenvalue of \begin{math}
F(\bar{\lambda } ,-q^{*})\end{math}.  If \begin{math}
v\end{math} is any meromorphic solution of \begin{displaymath}
\pi ^{*}(F(\lambda ,q))v=v\mu ,\end{displaymath}
then it is also a solution of \begin{displaymath}
\pi ^{*}(F^{-1}(\lambda ,q))v=v\frac{1}{\mu } ,\end{displaymath}
and \begin{math} \theta ^{*}(v)\end{math} is a
solution of \begin{displaymath}
\theta ^{*}(v)\theta ^{*}(\pi ^{*}(F^{-1}(\lambda
,q)))=\theta ^{*}(v)\pi ^{*}(F^{-1^{*}}(\bar{\lambda }
,q))=\end{displaymath} \begin{displaymath}
=\theta ^{*}(v)\pi ^{*}(F(\lambda
,-q^{*}))=\theta \left( \frac{1}{\mu }
\right) \theta ^{*}(v)=\mu \theta ^{*}(v).\end{displaymath}
If \begin{math} q^{*}=-q\end{math}, this
implies \begin{math} \theta ^{*}(v)=w\end{math}
and \begin{math} D^{t}(q)=\theta (D(q))\end{math}.
Now let \begin{math} D^{t}(q)\end{math}
be equal to \begin{math} \theta (D(q))\end{math}.
Since \begin{math} \theta ^{*}(\mu ^{x})=\mu ^{-x}\end{math},
Theorem~\ref{Theorem3.3} implies \begin{displaymath}
D^{t}(T_{x}q)=T_{-x}D^{t}(q)=T_{-x}\theta (D(q))
=\theta (T_{x}D(q))=\theta (D(T_{x})).\end{displaymath}
Then we have \begin{displaymath}
\theta ^{*}(v(T_{x}q))=w(T_{x}q).\end{displaymath}
Using the inverse of the map \begin{math}
q\mapsto D(q)\end{math}, constructed in
Section~\ref{Section3}, this implies \begin{math}
q^{*}=-q\end{math}. Now let \begin{math}
D\end{math} be any divisor of Riemann-Roch
type such that \begin{displaymath}
D+\theta (D)-b=(f)\end{displaymath} with some
meromorphic function \begin{math} f\end{math},
which is equal to 1
at all covering points of infinity. Due
to our assumption on \begin{math} Y,\ b\end{math}
is equal to \begin{displaymath}
D(q)+\theta (D(q))+\left( \theta ^{*}(v(q))v(q)\right)
\end{displaymath} with some \begin{math}
q\in {\cal H}^{\infty }_{\Bbb R}\end{math}.
This shows that \begin{displaymath}
\theta (b)=b\mbox{ and }
\theta ^{*}(f)=f.\end{displaymath}  Hence \begin{math}
f\end{math} is real on \begin{math}
\pi ^{-1}({\Bbb R})\end{math}. Then there
exists of course some large \begin{math}
\lambda _{0}\in {\Bbb R}\end{math}, such that \begin{math}
D\end{math} and \begin{math} b\end{math}
have no contributions at all covering
points of \begin{math} \lambda _{0}\end{math}
and the values of \begin{math} f\end{math}
at this covering points are positive.
Now let \begin{math} g\end{math} be any
cross section of \begin{math} {\cal
O}_{D-\pi^{-1}(\infty )}\end{math}. By definition \begin{math}
\frac{\theta ^{*}(g)gf}{\lambda -\lambda _{0}} d\lambda \end{math}
is a meromorphic differential form with
poles only at the covering points of \begin{math}
\lambda _{0}\end{math}. Hence the residue
of this differential form is equal to the
sum of the values of \begin{math}
2\pi \sqrt{-1} \theta ^{*}(g)gf\end{math} at the
covering points of infinity. Due to Theorem~\ref{Theorem2.1}
this total residue is zero and therefore
\begin{math} g\end{math} is zero at all
covering points of \begin{math} \lambda _{0}\end{math}.
The same is of course true for all \begin{math}
\lambda _{0}'\in [\lambda _{0}-\epsilon
,\lambda _{0}+\epsilon ]\end{math}
with some small \begin{math} \epsilon >0\end{math}.
Hence \begin{math} g\end{math} is equal
to zero. We proved that \begin{displaymath}
H^{0}(Y,{\cal O}_{D-\pi^{-1}(\infty )})\simeq H^{0}(Y,{\cal
O}_{\theta (D)-\pi^{-1}(\infty )})\simeq 0.\end{displaymath}
The asymptotic and total degrees of \begin{math}
D\end{math} and \begin{math} \theta (D)\end{math}
are equal to half the asymptotic and
total degree of \begin{math} b\end{math}. Hence they
are equal to \begin{math} (1,1,\ldots )\end{math}
in the sense of Remark~\ref{Remark2.1}.
Due to Serre duality this implies \begin{displaymath}
\tilde{H} ^{1}(Y,{\cal O}_{D-\pi^{-1}(\infty )})\simeq
\tilde{H} ^{1}(Y,{\cal O}_{\theta (D)-\pi^{-1}(\infty )})
\simeq 0.\end{displaymath}
Due to Lemma~\ref{Lemma5.1} this implies
that \begin{math} D\end{math} and \begin{math}
\theta (D)\end{math} are non-special in the
modified sense. As mentioned above, \begin{math}
\theta ^{*}(\mu ^{x})=\mu ^{-x}\end{math},
and therefore the action of \begin{math}
T_{x}\end{math} on \begin{math} \mbox{\it Jacobian} (Y)\end{math}
leaves the subspace \begin{math}
\mbox{\it Jacobian} _{\Bbb R}(Y)\end{math}
invariant and \begin{math} \mbox{\it Jacobian} _{\Bbb R}(Y)\end{math}
is contained in \begin{math} \mbox{\it Jacobian} _{0}(Y)\end{math}.
\hspace*{\fill } \begin{math} \Box \end{math}

\noindent
For compact Riemann surfaces the Jacobian
variety is isomorphic to the dual space
of all holomorphic forms modulo the
first homology group of the Riemann
surface. Here the first
homology group is embedded in the dual
space of all holomorphic forms by
integration of forms over
1-chains. On the other hand the Jacobian variety is
isomorphic to the connected component of
the Picard group and the dual space of
the holomorphic forms is isomorphic to
the Lie algebra of the Picard group.
 From this point of view the first
homology group is isomorphic to the
kernel of the exponential map from the
Lie algebras of the Picard group into the
connected component of the identity of
the Picard group. We will see that in
our case the same is true, if we
restrict ourself to the real part of the
Picard group. To do this let us impose
the reality condition on the first
homology group of our Riemann surface \begin{math}
Y\end{math}. We saw already that the
real part of the Lie algebra \begin{math}
H^{1}_{\mbox{\it \scriptsize modified} }(Y,{\cal O})
\end{math} of the
Picard group contains all cocycles \begin{math}
[f]\end{math}, which are equal to \begin{math}
[-\theta ^{*}(f)]\end{math}. The real part
of the Picard group \begin{math}
H^{1}_{\mbox{\it \scriptsize modified}
}(Y,{\cal O}^{*})\end{math} contains all
cocycles \begin{math} [f]\end{math},
which are equal to \begin{math} \left[ \frac{1}{\theta ^{*}(f)}
\right] \end{math}. The involution \begin{math}
\theta \end{math} induces an involution \begin{math}
\theta _{\#}\end{math} of \begin{math}
H_{1}(\pi ^{-1}({\Bbb C}),{\Bbb Z})\end{math}. Hence
the lattice of real periods of the
Jacobian variety is given by \begin{displaymath}
H_{1,{\Bbb R}}(\pi ^{-1}({\Bbb C}),{\Bbb Z})=\left\{ x\in
H_{1}(\pi ^{-1}({\Bbb C}),{\Bbb Z})|\theta _{\#}(x)=-x\right\}
.\end{displaymath} Since the branching
divisor \begin{math} b_{\mbox{\it \scriptsize effective} }\end{math} is
equal to the zeroes of \begin{math}
\theta ^{*}(v(q)))v(q)\end{math}, the support
of \begin{math} b_{\mbox{\it \scriptsize effective} }\end{math} is
contained in \begin{math} \pi ^{-1}({\Bbb C}
\setminus {\Bbb R})\end{math} and \begin{math}
\theta (b_{\mbox{\it \scriptsize effective} })=
b_{\mbox{\it \scriptsize effective} }\end{math}.  Hence we may
build the Riemann surface \begin{math}
\pi ^{-1}({\Bbb C})\end{math} out of \begin{math}
n\end{math} copies of \begin{math} {\Bbb C}\end{math}
by cutting and gluing along small lines,
which connect the points \begin{math}
(\lambda _{\iota },\mu _{\iota })\end{math} of \begin{math}
b_{\mbox{\it \scriptsize effective} }\end{math} with the corresponding
point \begin{math} (\bar{\lambda } _{\iota },\bar{\mu }
^{-1}_{\iota })\end{math} of \begin{math}
b_{\mbox{\it \scriptsize effective} }\end{math}. Hence we may choose
generators of \begin{math} H_{1,{\Bbb R}
}(\pi ^{-1}({\Bbb C}),{\Bbb Z})\end{math}, which are
in one to one correspondence with all pairs \begin{math}
(\lambda _{\iota },\mu _{\iota }),(\bar{\lambda } _{\iota }
,\bar{\mu } ^{-1}_{\iota })\end{math} of \begin{math}
b_{\mbox{\it \scriptsize effective} }\end{math}.
Since \begin{math} b_{\mbox{\it \scriptsize effective} }\end{math}
has asymptotic and total degree \begin{math}
(2,2,\ldots )\end{math}, we may label
such pairs with the index set \begin{math}
{\cal I}\end{math}. This shows that \begin{math}
H_{1,{\Bbb R}}(\pi ^{-1}({\Bbb C}),{\Bbb Z})\end{math} is
isomorphic to \begin{math} \oplus _{\iota \in
{\cal I}}{\Bbb Z}a_{\iota }\end{math}, where \begin{math}
a_{\iota }\end{math} are the (asymptotically
uniquely defined) generators in the
excluded domains with index \begin{math}
\iota \end{math}. Let us now define elements
\begin{math} \alpha _{\iota }\in \tilde{H}
^{1}_{\mbox{\it \scriptsize modified} }(Y,{\cal O})\end{math}, such that
\begin{displaymath}
\mbox{\it Res} (\alpha _{\iota },\omega )
=\int_{a_{\iota }}\omega \mbox{ for all }
\omega \in H_{\mbox{\it \scriptsize modified}
}^{0}(Y,\Omega ).\end{displaymath} Let us fix
some \begin{math} (\lambda _{0},\mu _{0})\in
\pi ^{-1}({\Bbb C})\end{math}, which is no
branchpoint of the covering map \begin{math}
\pi \end{math}. Then the function \begin{displaymath}
\frac{R(\lambda _{0},\mu )}{(\mu -\mu
_{0})(\lambda -\lambda _{0})R_{\mu }(\lambda _{0},\mu _{0})}
\end{displaymath} is meromorphic on \begin{math}
\pi ^{-1}({\Bbb C})\end{math} and has only
one pole at \begin{math} (\lambda _{0},\mu _{0})\end{math},
such that \begin{displaymath}
\frac{R(\lambda _{0},\mu )}{(\mu -\mu _{0})
(\lambda -\lambda _{0})R_{\mu }(\lambda _{0},\mu _{0})}
-\frac{1}{\lambda -\lambda _{0}} \end{displaymath}
is holomorphic near \begin{math}
(\lambda _{0},\mu _{0})\end{math}. Then the
function \begin{displaymath} \frac{1}{2\pi \sqrt{-1} }
\int_{a_{\iota }}
\frac{R(\lambda _{0},\mu )}{(\mu -\mu _{0})
(\lambda -\lambda _{0})R_{\mu }(\lambda _{0},\mu _{0})}
d\lambda _{0}\end{displaymath} is holomorphic on
the complement of the cycle \begin{math}
a_{\iota }\end{math} of \begin{math}
\pi ^{-1}({\Bbb C})\end{math}. For all \begin{math}
\omega \in H_{\mbox{\it \scriptsize modified}
}^{0}(Y,\Omega )\end{math} the sum of the
residues at all covering points of
infinity of the product of this function
with \begin{math} \omega \end{math} is equal
to \begin{math} \int_{a_{\iota }} \omega \end{math}.
Hence this function defines an element \begin{math}
\alpha _{\iota }\end{math} of
\begin{math} \tilde{H} ^{1}_{\mbox{\it \scriptsize modified}
}(Y,{\cal O})\end{math}, such that for all \begin{math}
\omega \in H_{\mbox{\it \scriptsize modified}
}^{0}(Y,\Omega )\end{math} \begin{displaymath}
\mbox{\it Res} (\alpha _{\iota },\omega )=
\int_{a_{\iota }} \omega .\end{displaymath} Since \begin{math}
\theta _{\#}(a_{\iota })=-a_{\iota }\end{math} we
claim that \begin{math}
\theta ^{*}(\alpha _{\iota })=\alpha _{\iota }\end{math}. In
fact, we have \begin{displaymath} \theta
^{*}(R_{\mu }(\lambda ,\mu ))=
-\frac{R_{\mu }(\lambda ,\mu )}{\mu ^{n-2}R(\lambda ,0)}
.\end{displaymath} Then
the cocycle \begin{math} \theta
^{*}(\alpha _{\iota })\end{math} can be represented by the
function \begin{displaymath} -\frac{1}{2\pi \sqrt{-1} }
\int_{a_{\iota }}\frac{\mu _{0}^{n-2}}{\mu
^{n}} \frac{R(\lambda _{0},0)}{R(\lambda ,0)}
\frac{R(\lambda _{0},\mu )}{(\mu ^{-1}-\mu
_{0}^{-1})(\lambda
-\lambda _{0})R_{\mu }(\lambda _{0},\mu _{0})}
d\lambda _{0}=\end{displaymath}
\begin{displaymath} =\frac{1}{2\pi \sqrt{-1} }
\int_{a_{\iota }}\left( \frac{\mu _{0}}{\mu } \right)
^{n-1}\frac{R(\lambda _{0},0)}{R(\lambda ,0)}
\frac{R(\lambda _{0},\mu )}{(\mu -\mu _{0})(\lambda
-\lambda _{0})R_{\mu }(\lambda _{0},\mu _{0})}
d\lambda _{0}.\end{displaymath} Since \begin{displaymath}
\left( \frac{R(\lambda _{0},0)}{R(\lambda ,0)} -1\right)
\frac{R(\lambda _{0},\mu )}{(\mu -\mu _{0})
(\lambda -\lambda _{0})} \end{displaymath}
is a holomorphic function on \begin{math}
\pi ^{-1}({\Bbb C})\end{math}, the cocycle \begin{math}
\theta ^{*}(\alpha _{\iota })\end{math} can be
represented by the function \begin{displaymath}
\frac{1}{2\pi \sqrt{-1} } \int_{a_{\iota }} \left( \frac{\mu _{0}}{\mu }
\right) ^{n-1}
\frac{R(\lambda _{0},\mu )}{(\mu -\mu _{0})
(\lambda -\lambda _{0})R_{\mu }(\lambda _{0},\mu _{0})}
d\lambda _{0}.\end{displaymath} Since \begin{displaymath}
\left( \left( \frac{\mu _{0}}{\mu } \right)
^{n-1}-1\right)
\frac{R(\lambda _{0},\mu )}{(\mu -\mu _{0})
(\lambda -\lambda _{0})} \end{displaymath}
is a holomorphic function on \begin{math}
\pi ^{-1}({\Bbb C})\end{math}, the cocycle \begin{math}
\theta ^{*}(\alpha _{\iota })\end{math} can be
represented by the function \begin{displaymath}
\frac{1}{2\pi \sqrt{-1} } \int_{a_{\iota }}
\frac{R(\lambda _{0},\mu )}{(\mu -\mu _{0})
(\lambda -\lambda _{0})R_{\mu }(\lambda _{0},\mu _{0})}
d\lambda _{0}.\end{displaymath} This shows that \begin{math}
\theta ^{*}(\alpha _{\iota })\end{math} is equal to \begin{math}
\alpha _{\iota }\end{math}. Since \begin{displaymath}
\frac{R(\lambda _{0},\mu )}{(\mu -\mu _{0})
(\lambda -\lambda _{0})R_{\mu }(\lambda _{0},\mu _{0})}
-\frac{1}{\lambda -\lambda _{0}} \end{displaymath}
is holomorphic near \begin{math}
(\lambda _{0},\mu _{0})\end{math}, the function \begin{displaymath}
\exp \left(
\int_{(\lambda _{1},\mu _{1})}^{(\lambda _{2},\mu _{2})}
\frac{R(\lambda _{0},\mu )}{(\mu -\mu _{0})(\lambda
-\lambda _{0})R_{\mu }(\lambda _{0},\mu _{0})}
d\lambda _{0}\right) \end{displaymath} is on
a small open set a solution of the divisor \begin{math}
(\lambda _{1},\mu _{1})-(\lambda _{2},\mu
_{2})\end{math},
if \begin{math}
(\lambda _{1},\mu _{1})\end{math} and \begin{math}
(\lambda _{2},\mu _{2})\end{math} are
elements of this small open set.
If we divide the path of integration
into small intervals we see that the
same is true for arbitrary \begin{math}
(\lambda _{1},\mu _{1}),(\lambda _{2},\mu _{2})\in
\pi ^{-1}({\Bbb C})\end{math}. Now let \begin{math}
A_{\iota }(t)\end{math} be the line bundle
over \begin{math} Y\end{math} defined by
the cocycle \begin{math} \exp
(2\pi \sqrt{-1} t\alpha _{\iota })\end{math} for all \begin{math}
t\in {\Bbb C}\end{math}. Then \begin{math}
A_{\iota }(t)\end{math} is isomorphic in the
modified sense to the trivial line bundle
for all \begin{math} t\in {\Bbb Z}\end{math}.
For all \begin{math} t\in {\Bbb C}\end{math}
it is an element of \begin{math}
\tilde{H} ^{1}_{\mbox{\it \scriptsize modified}
}(Y,{\cal O}^{*})\end{math}
and for all \begin{math} t\in {\Bbb R}\end{math}
it is even an element of the real part \begin{math}
\tilde{H} ^{1}_{\mbox{\it \scriptsize modified}
,{\Bbb R}}(Y,{\cal O}^{*})\end{math}.

The line bundles \begin{math} A_{\iota }(t)\end{math}
have another description, which has the
advantage that it admits infinite tensor
products. The exact sequence of sheaf
homomorphisms \begin{displaymath} 0\rightarrow
{\Bbb Z}\rightarrow {\cal O}\rightarrow {\cal
O}^{*}\rightarrow 1,\end{displaymath}
where the homomorphism \begin{math} {\cal
O}\rightarrow {\cal O}^{*}\end{math} is
defined by \begin{math} f\mapsto \exp
(2\pi \sqrt{-1} f)\end{math}, induces the long
exact sequence \begin{displaymath} 0\rightarrow
H^{0}(Y,{\Bbb Z})\rightarrow H^{0}(Y,{\cal O})
\rightarrow H^{0}(Y,{\cal O}^{*})\rightarrow
H^{1}(Y,{\Bbb Z})\rightarrow \end{displaymath}
\begin{displaymath} \rightarrow H^{1}(Y,{\cal O})
\rightarrow H^{1}(Y,{\cal O}^{*})\rightarrow
H^{2}(Y,{\Bbb Z})\rightarrow \ldots \end{displaymath}
This long exact sequence decomposes
into \begin{displaymath} 0\rightarrow {\Bbb Z}
\rightarrow {\cal O}\rightarrow {\cal
O}^{*}\rightarrow 1\mbox{ and } \end{displaymath}
\begin{displaymath} 0\rightarrow
H^{1}(Y,{\Bbb Z})\rightarrow H^{1}(Y,{\cal O})
\rightarrow H^{1}(Y,{\cal O}^{*})\rightarrow
H^{2}(Y,{\Bbb Z})\rightarrow \ldots \end{displaymath}
For compact Riemann surfaces, \begin{math}
H^{1}(Y,{\cal O})\end{math} is
isomorphic to the Lie algebra of the
Picard group \begin{math} H^{1}(Y,{\cal O}^{*})
\end{math}. From this point of
view the kernel of the exponential map
of the Lie algebra of the Picard group
into the Picard group is isomorphic to \begin{math}
H^{1}(Y,{\Bbb Z})\end{math}. Due to Poincar\'{e}
duality this first cohomology group is
isomorphic to the first homology group.
Now we use a special covering of \begin{math}
Y\end{math}, in order to define the
elements \begin{math} \alpha _{\iota }\end{math}
in \begin{math}
H^{1}_{\mbox{\it \scriptsize modified} }(Y,{\cal O})\end{math},
which correspond to \begin{math}
a_{\iota }\in H_{1}(\pi ^{-1}({\Bbb C}),{\Bbb Z})\end{math}.
Let \begin{math} U_{-}\end{math} be the
complement of all the circles \begin{math}
(a_{\iota })_{\iota \in {\cal I}}\end{math} in \begin{math}
Y\end{math} and let \begin{math}
(U_{\iota })_{\iota \in {\cal I}}\end{math} be
small pairwise disjoint open tubular
neighbourhoods of the circles \begin{math}
(a_{\iota })_{\iota \in {\cal I}}\end{math}. The
intersection \begin{math} U_{\iota }\cap
U_{-}\end{math} decomposes into two
connected components. Now let \begin{math}
\alpha _{\iota }\in
H^{1}_{\mbox{\it \scriptsize modified} }(Y,{\cal O})\end{math}
be the cocycle, which may be represented
by the element of \begin{math} C^{1}(\{
U_{-},U_{\iota }|\iota \in {\cal I}\} ,{\Bbb Z})\end{math},
which is zero on \begin{math}
U_{\iota '}\cap U_{-}\end{math} if \begin{math}
\iota '\neq \iota \end{math} and on one
component of \begin{math} U_{\iota }\cap
U_{-}\end{math} equal to 1
and zero on the other component. The
orientation of \begin{math} a_{\iota }\end{math}
determines on which component of \begin{math}
U_{\iota }\cap U_{-}\end{math} the
representative of \begin{math} \alpha _{\iota }\end{math}
vanishes and on which it is equal to \begin{math}
1\end{math}. Now \begin{math} \exp
(2\pi \sqrt{-1} \alpha _{\iota })\end{math} is of course a
representative of the trivial line
bundle in \begin{math}
H^{1}_{\mbox{\it \scriptsize modified}
}(Y,{\cal O}^{*})\end{math}. Moreover, for any
sequence \begin{math} (t_{\iota })_{\iota \in {\cal
I}}\in {\Bbb R}^{{\cal I}}\sum_{\iota \in {\cal
I}} t_{\iota }\alpha _{\iota }\end{math} defines an
element of \begin{math}
H^{1}_{\mbox{\it \scriptsize modified}
}(Y,{\cal O})\end{math} and \begin{math} \exp \left(
2\pi \sqrt{-1} \sum_{\iota \in {\cal I}}
t_{\iota }\alpha _{\iota }\right) \end{math} defines
a line bundle in \begin{math}
H^{1}_{\mbox{\it \scriptsize modified}
}(Y,{\cal O}^{*})\end{math}. Therefore
the map \begin{displaymath} \left( {\Bbb R}
/{\Bbb Z}\right) ^{{\cal I}}\rightarrow
H^{1}_{\mbox{\it \scriptsize modified}
}(Y,{\cal O}^{*}),(t_{\iota })_{\iota \in {\cal
I}}\mapsto \otimes _{\iota \in {\cal I}}
A_{\iota }(t_{\iota })\end{displaymath} defines a group
homomorphism from the compact abelian
group \begin{math} ({\Bbb R}/{\Bbb C})^{{\cal I}}\end{math}
into the real part of the Picard group
in the modified sense.
\begin{Proposition}
\label{Proposition8.1}
If the excluded domains of some \begin{math}
U_{l,\epsilon }\end{math} have asymptotically
no overlap, the actions of \begin{math}
{\Bbb R}/{\Bbb Z}\end{math} on \begin{math}
\mbox{\it Jacobian} _{\Bbb R}(Y)\end{math} defined by the tensor
product with \begin{math} A_{\iota }(t)\end{math}
fit together to a continuous action of
the compact group \begin{math} ({\Bbb R}/{\Bbb Z}
)^{{\cal I}}\end{math} on \begin{math}
\mbox{\it Jacobian} _{\Bbb R}(Y)\end{math}.
\end{Proposition}
Proof: We use the same arguments as in
the proof of Theorem~\ref{Theorem5.1}.
The action of the tensor product with \begin{math}
A_{\iota }(t)\end{math} does not change \begin{math}
g_{\iota '}\end{math} if \begin{math}
\iota '\neq \iota \end{math} and \begin{math}
|\iota |,|\iota '|\end{math}  are large enough.
On the cocycle \begin{math} g_{\iota }\end{math}
this action is given by \begin{displaymath}
g_{\iota }(t)=Ad\left( \exp \left( -2\pi t\sqrt{-1} \mbox{\it diagonal}
(n_{1},\ldots ,n_{n})\right) \right)
g_{\iota },\end{displaymath} where \begin{math}
n_{i}-n_{j}=1\end{math}, and all other \begin{math}
n\end{math}'s arbitrary, with \begin{math}
\iota =(i,j,k)\end{math}. In particular,
\begin{math} \| g_{\iota }(t)-\unity \|_{l,\epsilon ,\iota }\end{math}
does not depend on \begin{math} t\end{math}.
Due to Proposition~\ref{Proposition5.1}
the compact group \begin{math}
({\Bbb R}/{\Bbb Z})^{{\cal I}}\end{math} then acts
continuously on \begin{math}
\mbox{\it Jacobian} _{\Bbb R}(Y)\end{math}.
\hspace*{\fill } \begin{math} \Box \end{math}
\begin{Theorem} \label{Theorem8.3}
For all \begin{math} q,\tilde{q} \in
{\cal H}^{\infty }_{\Bbb R},\
H_{q,\mbox{\it \scriptsize modified} }^{1}(Y,{\cal O})
\end{math} is equal to \begin{math}
H^{1}_{\tilde{q} ,\mbox{\it \scriptsize
modified} }(Y,{\cal O})\end{math}.
Furthermore, there exists a dual basis \begin{math}
(\omega _{\iota })_{\iota \in {\cal I}}\end{math},
such that for all \begin{math}  \omega _{\iota }\in
H_{\mbox{\it \scriptsize modified} }^{0}(Y,\Omega )
\end{math} the following holds: \begin{description}
\item[(i)] \begin{math}
\int_{a_{\iota '}} \omega _{\iota }=\delta
_{\iota '\iota }=\mbox{\it Res} (\alpha _{\iota '},\omega
_{\iota })\end{math}.
\item[(ii)] \begin{math}
\mbox{\it Res} ([f],\omega _{\iota })
\end{math} exists for all \begin{math}
 [f]\in H_{q,\mbox{\it \scriptsize modified} }^{1}(Y,{\cal O})
\end{math}.
\item[(ii)] \begin{math} \omega =\sum_{\iota \in
{\cal I}} \mbox{\it Res} (\alpha _{i},\omega )\omega _{\iota }\end{math}
for all \begin{math} \omega \in
H_{\mbox{\it \scriptsize modified} }^{0}(Y,\Omega )
\end{math}.
\item[(iv)] \begin{math} [f]=\sum_{\iota \in
{\cal I}} \mbox{\it Res} ([f],\omega
_{\iota })\alpha _{\iota }\end{math}
for all \begin{math}
[f]\in H_{q,\mbox{\it \scriptsize modified} }^{1}(Y,{\cal O})
\end{math}. \end{description}
\end{Theorem}
Proof\footnote{Those readers, who are
not interested in this proof may skip to
Theorem~\ref{Theorem8.4}.}: For large \begin{math} |\iota |\end{math},
the Riemann surface \begin{math} Y_{\iota }\end{math}
decomposes into a two-fold covering of \begin{math}
{\Bbb P}_{1}\end{math} and \begin{math} (n-2)\end{math}
copies of \begin{math} {\Bbb P}_{1}\end{math}.
 Let \begin{math} \beta _{\iota }\end{math} and \begin{math}
\bar{\beta } _{\iota }\end{math} be the values
of \begin{math} \lambda \end{math} at the two
branchpoints of \begin{math} Y_{\iota }\end{math}.
Then the two-fold covering of \begin{math}
{\Bbb P}_{1}\end{math} may be described by \begin{displaymath}
\lambda =\frac{\beta _{\iota }-\bar{\beta }
_{\iota }}{4\sqrt{-1} }
\left( \kappa -\frac{1}{\kappa } \right)
+\frac{\beta _{\iota }+\bar{\beta } _{\iota }}{2}
,\end{displaymath}
where \begin{math} \kappa \in {\Bbb P}_{1}\end{math}
gives the parameterization of the two-fold
covering, which is equal to \begin{math}
0\end{math} and \begin{math} \infty \end{math}
at the two covering points of infinity
and equal to \begin{math} \pm \sqrt{-1} \end{math}
at the two branchpoints. Now let \begin{math}
\kappa _{\iota }\end{math} describe the divisor \begin{math}
D_{\iota }(q)\end{math}. Then we claim that
for all \begin{math} q\in {\cal H}^{\infty }_{\Bbb R}\end{math}
the sequences \begin{math} \left|
|\kappa _{\iota }|-1\right| \end{math} decreases
faster than every inverse power of \begin{math}
|\iota |\end{math}. This is an easy
consequence of the following fact: If \begin{math}
v(q)=\pi ^{*}(g_{+\iota })v_{\iota }(q)\end{math}
is the Birkhoff factorization of \begin{math}
h(D,\cdot )\end{math} around the
excluded domain with index \begin{math}
\iota \end{math} (see the proof of the
implication (i)\begin{math} \Rightarrow \end{math}
(ii) of Theorem~\ref{Theorem4.3}), \begin{math}
\| g_{+\iota }-\unity \| _{l,\epsilon ,\iota }\end{math}
decreases faster than every inverse
power of \begin{math} |\iota |\end{math},
and the meromorphic function \begin{displaymath}
\theta ^{*}(\pi ^{*}(g_{+\iota })v_{\iota }(q)))
\pi ^{*}(g_{+\iota })v_{\iota }(q)\end{displaymath}
is locally a solution of the divisor
\begin{math} b_{\mbox{\it \scriptsize effective} }-
D(q)-\theta (D(q))\end{math}.
In Lemma~\ref{Lemma7.1} we showed that
the elements of \begin{math}
H_{q,\mbox{\it \scriptsize modified} }^{1}(Y,{\cal O})\end{math}
are in one to one correspondence to
Mittag Leffler distributions, or more
precisely, global sections of the sheaf \begin{math}
{\cal O}_{D(q)}/{\cal O}\end{math}.
Asymptotically such sections have the
form \begin{math}
\frac{c_{\iota }}{\kappa -\kappa _{\iota }} \end{math} on
the Riemann surface \begin{math} Y\end{math}.
\begin{Lemma} \label{Lemma8.1} The
sequence \begin{math} (c_{\iota })_{\iota \in
{\cal I}}\end{math} corresponds to an element
of \begin{math}
H_{q,\mbox{\it \scriptsize modified} }^{1}(Y,{\cal O})\end{math}, if and
only if \begin{math} \left| (\beta _{\iota }-\bar{\beta }
_{\iota })c_{\iota }\right| \end{math} decreases
faster than every inverse power of \begin{math}
|\iota |\end{math}. \end{Lemma}
Proof: Let \begin{math} \underline{P} _{\iota }\end{math}
be the value of the function \begin{displaymath}
P\frac{d\lambda }{d\kappa } =P
\frac{(\beta _{\iota }-\bar{\beta }
_{\iota })(\kappa ^{2}+1)}{4\sqrt{-1} \kappa ^{2}}
\end{displaymath} at the point \begin{math}
\kappa _{\iota }\end{math} of the divisor \begin{math}
D\end{math}.  Since \begin{math} \frac{d\lambda }{d\kappa }
\end{math} is zero at the branchpoints,
this function is holomorphic in the
excluded domain with index \begin{math}
\iota \end{math}, and \begin{math} \underline{P} _{\iota }\end{math}
is well defined. Let us now calculate
the meromorphic function \begin{math}
P_{\iota }\end{math} on the
Riemann surface \begin{math} Y_{\iota }\end{math}
corresponding to the divisor \begin{math}
D_{\iota }: \end{math} \begin{math}
v_{\iota }\end{math} is
given by \begin{math} \left( \begin{array}{c}
1 \\
\frac{\kappa +\kappa _{\iota }}{\kappa -\kappa _{\iota }} \end{array}
\right) \end{math}. The transposed
divisor \begin{math} b_{\iota }-D_{\iota }\end{math}
admits two global sections \begin{math} \left(
\frac{(\kappa -\kappa _{\iota })(\kappa
-\kappa _{\iota }^{-1})}{2(\kappa ^{2}+1)}
,\frac{(\kappa -\kappa _{\iota })(\kappa
+\kappa _{\iota }^{-1})}{2(\kappa ^{2}+1)}
\right) \end{math}. Then \begin{math}
P_{\iota }\end{math} is equal to \begin{displaymath}
\frac{1}{2(\kappa ^{2}+1)} \left( \begin{array}{cc}
(\kappa -\kappa _{\iota })(\kappa -\kappa _{\iota }^{-1}) &
(\kappa -\kappa _{\iota })(\kappa +\kappa _{\iota }^{-1})\\
(\kappa +\kappa _{\iota })(\kappa -\kappa _{\iota }^{-1}) &
(\kappa +\kappa _{\iota })(\kappa +\kappa _{\iota }^{-1})
\end{array} \right) .\end{displaymath} Since
\begin{math} |\kappa _{\iota }|\end{math}
converges to 1
and since the restriction of \begin{math}
P\end{math} to the Riemann surface \begin{math}
Y_{\iota }\end{math} converges to \begin{math}
P_{\iota }\end{math}, the sequence \begin{math}
\left\| \frac{\underline{P} _{\iota }}{\beta _{\iota }-\bar{\beta }
_{\iota }} \right\| \end{math} is
asymptotically bounded by \begin{displaymath}
\frac{1}{c} \leq
\left\| \frac{\underline{P} _{\iota }}{\beta _{\iota }-\bar{\beta }
_{\iota }} \right\| \leq c\mbox{ with some }
c>1.\end{displaymath} From the proof of
Theorem~\ref{Theorem7.1} it follows that \begin{displaymath}
\delta v=\pi ^{*}\left( \sum_{\iota }\frac{\underline{P}
_{\iota }c_{\iota }}{\lambda -\lambda _{\iota }}
\right) v.\end{displaymath}
Hence \begin{math} (c_{\iota })_{\iota \in
{\cal I}}\end{math} corresponds to an element \begin{math}
[f]\end{math} of \begin{math}
H_{q,\mbox{\it \scriptsize modified} }^{1}(Y,{\cal O})\end{math}
if and only if \begin{math} \sum_{\iota } \frac{\underline{P}
_{\iota }c_{\iota }}{\lambda -\lambda _{\iota }} \end{math} is a
matrix valued meromorphic function on \begin{math}
X\end{math}.  Due to Example~\ref{Example2.1}
this is equivalent to the condition that
\begin{math} |c_{\iota }(\beta _{\iota }-\bar{\beta }
_{\iota })|\end{math} decreases faster than
every inverse power of \begin{math}
|\iota |\end{math}.
\hspace*{\fill } \begin{math} \Box \end{math}

\noindent
Continuation of the proof of
Theorem~\ref{Theorem8.3}: Let \begin{math}
([f_{\iota }(q)])_{\iota \in {\cal I}}\end{math}
denote the basis of \begin{math}
H_{q,\mbox{\it \scriptsize modified} }^{1}(Y,{\cal O})
\end{math} corresponding to the
Mittag Leffler distributions, which
vanish outside the excluded domain
with index \begin{math} \iota \end{math} and
in this excluded domain is given by \begin{math}
\frac{1}{\kappa -\kappa _{\iota }} \end{math} as used
in the foregoing lemma. There exists
also a dual basis of \begin{math}
H_{\mbox{\it \scriptsize modified} }^{0}(Y,\Omega ):\end{math}
Let \begin{math} \underline{v} _{\iota }\end{math}
be the value of the function \begin{math}
v\frac{d\kappa }{d\lambda } (\lambda -\lambda _{\iota })\end{math}
at the point \begin{math} \kappa _{\iota }\end{math}.
Then we claim that the forms \begin{displaymath}
\omega _{\iota }(q)=\frac{w\underline{v}
_{\iota }}{(\lambda -\lambda _{\iota })wv} d\lambda \end{displaymath}
are dual to the basis \begin{math}
([f_{\iota }(q)])_{\iota \in {\cal
I}}\end{math} of \begin{math}
H_{q,\mbox{\it \scriptsize modified} }^{1}(Y,{\cal O})
\end{math}. Every form \begin{math} \omega \in
H_{\mbox{\it \scriptsize modified} }^{0}(Y,\Omega )
\end{math} can be written as a
meromorphic function times \begin{math}
d\lambda \end{math}. On the other hand the
meromorphic functions \begin{math}
w_{1},\ldots ,w_{n}\end{math} are a basis
of the meromorphic functions on \begin{math}
Y\end{math} over the meromorphic
functions on \begin{math} X\end{math}.
Hence every element \begin{math} \omega \in
H_{\mbox{\it \scriptsize modified} }^{0}(Y,\Omega )
\end{math} can be written as a sum \begin{displaymath}
\omega =\sum_{i=1}^{n} \left( \frac{w_{i}}{wv}
\pi ^{*}(f_{i})\right) d\lambda ,\mbox{ with
meromorphic functions \begin{math}
f_{\iota }\end{math} on \begin{math} X\end{math}.
} \end{displaymath} Since \begin{math}
\omega \end{math} may only have poles at \begin{math}
b\end{math}, the vector valued function \begin{math}
(f_{1},\ldots ,f_{n})\end{math} can have
poles only at the base points of the
divisor \begin{math} D\end{math}. More
precisely, \begin{math} \omega \end{math}
must be equal to an infinite sum:\begin{displaymath}
\omega =\frac{1}{2\pi \sqrt{-1} } \sum_{\iota }
\mbox{\it Res} ([f_{\iota }(q)],\omega )
\omega _{\iota }(q).\end{displaymath}
On the other hand Lemma~\ref{Lemma7.1} shows
that \begin{displaymath} [f]=\frac{1}{2\pi \sqrt{-1} }
\sum_{\iota } \mbox{\it Res} ([f],\omega _{\iota }(q))[f_{\iota }(q)]
\mbox{for all } [f]\in
H_{q,\mbox{\it \scriptsize modified} }^{1}(Y,{\cal O}).
\end{displaymath} This proves the claim.
\begin{Lemma} \label{Lemma8.2} Let \begin{math}
q,\tilde{q} \in \mbox{\it Isospectral} _{\Bbb R}(Y)
\end{math} and let
the excluded domains of \begin{math}
U_{l,\epsilon }\end{math} have asymptotically
no overlap. Then the following estimates
hold:\begin{displaymath} \left|
\mbox{\it Res}  ([f_{\iota }(q)],\omega _{\tilde{\iota } }(\tilde{q}
))\right| \leq \frac{|\beta _{\iota }-\bar{\beta }
_{\iota }|}{|\lambda _{\iota }-\lambda
_{\tilde{\iota } |}} \end{displaymath}
with some \begin{math} c>0\end{math} for
\begin{math} |\iota |\end{math} and \begin{math}
|\tilde{\iota } |\end{math}  large enough and \begin{math}
\iota \neq \tilde{\iota } \end{math}.
For \begin{math} \iota =\tilde{\iota } \end{math}
we have \begin{displaymath} \frac{1}{c} \leq
\left| \mbox{\it Res} ([f_{\iota }(q)],\omega _{\iota }(\tilde{q}
))\right| \leq c.\end{displaymath}
\end{Lemma}
Proof: By similar arguments as before \begin{math}
\|\underline{v} _{\iota }\|\end{math} is
asymptotically bounded \begin{displaymath}
\frac{1}{c} \leq \|\underline{v} _{\iota }\|\leq c\mbox{ with some }
c>0.\end{displaymath} Then
on the excluded domain with index \begin{math}
\iota \end{math} the form \begin{math}
\omega _{\tilde{\iota } }(\tilde{q} )\end{math}
becomes nearly equal to \begin{displaymath}
(\beta -\bar{\beta } _{\iota })\frac{\alpha (\kappa -\tilde{\kappa }
_{\iota })(\kappa -\tilde{\kappa }
_{\iota }^{-1})+\beta (\kappa -\tilde{\kappa }
_{\iota })(\kappa -\tilde{\kappa }
_{\iota })(\kappa +\kappa _{\iota }^{-1})}{8\sqrt{-1}
\kappa ^{2}(\lambda -\lambda _{\tilde{\iota }
})} d\kappa \end{displaymath} with some \begin{math}
\alpha ,\beta \in {\Bbb C}\end{math} such that \begin{math}
\frac{1}{c} \leq \sqrt{\| \alpha \| ^{2}+\|
\beta \| ^{2}} \leq c\end{math}. This
implies \begin{displaymath} \left|
\mbox{\it Res} ([f_{\iota }(q)],\omega _{\tilde{\iota } }(\tilde{q}
))\right| \leq c\frac{|\beta _{\iota }-\bar{\beta }
_{\iota }|}{|\lambda -\lambda _{\tilde{\iota } }|} \end{displaymath}
if \begin{math} \iota \neq \tilde{\iota }
\end{math}. Finally, on the excluded
domain with index \begin{math} \iota \end{math}
the form \begin{math} \omega _{\iota }(\tilde{q}
)\end{math} becomes nearly equal to \begin{math}
\kappa ^{-1}d\kappa \end{math}. This implies \begin{displaymath}
\frac{1}{c} \leq \left|
\mbox{\it Res} ([f_{\iota }(q)],\omega _{\iota }(\tilde{q}
))\right| \leq c\mbox{ for some } c\geq
2\pi .\end{displaymath}
\hspace*{\fill } \begin{math} \Box \end{math}

\noindent
Continuation of the proof of
Theorem~\ref{Theorem8.3}: If \begin{math}
(c_{\iota })_{\iota \in {\cal I}}\end{math} describes
an element of \begin{math} H_{q,\mbox{\it
\scriptsize modified} }^{1}(Y,{\cal O})\end{math},
the corresponding element of \begin{math}
H^{1}_{\tilde{q} ,\mbox{\it \scriptsize
modified}}(Y,{\cal O})\end{math} is
given by \begin{displaymath} \tilde{c}
_{\tilde{\iota } }=\sum_{\iota \in {\cal I}} \frac{1}{2\pi \sqrt{-1} }
\mbox{\it Res} ([f_{\iota }(q)],\omega _{\tilde{\iota } }(\tilde{q}
))c_{\iota }.\end{displaymath} Then the last
two lemmata show that \begin{math}
H_{q,\mbox{\it \scriptsize modified} }^{1}(Y,{\cal O})\end{math}
and \begin{math} H^{1}_{\tilde{q} ,\mbox{\it \scriptsize
modified}}(Y,{\cal O})\end{math}  are
equal. \begin{Lemma} \label{Lemma8.3}
For all \begin{math} \tilde{q} \in {\cal H}^{\infty }_{\Bbb R}\end{math}
and \begin{math} \iota \neq \tilde{\iota } \end{math}
\begin{displaymath} \left| \int_{a_{\iota }}
\omega _{\tilde{\iota } }(\tilde{q} )\right| \leq
c\frac{|\beta _{\iota }-\bar{\beta }
_{\iota }|}{|\beta _{\iota }-\bar{\beta } _{\tilde{\iota }
}|} \end{displaymath} and finally, for \begin{math}
\iota =\tilde{\iota } \end{math}
\begin{displaymath} \left| \int_{a_{\iota }}
\omega _{\tilde{\iota } }(\tilde{q} )\right| \geq
\frac{1}{c} .\end{displaymath} \end{Lemma}
The proof is similar to the proof of the
lemma before.
\hspace*{\fill } \begin{math} \Box \end{math}

\noindent
Continuation of the proof of
Theorem~\ref{Theorem8.3}: With this
lemma it is obvious that there exists
some \begin{math} K\geq 0\end{math}, and
a matrix \begin{math} (M_{\iota \tilde{\iota }
})_{\iota ,\tilde{\iota } \in {\cal I},\ |\iota |,|\tilde{\iota }
|\geq K}\end{math} such that
\begin{description}
\item[(i)] \begin{math}
\sum_{|\tilde{\iota } |\geq K} M_{\iota \tilde{\iota }
} \int_{a_{\iota '}} \omega _{\tilde{\iota } }(\tilde{q}
)=\delta _{\iota \iota '}\end{math} for all \begin{math}
 |\iota |,|\iota '|\geq
K\end{math}.
\item[(ii)] \begin{math} |M_{\iota \tilde{\iota }
}|\leq c|\beta _{\iota }-\bar{\beta } _{\iota }|\end{math}
with some \begin{math}
c>0\end{math}. \end{description}
\begin{Lemma} \label{Lemma8.4}
Let \begin{math} \omega \in
H_{\mbox{\it \scriptsize modified} }^{0}(Y,\Omega )\end{math} be
an element such that \begin{math}
\int_{a_{\iota }} \omega =0\end{math} for all \begin{math}
\iota \in {\cal I}\end{math}. Then \begin{math}
\omega \end{math} is equal to zero.\end{Lemma}
Proof\footnote{Another proof of this lemma can be found
in \cite{MKT2}, which carries over to
our situation. In fact, due to our
assumption all elements of \begin{math}
H_{\mbox{\it \scriptsize modified} }^{0}(Y,\Omega )
\end{math} can be proven to be square
integrable.}: It suffices to proof the statement
for a form, which satisfies the relation
\begin{math} \theta ^{*}(\omega )=-\omega \end{math}.
In fact, \begin{math}
\omega =1/2(\omega +\theta ^{*}(\omega ))+
1/2(\omega -\theta ^{*}(\omega ))\end{math}
and both forms \begin{math}
\omega -\theta ^{*}(\omega )\end{math}  and \begin{math}
\sqrt{-1} (\omega +\theta ^{*}(\omega ))\end{math} fulfil this
condition. Moreover, we have \begin{displaymath}
\int_{a_{\iota }}\theta ^{*}(\omega )=
\overline{\int_{\theta _{\#}(a_{\iota })}
\omega } =-
\overline{\int_{a_{\iota }}\omega } .\end{displaymath}
If \begin{math}b\in H_{1}(\pi ^{-1}({\Bbb
C}),{\Bbb Z})\end{math},
such that \begin{math} \theta _{\#}(b)=b\end{math},
the assumption implies
\begin{displaymath} \int_{b}
\theta ^{*}(\omega )=
\overline{\int_{b}\omega } =-
\int_{b}\omega
.\end{displaymath} This shows that \begin{displaymath}
\Re\left( \int_{b} \omega \right) =0\mbox{ for all } b\in
H_{1}(\pi ^{-1}({\Bbb C}),{\Bbb Z}).\end{displaymath} The
condition that \begin{math} \int_{a_{\iota }}
\omega =0\end{math} for all \begin{math}
\iota \in {\cal I}\end{math} implies that the
residue of \begin{math} \omega \end{math} at
all covering points of infinity vanishes
and therefore \begin{math}
\omega \end{math} is holomorphic even
at all covering points of infinity.
Hence the function \begin{displaymath}
Y\rightarrow {\Bbb R},(\lambda ,\mu )
\mapsto \Re \left( \int_{(\infty ,\infty
)}^{(\lambda ,\mu )} \omega \right) \end{displaymath}
is a harmonic function on \begin{math} Y\end{math},
which vanishes at \begin{math} (\infty ,\infty
)\end{math}.  Due to the maximum modulus
Theorem for harmonic functions (see \cite{Co})
this function is equal to zero.
Hence \begin{math} \omega \end{math} is
zero too.
\hspace*{\fill } \begin{math} \Box \end{math}

\noindent
Completion of the proof of
Theorem~\ref{Theorem8.3}:
With the help of the matrices \begin{math}
M_{\iota \tilde{\iota } }\end{math} introduced
above and the last lemma it follows that
there exists a basis of holomorphic forms \begin{math}
(\omega _{\iota })_{\iota \in {\cal
I}},\omega _{\iota }\in H_{\mbox{\it
\scriptsize modified} }^{0}(Y,\Omega )\end{math},
which fulfil the conditions
(i)-(iv) of the theorem.
\hspace*{\fill } \begin{math} \Box \end{math}
\begin{Theorem} \label{Theorem8.4} If
the excluded domains of some \begin{math}
U_{l,\epsilon }\end{math} have
asymptotically no overlap, the action of
\begin{math} ({\Bbb R}/{\Bbb Z})^{{\cal I}}\end{math}
on any point \begin{math} [D]\in \mbox{\it Jacobian}
_{\Bbb R}(Y)\end{math}
induces a homeomorphism between \begin{math}
({\Bbb R}/{\Bbb Z})^{{\cal I}}\end{math} and \begin{math}
\mbox{\it Jacobian} _{\Bbb R}(Y)\end{math}. Moreover, the
holomorphic forms \begin{math} (\omega _{\iota })_{\iota \in
{\cal I}}\end{math} define an embedding of the
real part \begin{math}
H^{1}_{{\Bbb R},\mbox{\it \scriptsize modified}
}(Y,{\cal O})\end{math} of any \begin{math}
H_{q,\mbox{\it \scriptsize modified} }^{1}(Y,{\cal O})
\end{math} onto a subspace of \begin{math}
{\Bbb R}^{{\cal I}}\end{math}, which is mapped
under the exponential map onto \begin{math}
({\Bbb R}/{\Bbb Z})^{{\cal I}}\end{math}.
Finally, the following diagram commutes:\\
\begin{picture}(460,100)(-200,-50)
\put(-100,30){\begin{math} H^{1}_{{\Bbb
R},\mbox{\it \scriptsize modified} }(Y,{\cal
O})\end{math} }
\put(-100,-45){\begin{math} H^{1}_{{\Bbb
R},\mbox{\it \scriptsize modified} }(Y,{\cal
O^{*}})\end{math} }
\put(110,30){\begin{math} {\Bbb R}^{{\cal
I}}\end{math} }
\put(100,-45){\begin{math} \left( {\Bbb
R}/{\Bbb Z}\right) ^{{\cal I}}\end{math} }
\put(0,35){\vector(1,0){70} }
\put(70,-40){\vector(-1,0){70} }
\put(-80,20){\vector(0,-1){40} }
\put(-75,-5){\begin{math} \exp \end{math} }
\put(115,20){\vector(0,-1){40} }
\put(120,-5){\begin{math} \exp (2\pi \sqrt{-1}
\cdot )\end{math} }
\end{picture}
\end{Theorem}
\begin{Corollary} \label{Corollary8.1}
If the excluded domains of some \begin{math}
U_{l,\epsilon }\end{math} have asymptotically
no overlap, the group \begin{math} ({\Bbb R}
/{\Bbb Z})^{{\cal I}}\end{math} acts on \begin{math}
\mbox{\it Isospectral} _{\Bbb R}(Y)\end{math} transitively and freely.
The action on any potential \begin{math}
q\in \mbox{\it Isospectral} _{\Bbb R}(Y)\end{math} induces a
homeomorphism between \begin{math} ({\Bbb R}
/{\Bbb Z})^{{\cal I}}\end{math} and \begin{math}
\mbox{\it Isospectral} _{\Bbb R}(Y)\end{math}. Moreover, there
exists for any \begin{math} q\in
\mbox{\it Isospectral} _{\Bbb R}(Y)\end{math}
an embedding from \begin{math} {\cal L}_{q,{\Bbb R}
}\end{math} onto a subspace of \begin{math}
{\Bbb R}^{{\cal I}}\end{math}, which does not
depend on \begin{math} q\end{math}, such
that the flow induced by the Lie algebra
element in \begin{math} {\Bbb R}^{{\cal I}}\end{math}
corresponding to some \begin{math} \delta q\in
{\cal L}_{q,{\Bbb R}}\end{math} is smooth in \begin{math}
{\cal H}^{\infty }\end{math}, and the derivative at \begin{math}
t=0\end{math} is equal to \begin{math}
\delta q\end{math}.  This action of \begin{math}
{\cal L}_{q,{\Bbb R}}\end{math} on \begin{math}
\mbox{\it Isospectral} _{\Bbb R}(Y)\end{math} is also transitive.
\end{Corollary}
In order to prove this theorem we first
need a lemma:
\begin{Lemma} \label{Lemma8.5}
For any two divisors \begin{math} [D],[\tilde{D}
]\in \mbox{\it Jacobian} _{\Bbb R}(Y)\end{math}, there exists a
sequence of 1-chains \begin{math} (c_{\iota })_{\iota \in
{\cal I}}\end{math}, such that \begin{description}
\item[(i)] \begin{math} c_{\iota }\end{math}
lies inside the excluded domain with
index \begin{math} \iota \end{math} of some
neighbourhood of infinity.
\item[(ii)] \begin{math}
\sum_{\iota \in {\cal I}} \partial c_{\iota }=
\tilde{D} -D\end{math}.
\item[(iii)] \begin{math} \sum_{\iota \in
{\cal I}} \int_{c_{\iota }}\omega
_{\tilde{\iota } }\in {\Bbb R}\end{math}
for all \begin{math}
\tilde{\iota } \in {\cal I}\end{math}. \end{description}
\end{Lemma}
Proof: There exists a meromorphic function \begin{math}
f\end{math}, which is equal to \begin{math}
1\end{math} at all covering points of
infinity such that \begin{displaymath}
(f)=\tilde{D} +\theta (\tilde{D} )-D-\theta (D).\end{displaymath}
Asymptotically the equation \begin{math}
f=\frac{x}{x-1} \end{math} has for all \begin{math}
x\in [0,1]\end{math} in each excluded
domain two solutions. This defines a
sequence of 1-chains \begin{math} (c_{\iota })_{\iota \in
{\cal I}}\end{math}, which fulfills condition
(i) and \begin{displaymath} \sum_{\iota \in
{\cal I}} \partial c_{\iota }=\tilde{D} -D',\mbox{ with some }
[D']\in \mbox{\it Jacobian} _{\Bbb R}(Y),\end{displaymath} such
that \begin{math} D'+\theta (D')=D+\theta (D)\end{math}.
Moreover, the arguments of the proof of
Theorem~\ref{Theorem8.3} show that
condition (iii) is also fulfilled. It remains to prove
the statement for \begin{math} \tilde{D}
+\theta (\tilde{D} )=D+\theta (D)\end{math}. In
this case there exists a natural
sequence \begin{math} (c_{\iota })_{\iota \in
{\cal I}}\end{math} of 1-chains, which
fulfills condition (i)-(iii): this sequence is on
\begin{math} Y_{\iota }\end{math}
asymptotically given by \begin{displaymath}
c_{\iota }:[0,1]\rightarrow Y_{\iota },t\rightarrow
\kappa (t)=\kappa _{\iota }\left( \frac{\tilde{\kappa }
_{\iota }}{\kappa _{\iota }} \right) ^{t}=\kappa _{\iota }\exp\left(
t\ln \left( \frac{\tilde{\kappa }
_{\iota }}{\kappa _{\iota }} \right) \right) ,\end{displaymath}
where \begin{math} \ln (\tilde{\kappa }
_{\iota }/\kappa _{\iota })\end{math} is uniquely
defined by \begin{math} -\pi <\Im (\ln (\tilde{\kappa }
_{\iota }/\kappa _{\iota }))\leq \pi \end{math}. In
this case \begin{math} c_{\iota }\end{math}
can be chosen to obey the relation \begin{math}
\theta (c_{\iota })=-c_{\iota }\end{math}. This and
again the arguments of the proof of
Theorem~\ref{Theorem8.3} imply condition
(iii).
\hspace*{\fill } \begin{math} \Box \end{math}

\noindent
Proof of Theorem~\ref{Theorem8.4}: Let
us fix some \begin{math} [D]\in \mbox{\it Jacobian}
_{\Bbb R}(Y)\end{math}.
Due to Proposition~\ref{Proposition8.1} the
action of \begin{math} ({\Bbb R}/{\Bbb Z})^{{\cal I}}\end{math}
on \begin{math} [D]\end{math} defines a
continuous map \begin{math} ({\Bbb R}/{\Bbb Z}
)^{{\cal I}}\rightarrow \mbox{\it Jacobian} _{\Bbb R}(Y)\end{math}. On
the other hand the last lemma defines a
map \begin{displaymath} \mbox{\it Jacobian} _{\Bbb R}(Y)\rightarrow
{\Bbb R}^{{\cal I}},[\tilde{D} ]\mapsto \left( \sum_{\iota \in
{\cal I}}\int_{c_{\iota }} \omega _{\tilde{\iota } }\right)
_{\tilde{\iota } \in {\cal I}},\mbox{ with } \sum_{\iota \in
{\cal I}}\partial c_{\iota }=\tilde{D} -D.\end{displaymath}
Now we claim that the composition of the
map \begin{math} \mbox{\it Jacobian} _{\Bbb R}(Y)\rightarrow
{\Bbb R}^{{\cal I}}\end{math} with the natural map \begin{math}
{\Bbb R}^{{\cal I}}\rightarrow ({\Bbb R}/{\Bbb Z})^{{\cal I}}\end{math}
is the inverse of the group action on \begin{math}
[D]:({\Bbb R}/{\Bbb Z})^{{\cal I}}\rightarrow
\mbox{\it Jacobian} _{\Bbb R}(Y)\end{math}.
First we show that this map does not
depend on the choice of \begin{math}
(c_{\iota })_{\iota \in {\cal I}}\end{math}.  If \begin{math}
(\tilde{c} _{\iota })_{\iota \in {\cal I}}\end{math}
is another sequence of 1-chains, which fulfills the
conditions (i)-(iii) of
Lemma~\ref{Lemma8.5}, then \begin{displaymath}
\sum_{\iota \in {\cal I}}\left( \int_{c_{\iota }}
\omega _{\tilde{\iota }
}-\int_{\tilde{c} _{\iota }} \omega _{\tilde{\iota }
}\right) \in {\Bbb Z}.\end{displaymath} In
fact, \begin{math} \partial \sum_{\iota \in
{\cal I}}c_{\iota }-\tilde{c} _{\iota }=0\end{math},
and \begin{displaymath} \sum_{\iota \in {\cal I}}\left(
\int_{c_{\iota }} \omega -\int_{\tilde{c} _{\iota }}
\omega +\int_{\theta (c_{\iota })} \omega -\int_{\theta (\tilde{c}
_{\iota })} \omega \right) =0.\end{displaymath}
Hence \begin{math} \sum_{\iota \in
{\cal I}} c_{\iota }-\tilde{c} _{\iota }\end{math} defines an
element of \begin{math} H^{1}(Y,{\Bbb Z})\end{math}, which lies in
the eigen space with eigenvalue \begin{math}
-1\end{math}  of the involution \begin{math}
\theta ^{\#}\end{math}. But these elements
are of the form \begin{math} \prod_{\iota \in
{\cal I}} {\Bbb Z}a_{\iota }\end{math}. This proves
\begin{displaymath}
\sum_{\iota \in {\cal I}}\left(
\int_{c_{\iota }} \omega _{\tilde{\iota }
}-\int_{\tilde{c} _{\iota }} \omega _{\tilde{\iota }
}\right) \in {\Bbb Z}.\end{displaymath} Hence
the map \begin{math} \mbox{\it Jacobian}
_{\Bbb R}(Y)\rightarrow
{\Bbb R}^{{\cal I}}\rightarrow
({\Bbb R}/{\Bbb Z})^{{\cal I}}\end{math}
does not depend on the choice of \begin{math}
(c_{\iota })_{\iota \in {\cal I}}\end{math}. Then this
map is continuous.

Secondly we prove that the map \begin{displaymath}
({\Bbb R}/{\Bbb Z})^{{\cal I}}\rightarrow
\mbox{\it Jacobian} _{\Bbb R}(Y)\rightarrow
{\Bbb R}^{{\cal I}}\rightarrow
({\Bbb R}/{\Bbb Z})^{{\cal I}}\end{displaymath}
is the identity map. For any element of
\begin{math} ({\Bbb R}/{\Bbb Z})^{{\cal I}}\end{math}
there exists an element of \begin{math}
(-1/2,1/2]^{{\cal I}}\end{math}, which is
mapped onto this element under the natural map
\begin{displaymath}
(-1/2,1/2]^{{\cal I}}\hookrightarrow {\Bbb R}^{{\cal I}}\rightarrow
({\Bbb R}/{\Bbb Z})^{{\cal I}}.\end{displaymath} Now the
multiplication with \begin{math} t\in
[0,1]\end{math} defines a continuous map
from \begin{math} 0\end{math}  to this
element of \begin{math} ({\Bbb R}/{\Bbb Z})^{{\cal I}}\end{math}
and hence also a continuous map \begin{math}
[0,1]\rightarrow \mbox{\it Jacobian} _{\Bbb R}(Y)\end{math},
which is equal to \begin{math} [D]\end{math}
at \begin{math} t=0\end{math} and
equal to any \begin{math} [\tilde{D} ]\end{math}
in the image of \begin{math} ({\Bbb R}/{\Bbb Z}
)^{{\cal I}}\rightarrow \mbox{\it Jacobian} _{\Bbb R}(Y)
\end{math} at \begin{math} t=1\end{math}.
This defines a sequence \begin{math}
(c_{\iota })_{\iota \in {\cal I}}\end{math} of
1-chains, which fulfills
condition (i)-(iii) of
Lemma~\ref{Lemma8.5}. Lemma~\ref{Lemma8.1}
shows that the infinitesimal generator
of this flow can be considered as an element of
\begin{math}
H_{q,\mbox{\it \scriptsize modified} }^{1}(Y,{\cal O})
\end{math}, and Theorem~\ref{Theorem8.4}
(iv) shows that the integration over all
\begin{math} (\omega _{\tilde{\iota }
})_{\tilde{\iota }
\in {\cal I}}\end{math} along this path \begin{math}
(c_{\iota })_{\iota \in {\cal I}}\end{math} gives back
the original element of \begin{math} ({\Bbb R}
/{\Bbb Z})^{{\cal I}}\end{math}. Hence the map
\begin{displaymath}
({\Bbb R}/{\Bbb Z})^{{\cal I}}\rightarrow
\mbox{\it Jacobian} _{\Bbb R}(Y)\rightarrow
{\Bbb R}^{{\cal I}}\rightarrow
({\Bbb R}/{\Bbb Z})^{{\cal I}}\end{displaymath}
is the identity map.

Let us finally also prove that the map
\begin{displaymath} \mbox{\it Jacobian} _{\Bbb R}(Y)
\rightarrow {\Bbb R}^{{\cal I}}\rightarrow
({\Bbb R}/{\Bbb Z})^{{\cal I}}\rightarrow
\mbox{\it Jacobian} _{\Bbb R}(Y)\end{displaymath} is the
identity map\footnote{This is equivalent
to Abel's Theorem for those divisors,
which obey the reality condition. In
general the analogous statement is more
complicated. It can be proven by
methods similar to those used in this
section. One half can be proven with
the methods of Lemma~\ref{Lemma8.5}
(compare with \cite[Section~10]{MKT2}).}. Again we want to
distinguish between the two cases of the
proof of Lemma~\ref{Lemma8.5}. If \begin{math}
\tilde{D} +\theta (\tilde{D} )\end{math} is
equal to \begin{math} D+\theta (D)\end{math},
there exists of course a sequence \begin{math}
(D_{n})_{n\in {\Bbb N}}\end{math} of integral
divisors of asymptotic and total degree \begin{math}
(1,1,\ldots )\end{math} such that \begin{description}
\item[(i)] \begin{math}
D_{n}+\theta (D_{n})=D+\theta (D)\end{math}
for all \begin{math}
n\in {\Bbb N}\end{math}.
\item[(ii)] \begin{math} D-D_{n}\end{math}
is a finite divisor of degree zero.
\item[(iii)] \begin{math} ([D_{n}])_{n\in {\Bbb N}
}\end{math} converges to \begin{math} [\tilde{D}
]\end{math}. \end{description} The
divisor \begin{math} D-D_{n}\end{math}
corresponds to an element of \begin{math}
\tilde{H} ^{1}_{{\Bbb R},\mbox{\it \scriptsize modified}
}(Y,{\cal O}^{*})\end{math}.  Moreover
it corresponds to an element of the Lie
algebra \begin{math} \tilde{H} ^{1}_{{\Bbb R}
,\mbox{\it \scriptsize modified}
}(Y,{\cal O})\end{math}. Theorem~\ref{Theorem8.3}
(iv) implies that for all \begin{math}
n\in {\Bbb N},\ [D_{n}]\end{math} is mapped
onto \begin{math} [D_{n}]\end{math}
under the map given above. Hence the
same is true for \begin{math} [\tilde{D}
]\end{math}. In the other case there
exists a continuous map \begin{displaymath}
[0,1]\rightarrow \mbox{\it Jacobian} _{\Bbb R}(Y),t\mapsto
[D(t)],\end{displaymath} which is mapped onto \begin{math}
[D]\end{math} for \begin{math} t=0\end{math}
and which is mapped onto \begin{math} [\tilde{D}
]\end{math} for \begin{math} t=1\end{math}.
The derivative of this map,
which was constructed in the proof of
Lemma~\ref{Lemma8.5} is furthermore a
map \begin{math} [0,1]\rightarrow
\tilde{H} ^{1}_{{\Bbb R},\mbox{\it \scriptsize modified}
}(Y,{\cal O})\end{math}. This derivative
may be integrated due to Lemma~\ref{Lemma8.1}
 to a map \begin{math} [0,1]\rightarrow
\tilde{H} ^{1}_{{\Bbb R},\mbox{\it \scriptsize modified}
}(Y,{\cal O})\end{math}. The composition
of this map \begin{math} [0,1]\rightarrow
\tilde{H} ^{1}_{{\Bbb R},\mbox{\it \scriptsize modified}
}(Y,{\cal O})\end{math} with the
exponential map \begin{math}
\tilde{H} ^{1}_{{\Bbb R},\mbox{\it \scriptsize modified}
}(Y,{\cal O})\rightarrow
\tilde{H} ^{1}_{{\Bbb R},\mbox{\it \scriptsize modified}
}(Y,{\cal O}^{*})\end{math} gives the map
\begin{displaymath} [0,1]\rightarrow
\tilde{H} ^{1}_{{\Bbb R},\mbox{\it \scriptsize modified}
}(Y,{\cal O}^{*}),t\mapsto \mbox{ line bundle corresponding to }
D(t)-D.\end{displaymath}
Now again Theorem~\ref{Theorem8.3}
implies that for all \begin{math} t\in
[0,1],\ [D(t)]\end{math} is mapped onto \begin{math}
[D(t)]\end{math} under the above map.
Due to Lemma~\ref{Lemma8.5}  the
combination of these two cases cover the
general case and the group action on \begin{math}
[D]\end{math} defines a homeomorphism \begin{math}
({\Bbb R}/{\Bbb Z})^{{\cal I}}\simeq
\mbox{\it Jacobian} _{\Bbb R}(Y)\end{math}.
The rest of Theorem~\ref{Theorem8.4}
follows from Theorem~\ref{Theorem8.3}.
\hspace*{\fill } \begin{math} \Box \end{math}

\noindent
Corollary~\ref{Corollary8.1} is a direct
consequence of Theorem~\ref{Theorem8.4}.
\hspace*{\fill } \begin{math} \Box \end{math}

\noindent
The assumption in the last two
theorems that the excluded domains have
asymptotically no overlap, can be
weakened. In fact, the arguments we gave
can be used to prove these theorems in
more general cases. But let us indicate, why this
assumption must not be dropped. If \begin{math}
p_{1}(\lambda )-p_{2}(\lambda )\end{math} and \begin{math}
p_{1}(\lambda )-p_{3}(\lambda )\end{math} are nearly
elements of \begin{math} 2\pi \sqrt{-1} {\Bbb Z},\
p_{2}(\lambda )-p_{3}(\lambda )\end{math} of
course is also nearly an element of \begin{math}
2\pi \sqrt{-1} {\Bbb Z}\end{math}. In the case \begin{math}
n=3\end{math}, the Riemann surface \begin{math}
Y_{\iota }\end{math} corresponding to the
overlapping excluded domain is a
three-fold covering of \begin{math} {\Bbb P}
_{1}\end{math} with 6
branchpoints. In the \begin{math}
\lambda \end{math}-plane the cuts and
branchpoints may be chosen e.g. like in figure 2,
where \begin{math} \{ \lambda _{i,j},\bar{\lambda }
_{i,j}\} \end{math} are the branchpoints
between the \begin{math} i\end{math}-th
and \begin{math} j\end{math}-th sheet
over \begin{math} {\Bbb P}_{1}\end{math}.\\
\begin{picture}(460,220)(-200,-120)

\put(-100,70){\circle*{1}}
\put(-100,-70){\circle*{1}}
\put(-95,70){\begin{math} \lambda _{1,2}\end{math} }
\put(-95,-80){\begin{math} \bar{\lambda } _{1,2}\end{math} }
\put(-100,70){\line(0,-1){140} }

\put(-20,50){\circle*{1}}
\put(-20,-50){\circle*{1}}
\put(-15,50){\begin{math} \lambda _{1,3}\end{math} }
\put(-15,-60){\begin{math} \bar{\lambda } _{1,3}\end{math} }
\put(-20,50){\line(0,-1){100} }

\put(120,70){\circle*{1}}
\put(120,-70){\circle*{1}}
\put(125,70){\begin{math} \lambda _{2,3}\end{math} }
\put(125,-80){\begin{math} \bar{\lambda } _{2,3}\end{math} }
\put(120,70){\line(0,-1){140} }

\put(-24,51){\vector(-4,1){60} }
\put(-24,-51){\vector(-4,-1){60} }

\put(-50,-115){Figure 2.}

\end{picture}

\noindent
If for example \begin{math}
\lambda _{1,2}=\lambda _{1,3}\end{math} the Riemann
surface is still non-singular.
But if furthermore \begin{math} \lambda _{2,3}\end{math}
becomes equal to \begin{math}
\lambda _{1,2}=\lambda _{1,3}\end{math} as indicated
in figure 3, the Riemann
surface becomes singular. \\
\begin{picture}(460,220)(-200,-120)

\put(-100,70){\circle*{1}}
\put(-100,-70){\circle*{1}}
\put(-95,70){\begin{math}
\lambda _{1,2}=\lambda _{1,3}\end{math} }
\put(-95,-80){\begin{math} \bar{\lambda }
_{1,2}=\bar{\lambda } _{1,3}\end{math} }
\put(-100,70){\line(0,-1){140} }

\put(120,70){\circle*{1}}
\put(120,-70){\circle*{1}}
\put(125,70){\begin{math} \lambda _{2,3}\end{math} }
\put(125,-80){\begin{math} \bar{\lambda } _{2,3}\end{math} }
\put(120,70){\line(0,-1){140} }

\put(98,70){\vector(-1,0){130} }
\put(98,-70){\vector(-1,0){130} }

\put(-50,-115){Figure 3.}

\end{picture}

\noindent
It will be one
two-fold covering of \begin{math} {\Bbb P}_{1}\end{math}
and one copy of \begin{math} {\Bbb P}_{1}\end{math}
connected by two ordinary double points.
Such a surface is described by the
algebraic equation \begin{displaymath}
R(\lambda ,\mu )=(\mu -\lambda )(\mu ^{2}-2\mu \lambda -1)=
\mu ^{3}-2\mu ^{2}\lambda +\mu
(2\lambda ^{2}-1)+\lambda =0.\end{displaymath}
Indeed, \begin{math} \mu ^{2}-2\mu \lambda -1=0\end{math}
describes a two-fold covering \begin{math}
\lambda =\frac{\mu ^{2}-1}{2\mu } \end{math}, with
branchpoints at \begin{math}
(\lambda _{1},\mu _{1})=(\sqrt{-1}
,\sqrt{-1} ),(\lambda _{2},\mu _{2})=
(-\sqrt{-1} ,-\sqrt{-1} )\end{math}
and \begin{math} \frac{d\lambda }{d\mu }
=\frac{2(\mu ^{2}+1)}{\mu ^{2}}
\end{math}. The real part of the
completion of the generalized Jacobian
variety\footnote{This completion of the
generalized Jacobian variety is defined
in the next section.} of this Riemann surface turns
out to decompose into two components:
One non-compact three dimensional group
isomorphic to \begin{math} {\Bbb R}\times
S^{1}\times S^{1}\end{math}
corresponding to \begin{math}
b_{\mbox{\it \scriptsize effective} }=
3(\sqrt{-1} ,\sqrt{-1} )+3(-\sqrt{-1} ,-\sqrt{-1} )\end{math}
and a compact component isomorphic to \begin{math}
S^{1}\end{math} corresponding to \begin{math}
b_{\mbox{\it \scriptsize effective} }=
(\sqrt{-1} ,\sqrt{-1} )+(\sqrt{-1} ,\sqrt{-1} )\end{math}.
If the assumption that the excluded domains
have asymptotically no overlap is
dropped, this example illustrates that
the perturbation of the
different Riemann surfaces to each other
cannot be estimated. Furthermore, in that case
certain holomorphic forms are not square
integrable.

\section{The singular case} \label{Section9}
In this section we want to investigate,
how the statements of the foregoing sections
can be generalized to the singular case
in the sense of
Definition~\ref{Definition3.1}. Let us
fix some equation \begin{math}
R(\lambda ,\mu )=0\end{math}, which describes
the curve of eigenvalues corresponding
to some potential \begin{math} q\in {\cal H}^{\infty }\end{math}.
To this curve there corresponds a unique
normalization \begin{math} Y\end{math},
which is an \begin{math} n\end{math}-fold
covering of \begin{math} {\Bbb P}_{1}, \pi :Y\rightarrow
{\Bbb P}_{1}\end{math}. Due to the counting
lemma all zeroes of \begin{math}
R_{\mu }(\lambda ,\mu )\end{math} can be arranged such
that asymptotically and totally to each
element of \begin{math} {\cal I}\end{math}
there corresponds one pair of zeroes in
the excluded domain with index \begin{math}
i\in {\cal I}\end{math}. In general not all
zeroes of \begin{math} R_{\mu }(\lambda ,\mu )\end{math}
are indeed branchpoints of the
normalization \begin{math} \pi :Y\rightarrow
{\Bbb P}_{1}\end{math}. But there always
exists some asymptotically and totally
unique subset \begin{math}
{\cal I}_{\mbox{\it \scriptsize analytic} }\subset
{\cal I}\end{math}, such that to each element
of \begin{math} {\cal I}_{\mbox{\it \scriptsize analytic} }
\end{math} there
corresponds one pair of branch points of
the covering map \begin{math} \pi :Y\rightarrow
{\Bbb P}_{1}\end{math}.  In Section~\ref{Section3}
we already mentioned that the algebraic
curve defined by \begin{math} R(\lambda ,\mu )=0\end{math}
and the normalization \begin{math} Y\end{math}
 are the extreme cases of the curves,
which really correspond to some
potential \begin{math} \tilde{q} \in {\cal H}^{\infty }\end{math}
corresponding to the curve of
eigenvalues defined by the equation \begin{math}
R(\lambda ,\mu )=0\end{math}.  Let \begin{math}
{\cal O}_{\mu }\end{math} be the free sheaf of
rings on \begin{math} {\Bbb C}\subset {\Bbb P}_{1}\end{math}
\begin{displaymath} {\cal O}_{{\Bbb P}
_{1}}^{n}\simeq {\cal O}_{\mu },(f_{1},\ldots
,f_{n})\mapsto \sum_{i=1}^{n}
f_{i}\mu ^{i-1}.\end{displaymath} Due to
the equation \begin{math} R(\lambda ,\mu )=0,\
{\cal O}_{\mu }\end{math} is in fact a sheaf of
rings. On the other hand \begin{math}
\pi _{*}({\cal O}_{Y})\end{math} also defines a
sheaf of rings on \begin{math} {\Bbb C}
\subset {\Bbb P}_{1}\end{math}. Moreover,
since \begin{math} \mu \end{math} is a
holomorphic function on \begin{math}
\pi ^{-1}({\Bbb C}),\ {\cal O}_{\mu }\end{math} is a
subsheaf of \begin{math} \pi _{*}({\cal O}_{Y})\end{math}
and both sheaves contain \begin{math} {\cal O}_{{\Bbb P}
_{1}}\end{math} as a ring subsheaf.
Now let \begin{math} {\cal S}\end{math}
be the set of all ring subsheaves,
which contain \begin{math} {\cal O}_{\mu }\end{math}
as a ring subsheaf. To each element of \begin{math}
{\cal S}\end{math}, there corresponds a
singular Riemann surface \begin{math} Y'\end{math}
with normalization \begin{math} p:Y\rightarrow
Y'\end{math}, such that the following
diagram commutes and \begin{math}
\pi _{*}({\cal O}_{Y'})\end{math} is the
corresponding subsheaf in \begin{math} {\cal
S.}\end{math} \\
\begin{picture}(460,80)(0,0)
\put(210,55){\begin{math} Y\end{math} }
\put(280,55){\begin{math} Y'\end{math} }
\put(245,15){\begin{math} {\Bbb P}_{1}\end{math}
} \put(245,65){\begin{math} p\end{math} }
\put(215,35){\begin{math} \pi \end{math}
} \put(275,35){\begin{math} \pi '\end{math}
}\put(215,50){\vector(1,-1){25}}
\put(280,50){\vector(-1,-1){25}}
\put(230,60){\vector(1,0){40} }
\end{picture}

\noindent
By abuse of notation we
will sometimes identify \begin{math} Y'\end{math}
with the corresponding element of \begin{math}
{\cal S}\end{math}.
In general divisors are defined to be
global sections of the sheaf \begin{math}
{\cal M}^{*}/{\cal O}^{*}\end{math}. Hence
they define locally free submodules of
the meromorphic functions. By abuse of
notation we will use divisors to define
finitely generated submodules of the
meromorphic functions: For any singular
Riemann surface \begin{math} Y'\end{math}
with normalization \begin{math} p:Y\rightarrow
Y'\end{math} such that the foregoing
diagram commutes
let the branching divisor \begin{math}
b_{\mbox{\it \scriptsize effective} }\end{math} be defined by the
property that \begin{math}
{\cal O}_{b_{\mbox{\it \scriptsize effective} }}\end{math}
is the \begin{math} {\cal O}_{Y'}\end{math}
submodule of meromorphic functions
defined by:\begin{displaymath}
\pi _{*}'({\cal O}_{b_{\mbox{\it \scriptsize effective}
}})=\left\{
g\in \pi _{*}'({\cal M}) \left|
\sum_{\mbox{\it \scriptsize sheets of }
\pi '} gf\in {\cal O}_{{\Bbb P}_{1}} \mbox{ for all } f\in
\pi _{*}'({\cal O}_{Y'})\right.
\right\} .\end{displaymath} In
general \begin{math}
{\cal O}_{b_{\mbox{\it \scriptsize effective} }}\end{math}
is not locally free, but the two extreme
cases \begin{math}
{\cal O}_{b_{\mbox{\it \scriptsize analytic} }}\end{math}
as well as \begin{math}
{\cal O}_{b_{\mbox{\it \scriptsize algebraic} }}\end{math}
are locally free. It
is quite easy to see that \begin{math}
{\cal O}_{Y'}\end{math} can be reconstructed
from the sheaf \begin{math}
{\cal O}_{b_{\mbox{\it \scriptsize effective} }}:\end{math}
\begin{displaymath} {\cal O}_{Y'}=\left\{ f\in
p_{*}({\cal O}_{Y})| fg\in
{\cal O}_{b_{\mbox{\it \scriptsize effective} }} \mbox{ for all }
g\in {\cal O}_{b_{\mbox{\it \scriptsize effective} }}
\right\} .\end{displaymath}
We will later give an example where \begin{math}
{\cal S}\end{math} is not a
countable set, but for \begin{math} n=2\end{math}
the situation is quite simple:
\begin{Proposition} \label{Proposition9.1}
If \begin{math} n=2\end{math} locally
there are only two cases \begin{description}
\item[(i)] \begin{math}
b_{\mbox{\it \scriptsize analytic} }=y_{0},
b_{\mbox{\it \scriptsize algebraic} }=(2m+1)y_{0},m\in
{\Bbb N}\end{math}
\item[(ii)] \begin{math} b_{\mbox{\it \scriptsize analytic} }=0,
b_{\mbox{\it \scriptsize algebraic} } =my_{1}+my_{2},\mbox{ with }
\pi (y_{1})=\pi (y_{2}).\end{math}
\end{description} In both cases \begin{math}
{\cal O}_{b_{\mbox{\it \scriptsize effective} }}
\end{math} is locally free. \begin{description}
\item[{\rm \normalsize \em In case (i)}]
\begin{math} b_{\mbox{\it \scriptsize effective} }\end{math}
is of the form \begin{math}
b_{\mbox{\it \scriptsize effective} }=(2m'+1)y_{0}
\end{math}, for \begin{math}
0\leq m'\leq m\end{math}.
\item[{\rm \normalsize \em In case (ii)}] \begin{math}
b_{\mbox{\it \scriptsize effective} }\end{math}
is of the form \begin{math}
b_{\mbox{\it \scriptsize effective} }=m'(y_{1}+y_{2})
\end{math}, for \begin{math}
0\leq m'\leq m\end{math}. \end{description}
More generally, all finitely generated
submodules \begin{math} {\cal F}\end{math}
of \begin{math} p_{*}({\cal
M}_{Y})\end{math} are locally free, if
\begin{displaymath} \left\{ f\in
p_{*}({\cal O}_{Y})| fg\in
{\cal F} \mbox{ for all }
g\in {\cal F} \right\} ={\cal O}_{Y'}.\end{displaymath}
\end{Proposition}
Proof: We may assume that \begin{math}
R(\lambda ,\mu )\end{math} has the form \begin{math}
\mu ^{2}=a(\lambda )\end{math} with some entire
function \begin{math} a(\lambda )\end{math}.
Hence the equation \begin{math}
R(\lambda ,\mu )=0\end{math} is locally
equivalent to \begin{math}
\mu ^{2}=(\lambda -\lambda _{0})^{k}\end{math}.  Now
there are two cases: \begin{description}
\item[(i)] \begin{math}
\mu ^{2}=(\lambda -\lambda _{0})^{2m+1}\end{math}.
Then \begin{math} b_{\mbox{\it \scriptsize analytic} }\end{math} is
equal to \begin{math} \pi ^{-1}(\lambda _{0})\end{math}
and \begin{math} b_{\mbox{\it \scriptsize algebraic} }\end{math} equal
to \begin{math} (2m+1)\pi ^{-1}(\lambda _{0})\end{math}.
Moreover, in this case \begin{math} {\cal O}_{\mu }\end{math}
is generated as an \begin{math}
{\cal O}_{{\Bbb P}_{1}}\end{math} module by \begin{math}
1\end{math}  and \begin{math} \mu \end{math}
 and \begin{math} \pi _{*}({\cal O}_{Y})\end{math}
is generated as an \begin{math} {\cal O}_{{\Bbb P}
_{1}}\end{math} module by 1
and \begin{math} \mu (\lambda -\lambda _{0})^{-m}\end{math}.
 For any \begin{math} {\cal O}_{{\Bbb P}_{1}}\end{math}
module \begin{math} {\cal O}_{Y'}\end{math},
which is contained in \begin{math}
\pi _{*}({\cal O}_{Y})\end{math}  and contains \begin{math}
{\cal O}_{\mu }\end{math},  there exists an
integer \begin{math} 0\leq m'\leq m\end{math}
such that this module is generated by \begin{math}
1\end{math} and \begin{math}
\mu (\lambda -\lambda _{0})^{-m'}\end{math}.  The
corresponding module \begin{math}
{\cal O}_{b_{\mbox{\it \scriptsize effective} }}
\end{math} is the free \begin{math}
{\cal O}_{Y'}\end{math} module generated by \begin{math}
\mu (\lambda -\lambda _{0})^{m-m'-1}\end{math} and the
divisor \begin{math} b_{\mbox{\it \scriptsize effective} }\end{math} is
equal to \begin{math}
b_{\mbox{\it \scriptsize effective} }=(2m'+1)\pi
^{-1}(\lambda _{0})\end{math}.
\item[(ii)] \begin{math}
\mu ^{2}=(\lambda -\lambda _{0})^{2m+1}\end{math}.
Then \begin{math} b_{\mbox{\it \scriptsize analytic} }=0\end{math}
and \begin{math} \pi ^{-1}(\lambda _{0})\end{math}
consists of two points \begin{math}
y_{1}\end{math} and \begin{math} y_{2}\end{math}.
\begin{math} b_{\mbox{\it \scriptsize algebraic} }\end{math} is equal
to \begin{math} m(y_{1}+y_{2})\end{math}.
Again \begin{math} {\cal O}_{\mu }\end{math} and
\begin{math} \pi _{*}({\cal O}_{Y})\end{math}
are generated as \begin{math}
{\cal O}_{{\Bbb P}_{1}}\end{math} modules by \begin{math}
1, \mu \end{math}
 and \begin{math} 1, \mu (\lambda -\lambda _{0})^{-m}\end{math},
respectively. Any \begin{math} {\cal O}_{{\Bbb P}_{1}}\end{math}
module \begin{math} {\cal O}_{Y'}\end{math}
is again generated by \begin{math}
1\end{math} and \begin{math}
\mu (\lambda -\lambda _{0})^{-m'}\end{math}, with some
\begin{math} 0\leq m'\leq m\end{math}.  The
corresponding module \begin{math}
{\cal O}_{b_{\mbox{\it \scriptsize effective} }}
\end{math} is the free \begin{math}
{\cal O}_{Y'}\end{math} module generated by \begin{math}
\mu (\lambda -\lambda _{0})^{m-m'-1}\end{math} and the
divisor \begin{math} b_{\mbox{\it \scriptsize effective} }\end{math} is
equal to \begin{math}
b_{\mbox{\it \scriptsize effective} }=m'(y_{1}+y_{2})\end{math}.
\end{description}
The same arguments show that all
finitely generated submodules \begin{math}
{\cal F}\end{math} are locally of the
same form as \begin{math} {\cal
O}_{b_{\mbox{\it \scriptsize effective} }}
\end{math} with \begin{math}
m'\in {\Bbb Z}\end{math} if \begin{displaymath} \left\{ f\in
p_{*}({\cal O}_{Y})| fg\in {\cal F} \mbox{ for all }
g\in {\cal F} \right\} ={\cal O}_{Y'}.\end{displaymath}
\hspace*{\fill } \begin{math} \Box \end{math}

\noindent
Now we explain the modification of
Section~\ref{Section3} in case of
singular Riemann surfaces. In that
section we investigated the dual
eigen bundle of some potential \begin{math}
q\end{math}. Due to Definition~\ref{Definition3.1}
 the potential \begin{math} q\end{math}
completely determines the singular
Riemann surface, which corresponds to this
potential. In this context we will take
for granted that \begin{math} Y'\end{math}
corresponds to \begin{math} q\end{math},
or equivalently that \begin{math}
\varepsilon '_{q}:{\cal O}_{q}\rightarrow
\pi '_{*}({\cal O}_{Y'})\end{math}
is an isomorphism of sheaves. For
singular Riemann surfaces the sheaves \begin{math}
{\cal O}_{D(q)}\end{math} and \begin{math}
{\cal O}_{D^{t}(q)}\end{math} have to be
defined in a slightly different way: The
meromorphic functions on \begin{math} Y\end{math}
and \begin{math} Y'\end{math} coincide,
hence \begin{math} v\end{math} and \begin{math}
w\end{math}  are also meromorphic
functions on \begin{math} Y'\end{math}.
Now let \begin{math} {\cal O}_{D(q)}\end{math}
and \begin{math} {\cal O}_{D^{t}(q)}\end{math}
be the \begin{math} {\cal O}_{Y'}\end{math}
submodules of \begin{math} p_{*}({\cal M
}_{Y})\end{math}, which are generated by
\begin{math} v_{1},\ldots ,v_{n}\end{math}
and \begin{math} w_{1}\ldots ,w_{n}\end{math},
respectively. Then Theorem~\ref{Theorem3.1}
is still valid. Since the singular
Riemann surface \begin{math} Y'\end{math}
corresponds to \begin{math} q\end{math},
or equivalently \begin{math}
\varepsilon '_{q}:{\cal O}_{q}\rightarrow
\pi '_{*}({\cal O}_{Y'})\end{math} is an
isomorphism, \begin{math} f\end{math} is
locally an element of \begin{math}
\pi '({\cal O}_{Y'})\end{math} if and only if \begin{math}
\sum_{\mbox{\it \scriptsize sheets of
}\pi }fP\end{math} is holomorphic. This
implies that the entries of \begin{math}
P\end{math} generate the \begin{math}
{\cal O}_{Y'}\end{math} module \begin{math}
{\cal O}_{b_{\mbox{\it \scriptsize effective} }}
\end{math}. This generalizes
statement (iv) of
Lemma~\ref{Lemma3.3}. Lemma~\ref{Lemma3.3}
(i)-(iii) remains valid. Due to an easy
exercise the degree function extends to
a unique function \begin{displaymath}
\deg :\mbox{finitely generated submodules of }
p_{*}({\cal M}_{Y})\rightarrow {\Bbb Z},\end{displaymath}
which obeys the two properties \begin{description}
\item[(i)] If \begin{math} {\cal F}\subseteq {\cal
F}'\end{math} are two finitely generated
submodules of \begin{math} p_{*}({\cal
M}_{Y})\end{math} \begin{displaymath}
\deg ({\cal F}')-\deg ({\cal
F})=\dim H^{0}(Y',{\cal F}'/{\cal F}).\end{displaymath}
\item[(ii)] If \begin{math} D\end{math}
is a divisor of \begin{math} Y'\end{math}
in the correct sense, \begin{math} \deg
(D)=\deg ({\cal O}_{D})\end{math}. \end{description}
Moreover the Riemann-Roch Theorem can be
generalized to this context:

\noindent
In fact, let \begin{math} \pi ':Y'\rightarrow {\Bbb P}_{1}\end{math}
be a singular \begin{math}
n\end{math}-fold covering over \begin{math}
{\Bbb P}_{1}\end{math} with normalization \begin{math}
p:Y\rightarrow Y'\end{math}, such that
the following diagram commutes. \\
\begin{picture}(460,80)(0,0)
\put(210,55){\begin{math} Y\end{math} }
\put(280,55){\begin{math} Y'\end{math} }
\put(245,15){\begin{math} {\Bbb P}_{1}\end{math}
} \put(245,65){\begin{math} p\end{math} }
\put(215,35){\begin{math} \pi \end{math}
} \put(275,35){\begin{math} \pi '\end{math}
}\put(215,50){\vector(1,-1){25}}
\put(280,50){\vector(-1,-1){25}}
\put(230,60){\vector(1,0){40} }
\end{picture}

\noindent
Let \begin{math} {\cal F}\end{math}
be any finitely generated \begin{math}
{\cal O}_{Y'}\end{math} submodule of \begin{math}
p_{*}({\cal M}_{Y})\end{math} and \begin{math}
{\cal O}_{b_{\mbox{\it \scriptsize effective} }}
\end{math} be defined as above,
then the following relation holds:
\begin{displaymath} \dim H^{0}(Y',{\cal
F})-\dim H^{1}(Y',{\cal F})=n+\deg
({\cal F})-\frac{\deg ({\cal
O}_{b_{\mbox{\it \scriptsize effective} }})}{2} .\end{displaymath}
Since \begin{math}
\pi _{*}({\cal O}_{b_{\mbox{\it \scriptsize analytic} }})\subseteq
\pi _{*}({\cal O}_{b_{\mbox{\it \scriptsize effective} }})\subseteq
\pi _{*}({\cal O}_{b_{\mbox{\it \scriptsize algebraic} }})
\end{math} there exists
a subset \begin{math}
{\cal I}_{\mbox{\it \scriptsize effective} }\end{math}
of \begin{math} {\cal I}\end{math},
which contains \begin{math}
{\cal I}_{\mbox{\it \scriptsize algebraic} }\end{math},
such that to each element of \begin{math}
{\cal I}_{\mbox{\it \scriptsize effective} }
\end{math} there corresponds a pair
of degrees of \begin{math}
{\cal O}_{b_{\mbox{\it \scriptsize effective} }}\end{math}.
The degree of a finitely generated
submodule of \begin{math} p_{*}({\cal
M}_{Y})\end{math} is a sequence of
numbers indexed by \begin{math}
{\cal I}_{\mbox{\it \scriptsize effective} }\end{math}
in the sense of Remark~\ref{Remark2.1}.
With this modification Theorem~\ref{Theorem3.1}
is valid also in the general case and
relation (\ref{g3.2}) generalizes to the
relation \begin{displaymath} {\cal O}_{D(q)}\otimes
{\cal O}_{D^{t}(q)}\simeq
{\cal O}_{b_{\mbox{\it \scriptsize effective} }},\end{displaymath}
where the isomorphism is given by
multiplication with the function \begin{math}
wv\end{math}. The rest of Section~\ref{Section3}
can be carried over to the general case
in an obvious way.

In Section~\ref{Section4} the divisor \begin{math}
D\end{math} should be considered in the
general case as a finitely generated
submodule of \begin{math} p_{*}({\cal
M}_{Y})\end{math}. Moreover, the
branching divisor has to be specified as
the module \begin{math}
{\cal O}_{b_{\mbox{\it \scriptsize effective} }}\end{math}
defined above, and the degree of a
divisor is in the general case a
sequence indexed by \begin{math}
{\cal I}_{\mbox{\it \scriptsize effective} }\end{math}
and not by \begin{math} {\cal I}\end{math}. With
this modification Section~\ref{Section4}
is also true in the general case. In
particular equation (\ref{g4.5}) is also
true in the singular case \begin{math}
\lambda _{1}=\lambda _{2}\end{math}, and therefore
Corollary~\ref{Corollary4.1} remains valid too.

In the general case
Definition~\ref{Definition5.1} has to
be more precise:
\begin{Definition}
\label{Definition5.1'} The Jacobian
variety of a singular Riemann surface
\begin{math} Y'\in {\cal S}\end{math} is
defined to be the set of all
equivalence classes of finitely
generated submodules \begin{math} {\cal
O}_{D}\end{math} of \begin{math}
p_{*}({\cal M}_{Y})\end{math} such that
\begin{description}
\item[(i)] \begin{math} \{ f\in p_{*}({\cal
O}_{Y}) | f{\cal O}_{D}\subset {\cal
O}_{D}\} ={\cal O}_{Y'}.\end{math}
\item[(ii)] \begin{math} {\cal O}_{D}\end{math}
is of Riemann-Roch type.
\item[(iii)] \begin{math} {\cal O}_{D}\end{math}
is integral; this means that \begin{math}
{\cal O}_{D}\end{math} contains \begin{math}
{\cal O}_{Y'}\end{math}.
\item[(iv)] The asymptotic and total
degree of \begin{math} {\cal O}_{D}\end{math}
is equal to \begin{math} (1,1,\ldots
)\end{math}. \end{description}
\end{Definition}
We want to emphasize that this
definition is not analogous to the
generalized Jacobian variety defined in
\cite{Se}. In fact, we do not assume
that \begin{math} {\cal O}_{D}\end{math}
is locally free\footnote{Our definition
is similar to the compactification of
the generalized Jacobian variety with
the help of torsion free sheaves (see
\cite{DS}).}. We will give an
example of such modules, which are not
locally free. Such an integral divisor \begin{math}
{\cal O}_{D}\end{math} is called non-special
in the modified sense, if there
exists only one integral divisor, which
is equivalent in the modified sense to \begin{math}
{\cal O}_{D}\end{math}. Lemma~\ref{Lemma5.1}
is valid also in the general case.
Furthermore, Theorem~\ref{Theorem5.1}
holds even for the Jacobian variety \begin{math}
\mbox{\it Jacobian} (Y')\end{math} of a singular Riemann
surface. In the proof of this theorem
the assumption (i) on cocycles, which
correspond to line bundles on \begin{math}
Y'\end{math} has to be replaced in the
general case by the condition \begin{description}
\item[Cocycle (i)'] for all \begin{math} \iota \in
{\cal I}_{\mbox{\it \scriptsize effective} }
\end{math} and for all holomorphic
sections \begin{math} f\end{math}  of \begin{math}
{\cal O}_{Y'}\end{math} over \begin{math}
\pi ^{-1}(U_{\iota })\end{math} there exists a
holomorphic function \begin{math}
g_{+,\iota }:U_{\iota }\rightarrow n\times
n\end{math}-matrices, such that \begin{displaymath}
g_{+,\iota }g_{\iota }=g_{\iota }\mbox{\it diagonal}
(f_{1},\ldots ,f_{n}),\end{displaymath}
where \begin{math} f_{i}\end{math} is
the restriction of \begin{math} f\end{math}
to the \begin{math} i\end{math}-th sheet
of \begin{math} \pi ^{-1}(U_{\iota }\cap
U)\end{math}. \end{description}
In general let \begin{math} \mbox{\it Isospectral} (Y')\end{math}
be the subspace of all potentials, which
correspond to the singular Riemann
surface \begin{math} Y'\end{math} in
the sense of
Definition~\ref{Definition3.1}.
Theorem~\ref{Theorem5.2} holds in the
general case.

The space of all regular forms \begin{math}
H_{\mbox{\it \scriptsize modified} }^{0}(Y',\Omega )
\end{math} on the singular Riemann
surface \begin{math} Y'\end{math} is
equal to the vector space of all
sections of \begin{math} {\cal
O}_{b_{\mbox{\it \scriptsize effective} }-\pi ^{-1}(\infty )}\end{math}
times \begin{math} d\lambda \end{math}. This
is analogous to the definition of
regular forms in \cite{Se}. Now let \begin{math}
q\end{math}  correspond to the singular
Riemann surface \begin{math}
Y'\end{math}. Then the form \begin{math}
\frac{1}{\mu } \frac{\partial \mu }{\partial q}
(\delta q)d\lambda \end{math}
is an element of \begin{math}
H_{\mbox{\it \scriptsize modified} }^{0}(Y',\Omega )
\end{math} for all \begin{math} \delta q\in T_{q}{\cal
H}^{\infty }\end{math}.
In fact, let \begin{math} f\end{math} be
a local section of \begin{math} \pi _{*}'({\cal
O}_{Y'})\end{math}. Then \begin{math}
a_{f}=\sum_{\mbox{\it \scriptsize sheets of
}\pi '} v(q)fw(q)\end{math} is a holomorphic \begin{math}
n\times n\end{math}-matrix valued
function. Moreover, the following
relation holds: \begin{displaymath}
\sum_{\mbox{\it \scriptsize sheets of }
\pi '} \frac{1}{\mu } \frac{\partial \mu }{\partial q}
(\delta q)f=\mbox{\it tr} \left(
a_{f}F^{-1}(\cdot ,q)\frac{\partial F(\cdot ,\lambda )}{\partial q}
(\delta q)\right) .\end{displaymath} Hence
the left hand side is holomorphic for all
local sections of \begin{math} \pi _{*}'({\cal
O}_{Y'})\end{math}. This shows that in
general \begin{math} \Omega _{q}\end{math}
is a map from \begin{math} T_{q}{\cal H}^{\infty
}\end{math} into \begin{math}
H_{\mbox{\it \scriptsize modified} }^{0}(Y',\Omega )\end{math}.

The Darboux coordinates of singular
potentials are more complicated: For a
singular potential \begin{math} q\end{math}
let \begin{math} D(q)\end{math} be the
sum of the divisor, which is given by
the support of the sheaf \begin{math} {\cal
O}_{D(q)}/{\cal O}_{Y'}\end{math} with
multiplicity equal to the local
dimension of this sheaf, plus the
divisor, which is given by the support
of the sheaf \begin{math} {\cal
O}_{b_{\mbox{\it \scriptsize algebraic} }}/
{\cal O}_{b_{\mbox{\it \scriptsize effective} }}\end{math}
with multiplicity equal to half the
local dimension of this sheaf. Then
Lemma~\ref{Lemma6.1}  remains valid. But
the first statement of Theorem~\ref{Theorem6.1}
is true only if \begin{math} {\cal
O}_{D(q)}\end{math} is a locally free
sheaf. In fact, in this case the proof
of Theorem~\ref{Theorem6.1} carries
over. Now we want to
give an example, in which the first
statement of this theorem is false.
\begin{Example} \label{Example9.1}
Let \begin{math} n=3, p=\mbox{\it diagonal}
(1,0,-1)\end{math}, \begin{displaymath}
q=\left( \begin{array}{ccc}
0 & 0 & a\\
0 & 0 & b\\
0 & 0 & 0\end{array} \right) \end{displaymath}
with two constants \begin{math} a\end{math}
and \begin{math} b\end{math} not
depending on \begin{math} x\end{math}.
The corresponding spectral curve is
defined by the equation \begin{displaymath}
R(\lambda ,\mu )=(\mu -\exp (-\lambda ))(\mu -1)(\mu -\exp
(\lambda ))=0.\end{displaymath}  It is quite
obvious that the normalization \begin{math}
Y\end{math} of this
singular Riemann surface is the disjoint
union of three copies of \begin{math} {\Bbb P}
_{1}\end{math}. Moreover, \begin{math} \nu =\ln (\mu )\end{math}
is a holomorphic function on this
normalization \begin{math} Y\end{math}.
It is easy to see that the
singular Riemann surface \begin{math} Y'\end{math}
corresponding to \begin{math} q\end{math}
may be described as follows: \\
\begin{picture}(460,80)(0,0)
\put(110,55){\begin{math} {\Bbb P}
_{1}\cup {\Bbb P}_{1}\cup {\Bbb P}_{1}=
\end{math} }
\put(210,55){\begin{math} Y\end{math} }
\put(280,55){\begin{math} \tilde{Y} \end{math} }
\put(245,15){\begin{math} {\Bbb P}_{1}\end{math}
} \put(245,65){\begin{math} p\end{math} }
\put(215,35){\begin{math} \pi \end{math}
} \put(275,35){\begin{math} \tilde{\pi } \end{math}
}\put(215,50){\vector(1,-1){25}}
\put(280,50){\vector(-1,-1){25}}
\put(230,60){\vector(1,0){40} }
\end{picture}

\noindent
Let \begin{math} \tilde{Y} \end{math}
be the algebraic curve defined by the equation
\begin{displaymath} (\nu -\lambda )\nu
(\nu +\lambda )=0\end{displaymath}
and \begin{math} {\Bbb P}
_{1}\cup {\Bbb P}_{1}\cup {\Bbb P}_{1}\end{math}
the corresponding normalization. Now the
sheaf \begin{math} \pi _{*}({\cal O}_{Y})\end{math}
can be identified with the \begin{math} {\cal
O}_{{\Bbb P}_{1}}\end{math} module, which is
generated by \begin{math}
1,\nu /\lambda ,\nu ^{2}/\lambda ^{2}\end{math}. Then
the direct image of the structure sheaf
of \begin{math} Y'\end{math} is the \begin{math}
{\cal O}_{{\Bbb P}_{1}} \end{math} module
generated by \begin{math}
1,\nu ,\nu ^{2}/\lambda \end{math}. The degree of \begin{math}
{\cal O}_{b_{\mbox{\it \scriptsize effective} }}
\end{math} is equal to \begin{math}
4\end{math}. The sheaf \begin{math} {\cal
O}_{D(q)}\end{math} is generated as an \begin{math}
{\cal O}_{Y'}\end{math} module by \begin{displaymath}
1,b\frac{\nu }{\lambda } +(a+b)
\frac{\nu ^{2}}{\lambda ^{2}}
,\frac{\nu (\nu -\lambda )}{2\lambda -a-2b}
.\end{displaymath}
Hence the first  summand of the divisor \begin{math}
D(q)\end{math} defined above is equal to
\begin{displaymath} (0,1)+\left(
\frac{a+2b}{2}
,\exp \left( \frac{a+2b}{2} \right) \right)
.\end{displaymath} The second summand of
the divisor does not depend on \begin{math}
a\end{math}  and \begin{math}
b\end{math}. This shows that the
Darboux coordinates of these potentials
are the same, if the value of \begin{math}
a+2b\end{math} is the same.
Furthermore, the \begin{math} {\cal
O}_{Y'}\end{math} module \begin{math} {\cal
O}_{D(q)}\end{math} is not locally free
at the singular point with \begin{math}
\lambda =0\end{math}. \end{Example}

In the eighth section Lemma~\ref{Lemma7.1}
may be extended in the obvious way to
the general situation. Theorem~\ref{Theorem7.1}
and its proof generalizes to the case in which \begin{math} {\cal
O}_{b_{\mbox{\it \scriptsize effective} }}
\end{math} is equal to \begin{math}
{\cal O}_{b_{\mbox{\it \scriptsize algebraic} }}\end{math}. The
statements (ii) and (iii) of Theorem~\ref{Theorem7.1}
are valid for all potentials \begin{math}
q\end{math}  and the corresponding
possibly singular Riemann surfaces \begin{math}
Y'\end{math}. The other statements must
be replaced by \begin{description}
\item[(iv)'] Let \begin{math}
{\cal L}_{q}^{\perp }\end{math} be the
subspace of \begin{math} T_{q}{\cal H}^{\infty
}\end{math} of all elements \begin{math}
\delta q\end{math}, such that the symplectic
form of \begin{math} \delta q\end{math}  with
all elements of \begin{math} {\cal
L}_{q}\end{math} vanishes. Then \begin{math}
{\cal L}_{q}^{\perp }\end{math} is
contained in \begin{math} {\cal
L}_{q}\end{math}.
\item[(i)'] There exists a
homomorphism\footnote{If \begin{math}
n=2\end{math} this homomorphism can be
proven to be an isomorphism. In general
this is not true, because locally there may
exist additional flows (compare with
Example~\ref{Example9.2}.) Moreover, in some sense the codimension
of the image of the map \begin{math} d\Gamma _{q}\end{math}
in \begin{math} {\cal L}_{q}\end{math}
is equal to the degree of \begin{math} {\cal
O}_{b_{\mbox{\it \scriptsize algebraic} }}
\end{math} minus the degree of
\begin{math} {\cal O}_{b_{\mbox{\it \scriptsize effective} }}
\end{math}.}
of vector spaces \begin{math} d\Gamma _{q}:
H_{q,\mbox{\it \scriptsize modified} }^{1}(Y',{\cal O}) \rightarrow
{\cal L}_{q}^{\perp }\end{math} which is
uniquely determined by the property that
for all \begin{math} [f]\in
H_{q,\mbox{\it \scriptsize modified} }^{1}(Y',{\cal O})
\end{math} \begin{displaymath}
\frac{\partial v}{\partial q}
(d\Gamma _{q}([f])=\delta v,\end{displaymath}
where \begin{math} \delta v\end{math}  was
defined in Lemma~\ref{Lemma7.1}.
\end{description}

In the ninth section Theorem~\ref{Theorem8.1}
may be generalized in the obvious way.
\begin{Theorem} \label{Theorem9.1}
Let \begin{math} q\end{math} satisfy the
reality condition \begin{math}
q^{*}=-q\end{math}. Moreover,
assume that the excluded domains of some
\begin{math} U_{l,\epsilon }\end{math} of the
corresponding Riemann surface \begin{math}
Y'\end{math}  have asymptotically no
overlap. Then this Riemann surface \begin{math}
Y'\end{math} has asymptotically no
singularities; this means that the
effective branching divisor is
asymptotically equal to the analytic
branching divisor.
\end{Theorem}
Proof: Due to the assumption that the
excluded domains have asymptotically no
overlap, over some neighbourhood \begin{math}
U\end{math} of \begin{math} \lambda =\infty \end{math}
of \begin{math} {\Bbb P}_{1}\end{math} the
Riemann surface \begin{math} Y'\end{math}
can only have ordinary double points as
singular points. More precisely, two
points of \begin{math} Y\end{math} ,
which are identified in \begin{math} Y'\end{math}
must be covering points of the same \begin{math}
\lambda \in {\Bbb C}\end{math}. Since the effective
branching divisor \begin{math}
b_{\mbox{\it \scriptsize effective} }\end{math}
is a fixed point of \begin{math}
\theta \end{math}, the value of \begin{math}
\lambda \end{math} at such an ordinary double
point must be real. On the other hand
the equation \begin{displaymath}
b_{\mbox{\it \scriptsize effective} }
=D(q)+\theta (D(q))+(\theta ^{*}(v(q))v(q))\end{displaymath}
shows that on \begin{math} \pi ^{-1}({\Bbb R})\end{math}
all multiplicities of \begin{math}
b_{\mbox{\it \scriptsize effective} }\end{math}
are even. This proves that \begin{math}
Y'\end{math} has asymptotically no
singular points.
\hspace*{\fill }\begin{math} \Box \end{math}

\noindent
Due to this theorem in the rest of the
ninth section the singular
Riemann surface \begin{math} Y'\end{math}
may be considered as a non-singular Riemann
surface over some neighbourhood \begin{math}
U\end{math} of \begin{math} \lambda =\infty \end{math}
of \begin{math} {\Bbb P}_{1}\end{math}
together with finitely many
singularities over \begin{math} {\Bbb P}
_{1}\setminus U\end{math}.
Proposition~\ref{Proposition8.1} and
Theorem~\ref{Theorem8.3} remain valid,
if \begin{math} {\cal I}\end{math} is replaced
by \begin{math}
{\cal I}_{\mbox{\it \scriptsize effective} }\end{math}. In
general, all statements and proofs of
the ninth section may be extended to the
case \begin{math}
{\cal O}_{b_{\mbox{\it \scriptsize effective} }}\end{math}
being equal to \begin{math} {\cal
O}_{b_{\mbox{\it \scriptsize analytic} }}\end{math}. Then it is obvious
that the statements of Theorem~\ref{Theorem8.4}
and Corollary~\ref{Corollary8.1}
generalize to:
\begin{Theorem} \label{Theorem9.2} If
the excluded domains of some \begin{math}
U_{l,\epsilon }\end{math} have
asymptotically no overlap, the action of
\begin{math} ({\Bbb R}/{\Bbb Z})^{{\cal
I}_{\mbox{\it \scriptsize effective} }}\end{math}
on any point \begin{math}
[D]\in \mbox{\it Jacobian} _{\Bbb R}(Y')\end{math}
induces a homeomorphism between \begin{math}
({\Bbb R}/{\Bbb Z})^{{\cal I}_{\mbox{\it \scriptsize effective} } }
\end{math}
times finitely many copies of a finite dimensional abelian Lie
group and \begin{math}
\mbox{\it Jacobian} _{\Bbb R}(Y')\end{math}. Moreover, the
holomorphic forms \begin{math} (\omega _{\iota })_{\iota \in
{\cal I}}\end{math} define an embedding of the
real part \begin{math}
H^{1}_{{\Bbb R},\mbox{\it \scriptsize modified}
}(Y,{\cal O})\end{math} of any \begin{math}
H_{q,\mbox{\it \scriptsize modified}
}^{1}(Y,{\cal O})\end{math} onto a subspace of \begin{math}
{\Bbb R}^{{\cal I}_{\mbox{\it \scriptsize effective} }}
\end{math}, which is mapped
under the exponential map onto \begin{math}
({\Bbb R}/{\Bbb Z})^{{\cal I}_{\mbox{\it \scriptsize effective} }}
\end{math}. \end{Theorem}
\begin{Corollary} \label{Corollary9.1}
If the excluded domains of some \begin{math}
U_{l,\epsilon }\end{math} have asymptotically
no overlap the group \begin{math} ({\Bbb R}
/{\Bbb Z})^{{\cal I}_{\mbox{\it \scriptsize effective} }
}\end{math} times a finite dimensional
abelian Lie group acts on
each connected component of \begin{math}
\mbox{\it Isospectral} _{\Bbb R}(Y')\end{math}
transitively and freely. Moreover,
the action on any potential \begin{math}
q\in \mbox{\it Isospectral} _{\Bbb R}(Y)\end{math} induces a
homeomorphism between the connected
components of \begin{math}
\mbox{\it Isospectral} _{\Bbb R}(Y')\end{math}
and \begin{math} ({\Bbb R}
/{\Bbb Z})^{{\cal I}_{\mbox{\it \scriptsize effective} }}
\end{math} times
a finite dimensional abelian Lie group. Finally,
for any \begin{math} q\in
\mbox{\it Isospectral} _{\Bbb R}(Y)\end{math}
there exists an embedding from \begin{math} {\cal L}_{q,{\Bbb R}
}^{\perp }\end{math} onto a subspace of \begin{math}
{\Bbb R}^{{\cal I}_{\mbox{\it \scriptsize effective} }}
\end{math} times the
Lie algebra of the finite dimensional
abelian Lie group mentioned above, and which does not
depend on \begin{math} q\end{math}, such
that the flow induced by the Lie algebra
element in \begin{math} {\Bbb R}^{{\cal
I}_{\mbox{\it \scriptsize effective} }}\end{math} times the finite
dimensional abelian Lie algebra
corresponding to some \begin{math} \delta q\in
{\cal L}_{q,{\Bbb R}}^{\perp }\end{math} is smooth in \begin{math}
{\cal H}^{\infty }\end{math}, and the derivative at \begin{math}
t=0\end{math} is equal to \begin{math}
\delta q\end{math}.  This action of \begin{math}
{\cal L}_{q,{\Bbb R}}\end{math} on \begin{math}
\mbox{\it Isospectral} _{\Bbb R}(Y)\end{math} is also transitive
on the connected components.
\end{Corollary}
\hspace*{\fill }\begin{math} \Box \end{math}

\noindent
 From the point of view of Riemann
surfaces the union of all the Jacobian
varieties \begin{displaymath} \bigcup\limits_{Y'\in {\cal
S}}\mbox{\it Jacobi} (Y')\end{displaymath}
is the completion of the
Jacobian variety of the most singular
Riemann surface \begin{math}
Y_{\mbox{\it \scriptsize algebraic} }\end{math},
the algebraic curve
defined by the equation \begin{math}
R(\lambda ,\mu )=0\end{math}. In fact, in case
of Riemann surfaces of finite genus this
is a compactification of the generalized
Jacobian variety (compare with
\cite{DS}). The Jacobian variety of
Riemann surfaces of infinite genus is
not compact any more. Nevertheless the
generalized Jacobian variety of a
singular Riemann surface of infinite
genus is a metrizable space and has a
completion. The real part of the
Jacobian variety of a Riemann surface of
infinite genus is compact and the real
part of the completion of the
generalized Jacobian variety of a
singular Riemann surface of infinite
genus should be compact too. This space is
homeomorphic to the subspace of all
potentials \begin{math} \tilde{q}
\end{math}, such that \begin{math} \det
(\mu \unity -F(\lambda ,\tilde{q}
))=R(\lambda ,\mu )\end{math}.
 From the point of view of integrable
systems, all Lagrangian subspaces
decompose into the union of invariant
subspaces under the action of \begin{math}
\tilde{H} ^{1}_{\mbox{\it \scriptsize modified}
}(Y_{\mbox{\it \scriptsize algebraic} },{\cal
O}^{*})\end{math}, where \begin{math}
Y_{\mbox{\it \scriptsize algebraic} }\end{math}
is the most singular Riemann surface
described by the equation \begin{math}
R(\lambda ,\mu )=0\end{math}:
\begin{displaymath}
\bigcup\limits_{Y'\in {\cal S}} \mbox{\it Isospectral}
(Y')=\bigcup\limits_{Y'\in {\cal S}} \mbox{\it Jacobian}
_{0}(Y').\end{displaymath} Note also that the action of
these Lie groups on these subspaces is given by
hamiltonian flows. If this is a
countable union, there can't exist further
integrals of motion, which correspond to
additional flows. If \begin{math} n\end{math}
is equal to 2,
Proposition~\ref{Proposition9.1} proves
that this is the case. If we restrict
our attention to the real parts of these
isospectral sets, there exists a
group, which acts transitively on these
components. Furthermore, the real parts
of the decompositions of the Lagrangian
subspaces are equal to the real parts of
the decompositions of the
completion of the generalized Jacobian
variety of \begin{math}
Y_{\mbox{\it \scriptsize algebraic}
}\end{math}.
Finally let us
present an example, where \begin{math} {\cal
S}\end{math} is not a countable set such that
locally some more integrals exist. These
additional integrals of course do not
extend to any open set of the symplectic
space.
\begin{Example} \label{Example9.2}
Let \begin{math} n=4\end{math} \footnote{There
exist more complicated examples for the case \begin{math}
n=3\end{math}.} and \begin{math}
p=\mbox{\it diagonal}
(2,1,-1,-2)\end{math}. We consider the
constant potentials \begin{displaymath}
\left( \begin{array}{cccc}
0 & a & b & c\\
0 & 0 & d & e \\
0 & 0 & 0 & f\\
0 & 0 & 0 & 0\end{array} \right) .\end{displaymath}
The corresponding action of the finite
dimensional Lie group is given by
\begin{displaymath} {\Bbb C}^{6}\times {\Bbb C}
^{6}\rightarrow {\Bbb C}^{6},(t_{1},\ldots
,t_{6})\times (a,b,c,d,e,f)\mapsto (\tilde{a}
,\tilde{b} ,\tilde{c} ,\tilde{d} ,\tilde{e}
,\tilde{f} ),\mbox{ with } \end{displaymath}
\begin{displaymath} \begin{array}{l}
\tilde{a} =\exp (t_{1})a\\
\tilde{b} =\exp
(t_{1}+t_{2})(b+(3t_{4}+6t_{6})ad)\\
\tilde{c} =\exp
(t_{1}+t_{2}+t_{3})(c+4t_{4}(ae+bf)
+4t_{6}(ae-bf)+(12t_{4}^{2}+4t_{5}-12t_{6}^{2})adf)\\
\tilde{d} =\exp (t_{2})d\\
\tilde{e} =\exp
(t_{2}+t_{3})(e+(3t_{4}-6t_{6})df)\\
\tilde{f} =\exp (t_{3})f.\end{array}
\end{displaymath} This action decomposes
into several orbits with different
dimensions: \begin{enumerate}
\item If \begin{math} a\neq 0,d\neq
0,f\neq 0\end{math} there is a
six dimensional orbit, which is dense.
\item If \begin{math} a=0,d\neq 0,f\neq
0,b\neq 0\end{math} or \begin{math}
a\neq 0,d\neq 0,f=0,e\neq 0\end{math} there
is a five dimensional orbit.
In both cases the fifth
flow is trivial.
\item If \begin{math} a\neq 0,d=0,f\neq
0,b\neq 0,e\neq 0\end{math} there is a
one dimensional family of four
dimensional orbits. Again the fifth flow
is trivial and the fourth and sixth flows
may be transformed into each other.
The expression \begin{math} ae/bf\end{math}
is an additional integral of motion.
\item If \begin{math} a=0,d\neq
0,f=0,b\neq 0,c\neq 0,e\neq 0\end{math}
the fourth, the fifth and the sixth flows
are trivial. Hence there is a one
dimensional family of three dimensional
orbits. The value of \begin{math}
be/dc\end{math} is an additional
integral of motion.
\item Some more orbits of lower
dimensions.
\end{enumerate} Due to the third and
fourth case the set \begin{math} {\cal
S}\end{math} of singular curves
corresponding to this spectral curve
\begin{displaymath}
R(\lambda ,\mu )=(\mu -\exp (-2\lambda ))(\mu -\exp
(-\lambda ))(\mu -\exp (\lambda ))(\mu
-\exp (2\lambda ))\end{displaymath}
is not countable. Let us now consider the third case in
some more detail.
Like in Example~\ref{Example9.1} we can
define the singular Riemann surface with
the help of the singular curve \begin{math}
\tilde{Y} \end{math} defined by the
equation \begin{displaymath}
(\nu -2\lambda )(\nu -\lambda )(\nu +\lambda )(\nu +2\lambda )=0.
\end{displaymath} The normalization of
this curve is the four-fold covering
\begin{math} {\Bbb P}_{1}\cup {\Bbb P}_{1}\cup
{\Bbb P}_{1}\cup {\Bbb P}_{1}\end{math} of \begin{math}
{\Bbb P}_{1}\end{math}:\\
\begin{picture}(460,80)(0,0)
\put(110,55){\begin{math} {\Bbb P}
_{1}\cup {\Bbb P}_{1}\cup {\Bbb
P}_{1}\cup {\Bbb P}_{1}=\end{math} }
\put(210,55){\begin{math} Y\end{math} }
\put(280,55){\begin{math} \tilde{Y} \end{math} }
\put(245,15){\begin{math} {\Bbb P}_{1}\end{math}
} \put(245,65){\begin{math} p\end{math} }
\put(215,35){\begin{math} \pi \end{math}
} \put(275,35){\begin{math} \tilde{\pi } \end{math}
}\put(215,50){\vector(1,-1){25}}
\put(280,50){\vector(-1,-1){25}}
\put(230,60){\vector(1,0){40} }
\end{picture}

\noindent
The \begin{math} {\cal O}_{{\Bbb P}_{1}}\end{math}
module \begin{math} \pi _{*}({\cal
O}_{Y})\end{math} is generated by \begin{math}
1,\nu /\lambda ,\nu ^{2}/\lambda ^{2},\nu ^{3}/\lambda ^{3}\end{math}.
Furthermore, the submodule \begin{math} \pi '_{*}({\cal
O}_{Y'})\end{math} is generated by \begin{displaymath}
1,\nu ,\nu ^{2},\nu ^{3},
(ae-bf)\frac{\nu ^{2}}{\lambda }
-(ae+bf)\frac{\nu ^{3}}{\lambda ^{2}}
.\end{displaymath} There
exists an additional flow on \begin{math}
\mbox{\it Isospectral} (Y')\end{math} corresponding to the
additional integral of motion. Hence in this case
the image of the homomorphism \begin{math}
d\Gamma _{q}:H_{q,\mbox{\it \scriptsize modified}
}^{1}(Y',{\cal O})\rightarrow
{\cal L}_{q}^{\perp }\end{math} is not equal
to \begin{math} {\cal
L}_{q}^{\perp }\end{math}.
\end{Example}

\appendix
\section{Borel summability} \label{Appendixa}
In this Appendix we want to prove that
equations (\ref{g1.7a}) and
(\ref{g1.7b}) establish a one to one
correspondence between formal power
series \begin{displaymath}
q(x)=\sum_{l=0}^{\infty }
q_{l}x^{l}+b_{0},\mbox{ with } q_{n}
\mbox{ an offdiagonal matrix for all }
l\in {\Bbb N},\end{displaymath} and \begin{math}
b_{0}\end{math} due to
Assumption~\ref{Assumption1}
a fixed diagonal matrix, and formal
power series \begin{displaymath}
\unity +\sum_{m=1}^{\infty } a_{m}\lambda ^{-m},\mbox{ with }
a_{m} \mbox{ an offdiagonal matrix for all }
m\in {\Bbb N}.\end{displaymath} Furthermore,
we will show that the formal power
series \begin{math} q(x)\end{math}
defines a analytic function in some
neighbourhood of the point \begin{math}
x=0\end{math},  if and only if the power
series \begin{math} \sum_{m=1}^{\infty }
a_{m}\lambda ^{-m}/m!\end{math}
defines an analytic function in some
neighbourhood of \begin{math} \lambda ^{-1}=0\end{math}.
Hence the power series \begin{math}
\sum_{m=1}^{\infty } a_{m}(x)\lambda ^{-m}\end{math}
 of Theorem~\ref{Theorem1.2}  is Borel
summable at some point \begin{math} x\in
{\Bbb R}\end{math} if and only if \begin{math}
q(\cdot )\end{math} is analytic at this
point \begin{math} x\in {\Bbb R}\end{math}.
Moreover, the asymptotic expansion of \begin{math}
v\end{math} near infinity completely
determines the potential \begin{math} q\end{math},
if and only if \begin{math} q\end{math}
is analytic.

Now let \begin{math} a_{m}(x)=\sum_{l=0}^{\infty
} a_{m,l}x^{l}\end{math}  and \begin{math}
\sum_{l=0}^{\infty } b_{m,l}x^{l}\end{math}
 be for all \begin{math} m\in {\Bbb N}\end{math}
formal power series, such that \begin{math}
a_{m,l}\end{math} are offdiagonal
matrices and \begin{math} b_{m,l}\end{math}
are diagonal matrices. The equation \begin{displaymath}
[a_{M+1}(x),p]+b_{M}(x)=\end{displaymath}
\begin{equation} =\frac{da_{M}(x)}{dx}
+[a_{1}(x),p]a_{M}(x)+[b_{0},a_{M}(x)]-\sum_{m=1}^{M-1}
a_{m}(x)b_{M-m}(x)\label{gA1} \end{equation}
completely determines all the formal
power series \begin{math} a_{M+1}(x)\end{math}
and \begin{math} b_{M}(x)\end{math}  in
terms of the power series \begin{math}
a_{1}(x),\ldots
,a_{M}(x),b_{1}(x),\ldots ,b_{M-1}(x)\end{math}.
The inductive use of these equations
determines the formal power series \begin{math}
a_{M+1}(x)\end{math}  and \begin{math}
b_{M}(x)\end{math} for all \begin{math} M\in
{\Bbb N}\end{math} in terms of the power
series \begin{math} a_{1}(x)\end{math}.
Then equation (\ref{g1.7a}) \begin{displaymath}
q(x)=[a_{1}(x),p]+b_{0}\end{displaymath}
shows that the formal power series
\begin{displaymath} \unity +\sum_{m=0}^{\infty
}a _{m}\lambda ^{-m}=\unity +\sum_{m=0}^{\infty }
a_{m}(0)\lambda ^{-m}\end{displaymath} is
completely determined in terms of the
formal power series \begin{math} q(x)\end{math}.

On the other hand the equation \begin{displaymath}
\left( \frac{d}{dx} \right)
^{L+1}a_{M}(x)-\left( \frac{d}{dx} \right)
^{L}b_{M}(x)=\end{displaymath} \begin{equation}
=[a_{M+1}^{(L)}(x),p]-[b_{0},a_{M}^{(L)}(x)]
-\sum_{l=0}^{L} \left( \begin{array}{c}
L \\
l \end{array} \right) \left(
[a_{1}^{(l)}(x),p]a_{M}^{(L-l)}(x)+\sum_{M=1}^{M-1}
a_{m}^{(l)}(x)b_{M-m}^{(L-l)}\right) ,\label{gA2}
\end{equation} where the superscript
denotes formal derivatives with respect
to \begin{math} x\end{math},  determines
the \begin{math} (L+1)\end{math}-th
derivative of the power series \begin{math}
a_{M}(x)\end{math}  and the \begin{math}
L\end{math}-th derivative of the power
series \begin{math} b_{M}(x)\end{math}
in terms of derivatives of order at most
\begin{math} L\end{math} of \begin{math}
a_{1}(x),\ldots
,a_{M+1}(x),b_{1}(x),\ldots ,b_{M-1}(x)\end{math}.
The inductive use of these equations
determines all derivatives of the
power series \begin{math}
a_{1}(x),a_{2}(x),\ldots \end{math} and \begin{math}
b_{1}(x),b_{2}(x),\ldots \end{math} in
terms of the power series \begin{math}
a_{1}(x),a_{2}(x),\ldots \end{math}. In
particular, all derivatives of the
power series \begin{math} a_{1}(x)\end{math}
at the point \begin{math} x=0\end{math}
are completely determined by the formal
power series \begin{math} \unity +\sum_{m=1}^{\infty
} a_{m}(0)\lambda ^{-m}\end{math}.  Hence
the formal power series \begin{math}
q(x)=[a_{1}(x),p]+b_{0}\end{math}
is completely determined by the formal
power series \begin{math} \unity +\sum_{m=1}^{\infty
} a_{m}(0)\lambda ^{-m}\end{math}.  This shows
the one to one correspondence between
these two formal power series.

Now let us assume that \begin{math} q\end{math}
is a holomorphic function for all \begin{math}
|x|<R_{x}\end{math}  and bounded in
norm for all \begin{math} |x|\leq R_{x}\end{math}
by some constant \begin{math} C>0\end{math}.
Due to Cauchy's estimate \cite{Co} all
coefficients are bounded: \begin{displaymath}
\| q_{l}\| \leq \frac{C}{R_{x}^{l}}
\Longleftrightarrow \| q^{(l)}(0)\| \leq
\frac{Cl!}{R_{x}^{l}} .\end{displaymath}
If we define the degree of the \begin{math}
l\end{math}-th derivative of \begin{math}
q\end{math} to be equal to \begin{math}
l+1\end{math}, the recursion formula
(\ref{gA1}) shows that \begin{math}
a_{M+1}(x)\end{math}  and \begin{math}
b_{M}(x)\end{math} are homogenous
differential polynomials of degree \begin{math}
M+1\end{math}. Now let \begin{math} \alpha >1\end{math}
be a real number greater than \begin{math}
1/|p_{i}-p_{j}|\end{math} for all \begin{math}
n\geq i>j\geq 1\end{math}  and let the
numbers \begin{math} (\gamma _{m})_{m\in {\Bbb N}}\end{math}
be defined inductively by \begin{displaymath}
\gamma _{M+1}=\alpha \left( \frac{M}{R_{x}}
\gamma _{M}+\sum_{m=1}^{M} \gamma _{m}\gamma _{M+1-m}\right)
\mbox{ and } \gamma _{1}=C+\| ad(b_{0})\|
.\end{displaymath}
Then the recursion formula (\ref{gA1})
implies inductively that \begin{displaymath}
\| a_{M+1}(0)\| \leq \gamma _{M+1} \mbox{ and }
\| b_{M}(0)\| \leq \gamma _{M+1}.\end{displaymath}
It is easy to see that \begin{displaymath}
\gamma _{M+1}\leq M!(C+\| ad(b_{0})\| )\left(
\left( C+\| ad(b_{0})\| +\frac{1}{R_{x}}
\right) \alpha \right) ^{M}.\end{displaymath} This implies
that for all \begin{displaymath}
|\lambda ^{-1}|<\frac{R_{x}}{\alpha (R_{x}(C+\|
ad(b_{0})\| )+1)} \end{displaymath} the
power series \begin{math} \unity +\sum_{m=1}^{\infty
} a_{m}(0)\lambda ^{-m}/m!\end{math}
defines a holomorphic function.

Now let the real positive numbers \begin{math}
\alpha _{M}^{(L)}\end{math} and \begin{math}
\beta _{M}^{(L)}\end{math} satisfy the
recursion relations \begin{equation} \|
\alpha _{M}^{(L+1)}=\| ad(p)\|
\alpha _{M+1}^{(L)}+\| ad(b_{0})\|
\alpha _{M}^{(L)}+\beta _{M}^{(L)}\label{gR1} \end{equation}
\begin{equation}
\beta _{M}^{(L)}=\sum_{l=0}^{L} \left( \begin{array}{c}
L \\
l \end{array} \right)
\left( \| ad(p)\|
\alpha _{1}^{(l)}\alpha _{M}^{(L-l)}+\sum_{m=1}^{M-1}
\alpha _{m}^{(l)}\beta _{M-m}^{(L-l)}\right) .\label{gR2}
\end{equation} Then the
inductive use of equation (\ref{gA2})
implies that the following estimates hold \begin{displaymath}
\| a_{M}^{(L)}(0)\| \leq \alpha _{M}^{L} \mbox{ and }
\| b_{M}^{(L-1)}(0)\| \leq
\beta _{M}^{(L-1)} \mbox{ for all } M,L\in
{\Bbb N},\end{displaymath} whenever \begin{math}
\| a_{M}(0)\| \leq \alpha _{M}^{(0)}\end{math}
for all \begin{math} M\in {\Bbb N}\end{math}.
If \begin{math} \unity +\sum_{m=1}^{\infty }
a_{m}(0)\lambda ^{-m}\end{math} is Borel
summable, we may assume \begin{displaymath}
\| a_{m}(0)\| \leq \frac{C(m-1)!}{R_{\lambda }^{m-1}}
\end{displaymath} with some \begin{math}
R_{\lambda }>0\end{math} and \begin{math} C>0\end{math}.
Hence we set \begin{displaymath}
\alpha _{m}^{(0)}=\frac{C(m-1)!}{R_{\lambda }^{m-1}}
.\end{displaymath} Now we claim that due
to the recursion relations (\ref{gR1})
and (\ref{gR2}) the following estimates
hold: \begin{displaymath} \alpha _{M}^{(L+1)}
\leq \frac{C(M+L)!}{R_{\lambda }^{M-1}}
\left( \frac{\| ad(p)\| }{R_{\lambda }}
+\frac{\| ad(b_{0})\| }{C(2\| ad(p)\| +R_{\lambda })}
+1\right) ^{L+1}\left( C\left(
\| ad(p)\| +\frac{R_{\lambda }}{2} \right) \right)
^{M+L}\end{displaymath} \begin{displaymath}
\beta _{M}^{(L)}
\leq \frac{C(M+L)!}{R_{\lambda }^{M-1}}
\left( \frac{\| ad(p)\| }{R_{\lambda }}
+\frac{\| ad(b_{0})\| }{C(2\| ad(p)\| +R_{\lambda }}
+1\right) ^{L}\left( C\left(
\| ad(p)\| +\frac{R_{\lambda }}{2} \right) \right)
^{M+L}\end{displaymath}
for all \begin{math} M\in {\Bbb N},L\in {\Bbb N}
_{0}\end{math}.  For the proof of the
claim we first note that for all \begin{math}
L\in {\Bbb N}_{0}\end{math},  all \begin{math}
0\leq l\leq L\end{math} and all \begin{math}
M\in {\Bbb N}\setminus \{ 1\} \end{math}
\begin{displaymath}
\left( \begin{array}{c}
L \\
l \end{array} \right)
\sum_{m=1}^{M-1} (m+l-1)!(M+L-m-l)!\leq
\frac{(L-l+1)(M+L)!}{(L+1)(L+2)} .\end{displaymath}
For \begin{math} M=2\end{math}  this is
obvious, and for \begin{math} M>2\end{math}
it follows by induction in \begin{math}
M\end{math}.  This implies that \begin{displaymath}
\sum_{l=0}^{L} \left( \begin{array}{c}
L \\
l \end{array} \right)
\sum_{m=1}^{M-1} (m+l-1)!(M+L-m-l)!\leq
\frac{(M+L)!}{2} .\end{displaymath}
Furthermore, the following estimate is
obvious: \begin{displaymath}
\sum_{l=0}^{L} \left( \begin{array}{c}
L \\
l \end{array} \right)
l!(M+L-l-1)!\leq (M+L)!.\end{displaymath}
With these estimates the claim is an
easy calculation. This claim directly
shows that \begin{math}
q(x)=[a_{1}(x),p]+b_{0}\end{math}
defines a holomorphic function on some
neighbourhood of \begin{math} x=0\in {\Bbb C}\end{math},
if the formal power series \begin{math}
\unity +\sum_{m=1}^{\infty } a_{m}(0)\lambda ^{-m}\end{math}
is Borel summable. We want to remark
that this result cannot be proven with
the well known theorem of Watson
\cite{WW}, which gives an explicit
formula to reconstruct a holomorphic
function out of his asymptotic
expansion, if the asymptotic expansion
is Borel summable. But there is another
way to prove it. In fact, the method of
Segal and Wilson \cite{SW} to produce
solutions of the Korteweg-de Vries equation with the
help of the Birkhoff factorization can
be generalized to this situation. In \cite{HSS}
there was given a modification of the
usual Birkhoff factorization (see
e.g.\cite{PS}) in order to cover more
solutions of the Korteweg-de Vries and
nonlinear Schr{\"o}\-din\-ger equations.
It is possible to go further in this
direction:\begin{description}
\item[-] Let \begin{math} L^{-}GL(n,{\Bbb C})\end{math}
be the group of all formal power series \begin{displaymath}
g_{-}(\lambda )=\unity +\sum_{m=1}^{\infty }
a_{m}\lambda ^{-m},\mbox{ with } n\times n\mbox{-matrices }
a_{m}\mbox{ for all } m\in {\Bbb N},\end{displaymath}
such that \begin{math} \sum_{m=1}^{\infty
} a_{m}\lambda ^{-m}/m!\end{math}
defines a holomorphic entire function in \begin{math}
\lambda ^{-1}\end{math}.
\item[-] Let \begin{math}L^{+}GL(n,{\Bbb C})\end{math} be
the group of holomorphic
entire functions \begin{displaymath}
g_{+}:{\Bbb C}\rightarrow GL(n,{\Bbb C}),\lambda \mapsto
g_{+(\lambda )} \mbox{ of type } 1,\end{displaymath}
i.e. \begin{math} g_{+}\end{math} is
asymptotically bounded by \begin{displaymath}
\| g_{+}(\lambda )\| \leq \exp (\alpha |\lambda |)\mbox{ with
some } \alpha >0.\end{displaymath} \end{description}
Then the product of elements of \begin{math}
L^{-}GL(n,{\Bbb C})\end{math}  with elements
of \begin{math} L^{+}GL(n,{\Bbb C})\end{math}
is well defined. Moreover, let \begin{math}
LGL(n,{\Bbb C})\end{math} be the group of
invertible elements \begin{displaymath}
g(\lambda )=\sum_{m\in {\Bbb Z}}a_{m}\lambda ^{m},\end{displaymath}
such that \begin{math} \unity +\sum_{m=-1}^{-\infty
} a_{m}\lambda ^{m}\end{math} is an element of \begin{math}
L^{-}GL(n,{\Bbb C})\end{math}  and \begin{math}
\sum_{m=0}^{\infty } a_{m}\lambda ^{m}\end{math}
defines an entire function \begin{math}
{\Bbb C}\rightarrow n\times n\end{math}-matrices
of type 1. Then the Birkhoff
factorization (\cite[Theorem 8.1.2]{PS})
can be carried over to this modification
of the loop group of \begin{math}
GL(n,{\Bbb C})\end{math}. With the help of
this Birkhoff factorization one can
reconstruct the potential \begin{math}
q(\cdot )\end{math} out of the
asymptotic expansion \begin{math}
\unity +\sum_{m=1}^{\infty } a_{m}(0)\lambda ^{-m}\end{math}
defined in Theorem~\ref{Theorem1.2} (see
\cite{HSS}).

In \cite{Sch} it is shown that all
higher flows of the Korteweg-de Vries and nonlinear Schr{\"o}\-din\-ger
equations correspond to the series of
hamilton functions, given by the
asymptotic expansion of \begin{math}
\ln (\mu )\end{math}  in terms of \begin{math}
\lambda ^{-1}\end{math} of
Theorem~\ref{Theorem1.3}. Hence the
statement of this Appendix is related to a
statement \cite[Theorem~10.1]{MKT} of McKean and Trubowitz, which
gives a condition for the spanning of
the `tangent' space of the
isospectral sets in the sense of
Section~\ref{Section7} by all the local
flows.
\section{Another reality condition} \label{Appendixb}
In this Appendix we indicate, how to
treat other reality conditions in a
fashion similar to that of
Section~\ref{Section8}. In the case of the
nonlinear Schr{\"o}\-din\-ger equation two reality conditions are
known: the so called non-focussing and
the self focussing nonlinear Schr{\"o}\-din\-ger equation. We will
see that these two reality conditions
are related to the two covering maps
induced by \begin{math} \lambda \end{math}
and \begin{math} \mu \end{math},
respectively. In fact, let us now
consider the reality condition
corresponding to the covering map
induced by \begin{math} \mu \end{math}.
For the nonlinear Schr{\"o}\-din\-ger equation this corresponds to
the non-focussing case. The methods of \cite{MKT}
can be carried over to this case, but we
will give a different approach.

For this purpose we assume that the
matrix \begin{math} p\end{math} is
invertible. Otherwise the transformation
\begin{math} p\mapsto pa+\unity \end{math}
corresponds to the transformation \begin{math}
(\lambda ,\mu )\mapsto (\lambda ,\mu \exp (-a\lambda ))\end{math}
without change of the Riemann surface,
as mentioned in the proof of
Theorem~\ref{Theorem6.1}. In addition we
assume that \begin{math} p_{1},\ldots
,p_{n}\end{math} are imaginary numbers.
Then the Lax equation can be written as
an eigenvalue equation \begin{equation} \left(
-\frac{d}{dx} p^{-1}-q(x)p^{-1}\right)
\phi (x)=\lambda \phi (x)\label{ge1} \end{equation}
for a vector valued function \begin{math}
\phi (x)=\left( \begin{array}{c}
\phi _{1}(x)\\
. \\
. \\
\phi _{n}(x)\end{array} \right) .\end{math}
With the domain given as the set of
absolutely continuous
differentiable functions, such that
\begin{math} \phi (1)=\mu \phi
(0)\end{math}, the operator \begin{displaymath}
L(\mu )=-\left( \frac{d}{dx}
p^{-1}+q(x)p^{-1}\right) \end{displaymath}
extends to a closed unbounded operator
(see e.g. \cite{RS}) of the Hilbert space \begin{math}
L^{2}([0,1],{\Bbb C}^{n})\end{math} of square
integrable functions with the scalar
product \begin{displaymath}
\left\langle \phi ,\tilde{\phi } \right\rangle
=\int_{0}^{1}\phi ^{*}(x)\tilde{\phi } (x)dx.\end{displaymath}
Moreover, if \begin{math} |\mu |=1\end{math}
and \begin{math}
(q(x)p^{-1})^{*}=q(x)p^{-1}\end{math}
this operator is essentially self-adjoint.
Hence we introduce the reality condition
\begin{displaymath} pq+q^{*}p=0.\end{displaymath}
The solution of the eigenvalue equation
is given by \begin{displaymath}
\phi (x)=pg(x,\lambda ,q)p^{-1}\phi (0).\end{displaymath}
The boundary condition requires \begin{math}
\phi (0)\end{math} to be an eigen vector of
\begin{math} pg(1,\lambda ,q)p^{-1}\end{math}
with eigenvalue \begin{math}
\mu \end{math}. Hence the equation
\begin{displaymath} R(\lambda ,\mu )=\det
(\mu \unity -g(1,\lambda ,q))=0\end{displaymath}
describes the eigenvalues \begin{math}
\mu \end{math} of \begin{math} g(1,\lambda ,q)\end{math}
and \begin{math} F(\lambda ,q)\end{math}
depending on \begin{math} \lambda \end{math},
as well as the eigenvalues \begin{math} \lambda \end{math}
of the unbounded operator \begin{math}
L(\mu )\end{math} depending on \begin{math}
\mu \end{math}. From the second point of
view \begin{math} R(\lambda ,\mu )=0\end{math}
describes an infinite-fold covering over
\begin{math} \mu \in {\Bbb C}\setminus \{ 0\}
\end{math}, such that all covering
points of some \begin{math} \mu \in
{\Bbb C}\setminus \{ 0\} \end{math} fit
together to form the spectrum of \begin{math}
L(\mu )\end{math}. The reality condition
implies \begin{displaymath} L^{*}(\mu )=L\left(
\frac{1}{\bar{\mu } } \right) \mbox{ for all
\begin{math} \mu \in {\Bbb C}\setminus \{ 0\}
\end{math}, } \end{displaymath} since \begin{math}
L(\mu )\end{math} is an unbounded operator
valued holomorphic function. We leave
aside the analytic aspects of such
functions and let us just transform the
reality condition into a condition on \begin{math}
g(x,\lambda ,q)\end{math}. Let \begin{math}
U(p)\end{math} be the Lie subgroup of \begin{math}
GL(n,{\Bbb C})\end{math} which fixes the
hermitian form defined by \begin{math}
p\end{math}: \begin{displaymath} U(p)=\{
g\in GL(n,{\Bbb C})|g^{*}pg=p\} .\end{displaymath}
The corresponding Lie algebra is given
by \begin{displaymath} u(p)=\{ a\in
gl(n,{\Bbb C})|pa+a^{*}p=0\} .\end{displaymath}
Then the reality condition is
equivalent to the condition that \begin{math}
q(x)+p\lambda \end{math} is an element of \begin{math}
u(p)\end{math} for all real \begin{math}
\lambda \end{math}. This implies that \begin{math}
g(x,\lambda ,q)\end{math} is an element of \begin{math}
U(p)\end{math} for all real \begin{math}
\lambda \end{math}. More generally, the
following relation holds: \begin{displaymath}
g^{*}(x,\bar{\lambda } ,q)pg(x,\lambda ,q)=p \mbox{ for
all \begin{math} \lambda \in {\Bbb C}\end{math}.}
\end{displaymath} Hence we have \begin{displaymath}
\overline{R\left( \bar{\lambda } ,\frac{1}{\bar{\mu }
} \right) } =\overline{\det \left( \frac{1}{\bar{\mu }
} \unity -g(\bar{\lambda } )\right) } =\det \left( \frac{1}{\mu }
\unity -pg^{-1}(\lambda )p^{-1}\right) =\end{displaymath}
\begin{displaymath} =\det \left( \frac{1}{\mu }
\unity -g^{-1}(\lambda )\right)
=\frac{R(\lambda ,\mu )}{(-\mu )^{n}
\det (g(\lambda ))} =\frac{R(\lambda ,\mu )}{(\mu )^{n}
R(\lambda ,0)} .\end{displaymath}
Again we have an antilinear involution
of the Riemann surface \begin{math}
Y\end{math}: \begin{displaymath} \theta :Y\rightarrow
Y,(\lambda ,\mu )\mapsto \left( \bar{\lambda }
,\frac{1}{\bar{\mu }
} \right) .\end{displaymath} Let us now
find a condition on the divisors, which
is equivalent to the reality condition.
For this purpose we carry over equation
(\ref{g3.2}) to the case of the covering
map induced by \begin{math} \mu \end{math}.
The solution of the eigenvalue equation
with quasi-periodic boundary conditions
is given by \begin{displaymath}
\phi (x)=\pi ^{*}(pg(x,\cdot ,q)h_{0}^{-1})v.\end{displaymath}
For the transposed eigenvalue equation \begin{equation}
\frac{d}{dx} \psi (x)p^{-1}
-q(x)p^{-1}\psi (x)=\lambda \psi (x) \label{gte1}
\end{equation} for a vector valued function \begin{math}
\psi (x)=(\psi _{1}(x),\ldots ,\psi
_{n}(x))\end{math}
with boundary condition \begin{math}
\psi (1)=\frac{1}{\mu } \psi (0)\end{math}
we have the solution \begin{displaymath}
\psi (x)=w\pi ^{*}(h_{0}g^{-1}(x,\cdot
,q)).\end{displaymath}
Hence the analogous operator \begin{math}
P\end{math} of Lemma~\ref{Lemma3.3}
is given by \begin{displaymath}
P:L^{2}([0,1],{\Bbb C}^{n})\rightarrow L^{2}([0,1],{\Bbb C}
^{n}),\chi \mapsto
P\chi =\phi \frac{\int_{0}^{1} \psi (x)\chi (x)dx}{\int_{0}^{1}
\psi (x)\phi (x)dx} .\end{displaymath}
Similar to Lemma~\ref{Lemma3.3} we have:
\begin{Lemma} \label{Lemmab1}
\begin{description}
\item[(i)] \begin{math} P^{2}=P.\end{math}
\item[(ii)] \begin{math}
L(\mu )P=PL(\mu )=\lambda P.\end{math}
\item[(iii)] \begin{math}
\sum_{\mbox{\it \scriptsize
sheets of the covering map induced by }
\mu } P=\unity \end{math}.
\item[(iv)] The divisor of \begin{math}
P\end{math} is equal to \begin{math}
-b_{\mu }\end{math}, the branching divisor
of the covering map induced by \begin{math}
\mu \end{math}. \end{description}
\end{Lemma}
We only want to indicate the proof of
this lemma. The first two statements may
be verified directly. If \begin{math}
L(\mu )\end{math} is normal, the statement
(iii) is a consequence of the spectral
decomposition of \begin{math}
L(\mu )\end{math}. Since both sides are
meromorphic functions, this implies
(iii). In the proof of Lemma~\ref{Lemma6.2}
we used the fact that \begin{displaymath}
\int_{0}^{1}
\psi (x)\phi (x)dx=\int_{0}^{1}
w\pi ^{*}(h_{0}g^{-1}(x,\lambda ,q)pg(x,\lambda ,q)h_{0}^{-1})vdx=
\end{displaymath} \begin{displaymath}
=-wF^{-1}(\lambda ,q)\frac{\partial F(\lambda ,q)}{\partial \lambda }
v=-\frac{1}{\mu } \frac{d\mu }{d\lambda } wv.\end{displaymath}
The divisor of the meromorphic function \begin{math}
\mu ^{-1}\frac{d\mu }{d\lambda } \end{math} is
equal to the branching divisor of the
covering map induced by \begin{math} \mu \end{math}
minus the branching divisor of the
covering map induced by \begin{math}
\lambda \end{math}. A similar argument to that
in Lemma~\ref{Lemma3.3} now shows (iv).
\hspace*{\fill } \begin{math} \Box \end{math}

\noindent
For all matrix valued meromorphic
functions \begin{math} f\end{math} on \begin{math}
Y\end{math}, let \begin{math}
\theta ^{*}(f)\end{math} again be the function \begin{math}
\theta ^{*}(f)=(f\circ \theta )^{*}\end{math}. As
we saw above, the reality condition
implies \begin{displaymath} F^{-1^{*}}(\bar{\lambda }
,q)=h_{0}ph_{0}^{-1}F(\lambda ,q)h_{0}p^{-1}h_{0}^{-1}.\end{displaymath}
Then we have \begin{displaymath}
\theta ^{*}(v)h_{0}ph_{0}^{-1}\pi
^{*}(F(\cdot ,q))=\theta ^{*}(v)\theta ^{*}(\pi ^{*}(F^{-1}(\cdot
,q)))h_{0}ph_{0}^{-1}=\mu
\theta ^{*}(v)h_{0}ph_{0}^{-1}.\end{displaymath}
Thus the reality condition transforms to
the relation: \begin{math}
w=\theta ^{*}(v)h_{0}ph_{0}^{-1}\end{math}
is a solution of (\ref{g3.1b}), if and
only if \begin{math} v\end{math}  is a
solution of (\ref{g3.1a}). Of course
this is not compatible with the
normalization \begin{math}
v_{1}=1=w_{1}\end{math}. Hence we choose
another normalization: For every
potential \begin{math} q\end{math} let \begin{math}
\phi (x)\end{math} and \begin{math} \psi (x)\end{math}
be the unique solutions of
(\ref{ge1}) with boundary condition \begin{math}
\phi (1)=\phi (0)\mu \end{math} and
(\ref{gte1}) with boundary condition \begin{math}
\psi (0)=\mu ^{-1}\psi (1)\end{math}, such that
\begin{math} \sum_{i=1}^{n} \phi _{i}(0)=\sqrt{n}
=\sum_{i=1}^{n} \psi _{i}(0)\end{math},
respectively. Due to this normalization \begin{math}
\phi \end{math} and \begin{math} \psi \end{math}
 can be considered as \begin{math}
L^{2}([0,1],{\Bbb C}^{n})\end{math} valued
meromorphic functions on \begin{math} Y\end{math}
with no zeroes. If \begin{math} v\end{math}
and \begin{math} w\end{math} are the
solutions of (\ref{g3.1a}) and
(\ref{g3.1b}) with normalization \begin{math}
v_{1}=1=w_{1}\end{math}, respectively, \begin{math}
\phi \end{math} and \begin{math} \psi \end{math}
 are given by \begin{displaymath}
\phi (x)=\pi ^{*}(pg(x,\cdot ,q)h_{0}^{-1})v
\frac{1}{(h_{0}ph_{0}^{-1}v)_{1}}
\end{displaymath} \begin{displaymath}
\psi (x)=w\pi ^{*}(h_{0}g^{-1}(x,\cdot
,q)).\end{displaymath}
Hence the reality condition transforms
to the condition \begin{displaymath}
\phi (x)=\theta ^{*}(\psi (x))=
\theta ^{*}(w\pi ^{*}(h_{0}g^{-1}(x,\cdot
,q))).\end{displaymath} It is obvious
that the divisor of \begin{math} \psi \end{math}
is equal to \begin{math} -D^{t}(q)\end{math}
and the divisor of \begin{math} \phi \end{math}
is equal to \begin{math}
-D(q)-((h_{0}ph_{0}^{-1}v)_{1})\end{math}.
This function takes the values \begin{math}
p_{1}\sqrt{n} ,\ldots ,p_{n}\sqrt{n} \end{math}
at the covering points of infinity.
Therefore it does not define an
equivalence relation in the modified
sense. As a direct consequence of Lemma~\ref{Lemmab1}
we have \begin{equation} (\phi )+(\psi )+\left(
\int_{0}^{1} \psi (x)\phi (x)dx\right)
=b_{\mu }.\label{gb1} \end{equation}
With the normalization given above the
function \begin{math}
\int_{0}^{1}\psi (x)\phi (x)dx\end{math}
is equal to \begin{displaymath}
-\frac{1}{\mu } \frac{d\mu }{d\lambda }
\frac{wv}{(wh_{0}ph_{0}^{-1}v)_{1}}
.\end{displaymath}
This function takes the same value at
all covering points of infinity. Hence
it defines an equivalence relation in
the modified sense. Now we can state a
theorem analogous to
Theorem~\ref{Theorem8.1}:
\begin{Theorem} \label{Theoremb2}
In general the set of fixed points of the
involution \begin{math} \theta \end{math}
decomposes into several connected
components\footnote{In the hyperelliptic
case \begin{math}
n=2\end{math} it is more convenient to
use the antilinear involution \begin{math}
(\lambda ,\mu )\rightarrow (\bar{\lambda }
,\bar{\mu } )\end{math}. The
number of the connected components of
the set of fixed points of both
involutions is
equal to the genus of the Riemann
surface plus one. Such Riemann
surfaces are called \begin{math}
M\end{math}-curves (see e.g.
\cite{Kr2}). Then the  reality condition
simplifies to the condition, that the
divisor is invariant under this
involution.}. If a meromorphic function \begin{math}
f\end{math} gives an
equivalence relation in the modified
sense between \begin{math} D+\theta (D)\end{math}
and \begin{math} b_{\mu }\end{math},
then on each connected component
of the set of fixed points of the
involution \begin{math} \theta \end{math}
this function is either non-negative or
non-positive. Hence the real part
of the Jacobian variety \begin{displaymath}
\mbox{\it Jacobian} _{\Bbb R}(Y)=
\{ [D]\in \mbox{\it Jacobian} (Y)|D+\theta (D)\sim
b_{\mu }\mbox{ in the modified sense } \}\end{displaymath}
decomposes into several connected
components being characterized by the
sign of \begin{math} f\end{math} on each
connected component of the set of fixed
points of the involution \begin{math}
\theta \end{math}. Now let \begin{math}
\mbox{\it Jacobian} _{{\Bbb R},0}(Y)\end{math}
be the connected component of \begin{math}
\mbox{\it Jacobian} _{\Bbb R}(Y)\end{math}
corresponding to the case that \begin{math}
f\end{math} is non-negative on the whole set
of fixed points of the involution \begin{math}
\theta \end{math}. Then \begin{math}
\mbox{\it Jacobian} _{{\Bbb R},0}(Y)\end{math}
is contained in \begin{math}
\mbox{\it Jacobian} _{0}(Y)\end{math}, and \begin{math}
T_{x}\end{math} acts on this subspace.
Moreover, a potential satisfies the reality
condition \begin{math} pq+q^{*}p=0\end{math}
if and only if the inverse of the restriction of the
divisor of \begin{math}
\phi \end{math} to \begin{math} \pi
^{-1}({\Bbb C})\end{math} with the
normalization given above
is an element of \begin{math}
\mbox{\it Jacobian} _{{\Bbb R},0}(Y)\end{math}.
\end{Theorem}
The proof is quite similar to the proof
of Theorem~\ref{Theorem8.1}. We only indicate
the modifications. If \begin{math} \psi \end{math}
is equal to \begin{math} \theta
^{*}(\phi )\end{math} the function \begin{math}
\int_{0}^{1}\psi (x)\phi (x)dx\end{math}
is positive on the set of fixed points
of \begin{math} \theta \end{math}, and
the inverse of the restriction of the divisor of \begin{math}
\phi \end{math} to \begin{math} \pi ^{-1}({\Bbb C})\end{math}
is an element of \begin{math}
\mbox{\it Jacobian} _{{\Bbb R},0}(Y)\end{math}.
On the other hand let \begin{math} f\end{math}
be a function, such that \begin{math}
(f)\end{math} gives an
equivalence relation in the modified
sense between \begin{math} D+\theta (D)\end{math}
and \begin{math} b_{\mu }\end{math}. Then
\begin{math} \theta ^{*}(f)/f\end{math}
is a holomorphic function of \begin{math}
Y\end{math}, which is equal to \begin{math}
1\end{math} at all covering points of
infinity. Hence \begin{math} \theta
^{*}(f)\end{math} is equal to \begin{math}
f\end{math}  and \begin{math} f\end{math}
is real valued on the set of fixed
points of the involution \begin{math}
\theta \end{math}. Moreover, \begin{math}
f\end{math} can only have zeroes of even
order on this set. Then \begin{math} f\end{math}
is either non-negative or non-positive on each
connected component of this set.
Now let \begin{math} \mu _{0}\end{math}
be any value with \begin{math}
|\mu _{0}|=1\end{math}  and let \begin{math}
g\end{math}  be any cross section of \begin{math}
{\cal O}_{D-\pi ^{-1}(\infty )}\end{math}. Then
the total residue of the form \begin{math}
\theta ^{*}(g)gf(\mu -\mu _{0})^{-1}d\mu \end{math}
converges to zero. This again shows
that \begin{math} D\end{math}  is non-special
in the modified sense, if \begin{math} D\end{math}
is an element of \begin{math}
\mbox{\it Jacobian} _{{\Bbb R},0}(Y)\end{math}. The rest
of the proof is the same as the proof of
Theorem~\ref{Theorem8.1}.
\hspace*{\fill } \begin{math} \Box \end{math}

\noindent
Finally we want to mention that Theorem~\ref{Theorem8.1}
and Theorem~\ref{Theoremb2} are related
to the following fact: Let \begin{math}
LG,L^{-}G,L^{+}G\end{math} be the
subgroups of \begin{math} LGL(n,{\Bbb C}
),L^{-}GL(n,{\Bbb C}),L^{+}GL(n,{\Bbb C})\end{math}
of elements, which obey the relation
\begin{displaymath} g^{*}(\bar{\lambda }
)=g^{-1}(\lambda )\mbox{ for all }
\lambda \mbox{, with }|\lambda |=1.
\end{displaymath} Then the Birkhoff
factorization defines a diffeomorphism
onto the whole of \begin{math}
LG\end{math}: \begin{displaymath} L^{-}G\times
L^{+}G\rightarrow LG,(g_{-},g_{+})\mapsto
g_{-}g_{+}.\end{displaymath}

\noindent
{\em Acknowledgements}. The first part of
this work was performed while the author
was visiting the Forschungsinstitut
f{\"u}r Mathematik (ETH, Z{\"u}rich). He
would like to thank J. Moser for his
hospitality. Discussions with H.
Kn{\"o}rrer and E. Trubowitz have been
very stimulating and helpful.
Furthermore, the author thanks
F.~J.~Archer, P.~G.~Grinevich, J.~Mund and
R.~Schrader for comments and
corrections.


\begin{thebibliography}{MMM}
\bibitem[A-H-H]{AHH} Adams M. R.,
Harnad J., Hurtubise J.: Darboux
coordinates and Liouville-Arnold
integration in Loop algebras. Commun.
Math. Phys. {\bf 155}, 385-413 (1993).
\bibitem[A-vM]{AvM} Adler M., van
Moerbeke P.: Completely integrable systems,
Euclidian Lie algebras and curves. Adv.
Math. {\bf 38}, 267-317 (1980);
Lineraization of Hamiltonian systems,
Jacobi varieties and representation
theory. Adv. Math. {\bf 38}, 318-379
(1980).
\bibitem[B-S]{BS} Beals R., Sattinger D. H.:
On the complete integrability of
completely integrable systems. Commun.
Math. Phy. {\bf 138}, 409-436 (1991).
\bibitem[Bo]{Bo} Bourgain J.: Fourier
transform restriction phenomena for
certain lattice subsets and applications
to non-linear evolution equations.
Preliminary version.
\bibitem[Co]{Co} Conway J. B.: Functions of
one complex variable. Berlin,
Heidelberg, New York: Springer 1973.
\bibitem[D-K-N]{DKN} Dubrovin B. A.,
Krichever I. M., Novikov S. P.: Integrable
systems I. In: Arnold V. I., Novikov S. P.
(eds.) Dynamical Systems IV. Encyclopaedia of
Mathematical Sciences vol.4, pp.
173-280. Springer,(1990).
\bibitem[DS]{DS} D'Souza C.: Compactification
of generalized Jacobians. Pro. Indian Acad. Sci.
{\bf 88}, 419-457 (1979).
\bibitem [F-T]{FT} Faddeev L. D.,
Takhtajan L. A.: Hamiltonian methods in the
theory of solitons. Berlin,
Heidelberg, New York: Springer 1987.
\bibitem[F-K-T]{FKT} Feldman J.,
Kn{\"o}rrer H., Trubowitz E.: Riemann
surfaces of infinite genus. Preliminary
version.
\bibitem[F-K]{FK} Fordy A. P.,
Kulish P. P.: Nonlinear Schr{\"o}dinger equations and
simple Lie algebras: Commun. Math. Phys.
{\bf 89}, 427-443 (1983).
\bibitem[Fo]{Fo} Forster O.: Lectures
on Riemann surfaces. Berlin,
Heidelberg, New York: Springer 1981.
\bibitem[Fr]{Fr} Frenkel I. B.: Orbital
theory for affine Lie algebras. Invent.
Math. {\bf 77}, 301-352 (1984).
\bibitem[Gu]{Gu} Gunning R. C.:
Lectures on vector bundles over Riemann
surfaces. Princeton, New Jersey:
Princeton University Press 1967.
\bibitem[H-S-S]{HSS} Haak G., Schmidt M.,
Schrader R.: Group theoretic formulation of the
Segal-Wilson approach to integrable
systems with applications: Rev. Math.
Phys. {\bf 4}, 451-499 (1992).
\bibitem[Ho]{Ho} Hochstadt H.: On the
asymptotic spectrum of Hill's equation.
Archiv d. Mathematik {\bf 14}, 34-38 (1963).
\bibitem[I-M]{IM} Its A. R.,
Matveev V. B.: Hill's operator with finitely
many gaps. Functional Anal. Apll. {\bf
9}, 65-66 (1975).
\bibitem[Kr-1]{Kr} Krichever I. M.:
Methods of algebraic geometry in the
theory of non-linear equations.
Russian Math. Surveys {\bf 32} (6), 185-213
(1977).
\bibitem[Kr-2]{Kr2} Krichever I. M.:
Spectral theory of two-dimensional periodic
operators and its applications. Russian
Math. Surveys {\bf 44}, 145-225 (1989).
\bibitem[MK-T-1]{MKT} McKean H.,
Trubowitz E.: Hill's operator and hyperelliptic
function theory in the presence of
infinitely many branchpoints. Commun.
Pure Appl. Math. {\bf 29}, 143-226 (1976).
\bibitem[MK-T-2]{MKT2} McKean H.,
Trubowitz E.: Hill's surfaces and their theta
functions. American Math. Soc. {\bf 84}
(6), 1052-1085 (1978).
\bibitem[M-W]{MW} Magnus W., Winkler S.:
Hill's equation. New York, London,
Sydney: Interscience
Publishers 1966.
\bibitem[P-S]{PS} Pressley A., Segal G.:
Loop groups: Oxford: Clarendon Press
1986.
\bibitem[P-T]{PT} P{\"o}schel J.,
Trubowitz E.: Inverse spectral theory.
Orlando, Florida: Academic Press 1987.
\bibitem[R-S]{RS} Reed M., Simon B.: Functional
analysis. San Diego, California: Academic Press
1980.
\bibitem[R-S-T]{RST} Reyman A. G.,
Semenov-Tian-Shansky M. A.: Reduction of
Hamiltonian systems, affine Lie algebras
and Lax equations I and II. Inventiones math.
{\bf 54}, 81-100 (1979); Inventiones math.
{\bf 63}, 423-432 (1981).
\bibitem[Sch]{Sch} Schmidt M. U.: Solitonen
und Schleifengruppen. Berlin: Thesis 1990.
\bibitem[S-W]{SW} Segal G., Wilson G.:
Loop groups and equations of KdV type.
Publ. Math. I. H. E. S. {\bf 61}, 5-65 (1985).
\bibitem[Se]{Se} Serre J. P.: Algebraic
groups and class fields. Berlin,
Heidelberg, New York: Springer 1988.
\bibitem[W-W]{WW} Whittaker E. T.,
Watson G. N.: A course of modern analysis.
Cambridge: University Press 1927 (ed. 4).
\end{thebibliography}
\end{document}